\documentclass[preprint,superscriptaddress,tightenlines]{revtex4-1}
\usepackage{graphicx}
\usepackage{amsmath, amsfonts, bm}
\usepackage{mathrsfs}
\usepackage{tikz}
\usepackage{braket}
\usepackage{pgfplots}
\pgfplotsset{compat=1.17}

\newcommand{\vac}{\left\vert\text{vac}\right\rangle}
\newcommand{\bvac}{\left\langle\text{vac}\right\vert}

\newcommand{\dd}{\text{d}}

\newcommand{\ddt}[1]{\frac{\text{d}{#1}}{\text{d}t}}

\newcommand{\hc}{\text{H.c.}}
\newcommand{\cc}{\text{c.c.}}

\newcommand{\Hl}{H^{\text{L}}}
\newcommand{\Hnl}{H^{\text{NL}}}
\newcommand{\Hp}{H^{\text{phantom}}}
\newcommand{\Hr}{H_{\text{ring}}}
\newcommand{\Hrl}{H_{\text{ring}}^{\text{L}}}
\newcommand{\Hrnl}{H_{\text{ring}}^{\text{NL}}}
\newcommand{\Hrs}{H_{\text{ring}}^{\text{SPM}}}
\newcommand{\Hrx}{H_{\text{ring}}^{\text{XPM}}}
\newcommand{\Hcl}{H_{\text{chan}}^{\text{L}}}

\newcommand{\Hcs}{H_{\text{chan}}^{\text{SPM}}}
\newcommand{\Hcx}{H_{\text{chan}}^{\text{XPM}}}

\newcommand{\grs}{\gamma_{\text{ring}}^{\text{SPM}}}
\newcommand{\grx}{\gamma_{\text{ring}}^{\text{XPM}}}
\newcommand{\gcs}{\gamma_{\text{chan}}^{\text{SPM}}}
\newcommand{\gcx}{\gamma_{\text{chan}}^{\text{XPM}}}

\newcommand{\Ars}{A_{\text{ring}}^{PPPP}}
\newcommand{\Arx}{A_{\text{ring}}^{PSPS}}
\newcommand{\Acs}{A_{\text{chan}}^{PPPP}}
\newcommand{\Acx}{A_{\text{chan}}^{PSPS}}

\begin{document}

\title{Beyond photon pairs: Nonlinear quantum photonics in the high-gain regime}

\author{N.~Quesada}
\affiliation{Department of Engineering Physics, \'Ecole Polytechnique de Montr\'eal, Montr\'eal, QC, H3T 1JK, Canada}
\affiliation{Xanadu, Toronto, ON, M5G 2C8, Canada}
\author{L.~G.~Helt}
\affiliation{Xanadu, Toronto, ON, M5G 2C8, Canada}
\author{M.~Menotti}
\affiliation{Xanadu, Toronto, ON, M5G 2C8, Canada}
\author{M.~Liscidini}
\affiliation{Dipartimento di Fisica, Università degli Studi di Pavia, Via Bassi 6, 27100 Pavia, Italy}
\author{J.~E.~Sipe}
\email{sipe@physics.utoronto.ca}
\affiliation{Department of Physics, University of Toronto, 60 St. George Street, Toronto, Ontario M5S 1A7, Canada}




\begin{abstract}
Integrated optical devices will play a central role in the future development of nonlinear quantum photonics. Here we focus on the generation of high-gain nonclassical light within them. Starting from the solid foundation provided by Maxwell’s equations, we then move to applications by presenting a unified formulation that allows for a comparison of stimulated and spontaneous experiments in ring resonators and nanophotonic waveguides, and leads directly to the calculation of the quantum states of light generated in high-gain experiments.
\end{abstract}

\maketitle

\section{Introduction}
Spontaneous emission, where a single photon is emitted as an atom decays from an excited state to its ground state, was the earliest source of nonclassical light.  Within a decade of the invention of the laser, however, researchers began to explore the possibility of using parametric fluorescence to generate other nonclassical states \cite{harris67,magde1967study}.  Here the state of a material medium is left unchanged after the passage of a strong pump pulse, but because of optical nonlinearities it is possible for a pair of new photons to be generated. Much of the interest has focused on two processes: In one of them, called ``spontaneous parametric down-conversion" (SPDC), a pump photon splits into two daughter photons, usually referred to as ``signal'' and ``idler,'' in what can be thought of as a ``photon fission" process. This is sufficiently strong only in materials without inversion symmetry, where a $\chi_2$ nonlinear susceptibility is present. In the other process, called ``spontaneous four-wave mixing" (SFWM), what can be thought of as an ``elastic scattering'' between two pump photons occurs to make two new photons, again usually referred to as ``signal'' and ``idler,'' with one of the new photons at a frequency higher than that of the pump photons, and the other at a frequency lower than that. This process is governed by a $\chi_3$ nonlinear susceptibility, and so in principle can occur in any material medium. In both cases the pairs of photons produced are typically ``entangled,'' and thus of interest in quantum information processing protocols.

It was soon realized that not only could a strong enough pump pulse generate a pair of photons with significant probability, but that multiple pairs might also be generated. The quantum superposition of these states with different numbers of pairs of photons leads to a state of light called ``squeezed,'' in that the uncertainty in the amplitude of one quadrature is suppressed below the usual quantum limit, while the uncertainty in the amplitude of the other is correspondingly increased so as not to violate the uncertainty relation~\cite{lvovsky2015squeezed,andersen201630}.  These squeezed states are central to current efforts in quantum information processing based on continuous variables~\cite{weedbrook2012gaussian,braunstein2005squeezing,rudolph2017optimistic,bourassa2021blueprint,bromley2020applications,larsen21fault}.

One may think that the low efficiency of SPDC or SFWM is fundamentally related to their quantum nature. However, this is not the case, for it rather depends on the weakness of the nonlinear light-matter interaction. In this respect, there are at least two strategies to tackle this problem, either by increasing the nonlinear interaction length -- e.g., by working with bigger and bigger crystals -- or by enhancing the intensity of the electromagnetic field interacting with the matter. 

In the case of SFWM, which is the weaker of the two above-mentioned nonlinear processes, significant progress was made at the beginning of this century with the use optical fibers \cite{mosley2008heralded,Inoue2004}. These allow one to simultaneously enhance the electromagnetic field intensity by leveraging the  transverse light confinement and, at the same time, increasing the  nonlinear interaction lengths from a few millimeters to  few kilometers. 

More recently, studies of squeezed light and its generation have intersected with impressive improvements in the design and fabrication of integrated photonic structures involving channel waveguides and resonators. In these systems, the spatial and temporal confinement of light can enhance the nonlinear interaction strength up to ten orders of magnitudes with respect to what can be achieved in bulk systems, leading to efficient parametric fluorescence with continuous wave excitation at milliwatt pump powers. 

One important consequence is that the generation of nonclassical light ``beyond photon pairs'' is becoming commonplace~\cite{harder2013optimized,harder2016single,finger2015raman,chen2021photon,placke2020engineering,surya2018efficient,lu2020toward,cavanna2020progress,florez2020pump}.  A second consequence is that a more detailed comparison of theory and experiment is now possible, since implementations of well-characterized integrated structures do not suffer from the uncertainties that can plague experiments with bulk systems, such as details of beam shape and propagation.

From a theoretical point of view, the description of the nonlinear light-matter interaction in micro- and nanostructures is far more complex than that required in bulk systems, starting with the quantization of the electromagnetic field and moving on to the construction of the Hamiltonian describing a system that can be composed of several  optical elements. Finally, the strong enhancement of the nonlinear response of the system necessitates tools able to describe fast, non-trivial dynamics, which cannot always by approached using perturbation theories.  
In this tutorial we present an overview of the physics of nonlinear quantum optics ``beyond photon pairs.'' We begin in Sec.~\ref{sec:Quantization} with a treatment of the quantization of light in integrated photonic structures.  Although correct treatments of this subject can of course be found in the literature, there are pitfalls in adding to the linear Hamiltonian what one might think would be the obvious nonlinear contribution. We discuss the subtleties involved, and detail the mode expansion and nonlinear Hamiltonians for treating channel waveguides and microring resonators, the latter being the cavity structure we use as an example in this tutorial. In Sec.~\ref{sec:Ket} we consider general ``squeezing Hamiltonians'' that result when the usual classical approximation is made for the pump pulse; these Hamiltonians then only involve sums of products of pairs of operators.  We consider low-gain solutions, and then develop a general framework in which the solution of the Heisenberg equations of motion can be leveraged to build the full ket describing the squeezed state of light. This can be used to treat both spontaneous and stimulated parametric fluorescence, and we present results for both photon-number and homodyne statistics of the generated light.  In Sec.~\ref{sec:Waveguides} we focus on waveguides, and use a particular example to illustrate the low gain regime and the joint spectral amplitude of pairs of photons that can be generated, and how it is modified upon entering the high-gain region. In Sec.\ref{sec:Rings} we turn to microring resonators, and consider in detail both single- and dual-pump SFWM, as well as SPDC.  With all this as background we turn in Sec.~\ref{sec:Connections} to connections between classical and quantum nonlinear optics, and how the spontaneous processes we have considered can be understood as ``stimulated'' by vacuum power fluctuations.  In Sec.~\ref{sec:Tensor} we move beyond the world of coherent states and squeezed states, both of which fall in the category of Gaussian states. We present a leading strategy, based on ``matrix product states,'' to describe states more complicated than Gaussian.  Since moving beyond information processing that involves only Gaussian states and measurements is necessary to achieve universal quantum computing~\cite{bartlett2002universal,bartlett2002efficient}, the description and characterization of non-Gaussian states is a current area of active research, with many different approaches being explored~\cite{chabaud2020stellar,yanagimoto2021efficient,bourassa2021fast,walschaers2021non}.  Indeed, it can be hoped that at some point in the future a tutorial that could be considered a successor to this one, and might be called ``Beyond Gaussian states,'' will be available! Finally, we conclude in Sec.~\ref{sec:Conclusion}.

\section{Quantization in integrated photonics structures}\label{sec:Quantization}
In this section we develop the quantization of nonlinear optics in
an approach particularly suitable for treating integrated photonic
structures. In linear quantum optics the usual treatments are naturally
based on treating the electric and magnetic fields $\bm{E}(\bm{r},t)$
and $\bm{B}(\bm{r},t)$ as fundamental, but a straightforward
extension of this to nonlinear optics leads to error \cite{quesada2017why}. In fact, for
nonlinear quantum optics a treatment going back to the early work
of Born and Infeld \cite{born1934quantization}, which treats $\bm{D}(\bm{r},t)$
and $\bm{B}(\bm{r},t)$ as the fundamental fields,
seems the simplest way to develop the Hamiltonian treatment directly
without first introducing an underpinning Lagrangian framework. In Sec.~\ref{sec:TroubleWithE}
we illustrate the problems that can arise if a Hamiltonian
treatment of quantum nonlinear optics is based on $\bm{E}(\bm{r},t)$
and $\bm{B}(\bm{r},t)$ as the fundamental fields,
and as a preamble to nonlinear quantum optics we develop \emph{linear}
quantum optics based on $\bm{D}(\bm{r},t)$ and $\bm{B}(\bm{r},t)$
as the fundamental fields in Sec.~\ref{sec:LinearQO}. Of particular interest are
channel waveguide and ring resonator structures, and after introducing
the general quantization procedure we treat those special cases in
Sec.~\ref{sec:special_cases}. Initially we neglect dispersion, treating the relative
dielectric constant $\varepsilon_{1}(\bm{r})$ as independent
of frequency, but in Sec.~\ref{sec:dispersion} we indicate how dispersion can be
included in frequency regions where absorption can be neglected, such
as below the band-gap of a semiconductor, which is the usual regime
of interest for integrated photonic structures. In Sec.~\ref{sec:ring_channel_coupling} we present
a treatment of the coupling between channel waveguides and ring resonators
in the linear regime that will be useful in later sections, and in
Sec.~\ref{sec:nonlinear_intro} we introduce nonlinearities. The particular results for
the Hamiltonians describing self- and cross-phase modulation in channels
and rings are detailed in Secs.~\ref{sec:SXPM_channels} and \ref{sec:SXPM_rings} respectively; in later sections of this tutorial we introduce the appropriate Hamiltonians for other nonlinear processes.

Throughout this tutorial we generally assume that the relative dielectric
constant $\varepsilon_{1}(\bm{r})$---or, if dispersion
effects are included, $\varepsilon_{1}(\bm{r};\omega)$---can be treated as
a scalar. Structures of interest, such as those
involving lithium niobate, are clearly exceptions to this. The generalization
to a relative dielectric tensor is straightforward, but we do not
do it explicitly here since it complicates the formulas and, we believe, makes the physics we want to highlight in this tutorial less accessible.  

\subsection{The trouble with E}\label{sec:TroubleWithE}

The quantization of the electromagnetic field in vacuum is a standard
exercise in elementary quantum optics, and is naturally formulated
in terms of the electric and magnetic fields $\bm{E}(\bm{r},t)$
and $\bm{B}(\bm{r},t)$ \cite{gerry2005introductory}. The generalization
to include a uniform, frequency independent dielectric constant is
just as easy, and including even a position dependent relative dielectric
constant $\varepsilon_{1}(\bm{r})$, where the constitutive
relations are taken to be 
\begin{align}
 & \bm{D}(\bm{r},t)=\epsilon_{0}\varepsilon_{1}(\bm{r})\bm{E}(\bm{r},t)\label{eq:constitutive}\\
 & \bm{B}(\bm{r},t)=\mu_{0}\bm{H}(\bm{r},t),\nonumber 
\end{align}
with $\varepsilon_{1}(\bm{r})$ real, is straightforward.

\begin{figure}
    \centering
    \includegraphics[width=1.0\textwidth]{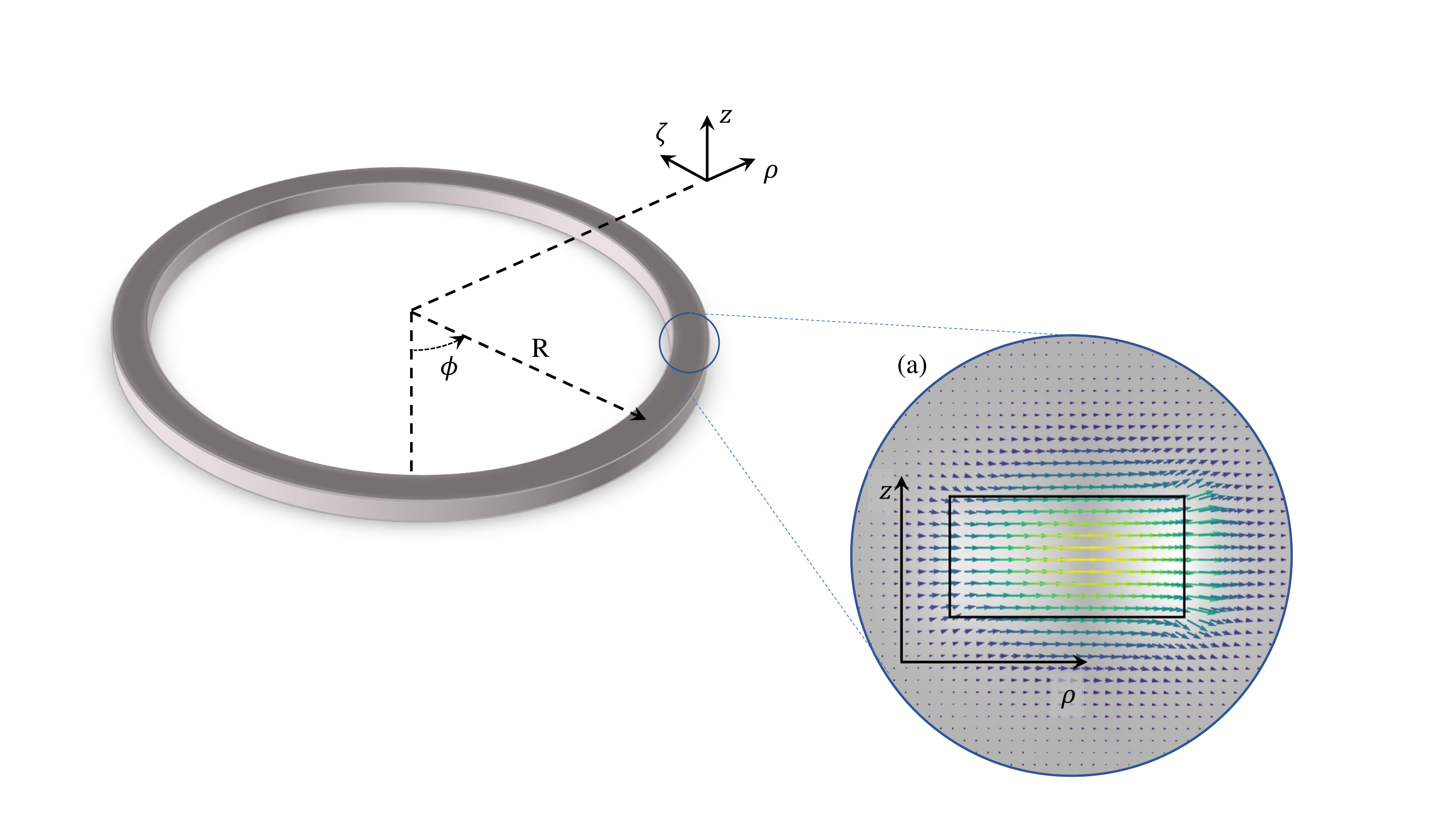}
    \caption{Schematic representation of an isolated ring resonator. The inset (a) reports the typical field profile of the fundamental transverse-electric mode supported by the bent waveguide. In the simulation the waveguide is 800 nm wide and 400 nm high, is made of silicon nitride fully clad in silica, and has a bending radius of 10 $\mu$m. The simulation is performed at a wavelength of 1550 nm. The contour plot accounts for the $E_{\zeta}$ component, while the arrow field represents $\bm{E}_{\bm{r}_\perp}$}
    \label{fig:isolated_ring}
\end{figure}

For our discussion in this section we just quote the results for a
dielectric ring shown in Fig.~\ref{fig:isolated_ring}; we give a derivation of the
results in Sec.~\ref{sec:special_cases}. Cylindrical coordinates $z$,
$\rho=\sqrt{x^{2}+y^{2}}$, and the angle $\phi$ are the natural
variables; it is convenient to introduce a nominal radius $R$ and
use $\zeta=R\phi$ in place of $\phi$. Denoting $\bm{r}_{\perp}=(\rho,z)$,
a volume element is $\dd\bm{r}=\dd\bm{r}_{\perp}\dd\zeta$,
where $\dd\bm{r}_{\perp}=R^{-1}\rho \dd\rho \dd z$, and $\zeta$
varies from $0$ to $\mathcal{L}=2\pi R$. With the relative dielectric
constant $\varepsilon_{1}(\bm{r})=\varepsilon_{1}(\bm{r}_{\perp})$
in the ring large enough to confine modes, the modes will be labeled
by integer quantum numbers $m_{J}$, where the wave number associated
with propagation around the ring is $\kappa_{J}=2\pi m_{J}/\mathcal{L}$;
for the moment we restrict ourselves to one relevant transverse field
structure for each $\kappa_{J}$, identify its frequency by $\omega_{J}$,
and consider only these bound modes. Then introducing raising and
lowering operators $c_{J}^{\dagger}$ and $c_{J}$ for each mode,
neglecting zero-point energies the Hamiltonian is 
\begin{align}
 & \Hrl=\sum_{J}\hbar\omega_{J}c_{J}^{\dagger}c_{J},\label{eq:Hring_first}
\end{align}
and the electric field operator is written in terms of them as 
\begin{align}
  \bm{E}(\bm{r})=\sum_{J}\sqrt{\frac{\hbar\omega_{J}}{2\mathcal{L}}}c_{J}\bm{e}_{J}(\bm{r}_{\perp})e^{i\kappa_{J}\zeta}+\hc,\label{eq:E_ring_use}
\end{align}
where the $\bm{e}_{J}(\bm{r}_{\perp})$
are determined so that the Heisenberg operator corresponding to \eqref{eq:E_ring_use} satisfies Maxwell's equations, and then they are appropriately normalized; the ``+H.c." indicates that the Hermitian conjugate of the preceding should be added. In writing (\ref{eq:E_ring_use}) we have assumed for simplicity that for the modes of interest the amplitudes $\bm{e}_{J}(\bm{r}_{\perp})$ lie in the $z$ direction; we will write down more general expressions later.   

This is all uncontroversial. All of the possible derivations of~(\ref{eq:Hring_first},~\ref{eq:E_ring_use})
rely on identifying the energy density $\mathfrak{h}(\bm{r})$
at a point $\bm{r}$ in space, and noting that for a change
in the fields at this point the energy density changes according to~\cite{born1999principles}
\begin{align}
 & d\mathfrak{h}=\bm{H}\cdot \dd\bm{B}+\bm{E}\cdot \dd\bm{D},\label{eq:dh}
\end{align}
and so for the constitutive relations of \eqref{eq:constitutive} we
find an expression for the classical energy density, 
\begin{align}
 & \mathfrak{h}(\bm{r},t)=\frac{1}{2\mu_{0}}\bm{B}(\bm{r},t)\cdot\bm{B}(\bm{r},t)+\frac{\epsilon_{0}\varepsilon_{1}(\bm{r})}{2}\bm{E}(\bm{r},t)\cdot\bm{E}(\bm{r},t),\label{eq:H_density}
\end{align}
with the volume integral of this equal to the total energy. In accordance
with the prescription for canonical quantization, the operator version
of that integral is set equal to $\Hrl$ after the subtraction
of the zero-point energy.

But what would seem to be a straightforward extension of this to include
nonlinear effects, where then 
\begin{align}
 & \Hr=\Hrl+\Hrnl,\label{eq:Hring_total}
\end{align}
leads to erroneous results. To see the difficulty, note that with the approximation that the $\bm{e}_{J}(\bm{r}_{\perp})$
of interest lie purely in the $z$ direction, then $\bm{E}(\bm{r})$
will be as well, $\bm{B}(\bm{r})$ can correspondingly
be approximated as lying purely in the radial direction, and we denote
those components by $e_{J}(\bm{r}_{\perp})$, $E(\bm{r})$,
and $B(\bm{r})$ respectively. Assuming the usual series expansion
for the nonlinear response up to some power $N$, 
\begin{align}
 & D(\bm{r},t)=\epsilon_{0}\sum_{n=1}^{N}\varepsilon_{n}(\bm{r})E^{n}(\bm{r},t),\\
 & B(\bm{r},t)=\mu_{0}H(\bm{r},t),\nonumber
\end{align}
from (\ref{eq:dh}) we find 
\begin{align}
 & \mathfrak{h}(\bm{r},t)=\frac{1}{2\mu_{0}}B^{2}(\bm{r},t)+\epsilon_{0}\sum_{n=1}^{N}\frac{n}{n+1}\varepsilon_{n}(\bm{r})E^{n+1}(\bm{r},t).\label{eq:hEexpansion}
\end{align}
Considering only a third order nonlinearity $(n=3)$ as well as the
usual linear ($n=1$) response, a natural argument would be to notice
that the magnetic term and the first electric term in (\ref{eq:hEexpansion})
give the usual linear terms (\ref{eq:Hring_first}), when the usual
expressions (\ref{eq:E_ring_use}) for the electric and magnetic fields
are used, and so taking the integral of (\ref{eq:hEexpansion}) over
all space we can identify 
\begin{align}
 & \Hrnl=\frac{3}{4}\epsilon_{0}\int\varepsilon_{3}(\bm{r})E^{4}(\bm{r})\dd\bm{r}.\label{eq:Hwrong}
\end{align}
Using the electric field expression (\ref{eq:E_ring_use}) in this
expression (\ref{eq:Hwrong}), and combining it with $\Hrl$
to form the total Hamiltonian (\ref{eq:Hring_total}), it would seem
we could obtain the nonlinear equations for the dynamics of the $c_{J}$.

But this leads to nonsense. To see that, consider just one mode $P$
and its nonlinear interaction with itself. Neglecting normal ordering
corrections in the nonlinear term in (\ref{eq:Hwrong}), we find that
the equation for $c_{P}$ is 
\begin{align}
 & \ddt{c_{P}}=-i\omega_{P}c_{P}-\frac{9i}{\hbar}\left(\frac{\hbar\omega_{P}}{2\mathcal{L}}\right)^{2}Kc_{P}^{\dagger}c_{P}c_{P},\label{eq:dynamics_wrong}
\end{align}
where 
\begin{align}
 & K=\epsilon_{0}\int\varepsilon_{3}\left(\bm{r}\right)\left[e_{P}^{*}\left(\bm{r}_{\perp}\right)\right]^{2}\left[e_{P}\left(\bm{r}_{\perp}\right)\right]^{2}\dd\bm{r}_{\perp}\dd\zeta.\label{eq:Kdef}
\end{align}
In the classical limit, where $c_{P}(t)$ and $c_{P}^{\dagger}(t)$
($=c_{P}^{*}(t))$ are just variables, we see that this leads to a
prediction of a shift in frequency 
\begin{align}
 & \omega_{P}\rightarrow\omega_{P}+\frac{9}{\hbar}\left(\frac{\hbar\omega_{P}}{2\mathcal{L}}\right)^{2}Kc_{P}^{\dagger}c_{P}.\label{eq:shift_wrong}
\end{align}
That is, for a positive $\varepsilon_{3}(\bm{r}$), for which
$K$ is positive, the frequency is predicted to \emph{increase} with increasing
intensity. Yet that is incorrect: Initially we can always write the
frequency $\omega_{P}$ as $c\kappa_{P}/n_{\text{eff}}$, where $n_{\text{eff}}$
is an effective index of refraction of the mode in the ring, and a
positive $\varepsilon_{3}(\bm{r})$ will certainly lead to
an increase with intensity of the effective index of the ring, and
thus to a frequency that \emph{decreases} with increasing intensity. In
fact, rather than (\ref{eq:dynamics_wrong}), the correct shift in
frequency is given by 
\begin{align}
 & \omega_{P}\rightarrow\omega_{P}-\frac{3}{\hbar}\left(\frac{\hbar\omega_{P}}{2\mathcal{L}}\right)^{2}Kc_{P}^{\dagger}c_{P},\label{eq:shift_right}
\end{align}
as we show in Sec.~\ref{sec:SXPM_rings}. So~\eqref{eq:shift_wrong}
fails to correctly describe both the size of the effect \text{and} its
sign, and it would lead to the wrong prediction, for example, of the
group velocity dispersion regime in which soliton propagation would
exist.

What has gone wrong? The Hamiltonian density (\ref{eq:Hwrong}) is
certainly equal to the energy density, as it should be for canonical
quantization. But the use of the form (\ref{eq:E_ring_use}) of the
electric and magnetic fields in that Hamiltonian, together with the
usual commutation relations for the operators $c_{P}$ and $c_{P}^{\dagger}$,
does \emph{not} lead to the correct equations---Maxwell's equations----in the nonlinear regime. This is apparent because, very generally,
the equation for the Heisenberg operator $\bm{B}(\bm{r},t)$
would be 
\begin{align}
 & \frac{\partial\bm{B}(\bm{r},t)}{\partial t}=\frac{1}{i\hbar}\left[\bm{B}(\bm{r},t),\Hrl\right]+\frac{1}{i\hbar}\left[\bm{B}(\bm{r},t),\Hrnl\right].
\end{align}
Keeping only the first commutator on the right we get Faraday's, law,
$\dot{\bm{B}}(\bm{r},t)=-\nabla\times\bm{E}(\bm{r},t)$;
the theory was built to give this in the linear regime. But it is
easy to see that the second commutator on the right is nonvanishing,
and then leads to a violation of Faraday's law when nonlinear effects
are included.

A hint of the problem can be gleaned from the differential form (\ref{eq:dh})
for the change in the energy density, which suggests that $\bm{B}$
and $\bm{D},$ rather than $\bm{B}$ and $\bm{E}$,
should be thought of as the fundamental fields. This is in contrast
to a Lagrangian approach, where the differential form for the change
in the Lagrangian density is given \cite{born1934quantization} by 
\begin{align}
 & \dd\mathfrak{L=}-\bm{D}\cdot \dd\bm{E}+\bm{H}\cdot \dd\bm{B}.
\end{align}
Indeed, one way to proceed is to return to the Lagrangian
formulation, include the nonlinear interactions, and ``rebuild''
the Hamiltonian formulation from that starting point. This has been done at a microscopic level involving two- or many-level atoms~\cite{hillery1985semiclassical, hillery1997quantized, drummond1999quantum}, and from a macroscopic perspective using susceptibilities~\cite{hillery1984quantization, abram1991quantum}.  For an overview, see Drummond and Hillery~\cite{drummond2014quantum}. Alternately, as was first pointed out by Born and
Infeld~\cite{born1934quantization}, one can begin with $\bm{B}$ and $\bm{D}$
as the fundamental fields and directly introduce the Hamiltonian formulation,
without relying on an earlier Lagrangian formulation. When the focus
is on the Hamiltonian framework, as it is often in quantum optics,
this is a simpler strategy; it is the one we follow here~\cite{sipe2004effective}. In the next section we implement this
in the linear regime, and in Sec.~\ref{sec:nonlinear_intro} we extend it
to nonlinear optics. For a recent tutorial on these issues, see Raymer~\cite{raymer2020quantum}.

\subsection{Linear quantum optics}\label{sec:LinearQO}

We begin by adopting the constitutive relations (\ref{eq:constitutive}),
assuming a real relative dielectric constant $\varepsilon_{1}(\bm{r})$
that is dispersionless but arbitrarily position dependent. In Sec.~\ref{sec:dispersion}
we will generalize to include material dispersion.

\subsubsection{Modes}

We begin with the classical Maxwell equations

\begin{align}
 & \frac{\partial\bm{B}(\bm{r},t)}{\partial t}=-\bm{\nabla}\times\bm{E}(\bm{r},t),\label{Maxwell}\\
 & \frac{\partial\bm{D}(\bm{r},t)}{\partial t}=\bm{\nabla}\times\bm{H}(\bm{r},t),\nonumber \\
 & \bm{\nabla}\cdot\bm{B}(\bm{r},t)=0,\nonumber \\
 & \bm{\nabla}\cdot\bm{D}(\bm{r},t)=0,\nonumber 
\end{align}
and look for a solutions in the linear regime where the constitutive
relations are (\ref{eq:constitutive}). Looking for solutions of the
form 
\begin{align}
 & \bm{D}(\bm{r},t)=\bm{D}_{\alpha}(\bm{r})e^{-i\omega_{\alpha}t}+\cc,\label{eq:modes}\\
 & \bm{B}(\bm{r},t)=\bm{B}_{\alpha}(\bm{r})e^{-i\omega_{\alpha}t}+\cc,\nonumber 
\end{align}
with $\omega_{\alpha}$ positive, we will have a solution of (\ref{Maxwell})
if 
\begin{align}
 & \bm{\nabla}\times\left[\frac{\bm{\nabla\times}\bm{B}_{\alpha}(\bm{r})}{\varepsilon_{1}(\bm{r})}\right]=\frac{\omega_{\alpha}^{2}}{c^{2}}\bm{B}_{\alpha}(\bm{r}),\label{eq:mode_equations}\\
 & \bm{\nabla\cdot}\bm{B}_{\alpha}(\bm{r})=0,\nonumber \\
 & \bm{D}_{\alpha}(\bm{r})=\frac{i}{\omega_{\alpha}\mu_{0}}\bm{\nabla}\times\bm{B}_{\alpha}(\bm{r}).\nonumber 
\end{align}
The first of these is the ``master equation'' familiar from work
on photonic crystals~\cite{joannopoulos2011photonic}. Whatever the functional form of $\varepsilon_{1}(\bm{r}),$
we refer to a solution $\left(\bm{D}_{\alpha}\left(\bm{r}\right),\bm{B}_{\alpha}\left(\bm{r}\right)\right)$ of \eqref{eq:mode_equations} as a ``mode'' of the system. Note
that for every such mode we can identify another mode 
\begin{equation}\label{eq:star_condition}
\left(\bm{D}_{\alpha'}\left(\bm{r}\right),\bm{B}_{\alpha'}\left(\bm{r}\right)\right)=\left(-\bm{D}_{\alpha}^{*}\left(\bm{r}\right),\bm{B}_{\alpha}^{*}\left(\bm{r}\right)\right)
\end{equation}
with $\omega_{\alpha'}=\omega_{\alpha}.$ The organization of all
modes into pairs, such as $\left(\bm{D}_{\alpha}\left(\bm{r}\right),\bm{B}_{\alpha}\left(\bm{r}\right)\right)$
and $\left(\bm{D}_{\alpha'}\left(\bm{r}\right),\bm{B}_{\alpha'}\left(\bm{r}\right)\right)$,
can be done just as was outlined earlier for acoustic modes~\cite{sipe2016hamiltonian}, but note that different sign conventions have been used in
the past for partner modes. Following the notation there we
refer to $\alpha'$ as the ``partner mode'' of $\alpha.$

We begin by looking at modes that are also chosen to satisfy periodic
boundary conditions over a normalization volume $V$. Then the operator
$\mathcal{M}$,
\begin{align}
 & \mathcal{M}(\cdots)\equiv\bm{\nabla}\times\left[\frac{\bm{\nabla\times}(\cdots)}{\varepsilon_{1}(\bm{r})}\right]\label{eq:M_op}
\end{align}
is Hermitian, in that 
\begin{align}
 & \int_{V}\bm{A}^{*}(\bm{r})\cdot\mathcal{M}(\bm{C}(\bm{r}))\dd\bm{r}=\int_{V}\left(\mathcal{M}(\bm{A}(\bm{r}))\right)^{*}\cdot\bm{C}(\bm{r})\dd\bm{r},
\end{align}
for any such periodic functions $\bm{A}(\bm{r})$
and $\bm{C}(\bm{r}).$ From this it follows immediately
that 
\begin{align}
 & \int_{V}\bm{B}_{\alpha}^{*}(\bm{r})\cdot\bm{B}_{\beta}(\bm{r})\dd\bm{r}=0,\label{eq:orthogonality}\\
 & \omega_{\alpha}\neq\omega_{\beta},\nonumber 
\end{align}
with the orthogonality also holding if $\beta=\alpha'$ and the pairs
of modes are properly chosen~\cite{sipe2016hamiltonian}. We normalize the modes
according to 
\begin{align}
 & \int_{V}\frac{\bm{B}_{\alpha}^{*}(\bm{r})\cdot\bm{B}_{\alpha}(\bm{r})}{\mu_{0}}\dd\bm{r}=1,\label{eq:norm_B}
\end{align}
and it then follows from (\ref{eq:mode_equations}) that 
\begin{align}
 & \int_{V}\frac{\bm{D}_{\alpha}^{*}(\bm{r})\cdot\bm{D}_{\alpha}(\bm{r})}{\epsilon_{0}\varepsilon_{1}(\bm{r})}\dd\bm{r}=1.\label{eq:norm_D}
\end{align}

\subsubsection{Canonical formulation}

To construct a canonical formulation, one traditionally begins
with a set of functions of the canonical coordinates and momenta,
$q_{i}$ and $p_{j}$ respectively, and with the Poisson bracket of
two such functions $f(q_{i},p_{j})$, $g(q_{i},p_{j})$ defined according
to 
\begin{align}
 & \left\{ f,g\right\} =\sum_{k}\left(\frac{\partial f}{\partial q_{k}}\frac{\partial g}{\partial p_{k}}-\frac{\partial f}{\partial p_{k}}\frac{\partial g}{\partial q_{k}}\right),\label{eq:usualPB}
\end{align}
from which follows the properties 
\begin{align}
 & \left\{ f,g\right\} =-\left\{ g,f\right\} ,\label{eq:Lie_conditions}\\
 & \left\{ f+g,b\right\} =\left\{ f,b\right\} +\left\{ g,b\right\} \nonumber \\
 & \left\{ fg,b\right\} =f\left\{ g,b\right\} +g\left\{ f,b\right\} \nonumber \\
 & \left\{ f,\left\{ g,b\right\} \right\} +\left\{ g,\left\{ b,f\right\} \right\} +\left\{ b,\left\{ f,g\right\} \right\} =0.\nonumber 
\end{align}
We then find the familiar results 
\begin{align}
 & \left\{ q_{i},q_{j}\right\} =\left\{ p_{i},p_{j}\right\} =0,\label{eq:canPB}\\
 & \left\{ q_{i},p_{j}\right\} =\delta_{ij}.\nonumber 
\end{align}
The dynamical equations are then taken to be given by 
\begin{align}
 & \ddt{f}=\left\{ f,H\right\} .\label{eq:can_dynamics}
\end{align}

However, one can take a more general approach. Suppose we have a set
of quantities $(f,g,b,....)$ and a definition of their Poisson brackets
satisfying~\eqref{eq:Lie_conditions} such that the set is closed
under the operations specified by~\eqref{eq:Lie_conditions}; then
those quantities together with their Poisson brackets form a representation
of a Lie algebra. If there is a Hamiltonian $H$ that is an element
of that set, and the dynamics of the quantities are given by (\ref{eq:can_dynamics}),
we take this to constitute a canonical formulation.

As an example, suppose that instead of \emph{deriving}~\eqref{eq:canPB}
from~\eqref{eq:usualPB} we simply \emph{assert}~\eqref{eq:canPB} for
the $q_{i}$ and the $p_{j}$, and take our set of quantities to be
the $q_{i}$, the $p_{j}$ , and all polynomial functions of them.
Then, together with the definition that the Poisson bracket of two
numbers vanishes, and the definition that the Poisson bracket of a
number with any of the quantities vanishes, the Poisson brackets of
all such quantities can be determined from~\eqref{eq:Lie_conditions} and~\eqref{eq:canPB},
and the set of quantities is closed under those operations. If the
Hamiltonian is taken as a polynomial function of the $q_{i}$ and
the $p_{j}$, then the dynamics given by~\eqref{eq:can_dynamics}
are precisely what we would expect from the more traditional approach.
This more general approach is often useful. One example is if one
wants to treat the angular momentum components $J_{i}$ of an object
as fundamental, as one does for the spin of a particle; then there
are no canonical position and momentum in the description, but the
Lie algebra involves the Poisson bracket 
\begin{align}
 & \left\{ J_{x},J_{y}\right\} =J_{z},
\end{align}
and the brackets that follow from cyclic permutations.

That approach is also useful here. Returning to our equations (\ref{Maxwell})
we note that the divergence equations can be taken as initial conditions,
for if they are satisfied at an initial time and the curl equations
are satisfied at all later times, the divergence equations will also
be satisfied at all later times. So we need only construct dynamical
equations for the curl equations, and for ultimate use in quantization
we seek to do this with a Hamiltonian that is numerically equal to
the energy. From (\ref{eq:H_density}) we see that the Hamiltonian
should then be
\begin{align}
 & \Hl=\int_{V}\left(\frac{\bm{D}(\bm{r},t)\cdot\bm{D}(\bm{r},t)}{2\epsilon_{0}\varepsilon_{1}(\bm{r})}+\frac{\bm{B}(\bm{r},t)\cdot\bm{B}(\bm{r},t)}{2\mu_{0}}\right)\dd\bm{r},\label{eq:linear_Hamiltonian}
\end{align}
where $V$ is the normalization volume.

As first pointed out by Born and Infeld~\cite{born1934quantization}, to get the correct
dynamical equations the fundamental Poisson brackets should be taken
as 
\begin{align}
 & \left\{ D^{i}(\bm{r},t),D^{j}(\bm{r'},t)\right\} =\left\{ B^{i}(\bm{r},t),B^{j}(\bm{r'},t)\right\} =0,\label{eq:fieldPB}\\
 & \left\{ D^{i}(\bm{r},t),B^{j}(\bm{r'},t)\right\} =\epsilon^{ilj}\frac{\partial}{\partial r^{l}}\delta(\bm{r}-\bm{r'}).\nonumber 
\end{align}
for using these with (\ref{eq:linear_Hamiltonian}) we recover our
desired dynamics, 
\begin{align}
 & \frac{\partial\bm{D}(\bm{r},t)}{\partial t}=\left\{ \bm{D}(\bm{r},t),\Hl\right\} =\frac{1}{\mu_{0}}\bm{\nabla\times}\bm{B}(\bm{r},t),\\
 & \frac{\partial\bm{B}(\bm{r},t)}{\partial t}=\left\{ \bm{B}(\bm{r},t),\Hl\right\} =-\bm{\nabla\times}\left(\frac{\bm{D}(\bm{r},t)}{\epsilon_{0}\varepsilon_{1}(\bm{r})}\right).\nonumber
\end{align}

\subsubsection{Quantization}

We now quantize in the usual way, by taking 
\begin{align}
 & \left\{ \;\right\} \Rightarrow\frac{1}{i\hbar}\left[\;\right],\label{eq:quantization}
\end{align}
where $\left[\;\right]$ indicates the commutator. The equal time
commutation relations for Heisenberg operators are then given by 
\begin{align}
 & \left[D^{i}(\bm{r},t),D^{j}(\bm{r'},t)\right]=\left[B^{i}(\bm{r},t),B^{j}(\bm{r'},t)\right]=0,\label{eq:fieldCR}\\
 & \left[D^{i}(\bm{r},t),B^{j}(\bm{r'},t)\right]=i\hbar\epsilon^{ilj}\frac{\partial}{\partial r^{l}}\delta(\bm{r}-\bm{r'}),\nonumber 
\end{align}
and to satisfy the initial conditions we expand our fields in modes,
\begin{align}
 & \bm{D}(\bm{r},t)=\sum_{\alpha}\mathcal{C}_{\alpha}^{(1)}(t)\bm{D}_{\alpha}(\bm{r}),\\
 & \bm{B}(\bm{r},t)=\sum_{\alpha}\mathcal{C}_{\alpha}^{(2)}(t)\bm{B}_{\alpha}(\bm{r}),\nonumber
\end{align}
where here the $\mathcal{C}_{\alpha}^{(1)}(t)$ and the $\mathcal{C}_{\alpha}^{(2)}(t)$
are operators. Whatever their time dependence, the field operators
$\bm{D}(\bm{r},t)$ and $\bm{B}(\bm{r},t)$
will be divergenceless at all times, since $\bm{D}_{\alpha}(\bm{r})$
and $\bm{B}_{\alpha}(\bm{r})$ are divergenceless.

However, since those field operators must be Hermitian, the operators
$\mathcal{C}_{\alpha}^{(1)}(t)$ and $\mathcal{C}_{\alpha}^{(2)}(t)$
are not all independent; we easily find the conditions 
\begin{align}
 & \mathcal{C}_{\alpha'}^{(1)}(t)=-\left(\mathcal{C}_{\alpha}^{(1)}(t)\right)^{\dagger},\\
 & \mathcal{C}_{\alpha'}^{(2)}(t)=+\left(\mathcal{C}_{\alpha}^{(2)}(t)\right)^{\dagger},\nonumber
\end{align}
where we have used (\ref{eq:star_condition}). We can ensure these
conditions are satisfied by introducing new operators $a_{\alpha}(t)$
with no restrictions, such that 
\begin{align}
 & \mathcal{C}_{\alpha}^{(1)}(t)=\sqrt{\frac{\hbar\omega_{\alpha}}{2}}\left(a_{\alpha}(t)-a_{\alpha'}^{\dagger}(t)\right),\\
 & \mathcal{C}_{\alpha}^{(2)}(t)=\sqrt{\frac{\hbar\omega_{\alpha}}{2}}\left(a_{\alpha}(t)+a_{\alpha'}^{\dagger}(t)\right),\nonumber
\end{align}
where the factor $\sqrt{\left(\hbar\omega_{\alpha}\right)/2}$ is
introduced for later convenience; here $\alpha'$ is the partner mode
of $\alpha$. We then have 
\begin{align}
 & \bm{D}(\bm{r},t)=\sum_{\alpha}\sqrt{\frac{\hbar\omega_{\alpha}}{2}}\left(a_{\alpha}(t)\bm{D}_{\alpha}(\bm{r})+a_{\alpha}^{\dagger}(t)\bm{D}_{\alpha}^{*}(\bm{r})\right),\label{eq:field_expansions}\\
 & \bm{B}(\bm{r},t)=\sum_{\alpha}\sqrt{\frac{\hbar\omega_{\alpha}}{2}}\left(a_{\alpha}(t)\bm{B}_{\alpha}(\bm{r})+a_{\alpha}^{\dagger}(t)\bm{B}_{\alpha}^{*}(\bm{r})\right),\nonumber 
\end{align}
where we have again used (\ref{eq:star_condition}). The equal time
commutation relations (\ref{eq:fieldCR}) are then satisfied by taking
\begin{align}
 & \left[a_{\alpha}(t),a_{\beta}(t)\right]=0,\\
 & \left[a_{\alpha}(t),a_{\beta}^{\dagger}(t)\right]=\delta_{\alpha\beta},\nonumber
\end{align}
which are indeed required if the set of modes is complete. Using the
field expansions (\ref{eq:field_expansions}) in the expression (\ref{eq:linear_Hamiltonian})
for the Hamiltonian, we find 
\begin{align}
 \Hl &=\sum_{\alpha}\frac{\hbar\omega_{\alpha}}{2}\left(a_{\alpha}^{\dagger}(t)a_{\alpha}(t)+a_{\alpha}(t)a_{\alpha}^{\dagger}(t)\right)\label{eq:basic_Hamiltonian}\\
 & =\sum_{\alpha}\hbar\omega_{\alpha}\left(a_{\alpha}^{\dagger}(t)a_{\alpha}(t)+\frac{1}{2}\right)\nonumber \\
 & \rightarrow\sum_{\alpha}\hbar\omega_{\alpha}a_{\alpha}^{\dagger}(t)a_{\alpha}(t),\nonumber 
\end{align}
where in the last expression we have neglected the zero-point energy,
as we do henceforth. The equation for any Heisenberg operator $\mathcal{O}(t)$,
\begin{align}\label{eq:Heisenberg_O}
 & i\hbar\ddt{\mathcal{O}(t)}=\left[\mathcal{O}(t),\Hl\right],
\end{align}
[compare (\ref{eq:can_dynamics})], leads to 
\begin{align}
 & \ddt{a_{\alpha}(t)}=-i\omega_{\alpha}a_{\alpha}(t),
\end{align}
as expected.

\subsubsection{Special cases}\label{sec:special_cases}

\begin{figure}
    \centering
    \includegraphics[width=1.0\textwidth]{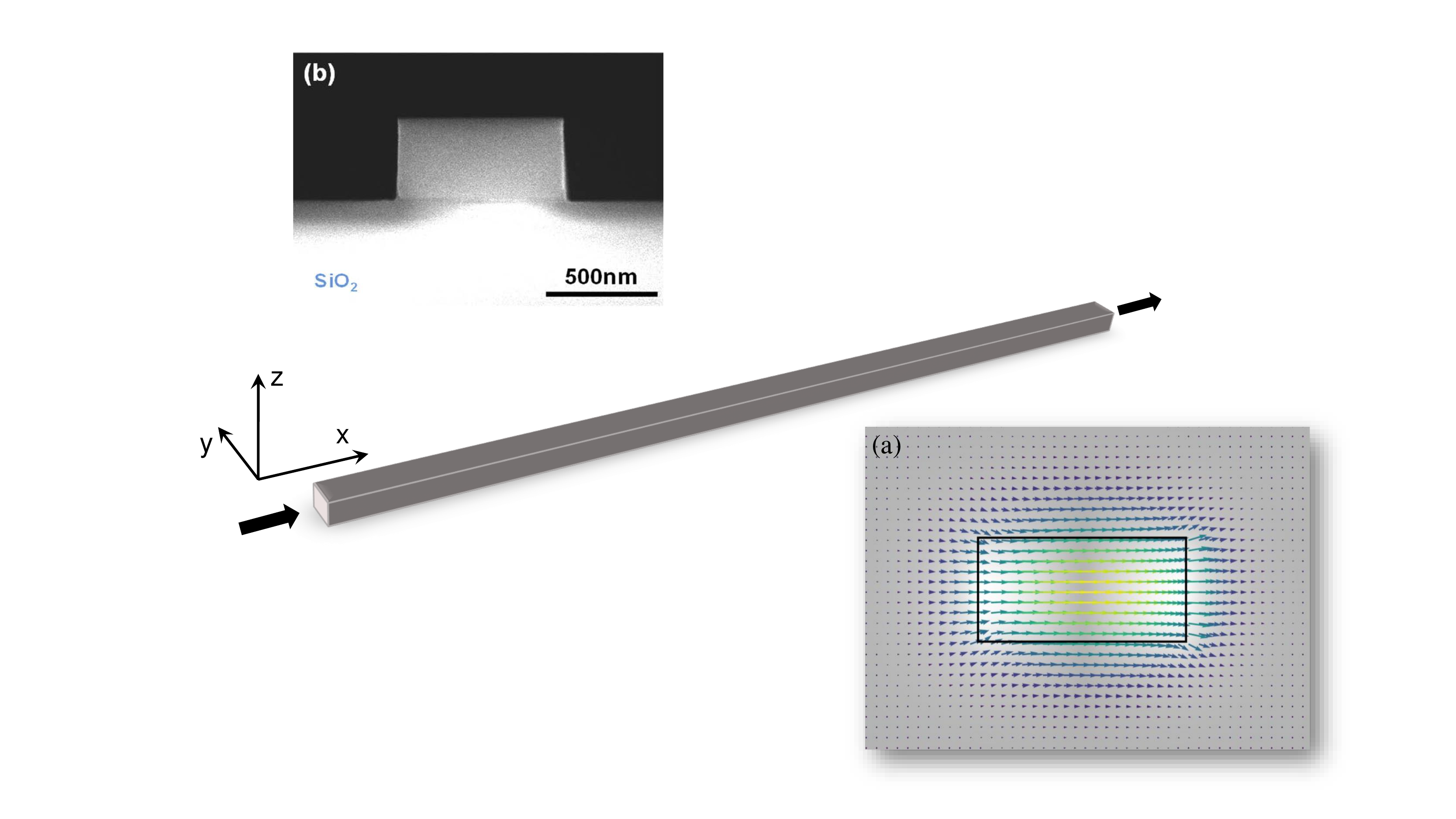}
    \caption{Schematic representation of a channel waveguide. (a) Typical field intensity profile of the fundamental transverse-electric mode, computed by an eigenmode solver. The waveguide and simulation parameters are the same as in Fig. \ref{fig:isolated_ring}, but with no curvature. By comparison with Fig. \ref{fig:isolated_ring} it is possible to appreciate the different field distribution produced by the waveguide curvature. The contour plot accounts for the $E_{x}$ component, while the arrow field represents $\bm{E}_{y,z}$; (b) Scanning Electron Microscope (SEM) image of a typical high index contrast waveguide cross-section. Image from Xie et al.~\cite{xie2020ultrahigh}.}
    \label{fig:isolated_channel}
\end{figure}

The usual first case of interest is a uniform medium. We treat that
for completeness in Appendix~\ref{sec:Normalization}. Here we consider the form of the field
operators for two simple structures from integrated optics, the first
a channel waveguide as indicated schematically in Fig.~\ref{fig:isolated_channel},
and the second a ring resonator, initially considered isolated and
indicated schematically in Fig.~\ref{fig:isolated_ring}; of interest in itself, the ring
resonator can also be taken as a simple example of an integrated optical
cavity.

\subsubsection*{Channel waveguide}

Looking first at the waveguide structure, we take $\bm{\hat{z}}$
normal to the substrate and $\bm{\hat{x}}$ the direction
along which the waveguide runs; then $\varepsilon_{1}(\bm{r})=\varepsilon_{1}(y,z)$
here. 

We will be interested in modes confined to the waveguide, and
so we need only introduce periodic boundary conditions in the $\bm{\hat{x}}$
direction, with periodicity over a length $L$. Then because of the
translational symmetry in the $x$ direction we can choose the modes
$(\bm{D}_{\alpha}(\bm{r}),\bm{B}_{\alpha}(\bm{r}))$
satisfying (\ref{eq:mode_equations}) to be of the form 
\begin{align}
 & \bm{D}_{Ik}(\bm{r})=\frac{\bm{d}_{Ik}(y,z)}{\sqrt{L}}e^{ikx},\label{eq:mode_form_periodic}\\
 & \bm{B}_{Ik}(\bm{r})=\frac{\bm{b}_{Ik}(y,z)}{\sqrt{L}}e^{ikx},\nonumber 
\end{align}
where $k=2\pi m/L,$ with $m$ an integer; the index $I$ labels the
different modes, which in their polarization profile will generally
be more complicated than either circular or linear polarization; we
label the frequency of such a mode label by $\omega_{Ik}$, and the
partner mode of a mode characterized by $k$ is one characterized
by $-k$. In the normalization conditions~\eqref{eq:norm_B}, and~\eqref{eq:norm_D}
we can let $y$ and $z$ vary over all values, with only $x$ limited
to a region of length $L$, yielding 
\begin{align}\label{eq:norm_B_waveguide_no_disp}
 & \int\frac{\bm{b}_{Ik}^{*}(y,z)\cdot\bm{b}_{Ik}(y,z)}{\mu_{0}}\dd y\dd z=1,
\end{align}
and
\begin{align}\label{eq:norm_D_waveguide_no_disp}
 & \int\frac{\bm{d}_{Ik}^{*}(y,z)\cdot\bm{d}_{Ik}(y,z)}{\epsilon_{0}\varepsilon_{1}(y,z)}\dd y\dd z=1,
\end{align}
as the normalization conditions of $\bm{b}_{Ik}(y,z)$ and
$\bm{d}_{Ik}(y,z).$ The general field expansions (\ref{eq:field_expansions})
then take the form
\begin{align}
 & \bm{D}(\bm{r},t)=\sum_{I,k}\sqrt{\frac{\hbar\omega_{Ik}}{2L}}a_{Ik}(t)\bm{d}_{Ik}(y,z)e^{ikx}+\hc,\label{eq:field_expansions_channel_periodic}\\
 & \bm{B}(\bm{r},t)=\sum_{I,k}\sqrt{\frac{\hbar\omega_{Ik}}{2L}}a_{Ik}(t)\bm{b}_{Ik}(y,z)e^{ikx}+\hc,\nonumber 
\end{align}
where $\omega_{Ik}$ is the frequency of mode $I,k,$
\begin{align}
 & \left[a_{Ik}(t),a_{I'k'}(t)\right]=0,\label{eq:comm_channel_periodic}\\
 & \left[a_{Ik}(t),a_{I'k'}^{\dagger}(t)\right]=\delta_{II'}\delta_{kk'},\nonumber 
\end{align}
and the Hamiltonian is 
\begin{align}
 & \Hl=\sum_{I,k}\hbar\omega_{Ik}a_{Ik}^{\dagger}(t)a_{Ik}(t).\label{eq:Hamiltonian_channel_periodic}
\end{align}

We consider passing to the limit of an infinite normalization length
$L$, which will involve a continuously varying $k$. We do this in
the usual heuristic way by replacing the sum over $k$ by an integral
over $k$, taking into account the density of points in $k$ space.
That is, we take 
\begin{align}
 & \sum_{k}\rightarrow\frac{\int \dd k}{\left(\frac{2\pi}{L}\right)}=\frac{L}{2\pi}\int \dd k.\label{eq:1Ddtoc}
\end{align}
We also have to replace the operators $a_{Ik}(t)$ associated with
discrete modes by operators associated with modes labeled by a continuous
varying $k;$ those latter operators we label $a_{I}(k,t).$ The connection
between the two can be identified by recalling that for the discrete
modes we have (\ref{eq:comm_channel_periodic}), and so in particular
\begin{align}
 & \sum_{k'}\left[a_{Ik}(t),a_{I'k'}^{\dagger}(t)\right]=\delta_{II'},\\
 & \sum_{k'}\left[a_{Ik}(t),a_{I'k'}(t)\right]=0.\nonumber
\end{align}
Recalling (\ref{eq:1Ddtoc}), we see that if we take 
\begin{align}
 & a_{Ik}(t)\rightarrow\sqrt{\frac{2\pi}{L}}a_{I}(k,t),\label{eq:1Ddtoc_second}
\end{align}
we arrive at 
\begin{align}
 & \int\left[a_{I}(k,t),a_{I}^{\dagger}(k',t)\right]\dd k'=\delta_{II'},\\
 & \int\left[a_{I}(k,t),a_{I}(k',t)\right]\dd k'=0,\nonumber
\end{align}
and since the operators at different $k$ are to be independent this
yields 
\begin{align}
 & \left[a_{I}(k,t),a_{I'}(k',t)\right]=0,\label{eq:comm_channel_infinite}\\
 & \left[a_{I}(k,t),a_{I'}^{\dagger}(k',t)\right]=\delta_{II'}\delta(k-k').\nonumber 
\end{align}
So the $a_{I}(k,t)$ are convenient operators for the expansion of
the electromagnetic field. Using~\eqref{eq:1Ddtoc} and~\eqref{eq:1Ddtoc_second}
in~\eqref{eq:field_expansions_channel_periodic} we find 
\begin{align}
 & \bm{D}(\bm{r},t)=\sum_{I}\int\sqrt{\frac{\hbar\omega_{Ik}}{4\pi}}a_{I}(k,t)\bm{d}_{Ik}(y,z)e^{ikx}\dd k+\hc,\label{eq:field_expansions_channel_infinite}\\
 & \bm{B}(\bm{r},t)=\sum_{I}\int\sqrt{\frac{\hbar\omega_{Ik}}{4\pi}}a_{I}(k,t)\bm{b}_{Ik}(y,z)e^{ikx}\dd k+\hc,\nonumber 
\end{align}
and using them in (\ref{eq:Hamiltonian_channel_periodic}) we find
\begin{align}
 & \Hcl=\sum_{I}\int\hbar\omega_{Ik}a_{I}^{\dagger}(k,t)a_{I}(k,t)\dd k.\label{eq:Hamiltonian_channel_infinite}
\end{align}
Here we have not changed the notation for $\omega_{Ik}$ nor
for $\bm{d}_{Ik}(y,z)$ and $\bm{b}_{Ik}(y,z)$, although
$k$ now various continuously; the normalization conditions remain~\eqref{eq:norm_B_waveguide_no_disp} and~\eqref{eq:norm_D_waveguide_no_disp}.
Of course, the integration over $k$ here only ranges over the values
for which the modes are confined to the waveguide structure; that
is the part of the electromagnetic field we include in (\ref{eq:field_expansions_channel_infinite}).

\subsubsection*{Ring resonators}

Next we turn to the ring structure shown in Fig.~\ref{fig:isolated_ring}, and consider the
form of the modes $\left(\bm{D}_{\alpha}\left(\bm{r}\right),\bm{B}_{\alpha}\left(\bm{r}\right)\right)$
that can at least be approximated as confined to the ring. Here we
use indices $J$ to denote the modes, and $\omega_{J}$ their frequencies;
the modes can be taken to be of the form 
\begin{align}
 & \bm{D}_{J}(\bm{r})=\frac{\mathtt{d}_{J}(\bm{r}_{\perp};\zeta)}{\sqrt{\mathcal{L}}}e^{i\kappa_{J}\zeta},\label{eq:ring_modes}\\
 & \bm{B}_{J}(\bm{r})=\frac{\mathsf{b}_{J}(\bm{r}_{\perp};\zeta)}{\sqrt{\mathcal{L}}}e^{i\kappa_{J}\zeta},\nonumber 
\end{align}
where, as in Sec.\ref{sec:TroubleWithE}, $\kappa_{J}=2\pi m_{J}/\mathcal{L}$
where $m_{J}$ is an integer; we use $J$ to denote the type of
the mode. The dependence of $\mathtt{d}_{J}(\bm{r}_{\perp};\zeta)$ and $\mathsf{b}_{J}(\bm{r}_{\perp};\zeta)$ on $\zeta$ arises because of components of the vectors in the $xy$ plane. As $\zeta$ varies the direction of those vectors will change, although $\mathtt{b}_{J}^{*}(\bm{r}_{\perp};\zeta)\cdot\mathtt{b}_{J}(\bm{r}_{\perp};\zeta)$ and $\mathtt{d}_{J}^{*}(\bm{r}_{\perp};\zeta)\cdot\mathtt{d}_{J}(\bm{r}_{\perp};\zeta)$ will be independent of $\zeta$ and we can put $\mathtt{b}_{J}^{*}(\bm{r}_{\perp};\zeta)\cdot\mathtt{b}_{J}(\bm{r}_{\perp};\zeta)=\mathtt{b}_{J}^{*}(\bm{r}_{\perp};0)\cdot\mathtt{b}_{J}(\bm{r}_{\perp};0)$ and $\mathtt{d}_{J}^{*}(\bm{r}_{\perp};\zeta)\cdot\mathtt{d}_{J}(\bm{r}_{\perp};\zeta)=\mathtt{d}_{J}^{*}(\bm{r}_{\perp};0)\cdot\mathtt{d}_{J}(\bm{r}_{\perp};0)$.

The electromagnetic field operators associated with the
modes (\ref{eq:ring_modes}) can then be written, using (\ref{eq:field_expansions}),
as 
\begin{align}
 & \bm{D}(\bm{r},t)=\sum_{J}\sqrt{\frac{\hbar\omega_{J}}{2\mathcal{L}}}c_{J}(t)\mathtt{d}_{J}(\bm{r}_{\perp},\zeta)e^{i\kappa_{J}\zeta}+\hc,\label{eq:field_expansions_ring}\\
 & \bm{B}(\bm{r},t)=\sum_{J}\sqrt{\frac{\hbar\omega_{J}}{2\mathcal{L}}}c_{J}(t)\mathtt{b}_{J}(\bm{r}_{\perp},\zeta)e^{i\kappa_{J}\zeta}+\hc,\nonumber 
\end{align}
where $J$ indicates both $\kappa_{J}$ and the type of mode, 
\begin{align}
 & \left[c_{J}(t),c_{J'}(t)\right]=0,\label{eq:comm_ring}\\
 & \left[c_{J}(t),c_{J'}^{\dagger}(t)\right]=\delta_{JJ'},\nonumber 
\end{align}
and the Hamiltonian is 
\begin{align}
 & \Hrl=\sum_{J}\hbar\omega_{J}c_{J}^{\dagger}c_{J},\label{Hamiltonian_ring}
\end{align}
as in (\ref{eq:Hring_first}).

The general normalization conditions (\ref{eq:norm_B},\ref{eq:norm_D})
here become 
\begin{align}
 & \int\frac{\mathtt{b}_{J}^{*}(\bm{r}_{\perp};0)\cdot\mathtt{b}_{J}(\bm{r}_{\perp};0)}{\mu_{0}}\dd\bm{r_{\perp}}=1,\label{eq:norm_B_ring_no_disp}
\end{align}
\begin{align}
 & \int\frac{\mathtt{d}_{J}^{*}(\bm{r}_{\perp};0)\cdot\mathtt{d}_{J}(\bm{r}_{\perp};0)}{\epsilon_{0}\varepsilon_{1}(\bm{r_{\perp}})}\dd\bm{r}_{\perp}=1,\label{eq:norm_D_no_disp}
\end{align}
The second of these gives the normalization condition for $\bm{e}_{J}(\bm{r}_{\perp};\zeta$),
since comparing (\ref{eq:E_ring_use}) with the first of (\ref{eq:field_expansions_ring})
we see that $\mathsf{d}_{J}(\bm{r}_{\perp};\zeta)=\epsilon_{0}\varepsilon_{1}(\bm{r})\mathsf{e}_{J}(\bm{r}_{\perp};\zeta).$

For the curvature of the ring not too great the fields $(\mathtt{d}_{J}(\bm{r}_{\perp};\zeta),\mathtt{b}_{J}(\bm{r}_{\perp};\zeta))$
will be related to those fields $(\bm{d}_{Ik}(y,z),\bm{b}_{Ik}(y,z))$
of a channel structure of the same cross-section according to 
\begin{align}
 & \mathtt{d}_{J}(\rho,z;0)\approx\bm{d}_{I_{\alpha}k_{\alpha}}(R-\rho,z),\label{eq:ring_channel_compare}\\
 & \mathtt{b}_{J}(\rho,z;0)\approx\bm{b}_{I_{\alpha}k_{\alpha}}(R-\rho,z).\nonumber 
\end{align}
[and compare~\eqref{eq:norm_B_waveguide_no_disp} and~\eqref{eq:norm_D_waveguide_no_disp}
with~\eqref{eq:norm_B_ring_no_disp} and \eqref{eq:norm_D_no_disp}].

\subsubsection{Including dispersion}\label{sec:dispersion}

Up to this point we have assumed that the relative dielectric constant
$\varepsilon_{1}(\bm{r})$ is independent of the frequency
of the optical field. More generally, however, if we have electric
and displacement fields 
\begin{align}
 & \bm{E}(\bm{r},t)=\bm{E}(\bm{r})e^{-i\omega t}+\cc,\label{eq:disp_fields}\\
 & \bm{D}(\bm{r},t)=\bm{D}(\bm{r})e^{-i\omega t}+\cc,\nonumber 
\end{align}
which for the moment we consider classical, they are related by a
frequency dependent $\varepsilon_{1}(\bm{r},\omega),$
\begin{align}
 & \bm{D}(\bm{r})=\epsilon_{0}\varepsilon_{1}(\bm{r};\omega)\bm{E}(\bm{r}).\label{eq:disp_constitutive}
\end{align}
The quantity $\varepsilon_{1}(\bm{r};\omega)$ is in general
complex, with its real part describing dispersive effects and its
imaginary part describing absorption; they are linked by Kramers-Kronig
relations~\cite{jackson1999classical}. This more complicated electrodynamics arises,
of course, because of the involvement of degrees of freedom of the
underlying material medium. Quantizing the dynamics then involves
quantizing both the ``bare'' electromagnetic field and those material
degrees of freedom to which it is coupled. Here we follow a strategy
introduced earlier~\cite{bhat2006hamiltonian}. Although the dynamics over a
wide frequency range was considered there, for integrated photonic
structures one is mainly interested in frequencies below the bandgap
of all materials, for which $\varepsilon_{1}(\bm{r};\omega)$
may be a strong function of frequency but is purely real. That is
the regime we consider here; the treatment of quantum electrodynamics
in the regime where absorption is presented has been investigated
using this strategy elsewhere~\cite{judge2013canonical}. 

We begin by noting that for $\varepsilon_{1}(\bm{r};\omega)$
purely real we can still consider ``modes'' for $\bm{D}(\bm{r},t)$
and $\bm{B}(\bm{r},t)$, as we did when dispersion
was neglected. Seeking solutions of Maxwell's equations (\ref{Maxwell})
of the form (\ref{eq:modes}), but now using the constitutive relations
\begin{align}
 & \bm{D}_{\alpha}(\bm{r})=\epsilon_{0}\varepsilon_{1}(\bm{r},\omega_{\alpha})\bm{E}_{\alpha}(\bm{r}),\label{eq:condisp}\\
 & \bm{B}_{\alpha}(\bm{r})=\mu_{0}\bm{H}_{\alpha}(\bm{r}),\nonumber 
\end{align}
we find we require 
\begin{align}
 & \bm{\nabla}\times\left[\frac{\bm{\nabla\times}\bm{B}_{\alpha}(\bm{r})}{\varepsilon_{1}(\bm{r};\omega_{\alpha})}\right]=\frac{\omega_{\alpha}^{2}}{c^{2}}\bm{B}_{\alpha}(\bm{r}),\label{eq:mode_equations_dispersion}\\
 & \bm{\nabla\cdot}\bm{B}_{\alpha}(\bm{r})=0,\nonumber \\
 & \bm{D}_{\alpha}(\bm{r})=\frac{i}{\omega_{\alpha}\mu_{0}}\bm{\nabla}\times\bm{B}_{\alpha}(\bm{r}).\nonumber 
\end{align}
The first of these is more complicated than the first of (\ref{eq:mode_equations}),
with $\omega_{\alpha}$ here appearing on both sides of the eigenvalue
equation. In general, for example for a photonic crystal, determining
the allowed $\omega_{\alpha}$ must be done self-consistently, with
some iteration. For the simpler structures we consider here, however,
that will not be a problem. Note also that the mode fields $\bm{B}_{\alpha}(\bm{r})$
and $\bm{B}_{\beta}(\bm{r})$ associated with different
frequencies $\omega_{\alpha}\neq\omega_{\beta}$ are in general \emph{not}
orthogonal; the condition (\ref{eq:orthogonality}) does not hold
here. Mathematically, this arises because such a $\bm{B}_{\alpha}(\bm{r})$
and a $\bm{B}_{\beta}(\bm{r})$ are eigenfunctions
of \emph{different} Hermitian operators, since in (\ref{eq:M_op}) $\varepsilon_{1}(\bm{r})$
must be replaced by $\varepsilon_{1}(\bm{r},;\omega)$ when it
acts on $\bm{B}_{\alpha}(\bm{r})$, and by $\varepsilon_{1}(\bm{r};\omega_{\beta})$
when it acts on $\bm{B}_{\beta}(\bm{r})$. Physically,
it arises because the true modes of the system involve both the electromagnetic
field and the degrees of freedom of the material medium; when these
full ``polariton modes'' are considered orthogonality reappears~\cite{bhat2006hamiltonian}.

Nonetheless, one finds that one can proceed with just $(\bm{D}_{\alpha}(\bm{r}),\bm{B}_{\alpha}(\bm{r}))$,
satisfying (\ref{eq:mode_equations_dispersion}), in the equations
(\ref{eq:field_expansions}) if the normalization conditions~\eqref{eq:norm_B} and~\eqref{eq:norm_D}
are modified; one formally introduces the material degrees of freedom
to show this, and then one need not explicitly deal with them again.
Once $\bm{D}_{\alpha}(\bm{r})$ and $\bm{B}_{\alpha}(\bm{r})$
are found from (\ref{eq:mode_equations_dispersion}), for future use
the appropriate normalization condition is most usefully written in
terms of $\bm{D}_{\alpha}(\bm{r})$, and for the case
of periodic boundary conditions is given by~\cite{bhat2006hamiltonian, sipe2009photons}
\begin{align}
 & \int_{V}\frac{\bm{D}_{\alpha}^{*}(\bm{r})\cdot\bm{D}_{\alpha}(\bm{r})}{\epsilon_{0}\varepsilon_{1}(\bm{r};\omega_{\alpha})}\frac{v_{p}(\bm{r};\omega_{\alpha})}{v_{g}(\bm{r};\omega_{\alpha})}\dd\bm{r}=1.\label{eq:norm_D-3}
\end{align}
instead of (\ref{eq:norm_D}). Here 
\begin{align}
 & v_{p}(\bm{r};\omega)=\frac{c}{n(\bm{r};\omega)}\label{eq:vphase}
\end{align}
is the local phase velocity at frequency $\omega$, where $n(\bm{r};\omega)=\sqrt{\varepsilon_{1}(\bm{r};\omega)}$
is the local index of refraction at that frequency, and 
\begin{align}
 & v_{g}(\bm{r};\omega)=\frac{v_{p}(\bm{r};\omega)}{1+\frac{\omega}{n(\bm{r};\omega)}\frac{\partial n(\bm{r};\omega)}{\partial\omega}}\label{eq:vgroup}
\end{align}
is the local group velocity at frequency $\omega.$ 

In Appendix~\ref{sec:Normalization} we consider the inclusion of dispersion in a description
of light in a uniform medium. Here we begin with the inclusion of
dispersion in the description of channel waveguide structures, first
considering periodic boundary conditions. The mode frequencies $\omega_{Ik}$
must be found by solving (\ref{eq:mode_equations_dispersion}) with
modes of the form (\ref{eq:mode_form_periodic}); following the discussion
above, it is easy to show from (\ref{eq:norm_D-3}) that the normalization condition (\ref{eq:norm_D_waveguide_no_disp})
is then replaced by 
\begin{align}
 & \int\frac{\bm{d}_{Ik}^{*}(y,z)\cdot\bm{d}_{Ik}(y,z)}{\epsilon_{0}\varepsilon_{1}(y,z;\omega_{Ik})}\frac{v_{p}(y,z;\omega_{Ik})}{v_{g}(y,z;\omega_{Ik})}\dd y\dd z=1.\label{eq:norm_D_waveguide_disp}
\end{align}
With this new normalization and the corresponding scaling for $\bm{b}_{Ik}(y,z;)$,
the expressions~\eqref{eq:field_expansions_channel_periodic},~\eqref{eq:comm_channel_periodic}, and~\eqref{eq:Hamiltonian_channel_periodic}
for the field expansions, the commutation relations, and the Hamiltonian
still all hold. Forming the Poynting vector $\bm{S}(\bm{r},t)=\bm{E}(\bm{r},t)\times\bm{H}(\bm{r},t)$,
if we look at a one-photon state 
\begin{align}
 & \left|\Psi\right\rangle =a_{Ik}^{\dagger}(0)\vac
\end{align}
we find 
\begin{align}
 & \int\left\langle \Psi|\bm{S}(\bm{r},t)|\Psi\right\rangle \cdot\bm{\hat{x}}\dd y\dd z=\frac{\hbar\omega_{Ik}}{L}v_{Ik},
\end{align}
where of course $(\hbar\omega_{Ik}/L)$ is the photon energy per unit
length in the waveguide, and 
\begin{align}
 & v_{Ik}\equiv\frac{\partial\omega_{Ik}}{\partial k}=\frac{1}{2}\int\left(\bm{e}_{Ik}(y,z)\times\bm{h}_{Ik}^{*}(y,z)+\bm{e}_{Ik}^{*}(y,z)\times\bm{h}_{Ik}(y,z)\right)\cdot\bm{\hat{x}}\dd y\dd z\label{eq:vIk}
\end{align}
is the group velocity of the mode, including both modal and material
dispersion; here 
\begin{align}
 & \bm{e}_{Ik}(y,z)=\frac{\bm{d}_{Ik}(y,z)}{\epsilon_{0}\varepsilon_{1}(y,z;\omega_{Ik})},\label{eq:little_e_and_h}\\
 & \bm{h}_{Ik}(y,z)=\frac{\bm{b}_{Ik}(y,z)}{\mu_{0}},\nonumber 
\end{align}
and (\ref{eq:vIk}) can be derived from the normalization condition
(\ref{eq:norm_D_waveguide_disp}), which we do in Appendix~\ref{sec:Group_velocity}. If
we pass to an infinite length waveguide structure, then~\eqref{eq:comm_channel_infinite},~\eqref{eq:field_expansions_channel_infinite}, and~\eqref{eq:Hamiltonian_channel_infinite}
all hold, with the $\bm{d}_{Ik}(y,z)$ and $\bm{b}_{Ik}(y,z)$
found as in the periodic boundary condition analysis, normalized according
to~\eqref{eq:norm_D_waveguide_disp}. 

Finally, for the isolated ring the inclusion of dispersion leads to
using the mode forms (\ref{eq:ring_modes}) in (\ref{eq:mode_equations_dispersion})
to find the mode frequencies $\omega_{J}$ and the fields $\mathtt{d}_{J}(\bm{r}_{\perp};\zeta)$
and $\mathtt{b}_{J}(\bm{r}_{\perp};\zeta)$. Instead of the normalization
condition (\ref{eq:norm_D_no_disp}) we use
\begin{align}
 & \int\frac{\mathtt{d}_{J}^{*}(\bm{r}_{\perp};0)\cdot\mathtt{d}_{J}(\bm{r}_{\perp};0)}{\epsilon_{0}\varepsilon_{1}(\bm{r_{\perp}};\omega_{J})}\frac{v_{p}(\bm{r}_{\perp};\omega_{J})}{v_{g}(\bm{r}_{\perp};\omega_{J})}\dd\bm{r}_{\perp}=1.\label{eq:norm_D_disp}
\end{align}
The field expansion (\ref{eq:field_expansions_ring}), the commutation
relations (\ref{eq:comm_ring}), and the form of the Hamiltonian (\ref{Hamiltonian_ring})
are all unchanged.

\subsection{Channel fields and ring-channel coupling}\label{sec:ring_channel_coupling}

\begin{figure}
    \centering
    \includegraphics[width=1.0\textwidth]{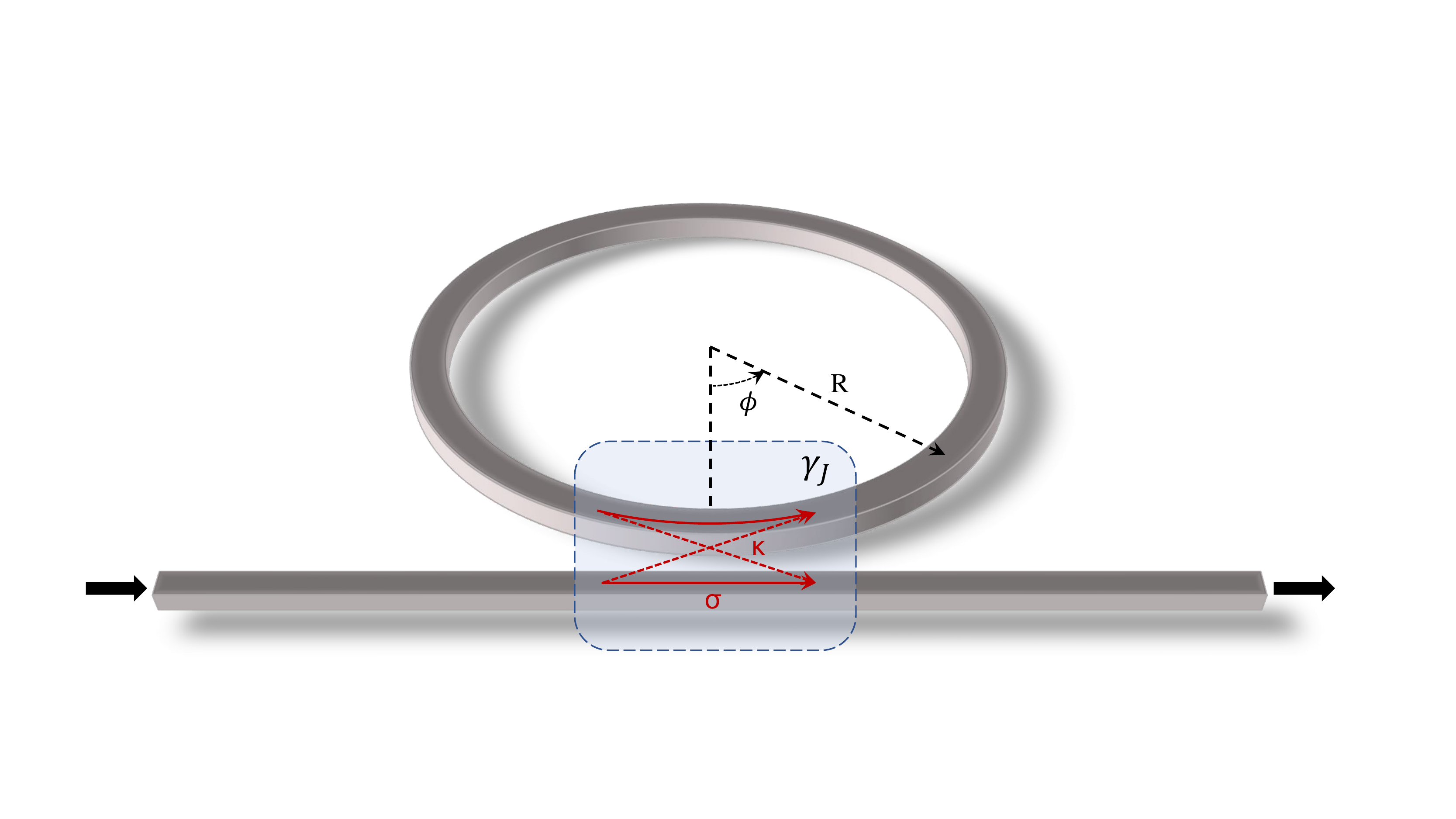}
    \caption{Schematic representation of a point-coupled ring resonator. The coupling is modelled by a simple $\begin{pmatrix}
    \sigma  & i\kappa \\
    i\kappa  & \sigma \\
    \end{pmatrix}$ matrix, where $(\sigma, \kappa) \in \mathbb{R}$ are the self-coupling and cross-coupling coefficients, respectively. The coupler is assumed lossless hence yielding $\sigma^2 + \kappa^2 = 1$.}
    \label{fig:coupled_ring}
\end{figure}

We now turn to the coupling between the channel and ring, in a structure
such as in Fig.~\ref{fig:coupled_ring}. 

In the frequency range indicated there
we identify the isolated ring resonances $\omega_{J}$, restricting
ourselves here to resonances associated with one transverse mode;
recall that the isolated ring fields are given by (\ref{eq:field_expansions_ring}),
with in general the normalization condition (\ref{eq:norm_D_disp}).
Our electromagnetic fields associated with the channel are expanded
as (\ref{eq:field_expansions_channel_infinite}), with commutation
relations (\ref{eq:comm_channel_infinite}) and a Hamiltonian (\ref{eq:Hamiltonian_channel_infinite});
the mode normalization condition is generally given by (\ref{eq:norm_D_waveguide_disp}).
In those expressions we have used the dummy index $I$ to indicate
the mode type. For the channel we now just consider one type, and
use the dummy index $J$ to label a range of $k$ (with a corresponding
range of $\omega_{Jk}$), introducing different wavenumber ranges,
each centered at a ring mode resonance. 

\begin{figure}
    \centering
    \includegraphics[width=0.75\textwidth]{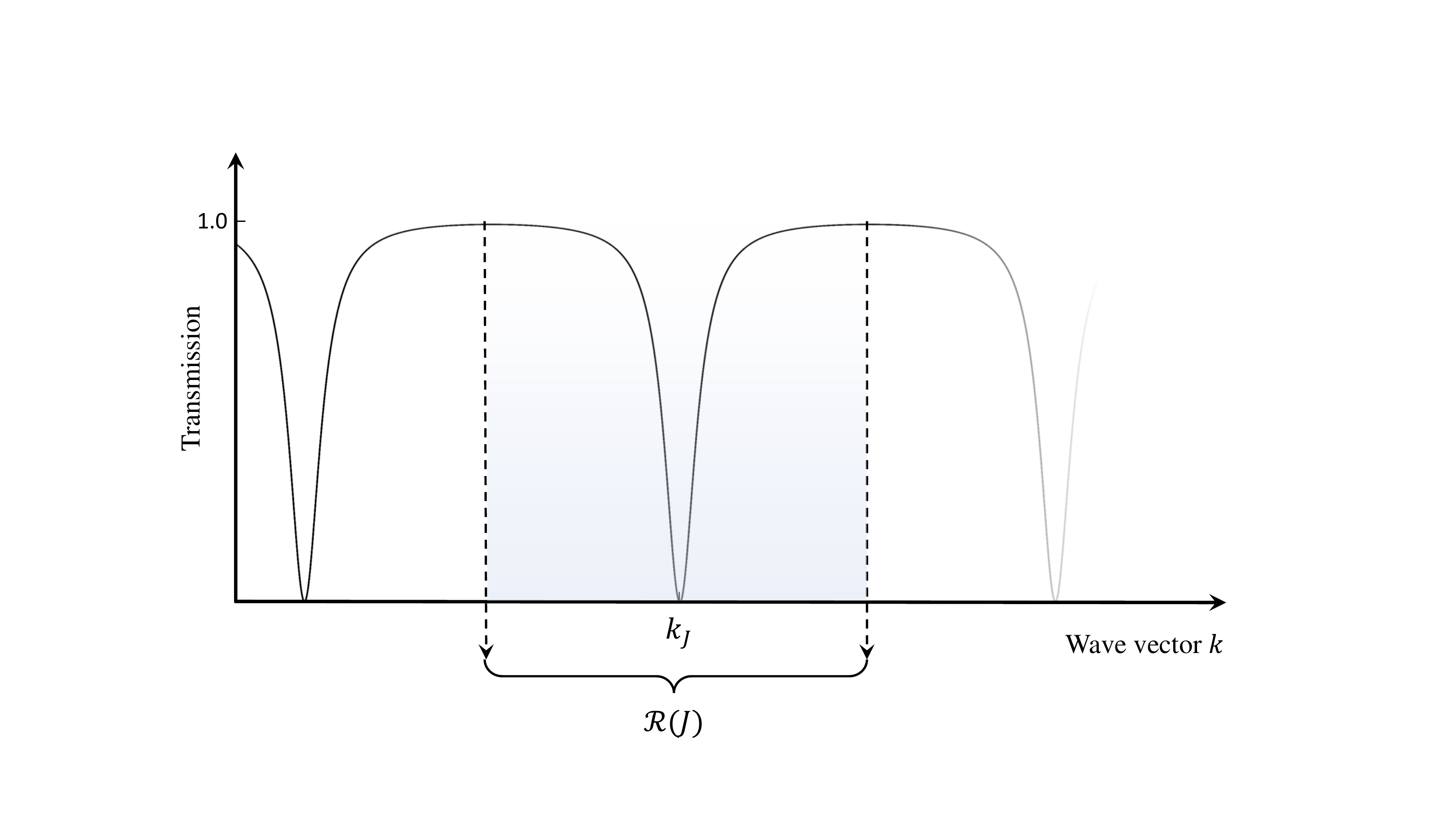}
    \caption{Typical transmission of a point-coupled ring resonator. Here $\mathscr{R}(J)$ represents the wave-vector range associated with each resonance $J$, centered at $k_J$.}
    \label{fig:k_range}
\end{figure}

If we denote the range of $k$ allotted to $J$ by $\mathscr{R}(J)$, (see Fig.~\ref{fig:k_range})
we can then write 
\begin{align}
 & \bm{D}(\bm{r},t)=\sum_{J}\int_{k\in\mathscr{R}(J)}\sqrt{\frac{\hbar\omega_{Jk}}{4\pi}}a_{J}(k,t)\bm{d}_{Jk}(y,z)e^{ikx}\dd k+\hc,\label{eq:field_expansions_channel_work}\\
 & \bm{B}(\bm{r},t)=\sum_{J}\int_{k\in\mathscr{R}(J)}\sqrt{\frac{\hbar\omega_{Jk}}{4\pi}}a_{J}(k,t)\bm{b}_{Jk}(y,z)e^{ikx}\dd k+\hc,\nonumber 
\end{align}
where 
\begin{align}
 & \left[a_{J}(k,t),a_{J'}(k',t)\right]=0,\label{eq:comm_channel_work}\\
 & \left[a_{J}(k,t),a_{J'}^{\dagger}(k',t)\right]=\delta_{JJ'}\delta(k-k'),\nonumber 
\end{align}
and the Hamiltonian is 
\begin{align}
 & \Hcl=\sum_{J}H_{J}^{\text{L}},\label{eq:Hamiltonian_channel_decomposition}
\end{align}
where 
\begin{align}
 & H_{J}^{\text{L}}=\int_{k\in\mathscr{R}(J)}\hbar\omega_{Jk}a_{J}^{\dagger}(k,t)a_{J}(k,t)\dd k.\label{eq:channel_Hamiltonian}
\end{align}
For some applications it is convenient to introduce ``channel field
operators,'' 
\begin{align}
 & \psi_{J}(x,t)\equiv\int_{k\in\mathscr{R}(J)}\frac{\dd k}{\sqrt{2\pi}}a_{J}(k,t)e^{i(k-k_{J})x},\label{eq:channel_field}
\end{align}
where we use $k_{J}$ to denote a reference $k$ in the center of
the range $\mathscr{R}(J)$, and before considering the coupling of
the channel to the ring we discuss these operators.

From the Hamiltonian (\ref{eq:channel_Hamiltonian}) their dynamics
are given by 
\begin{align}
 \frac{\partial\psi_{J}(x,t)}{\partial t} & =-i\int_{k\in\mathscr{R}(J)}\frac{\dd k}{\sqrt{2\pi}}\omega_{Jk}a_{J}(k,t)e^{i(k-k_{J})x}\label{eq:psi_dynamics1}\\
 & =-i\int_{k\in\mathscr{R}(J)}\frac{\dd k}{\sqrt{2\pi}}\left(\omega_{J}+v_{J}(k-k_{J})+\frac{1}{2}v'_{J}(k-k_{J})^{2}+...\right)a_{J}(k,t)e^{i(k-k_{J})x},\nonumber 
\end{align}
where in the second line we have put $\omega_{J}\equiv\omega_{Jk_{J}}$
and assumed that in the range $k\in\mathscr{R}(J)$ the full modal
dispersion is small enough that an expansion of the dispersion relation
is reasonable; we have put $v_{J}\equiv v_{Jk_{J}}$ and $v'_{J}\equiv v'_{Jk_{J}}$,
where 
\begin{align}
 & v'_{Jk}=\frac{\partial^{2}\omega_{Jk}}{\partial k^{2}}.
\end{align}
Then we can write (\ref{eq:psi_dynamics1}) as 
\begin{align}
 & \frac{\partial\psi_{J}(x,t)}{\partial t}=-i\omega_{J}\psi_{J}(x,t)-v_{J}\frac{\partial\psi_{J}(x,t)}{\partial x}+\frac{1}{2}iv'_{J}\frac{\partial^{2}\psi_{J}(x,t)}{\partial x^{2}}+...\label{eq:psi_dynamics2}
\end{align}
For many applications the group velocity dispersion within each channel
range $\mathscr{R}(J)$ can be neglected, and we can take simply 
\begin{align}
 & \frac{\partial\psi_{J}(x,t)}{\partial t}=-i\omega_{J}\psi_{J}(x,t)-v_{J}\frac{\partial\psi_{J}(x,t)}{\partial x}.\label{eq:psi_dynamics3}
\end{align}

Under certain conditions the Hamiltonian (\ref{eq:channel_Hamiltonian})
and the derivation of the dynamics of $\psi_{J}(x,t)$ that follow
from it can be simplified. We clearly have 
\begin{align}
 & \left[\psi_{J}(x,t),\psi_{J'}(x',t)\right]=0,
\end{align}
and 
\begin{align}\label{eq:channel_comm_1}
 & \left[\psi_{J}(x,t),\psi_{J'}^{\dagger}(x',t)\right]=0,
\end{align}
for $J\neq J'$, since different frequency ranges are involved. As
well, 
\begin{align}
 & \left[\psi_{J}(x,t),\psi_{J}^{\dagger}(x',t)\right]=\int_{k\in\mathscr{R}(J)}\frac{\dd k}{2\pi}e^{i(k-k_{J})(x-x')},
\end{align}
where we have used the commutation relations~\eqref{eq:comm_channel_work}.
Now suppose, for example, $k\in\mathscr{R}(J)$ for $-\Delta\kappa/2<k-k_{J}<\Delta\kappa/2;$
then 
\begin{align}
 & \left[\psi_{J}(x,t),\psi_{J}^{\dagger}(x',t)\right]=\frac{1}{2\pi}\frac{\sin\left(\Delta\kappa\frac{x-x'}{2}\right)}{\frac{x-x'}{2}}.
\end{align}
We will usually be interested in incident fields and generated fields
near the frequencies of ring resonances; if field excitations associated
with the channel field $\psi_{J}(x,t)$ are centered at $k=k_{J}$
and extend over a range $\Delta k\ll\Delta\kappa$, then for calculations
involving those fields and other independent fields we can take
\begin{align}
 & \left[\psi_{J}(x,t),\psi_{J}^{\dagger}(x',t)\right]\approx\delta(x-x'),\label{eq:approx_delta_comm}
\end{align}
and if we neglect group velocity dispersion for each channel field,
an expansion of the Hamiltonian $H_{J}^{\text{L}}$ (\ref{eq:channel_Hamiltonian})
along the lines used in (\ref{eq:psi_dynamics1}) leads to 
\begin{align}\label{eq:channel_linear_Hamiltonian_J}
 H_{J}^{\text{L}}&=\hbar\omega_{J}\int\psi_{J}^{\dagger}(x,t)\psi_{J}(x,t)\dd x\\
 & \quad -\frac{1}{2}i\hbar v_{J}\int\left(\psi_{J}^{\dagger}(x,t)\frac{\partial\psi_{J}(x,t)}{\partial x}-\frac{\partial\psi_{J}^{\dagger}(x,t)}{\partial x}\psi_{J}(x,t)\right)\dd x,\nonumber 
\end{align}
with the neglect of group velocity dispersion terms. With the full
Hamiltonian (\ref{eq:Hamiltonian_channel_decomposition}) and the
commutation relations 
\begin{align}
 & \left[\psi_{J}(x,t),\psi_{J'}^{\dagger}(x',t)\right]=\delta_{JJ'}\delta(x-x'),\label{eq:channel_comm}\\
 & \left[\psi_{J}(x,t),\psi_{J'}(x',t)\right]=0,\nonumber 
\end{align}
we indeed find the equations (\ref{eq:psi_dynamics3}).

To lowest order in $\Delta k/\Delta\kappa$ we can write the electromagnetic
field operators (\ref{eq:field_expansions_channel_work}) in terms
of the channel field operators by neglecting the variation of $\omega_{Jk}$,
and $\bm{b}_{Jk}(y,z)$ and $\bm{d}_{Jk}(y,z)$, over
the $\Delta k$ characterizing the excitation bandwidth. Putting $\bm{b}_{J}(y,z)\equiv\bm{b}_{Jk_{J}}(y,z)$
and $\bm{d}_{J}(y,z)\equiv\bm{d}_{Jk_{J}}(y,z)$,
we have 
\begin{align}
 & \bm{D}(\bm{r},t)\approx\sum_{J}\sqrt{\frac{\hbar\omega_{J}}{2}}\bm{d}_{J}(y,z)\psi_{J}(x,t)e^{ik_{J}x}+\hc,\label{eq:field_expansions_channel_work-1}\\
 & \bm{B}(\bm{r},t)\approx\sum_{J}\sqrt{\frac{\hbar\omega_{J}}{2}}\bm{b}_{J}(y,z)\psi_{J}(x,t)e^{ik_{J}x}+\hc\nonumber 
\end{align}
At the same level of approximation we have 
\begin{align}
 & \bm{E}(\bm{r},t)\approx\sum_{J}\sqrt{\frac{\hbar\omega_{J}}{2}}\bm{e}_{J}(y,z)\psi_{J}(x,t)e^{ik_{J}x}+\hc,\label{eq:field_expansions_channel_work-1-1}\\
 & \bm{H}(\bm{r},t)\approx\sum_{J}\sqrt{\frac{\hbar\omega_{J}}{2}}\bm{h}_{J}(y,z)\psi_{J}(x,t)e^{ik_{J}x}+\hc,\nonumber 
\end{align}
where 
\begin{align}\label{eq:eh_from_db}
 & \bm{e}_{J}(y,z)=\frac{\bm{d}_{J}(y,z)}{\epsilon_{0}\varepsilon_{1}(y,z;\omega_{J})},\\
 & \bm{h}_{J}(y,z)=\frac{\bm{b}_{J}(y,z)}{\mu_{0}},\nonumber 
\end{align}
[cf.~\eqref{eq:little_e_and_h}]. The total power flow in the waveguide
is 
\begin{align}
 & P(x,t)=\int \dd y\dd z\bm{S}(\bm{r},t)\cdot\bm{\hat{x}},
\end{align}
with $\bm{S}(\bm{r},t)=\bm{E}(\bm{r},t)\times\bm{H}(\bm{r},t)$.
Even a classical field oscillating at a single frequency will have
terms in the Poynting vector that vary as twice that frequency; as
usual we are interested in the slowly-varying part of the total power
flow, and so we omit terms in $\bm{S}(\bm{r},t)$
that will arise from the terms $\psi_{J}(x,t)\psi_{J'}(x,t)$ and
$\psi_{J}^{\dagger}(x,t)\psi_{J'}^{\dagger}(x,t)$. We use the commutation
relations (\ref{eq:channel_comm}) to write $\psi_{J}(x,t)\psi_{J'}^{\dagger}(x,t)$
as $\psi_{J}^{\dagger}(x,t)\psi_{J'}(x,t)$ plus a formally divergent
[see the discussion before~\eqref{eq:approx_delta_comm}] term proportional
to $\delta_{JJ'}$ that will, however, give a vanishing contribution
when channel modes going in the $+x$ and $-x$ directions are considered.
For the terms remaining proportional involving $\psi_{J}^{\dagger}(x,t)\psi_{J'}(x,t)$,
we neglect the contributions for $J\neq J'$ under the assumption
that any distinction frequencies $\omega_{J}$ and $\omega_{J'}$
are far enough apart that interference terms between them will be
rapidly varying. The final result is then~\cite{sipe2016hamiltonian}
\begin{align}
 & P(x,t)\rightarrow\sum_{J}\hbar\omega_{J}v_{J}\psi_{J}^{\dagger}(x,t)\psi_{J}(x,t),\label{eq:Pav}
\end{align}
where $v_{J}$ is that of (\ref{eq:psi_dynamics3}), and includes
both material and modal dispersion; corrections to (\ref{eq:Pav})
would arise if we considered variations in $\psi_{J}(x,t)$ more rapid
than could be well described by (\ref{eq:psi_dynamics3}).

\begin{figure}
    \centering
    \includegraphics[width=1.0\textwidth]{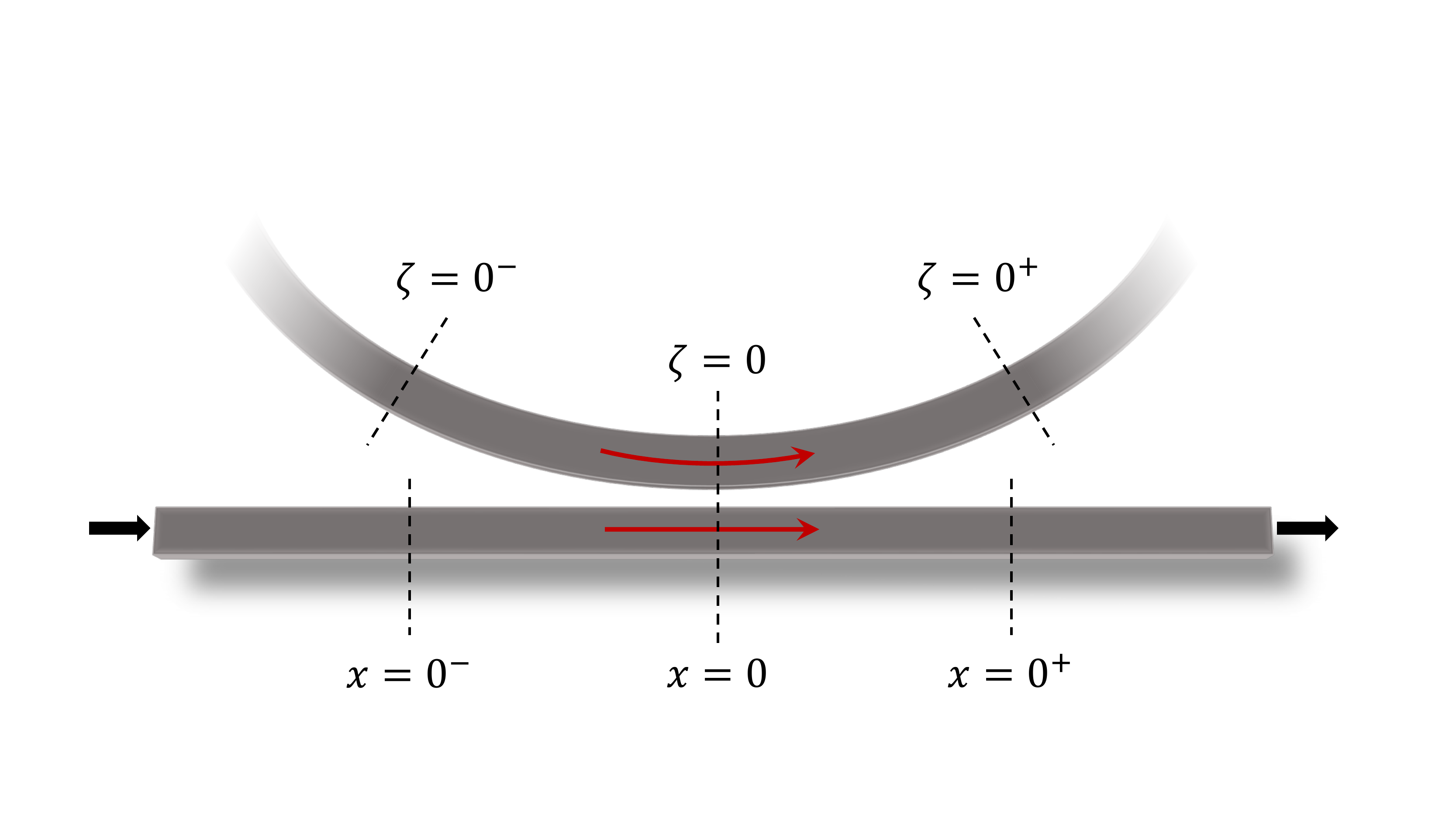}
    \caption{Zoom in on the point coupler of Fig.~\ref{fig:coupled_ring} between a channel and a ring, with particular emphasis on the conventional coordinate systems used in the two waveguide.}
    \label{fig:coupling}
\end{figure}

We now consider the channel and ring close enough that coupling can
occur. The coupling arises because the evanescent fields associated
with the ring and the channel overlap, and light can
move from the channel to the ring and vice-versa. The 
standard phenomenological model for this coupling~\cite{john2003optical}
begins by treating 
the field in the ring in the same manner as one treats the field in the channel; in our notation this involves introducing a field $\psi_r(\zeta,t)$, where $\zeta=0$ at the nominal coupling point, to describe the propagation in the ring. If we let $x=0$ denote the nominal coupling point for the channel field $\psi_J(x,t)$ associated with the ring resonance of interest (see Fig.~\ref{fig:k_range}), we then take the Fourier components of $\psi_J(x,t)$ and $\psi_r(\zeta,t)$,  \begin{align}
 & \psi_{J}(x,\omega)=\int\frac{\dd\omega}{\sqrt{2\pi}}\psi_{J}(x,t)e^{i\omega t},
 & \psi_{r}(\zeta,\omega)=\int\frac{\dd\omega}{\sqrt{2\pi}}\psi_{r}(\zeta,t)e^{i\omega t},
 \label{eq:single_omega}
\end{align}
and introduce self- and cross-coupling coefficients, $\sigma$ and $\kappa$ respectively, that relate the fields near the nominal coupling point,
\begin{align}
\begin{pmatrix}
\psi_{J}(0^{+},\omega) \\
\psi_{r}(0^{+},\omega)
\end{pmatrix} =  
\left(
\begin{array}{cc}
\ \sigma & i\kappa \\
i\kappa & \sigma \\
\end{array}
\right) \begin{pmatrix}
\psi_{J}(0^{-},\omega) \\ \psi_{r}(0^{-},\omega)
\end{pmatrix},
 \label{eq:coupling_matrix}
\end{align}
(see Fig.~\ref{fig:coupling}).

Here we assume that the waveguide properties of the ring and channel are the same, although the more general case can easily be considered.  Conventionally both $\sigma$ and $\kappa$ are taken to be real and positive; then energy conservation yields 
\begin{align}
 & \sigma^{2}+\kappa^{2}=1.
 \label{eq:energy_cons}
\end{align}
Generalizations to this are also straightforward, including the introduction of different $\sigma$ and $\kappa$ for frequencies close to different ring resonance frequencies. Now because light propagates freely from $\zeta=0^{+}$ to $\zeta=0^{-}$ within the ring, for fields of the form (\ref{eq:single_omega}) we have 
\begin{align}
    &\psi_{r}(0^{-},\omega)=\psi_{r}(0^{+},\omega) e^{ik\mathcal{L}},
\end{align}
where if the frequency $\omega$ of interest is close to the frequency $\omega_{J}$ of the $J^{th}$ resonance of the ring we can write 
\begin{align}
  & k=\frac{2\pi n_{J}}{\mathcal{L}}+(\omega-\omega_{J}) \left (\frac{dk}{d\omega} \right)_{\omega_{J}}+...
  \\ & = \frac{2\pi n_{J}}{\mathcal{L}}+\frac{\omega-\omega_{J}}{v_{J}}+... \label{eq:free_prop}
\end{align}
Stopping the expansion at this point we have $e^{ik\mathcal{L}}=e^{i\eta}$, where 
\begin{align}
    & \eta=\left(\frac{\omega-\omega_J}{v_J} \right) \mathcal{L}.
\end{align}
This can also be derived directly from the version of 
(\ref{eq:psi_dynamics3}) that would hold for $\psi_{r}(\zeta,t)$ in the ring. With this result used in (\ref{eq:free_prop}), when that equation is inserted in (\ref{eq:coupling_matrix}) we can solve for $\psi_{J}(0^{+},\omega)$ in terms of $\psi_{J}(0^{-},\omega)$,
\begin{align}
    &\psi_{J}(0^{+},\omega)=\left( \frac{\sigma-e^{i\eta}}{1-\sigma e^{i\eta}} \right) \psi_{J}(0^{-},\omega),
    \label{eq:exact_point}
\end{align}
where we have used (\ref{eq:energy_cons}). Recalling that we have taken $\sigma$ to be real, from (\ref{eq:exact_point}) we find that $|\psi_{J}(0^{+},\omega)|^2=|\psi_{J}(0^{-},\omega)|^2$. That is, the coupled channel-ring structure functions as an ``all pass filter," as would be expected.

We are typically interested in frequencies $\omega$ such that $|\omega-\omega_{J}|$ is much less than the frequency difference between neighboring resonances; then we have $|\eta| \ll 1$, and we can approximate $e^{i\eta} \approx 1+i\eta$. We are also typically interested in small ring-channel coupling, such that $\kappa \ll \sigma$; then since $|\eta|\ll1 $ we can take $\eta\sigma\approx\eta$ , and using these approximations in (\ref{eq:exact_point}) we have
\begin{align}
& \psi_{J}(0^{+},\omega) \approx \frac{\omega-\omega_{J}-i \left( \frac{(1-\sigma) v_{J}}{ \mathcal{L}} \right)} {\omega-\omega_{J}+i \left( \frac{(1-\sigma) v_{J}}{ \mathcal{L}} \right)} \psi_{J}(0^{-},\omega).
\label{eq:approx_point_result}
\end{align}
Note that the use of these two approximations respects the all-pass nature of the coupled channel-ring structure. 

While a whole range of linear and nonlinear optical phenomena in channel-ring structures can be considered at this level~\cite{john2003optical}, for quantum nonlinear optics we want to construct a quantum treatment of the channel-ring coupling that can be built into a Hamiltonian framework.  Beginning with the linear regime, we can do this by formally introducing a point coupling model ~\cite{vernon2015spontaneous}, with a coupling
constant $\gamma_{J}$. The full linear Hamiltonian is taken to be
\begin{align}
 \Hl &=\sum_{J}\left(\hbar\omega_{J}\int\psi_{J}^{\dagger}(x)\psi_{J}(x)\dd x-\frac{1}{2}i\hbar v_{J}\int\left(\psi_{J}^{\dagger}(x)\frac{\partial\psi_{J}(x)}{\partial x}-\frac{\partial\psi_{J}^{\dagger}(x)}{\partial x}\psi_{J}(x)\right)\dd x\right)\label{eq:ring+channel_H}\\
 & \quad +\sum_{J}\hbar\omega_{J}c_{J}^{\dagger}c_{J}+\sum_{J}\left(\hbar\gamma_{J}^*c_{J}^{\dagger}\psi_{J}\left(0\right)+\hc\right).\nonumber 
\end{align}
The first term on the right-hand-side is the Hamiltonian (\ref{eq:Hamiltonian_channel_decomposition})
for the channel fields, the second term is the Hamiltonian (\ref{Hamiltonian_ring})
for the modes in the ring, and the third terms represent a coupling
that can either remove a photon from the channel and create one in
the ring, or vice-versa. Here, as in the above, we consider only modes with $v_{J}>0.$
If we take $\zeta=0$ to identify the point on the ring closest to
the channel, then since the coupling involves the fields in the ring
and channel near that point we can expect the $\gamma_{J}$ to be
only weakly dependent on $J$; were $\zeta=\pi$ taken to identify
that point we would expect $\gamma_{J}\propto(-1)^{J}$; in SCISSOR
structures (``side-coupled integrated spaced sequences of optical resonators"), for example, both types of behavior arise~\cite{chak2006coupled}.

The Heisenberg equations of motion following from the Hamiltonian
(\ref{eq:ring+channel_H}) then are 
\begin{align}
 & \left(\frac{\partial}{\partial t}+v_{J}\frac{\partial}{\partial x}+i\omega_{J}\right)\psi_{J}(x,t)=-i\gamma_{J}c_{J}(t)\delta(x),\label{eq:coupling_dynamics}\\
 & \left(\frac{d}{dt}+i\omega_{J}\right)c_{J}(t)=-i\gamma_{J}^{*}\psi_{J}(0,t).\nonumber 
\end{align}
The formal solution of the first of these is
\begin{align}
 & \psi_{J}(x,t)=\psi_{J}(x-v_Jt,0)e^{-i\omega_{J}t}-\frac{i\gamma_{J}}{v_{J}}c_{J}(t-\tfrac{x}{v_{J}})\left[\theta(x)-\theta(x-v_{J}t)\right]e^{-i\omega_{J}x/v_{J}}.\label{eq:propagator}
\end{align}
We will then substitute this expression, evaluated at $x=0$, back into
the second of (\ref{eq:coupling_dynamics}). Note that the expression (\ref{eq:propagator}) respects the discontinuity of $\psi_{J}(x,t)$ across $x=0$ that was apparent in the standard phenomenological model discussed above. As was implicitly done there, we introduce one-sided limits
\begin{align}
 & \psi_{J}(0^{\pm},t)=\lim_{x\rightarrow0^{\pm}}\psi_{J}(x,t),\label{eq:PClimits}
\end{align}
and define $\psi_{J}(x,t)$ at $x=0$ via 
\begin{align}
 & \psi_{J}(0,t)\rightarrow\frac{1}{2}\left(\psi_{J}(0^{-},t)+\psi_{J}(0^{+},t)\right),\label{eq:midpoint}
\end{align}
~\cite{vernon2015spontaneous}. Using the limits (\ref{eq:PClimits}) in (\ref{eq:propagator}) we
find 
\begin{flalign}
 & \psi_{J}(0^{+},t)=\psi_{J}(0^{-},t)-\frac{i\gamma_{J}}{v_{J}}c_{J}(t),\label{eq:across_jump}
\end{flalign}
and then with $\psi_{J}(0,t)$ constructed from (\ref{eq:midpoint})
and used in the second of (\ref{eq:coupling_dynamics}) we find 
\begin{align}
 & \left(\ddt{}+\Gamma_{J}+i\omega_{J}\right)c_{J}(t)=-i\gamma_{J}^{*}\psi_{J}(0^{-},t),\label{eq:cdynamics}
\end{align}
where 
\begin{align}\label{eq:Gamma_J}
 & \Gamma_{J}=\frac{\left|\gamma_{J}\right|^{2}}{2v_{J}},
\end{align}
appears as an effective damping term in (\ref{eq:cdynamics}) characterizing
the rate at which energy in the ring can be lost to the channel. The
full solution of the equations (\ref{eq:coupling_dynamics}) then
follows from solving for $c_{J}(t)$ in terms of $\psi_{J}(0^{-},t)$
from (\ref{eq:cdynamics}), and then the determination of $\psi_{J}(0^{+},t)$
from $\psi_{J}(0^{-},t)$ from (\ref{eq:across_jump}). 

If we then introduce
Fourier components of $\psi_{J}(0^{\pm},t)$ as in the first of (\ref{eq:single_omega}), we find~\cite{vernon2015spontaneous} 
\begin{align}
 & \psi_{J}(0^{+},\omega)=\left[\frac{\omega_{J}-\omega+i\Gamma_{J}}{\omega_{J}-\omega-i\Gamma_{J}}\right]\psi_{J}(0^{-},\omega),\label{eq:IO_no_loss}
\end{align}
This can be compared with (\ref{eq:approx_point_result}), and the coupling constant $\gamma_{J}$ of the Hamiltonian model can be related to the self-coupling constant $\sigma$ of the standard phenomenological model,
\begin{align}
    &\frac{\left|\gamma_{J}\right|^{2}}{2v_{J}}=\frac{(1-\sigma)v_J}{\mathcal{L}}. 
    \label{eq:gamsig}
\end{align}

\begin{figure}
    \centering
    \includegraphics[width=1.0\textwidth]{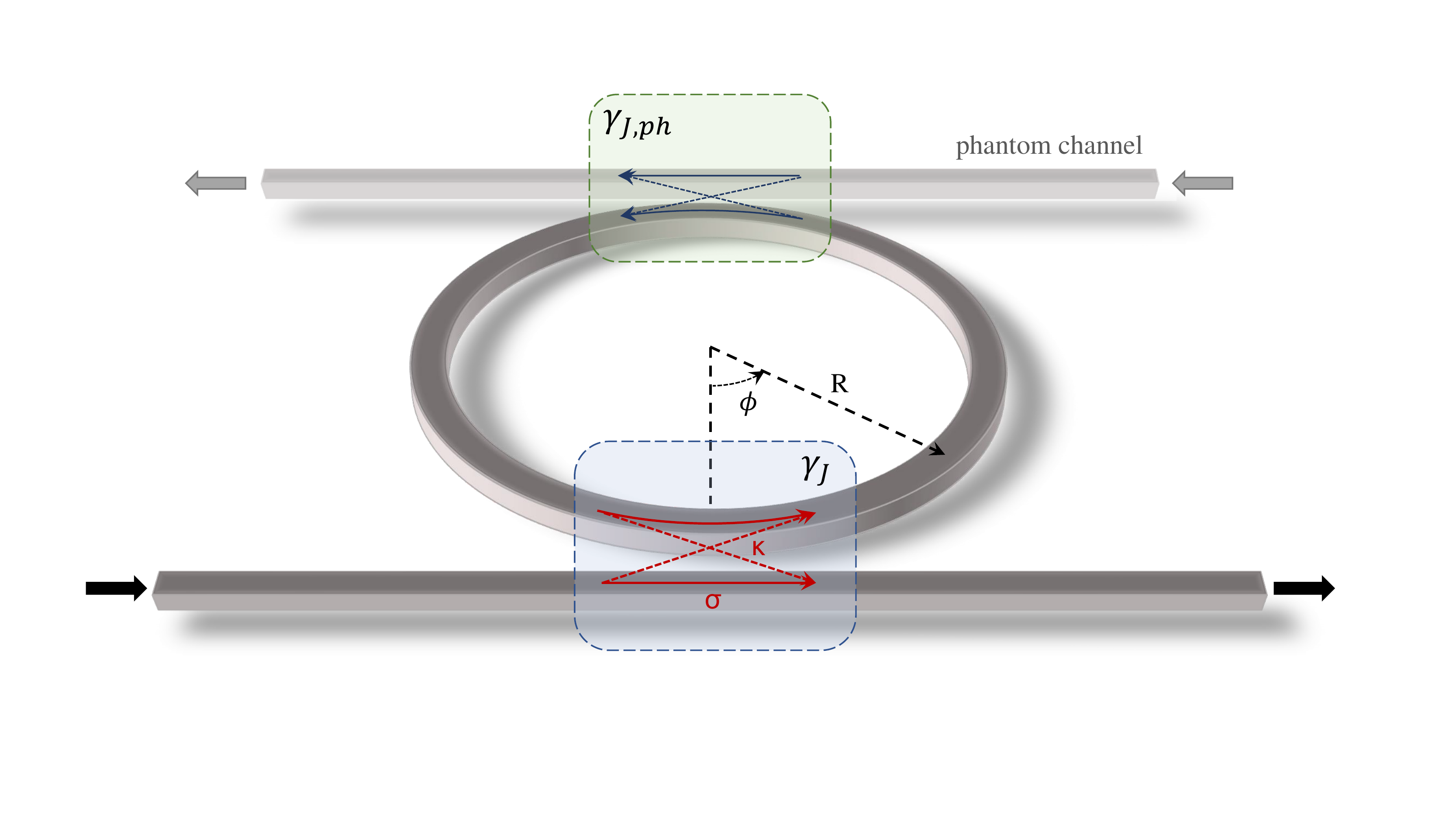}
    \caption{Schematic representation of a point-coupled ring resonator with and additional fictitious coupler and waveguide, referred to as ``phantom channel''.}
    \label{fig:phantom_channel}
\end{figure}

Ubiquitous scattering losses lead to the break-down of the all-pass
nature of this structure. We can model these by introducing a ``phantom channel''~\cite{vernon2015spontaneous}
(see Fig.~\ref{fig:phantom_channel}) that models the loss of light into
the environment, and more generally the possible scattering of light
into the system from the environment. 
Note that we indicate the positive $x$ direction in the phantom channel in the opposite direction of that in the real channel. Including the phantom channel and the coupling
of light into and out of it by taking 
\begin{align}
 & \Hl\rightarrow \Hl+\Hp,
\end{align}
where 
\begin{align}\label{eq:H_phantom}
 \Hp &=\sum_{J}\left(\hbar\omega_{J}\int\psi_{J\text{ph}}^{\dagger}(x)\psi_{J\text{ph}}(x)\dd x - \frac{1}{2}i\hbar v_{J\text{ph}}\int\left(\psi_{J\text{ph}}^{\dagger}(x)\frac{\partial\psi_{J\text{ph}}(x)}{\partial x}-\frac{\partial\psi_{J\text{ph}}^{\dagger}(x)}{\partial x}\psi_{J\text{ph}}(x)\right)\dd x\right)\\
 & \quad +\sum_{J}(\hbar \gamma_{J\text{ph}}c_{J}^{\dagger}\psi_{J\text{ph}}(0)+\hc),\nonumber
\end{align}
Here the $\psi_{J\text{ph}}(x,t)$ are the fields in the phantom channel,
taken to satisfy 
\begin{align}
 & \left[\psi_{J\text{ph}}(x,t),\psi_{J'\text{ph}}^{\dagger}(x',t)\right]=\delta_{JJ'}\delta(x-x'),\\
 & \left[\psi_{J\text{ph}}(x,t),\psi_{J'\text{ph}}(x',t)\right]=0,\nonumber
\end{align}
with an obvious generalization of the real channel notation for the phantom
channel parameters. In place of (\ref{eq:coupling_dynamics}) we have
the more general equations 
\begin{align}
 & \left(\frac{\partial}{\partial t}+v_{J}\frac{\partial}{\partial x}+i\omega_{J}\right)\psi_{J}(x,t)=-i\gamma_{J}c_{J}(t)\delta(x),\\\label{eq:gen_lin_dyn}
 & \left(\frac{\partial}{\partial t}+v_{J\text{ph}}\frac{\partial}{\partial x}+i\omega_{J}\right)\psi_{J\text{ph}}(x,t)=-i\gamma_{J\text{ph}}c_{J}(t)\delta(x),\nonumber\\
 & \left(\ddt{}+i\omega_{J}\right)c_{J}(t)=-i\gamma_{J}^{*}\psi_{J}(0,t)-i\gamma_{J\text{ph}}^{*}\psi_{J\text{ph}}(0,t).\nonumber
\end{align}
Following the same procedure as used above but taking both the real
and phantom channel into account, and introducing
\begin{align}
 & \psi_{J\text{ph}}(0^{\pm},\omega)=\int\frac{\dd t}{\sqrt{2\pi}}\psi_{J\text{ph}}(0^{\pm},t)e^{i\omega t},\\
 & c_{J}(\omega)=\int\frac{\dd t}{\sqrt{2\pi}}c_{J}(t)e^{i\omega t},\nonumber
\end{align}
we find 
\begin{align}
    & c_{J}(\omega)= - \frac{\gamma_{J}^{*}\psi_{J}(0^{-},\omega)+\gamma_{J\text{ph}}^{*}\psi_{J\text{ph}}(0^{-},\omega)}{\omega_{J}-\omega-i\overline{\Gamma}_{J}},\label{eq:cJomeg}
\end{align} 
where 
\begin{align}\label{eq:Gammabar_J}
 & \overline{\Gamma}_{J}=\Gamma_{J}+\Gamma_{J\text{ph}},
\end{align}
and
\begin{align}\label{eq:Gamma_Jph}
 & \Gamma_{J\text{ph}}=\frac{\left|\gamma_{J\text{ph}}\right|^{2}}{2v_{J\text{ph}}},
\end{align}
describes the scattering losses. For the usual situation
where $\left\langle \psi_{J\text{ph}}(0^{-},\omega)\right\rangle =0$ we have \begin{align}
    &|\left\langle c_{J}(\omega)\right\rangle|^{2}=\frac{|\gamma_{J}^{*}\left\langle\psi_{J}(0^{-},\omega)\right\rangle|^{2}}{(\omega_{J}-\omega)^{2}+\overline{\Gamma}_{J}^{2}}.
\end{align}
From this we can identify the full-width at half-maximum as  $\Delta\omega_{J}=2\overline{\Gamma}_{J}$, and so the "loaded" quality factor $Q$ for the resonance is given by \begin{align}
    & Q_{J}=\frac{\omega_{J}}{\Delta\omega_{J}}=\frac{\omega_{J}}{2\overline{\Gamma}_{J}}.
\end{align}
Defining an intrinsic $Q_{J\text{int}}$ due to scattering losses and an extrinsic $Q_{J\text{ext}}$ due to the coupling of the ring to the channel, we have
\begin{align}
    & \frac{1}{Q_{J}}=\frac{1}{Q_{J\text{int}}}+\frac{1}{Q_{J\text{ext}}},
\end{align}
where 
\begin{align}
    & Q_{J\text{ext}}=\frac{\omega_{J}}{2{\Gamma}_{J}}
\end{align}
and
\begin{align}
    & Q_{J\text{int}}=\frac{\omega_{J}}{2{\Gamma}_{J\text{ph}}}.
\end{align}

The solution of our general equations (\ref{eq:gen_lin_dyn}), again following the same procedure as used above, gives
\begin{align}
 & \psi_{J}(0^{+},\omega)=\left[\frac{\omega_{J}-\omega+i(\Gamma_{J}-\Gamma_{J\text{ph}})}{\omega_{J}-\omega-i\overline{\Gamma}_{J}}\right]\psi_{J}(0^{-},\omega)+\left[\frac{2i\Gamma_{J\text{ph}}}{\omega_{J}-\omega-i\overline{\Gamma}_{J}}\right]\psi_{J\text{ph}}(0^{-},\omega)\label{eq:IO_loss}
\end{align}
where the term proportional to $\psi_{J\text{ph}}(0^{-},\omega)$ in (\ref{eq:IO_loss})
describes the input noise from the loss channel; and again for the usual situation
where $\left\langle \psi_{J\text{ph}}(0^{-},\omega)\right\rangle =0$ we have
\begin{align}
 & \left\langle \psi_{J}(0^{+},\omega)\right\rangle =\left[\frac{\omega_{J}-\omega+i(\Gamma_{J}-\Gamma_{J\text{ph}})}{\omega_{J}-\omega-i\overline{\Gamma}_{J}}\right]\left\langle \psi_{J}(0^{-},\omega)\right\rangle ,
\end{align}
and so-called ``critical coupling'' is achieved if $\Gamma_{J}=\Gamma_{J\text{ph}}$,
for then $\left\langle \psi_{J}(0^{+},\omega_{J})\right\rangle =0.$

To avoid dealing with the fields $\psi_J(x,t)$ that have a discontinuity at $x=0$ it is convenient to introduce fields $\psi_{J<}(x,t)$ and $\psi_{J>}(x,t)$ associated with each $\psi_J(x,t)$. The first of these is equal to $\psi_J(x,t)$ for $x<0$ and, for $x\ge0$, follows from the evolution equation (\ref{eq:psi_dynamics3}) that would hold were there no coupling to the ring. Similarly, $\psi_{J>}(x,t)$ is given by $\psi_J(x,t)$ for $x>0$ and, for $x\le0$, follows from that evolution equation. Thus the fields  $\psi_{J<}(x,t)$ and $\psi_{J>}(x,t)$ are continuous even at $x=0$, and we have $\psi_J(0^-,t)=\psi_{J<}(0,t)$ and $\psi_J(0^+,t)=\psi_{J>}(0,t)$.

Finally, since the free spectral range (FSR) in the neighbourhood of the $J^{\text{th}}$ resonance is given by
\begin{align}
    & \delta\omega_{J}=v_{J}\delta\kappa_{J}=v_J\frac{2\pi}{\mathcal{L}},
\end{align}
the finesse in the neighborhood of the $J^\text{{th}}$ resonance is 
\begin{align}\label{eq:Finesse}
    & \mathcal{F}=\frac{\delta\omega_{J}}{\Delta\omega_{J}}=\frac{\pi v_{J}}{\overline{\Gamma}_{J} \mathcal{L}}.
\end{align}

From  this expression one can see that the finesse is a measurement of the effect of the resonant spatial and temporal confinement of light inside the ring. Indeed, it is inversely proportional to the resonator length and is directly proportional to the photon dwelling time in the resonator.  This quantity can be very useful, because it is  experimentally accessible from the structure transmission spectrum, which gives information about the FSR and the quality factor, and yet is also simply related to the on-resonance field  enhancement, $|\text{FE}(\omega_J)|$. This can be defined as the modulus of the ratio between the value of the displacement field measured in the center of the waveguide in any point of the ring and the value measured in the center of the channel waveguide before the point coupling. One can show that, at critical coupling (i.e., $\Gamma_J=\Gamma_{J\text{ph}}$),
\begin{equation}
  \left\vert\text{FE}_\text{crit}\left(\omega_J\right)\right\vert^{2}\simeq\frac{\mathcal{F}}{\pi}, 
\end{equation}
which means that at critical coupling the intensity inside the ring is increased by a factor approximately equal to a third of the finesse, with respect to that in the input channel.
As similar expression can be derived in the case of over coupling (i.e. $\Gamma_J\gg\Gamma_{J\text{ph}}$)
\begin{equation}
  \left\vert\text{FE}_\text{over}\left(\omega_J\right)\right\vert^{2}\simeq\frac{2}{\pi}\mathcal{F}. 
\end{equation}
\subsection{Including nonlinearities}\label{sec:nonlinear_intro}

We now turn to including the effects of nonlinearities. Considering
a very general dielectric structure, and beginning in the classical
regime, the usual description of the nonlinear response is in terms
of nonlinear response coefficients $\chi_{2}^{ijk}(\bm{r})$
and $\chi_{3}^{ijkl}(\bm{r}),$ etc., with the components
of the nonlinear polarization $\bm{P}_{\text{NL}}(\bm{r},t)$
given by 
\begin{align}
 & P_{\text{NL}}^{i}(\bm{r},t)=\epsilon_{0}\chi_{2}^{ijk}(\bm{r})E^{j}(\bm{r},t)E^{k}(\bm{r},t)+\epsilon_{0}\chi_{3}^{ijkl}(\bm{r})E^{j}(\bm{r},t)E^{k}(\bm{r},t)E^{l}(\bm{r},t)+\cdots.
\end{align}
The displacement field $\bm{D}(\bm{r},t)$ is then
given by $\bm{D}(\bm{r},t)=\epsilon_{0}\bm{E}(\bm{r},t)+\bm{P}(\bm{r},t),$
where $\bm{P}(\bm{r},t)$ is the full polarization,
so 
\begin{align}
 & D^{i}(\bm{r},t)=\epsilon_{0}\varepsilon_{1}(\bm{r})E^{i}(\bm{r},t)+\epsilon_{0}\chi_{2}^{ijk}(\bm{r})E^{j}(\bm{r},t)E^{k}(\bm{r},t)+\epsilon_{0}\chi_{3}^{ijkl}(\bm{r})E^{j}(\bm{r},t)E^{k}(\bm{r},t)E^{l}(\bm{r},t)+\cdots.\label{eq:DEwork1}
\end{align}
Here we have neglected material dispersion, and will formally do so
in the nonlinear components of the Hamiltonian in the rest of this
tutorial. It many cases the effects of material dispersion on the
nonlinear components can be reasonably supposed to be negligible.
Of course, when particular nonlinear terms are calculated the appropriate
$\chi_{n}$ for the frequencies of the fields involved can be employed;
we implicitly assume that below. And as well, in equations below that
involve $\varepsilon_{1}(\bm{r})$ we use the $\varepsilon_{1}(\bm{r};\omega)$
for the frequencies involved. But the full inclusion of dispersive
effects in the $\chi_{n}$ themselves, for $n>1$, is an outstanding
problem even at frequencies where absorption is not present.

Equation \eqref{eq:DEwork1} is a usual starting point for investigating
the propagation of light in nonlinear media. However, since our fundamental
fields are taken to be $\bm{D}(\bm{r},t)$ and $\bm{B}(\bm{r},t)$,
we want to rewrite this as an expression for $\bm{E}(\bm{r},t)$
in terms of $\bm{D}(\bm{r},t).$ To do that, write
(\ref{eq:DEwork1}) as 
\begin{align}
 & E^{i}(\bm{r},t)=\frac{D^{i}(\bm{r},t)}{\epsilon_{0}\varepsilon_{1}(\bm{r})}-\frac{\chi_{2}^{ijk}(\bm{r})}{\varepsilon_{1}(\bm{r})}E^{j}(\bm{r},t)E^{k}(\bm{r},t)-\frac{\chi_{3}^{ijkl}(\bm{r})}{\varepsilon_{1}(\bm{r})}E^{j}(\bm{r},t)E^{k}(\bm{r},t)E^{l}(\bm{r},t)+\cdots.
\end{align}
In the absence of any nonlineaerity $E^{i}(\bm{r},t)=D^{i}(\bm{r},t)/[\epsilon_{0}\varepsilon_{1}(\bm{r})]$,
of course, and we can iterate to find 
\begin{align}\label{eq:DEwork2}
 & E^{i}(\bm{r},t)=\frac{D^{j}(\bm{r},t)}{\epsilon_{0}\varepsilon_{1}(\bm{r})}-\frac{\Gamma_{2}^{ijk}(\bm{r})}{\epsilon_{0}}D^{j}(\bm{r},t)D^{k}(\bm{r},t)-\frac{\Gamma_{3}^{ijkl}(\bm{r})}{\epsilon_{0}}D^{j}(\bm{r},t)D^{k}(\bm{r},t)D^{l}(\bm{r},t)+\cdots,
\end{align}
where 
\begin{align}\label{eq:big_gammas}
 & \Gamma_{2}^{ijk}(\bm{r})=\frac{\chi_{2}^{ijk}(\bm{r})}{\epsilon_{0}\varepsilon_{1}^{3}(\bm{r})},\\
 & \Gamma_{3}^{ijkl}(\bm{r})=\frac{\chi_{3}^{ijkl}(\bm{r})}{\epsilon_{0}^{2}\varepsilon_{1}^{4}(\bm{r})}-2\frac{\chi_{2}^{ijm}(\bm{r})\chi_{2}^{mkl}(\bm{r})}{\epsilon_{0}^{2}\varepsilon_{1}^{5}(\bm{r})},\nonumber 
\end{align}
and we have used the fact that the $\chi_{n}(\bm{r})$ are
invariant under permutation of their Cartesian components. Note that
the $\Gamma_{n}$ , which are dimensionless, are then also invariant
under permutation of their Cartesian components; they are typically
positive. Using (\ref{eq:DEwork2}) in the expression (\ref{eq:dh})
we find 
\begin{align}
 H & =\int\left(\frac{\bm{D}(\bm{r},t)\cdot\bm{D}(\bm{r},t)}{2\epsilon_{0}\varepsilon_{1}(\bm{r})}+\frac{\bm{B}(\bm{r},t)\cdot\bm{B}(\bm{r},t)}{2\mu_{0}}\right)\dd\bm{r}\\
 & \quad -\frac{1}{3\epsilon_{0}}\int\Gamma_{2}^{ijk}(\bm{r})D^{i}(\bm{r},t)D^{j}(\bm{r},t)D^{k}(\bm{r},t)\dd\bm{r}\nonumber\\
 & \quad -\frac{1}{4\epsilon_{0}}\int\Gamma_{3}^{ijkl}(\bm{r})D^{i}(\bm{r},t)D^{j}(\bm{r},t)D^{k}(\bm{r},t)D^{l}(\bm{r},t)\dd\bm{r}+\cdots,
\end{align}
for the total energy. Adopting this as the Hamiltonian, and using
either the Poisson bracket expressions (\ref{eq:fieldPB}) (in the
classical regime), or the corresponding commutation relations (\ref{eq:fieldCR})
(in the quantum regime), we find that the Hamiltonian leads to the
dynamics 
\begin{align}
 & \frac{\partial\bm{D}(\bm{r},t)}{\partial t}=\frac{1}{\mu_{0}}\bm{\nabla\times}\bm{B}(\bm{r},t),\\
 & \frac{\partial\bm{B}(\bm{r},t)}{\partial t}=-\bm{\nabla\times}\bm{E}(\bm{r},t),\nonumber
\end{align}
where the components of $\bm{E}(\bm{r},t)$ are given
by (\ref{eq:DEwork2}). For the very general mode expansion (\ref{eq:field_expansions})
we then find the Schrödinger Hamiltonian 
\begin{align}
 & H=\sum_{\alpha}\hbar\omega_{\alpha}c_{\alpha}^{\dagger}c_{\alpha}+\Hnl,\label{eq:Hfull}
\end{align}
where 
\begin{align}\label{eq:HNL_general}
 \Hnl &=-\frac{1}{3\epsilon_{0}}\int\Gamma_{2}^{ijk}(\bm{r}):\left(D^{i}(\bm{r})D^{j}(\bm{r})D^{k}(\bm{r})\right):\dd\bm{r}\\
 & \quad -\frac{1}{4\epsilon_{0}}\int\Gamma_{3}^{ijkl}(\bm{r}):\left(D^{i}(\bm{r})D^{j}(\bm{r})D^{k}(\bm{r})D^{l}(\bm{r})\right):\dd\bm{r}+\cdots.\nonumber
\end{align}
In (\ref{eq:Hfull}) we have neglected the zero point energy in the
linear regime, and in (\ref{eq:HNL_general}) the notation ``$:(\quad):$''
indicates that the operator expressions inside should be normal ordered.
The corrections that result from this will lead to new zero point
energy terms, and as well frequency shifts and mixing in the linear
regime; these we neglect, considering the effects of the latter to
be included in the linear regime parameters.

Using an expansion of $\bm{D}(\bm{r})$ in terms of
the linear modes of the system the form of $\Hnl$ in terms of raising
and lowering operators can be explicitly constructed. We do this later
in this tutorial for spontaneous parametric down-conversion and spontaneous
four-wave mixing. In the rest of this section we treat the special
cases of self- and cross-phase modulation. For simplicity we assume
that $\Gamma_{3}^{ijkl}$ receives significant contributions only
from $\chi_{3}^{ijkl}$, and not from $\chi_{2}^{ijk}$ [see (\ref{eq:big_gammas})];
that usually holds, but if it doesn't the generalization of the formulas
given below is straightforward.

\subsubsection{Self- and cross-phase modulation in channels}\label{sec:SXPM_channels}

Considering an isolated channel, we still assume that we can take
our fields to be of the form (\ref{eq:field_expansions_channel_work-1}),
although the frequencies $\omega_{J}$ here are not linked to any
ring resonances, but merely identify the center frequencies of different
components of light that are well-separated from each other in frequency.
Assuming that we have just one such component; taking $J=P$, its self-interaction
is described by $\chi_{3}^{ijkl}(\bm{r})$, and using (\ref{eq:field_expansions_channel_work-1})
in (\ref{eq:HNL_general}) the term describing the resulting self-phase
modulation is
\begin{align}
 \Hcs &=-\frac{1}{4\epsilon_{0}}\left(\frac{\hbar\omega_{P}}{2}\right)^{2}\frac{4!}{2!2!}\left(\int \dd y\dd z\Gamma_{3}^{ijkl}(y,z)\left(d_{P}^{i}(y,z)d_{P}^{j}(y,z)\right)^{*}d_{P}^{k}(y,z)d_{P}^{l}(y,z)\right)\label{eq:HSPMwork}\\
 & \quad\times\int \dd x\psi_{P}^{\dagger}(x)\psi_{P}^{\dagger}(x)\psi_{P}(x)\psi_{P}(x),\nonumber 
\end{align}
where the combinatorial factor $4!/(2!2!)$ counts the number of ways,
after normal ordering, that the indicated combination of fields $\psi_{P}(x)$
and $\psi_{P}^{\dagger}(x)$ arise. We use (\ref{eq:big_gammas})
to write $\Gamma_{3}^{ijkl}(y,z)$ in terms of $\chi_{3}^{ijkl}(y,z)$,
and use (\ref{eq:eh_from_db}) to write the fields $\bm{d}_{P}(y,z)$
in terms of $\bm{e}_{P}(y,z)$. Then (\ref{eq:HSPMwork})
can be written as 
\begin{align}
 \Hcs &=-\gcs\frac{\hbar^{2}\omega_{P}}{2}v_{P}^{2}\int \dd x\psi_{P}^{\dagger}(x)\psi_{P}^{\dagger}(x)\psi_{P}(x)\psi_{P}(x),\label{eq:Hchannel_SPM}
\end{align}
where
\begin{align}
 & \gcs=\frac{3\omega_{P}\epsilon_{0}}{4v_{P}^{2}}\int \dd y\dd z\chi_{3}^{ijkl}(y,z)\left(e_{P}^{i}(y,z)e_{P}^{j}(y,z)\right)^{*}e_{P}^{k}(y,z)e_{P}^{l}(y,z).\label{eq:gamma_SPM_work1}
\end{align}
Taking as our Hamiltonian the sum of the linear Hamiltonian (\ref{eq:channel_linear_Hamiltonian_J})
and the nonlinear contribution (\ref{eq:Hchannel_SPM}) we find the
Heisenberg equations of motion yield
\begin{align}
 & \frac{\partial\psi_{P}(x,t)}{\partial t}=-i\omega_{P}\psi_{P}(x,t)-v_{P}\frac{\partial\psi_{P}(x,t)}{\partial x}+i\gcs\hbar\omega_{P}v_{P}^{2}\psi_{P}^{\dagger}(x,t)\psi_{P}(x,t)\psi_{P}(x,t).\label{eq:check_coeff_1}
\end{align}
In the classical limit, letting 
\begin{align}
 & \psi_{P}(x,t)\rightarrow\phi_{P}(x,t),\label{eq:check_coeff_2}\\
 & \psi_{P}^{\dagger}(x,t)\rightarrow\phi_{P}^{*}(x,t),\nonumber 
\end{align}
where $\phi_{P}(x,t)$ is a classical field, we find a solution of
the form $\phi_{P}(x,t)=\phi_{P}(x)\exp(-i\omega_{P}t)$, where 
\begin{align}\label{eq:check_coeff_3}
 \frac{\dd\phi_{P}(x)}{\dd x}&=i\gcs\hbar\omega_{P}v_{P}\left|\phi_{P}(x)\right|^{2}\phi_{P}(x)\\
 & =i\gcs P_{P}(x)\phi_{P}(x),\nonumber 
\end{align}
where $P_{P}(z)$ is the power in the channel [cf. (\ref{eq:Pav})].
Thus we can identify $\gcs$ as the self-phase modulation
coefficient for the channel~\cite{boyd2008nonlinear}.

To write $\gcs$ in a more transparent form, divide
the expression (\ref{eq:gamma_SPM_work1}) for it by the square
of 
\begin{align}
 & \epsilon_{0}c\int\frac{n(y,z;\omega_{P})}{v_{g}(y,z;\omega_{P})}\bm{e}_{P}^{*}(y,z)\cdot\bm{e}_{P}(y,z)\dd y\dd z=1,\label{eq:gamma_SPM_norm}
\end{align}
where the identify follows from the normalization condition (\ref{eq:norm_D_waveguide_disp})
and the relation (\ref{eq:eh_from_db}) between $\bm{d}_{P}(y,z)$
and $\bm{e}_{P}(y,z)$, as well as the expression (\ref{eq:vphase})
for the local phase velocity; we have also introduced the local index
of refraction $n(y,z;\omega)=\sqrt{\varepsilon_{1}(y,z;\omega)}$.
Then taking $\overline{n}_{P}$ to be a characteristic value of that index
in the region where $\bm{e}_{P}(y,z)$ is nonvanishing, and
$\overline{\chi}_{3}$ to be a characteristic value of the nonlinear
susceptibility tensor components in that region, we can combine~\eqref{eq:gamma_SPM_work1} and~\eqref{eq:gamma_SPM_norm}
to write 
\begin{align}
 & \gcs=\frac{3\omega_{P}\overline{\chi}_{3}}{4\epsilon_{0}\overline{n}_{P}^{2}c^{2}}\frac{1}{\Acs},\label{eq:gamma_SPM_chan_result}
\end{align}
where the effective area $\Acs$ is given by 
\begin{align}
 & \frac{1}{\Acs}=\frac{\int \dd y\dd z\left(\chi_{3}^{ijkl}(y,z)/\overline{\chi}_{3}\right)\left(e_{P}^{i}(y,z)e_{P}^{j}(y,z)\right)^{*}e_{P}^{k}(y,z)e_{P}^{l}(y,z)}{\left(\int\frac{n(y,z;\omega_{P})/\overline{n}_{P}}{v_{g}(y,z;\omega_{P})/v_{P}}\bm{e}_{P}^{*}(y,z)\cdot\bm{e}_{P}(y,z)\dd y\dd z\right)^{2}}.\label{eq:A_SPM_chan_result}
\end{align}
Note that if we define an effective nonlinear index $n_{2}$ in terms
of $\overline{\chi}_{3}$ and $\overline{n}_{P}$ in the usual way~\cite{boyd2008nonlinear}
\begin{align}
 & n_{2}=\frac{3\overline{\chi}_{3}}{4\epsilon_{0}c\overline{n}_{P}^{2}},
\end{align}
as is appropriate for a uniform medium, we have 
\begin{align}
 & \gcs=\frac{\omega_{P}}{c}n_{2}\frac{1}{\Acs},
\end{align}
the usual expression for $\gcs$ in terms of an effective
area~\cite{boyd2008nonlinear}. Further, if we consider an electric field with only
one component, put $\overline{\chi}_{3}$ equal to a typical value of the
relevant component of $\chi_{3}^{ijkl}(y,z)$ in the region where
$\bm{e}_{P}(y,z)$ is nonvanishing, and set $n(y,z;\omega_{P})/\overline{n}_{P}\approx v_{g}(y,z;\omega_{P})/v_{P}\approx1$
as well in that region, we have 
\begin{align}
 & \frac{1}{\Acs}\approx\frac{\int\left|e_{P}(y,z)\right|^{4}\dd y\dd z}{\left(\int\left|e_{P}(y,z)\right|^{2}\dd y\dd z\right)^{2}},
\end{align}
the usual expression for the effective area in that limit~\cite{boyd2008nonlinear}.

We now consider cross-phase modulation, a consequence of the nonlinear
interaction between a field $\psi_{S}(x,t)$ with frequency centered
at $\omega_{S}$ and a field $\psi_{P}(x,t)$ with frequency centered
at $\omega_{P}$.
The contribution to the nonlinear Hamiltonian responsible for cross-phase
modulation is 
\begin{align}
 \Hcx &=-\frac{1}{4\epsilon_{0}}\left(\frac{\hbar\omega_{P}}{2}\right)\left(\frac{\hbar\omega_{S}}{2}\right)4!\left(\int \dd y\dd z\Gamma_{3}^{ijkl}(y,z)\left(d_{P}^{i}(y,z)d_{S}^{j}(y,z)\right)^{*}d_{P}^{k}(y,z)d_{S}^{l}(y,z)\right)\label{eq:H_XPM_work}\\
 & \quad\times\int \dd x\psi_{P}^{\dagger}(x)\psi_{S}^{\dagger}(x)\psi_{P}(x)\psi_{S}(x),\nonumber 
\end{align}
where here the combinatorial factor arising from normal ordering is
$4!$, since there are four distinct field operators. In many applications
one of the fields is strong, say that centered at $\omega_{P}$, and
we are interested in describing its effect on the phase of the weaker
field, say that centered at $\omega_{S}$. We introduce a cross-phase
modulation coefficient $\gcx$ that is appropriate
in that scenario, 
\begin{align}
 & \gcx=\frac{3\epsilon_{0}\omega_{S}}{4v_{P}v_{S}}\int \dd y\dd z\chi_{3}^{ijkl}(y,z)\left(e_{P}^{i}(y,z)e_{S}^{j}(y,z)\right)^{*}e_{P}^{k}(y,z)e_{S}^{l}(y,z),\label{eq:gamma_XPM_work1}
\end{align}
[cf. the expression (\ref{eq:gamma_SPM_work1}) for $\gcs$],
in terms of which we can write 
\begin{align}
 & \Hcx=-2\gcx\hbar^{2}\omega_{P}v_{P}v_{S}\int \dd x\psi_{P}^{\dagger}(x)\psi_{S}^{\dagger}(x)\psi_{P}(x)\psi_{S}(x).\label{eq:H_XPM_channel_use}
\end{align}
Following the same strategy used above \eqref{eq:check_coeff_1}, \eqref{eq:check_coeff_2}, and~\eqref{eq:check_coeff_3}
in identifying the physical significance of $\gcs$,
taking the Hamiltonian to be the sum of the linear terms (\ref{eq:channel_linear_Hamiltonian_J})
and the nonlinear contribution (\ref{eq:H_XPM_channel_use}), in the
classical limit we find 
\begin{align}
 \frac{\dd\phi_{S}(x)}{\dd x} &=2i\gcx\hbar\omega_{P}v_{P}\left|\phi_{P}(x)\right|^{2}\phi_{S}(x)\\
 & =2i\gcx P_{P}(x)\phi_{S}(x),\nonumber
\end{align}
with the usual factor of $2$ for cross-phase modulation explicit
[cf. (\ref{eq:check_coeff_3})]. Dividing (\ref{eq:gamma_XPM_work1})
by unity using the left-hand side of (\ref{eq:gamma_SPM_norm}), once
at $\omega_{P}$ and once at $\omega_{S}$, we find 
\begin{align}
 \gcx=\frac{3\omega_{S}\overline{\chi}_{3}}{4\epsilon_{0}\overline{n}_{P}\overline{n}_{S}c^{2}}\frac{1}{\Acx},\label{eq:gamma_XPM_chan_result}
\end{align}
where 
\begin{align}
 \frac{1}{\Acx}=\frac{\int \dd y\dd z\left(\chi_{3}^{ijkl}(y,z)/\overline{\chi}_{3}\right)\left(e_{P}^{i}(y,z)e_{S}^{j}(y,z)\right)^{*}e_{P}^{k}(y,z)e_{S}^{l}(y,z)}{\left(\int\frac{n(y,z;\omega_{P})/\overline{n}_{P}}{v_{g}(y,z;\omega_{P})/v_{P}}\bm{e}_{P}^{*}(y,z)\cdot\bm{e}_{P}(y,z)\dd y\dd z\right)\left(\int\frac{n(y,z;\omega_{S})/\overline{n}_{S}}{v_{g}(y,z;\omega_{S})/v_{S}}\bm{e}_{S}^{*}(y,z)\cdot\bm{e}_{S}(y,z)\dd y\dd z\right)},\label{eq:A_XPM_chan_result}
\end{align}
with $\overline{n}_{P}$ and $\overline{n}_{S}$ here typical values of $n(y,z;\omega_{P})$
and $n(y,z;\omega_{S})$, respectively, in regions where the associated
electric fields are nonvanishing. Comparing these equations with the
corresponding expressions~\eqref{eq:gamma_SPM_chan_result} and~\eqref{eq:A_SPM_chan_result}
for self-phase modulation, we see that for $\omega_{P}$ and $\omega_{S}$
close we have $\gcs\approx\gcx$, as
expected.

\subsubsection{Self- and cross-phase modulation in rings}\label{sec:SXPM_rings}

In a ring coupled to a channel, the usual approximation is to assume
that the important nonlinear interaction is in the ring, since in
usual applications the field is concentrated there. So to construct
the appropriate Hamiltonian for self-phase moulation here we use the
expression (\ref{eq:field_expansions_ring}) for $\bm{D}(\bm{r})$
in the ring, keeping the contribution from just one mode $P$ with
frequency $\omega_{P}$, in the general nonlinear Hamiltonian (\ref{eq:HNL_general}).
Then keeping the contribution from $\chi_{3}$, the form of the
nonlinear Hamiltonian closely follows that for a channel, and we have
\begin{align}
 \Hrs &=-\frac{1}{4\epsilon_{0}}\left(\frac{\hbar\omega_{P}}{2\mathcal{L}}\right)^{2}\frac{4!}{2!2!}\left(\int \dd\bm{r}_{\perp}d\zeta\Gamma_{3}^{ijkl}(\bm{r}_{\perp})\left(\mathsf{d}_{P}^{i}(\bm{r}_{\perp};\zeta)\mathsf{d}_{P}^{j}(\bm{r}_{\perp};\zeta)\right)^{*}\mathsf{d}_{P}^{k}(\bm{r}_{\perp};\zeta)\mathsf{d}_{P}^{l}(\bm{r}_{\perp};\zeta)\right)\label{eq:HSPMwork-1}\\
 & \quad\times c_{P}^{\dagger}c_{P}^{\dagger}c_{P}c_{P}.\nonumber 
\end{align}
We can write this as 
\begin{align}
 & \Hrs=-\frac{\hbar^{2}\omega_{P}v_{P}^{2}\grs}{2\mathcal{L}}c_{P}^{\dagger}c_{P}^{\dagger}c_{P}c_{P},\label{eq:HringSPM}
\end{align}
where for convenience later we have introduced $v_{P}$ as the group
velocity at $\omega_{P}$ for light propagating in a channel with
a structure matching that of the ring, and 
\begin{align}
 & \grs=\frac{3\omega_{P}\epsilon_{0}}{4v_{P}^{2}}\frac{1}{\mathcal{L}}\int\chi_{3}^{ijkl}(\bm{r}_{\perp})\left(\mathsf{e}_{P}^{i}(\bm{r}_{\perp};\zeta)\mathsf{e}_{P}^{j}(\bm{r}_{\perp};\zeta)\right)^{*}\mathsf{e}_{P}^{k}(\bm{r}_{\perp};\zeta)\mathsf{e}_{P}^{l}(\bm{r}_{\perp};\zeta)\dd\bm{r}_{\perp}\dd\zeta.\label{eq:gamma_SPMring_work1}
\end{align}
where we have followed the procedure used for the channel and written $\mathsf{d}_{P}(\bm{r}_{\perp};\zeta)=\epsilon_{0}n^{2}(\bm{r}_{\perp};\omega_{P})\mathsf{e}_{P}(\bm{r}_{\perp};\zeta),$
and we have put $n(\bm{r}_{\perp};\omega)=\sqrt{\varepsilon_{1}(\bm{r}_{\perp};\omega)}$.
We then use the normalization condition (\ref{eq:norm_D_disp}) to
write 
\begin{align}
 & \epsilon_{0}c\int\frac{n(\bm{r}_{\perp};\omega_{P})}{v_{g}(\bm{r}_{\perp};\omega_{P})}\mathsf{e}_{P}^{*}(\bm{r}_{\perp};0)\cdot\mathsf{e}_{P}(\bm{r}_{\perp};0)\dd\bm{r}_{\perp}=1,\label{eq:ring_norm_use}
\end{align}
and again introducing characteristic values $\overline{n}_{P}$ and $\overline{\chi}_{3}$
for, respectively, $n(\bm{r}_{\perp};\omega_{P})$ and the
nonvanishing components of $\chi_{3}^{ijkl}(\bm{r}_{\perp})$
in the region where these quantities are important in the integrals
appearing in~\eqref{eq:gamma_SPMring_work1} and~\eqref{eq:ring_norm_use},
we can write $\grs$ as 
\begin{align}
 & \grs=\frac{3\omega_{P}\overline{\chi}_{3}}{4\epsilon_{0}\overline{n}_{P}^{2}c^{2}}\frac{1}{\Ars},\label{eq:gamma_SPM_ring_result}
\end{align}
where 
\begin{align}
 & \frac{1}{\Ars}=\frac{\mathcal{L}^{-1}\int\left(\chi_{3}^{ijkl}(\bm{r}_{\perp})/\overline{\chi}_{3}\right)\left(\mathsf{e}_{P}^{i}(\bm{r}_{\perp};\zeta)\mathsf{e}_{P}^{j}(\bm{r}_{\perp};\zeta)\right)^{*}\mathsf{e}_{P}^{k}(\bm{r}_{\perp};\zeta)\mathsf{e}_{P}^{l}(\bm{r}_{\perp};\zeta)\dd\bm{r}_{\perp}\dd\zeta}{\left(\int\frac{n(\bm{r}_{\perp};\omega_{P})/\overline{n}_{P}}{v_{g}(\bm{r}_{\perp};\omega_{P})/v_{P}}\mathsf{e}_{P}^{*}(\bm{r}_{\perp};0)\cdot\mathsf{e}_{P}(\bm{r}_{\perp};0)\dd\bm{r}_{\perp}\right)^{2}}.\label{eq:A_SPM_ring_result}
\end{align}
Comparing the results for a ring with those~\eqref{eq:gamma_SPM_chan_result},~\eqref{eq:A_SPM_chan_result}
for a channel, we see that if the ring structure is just the channel
structure formed in a ring, and if the ring is large enough that the
mode fields are approximately the same (\ref{eq:ring_channel_compare})
then we can expect $\grs\approx\gcs$
if the electric field amplitude is largely in the $\bm{\hat{z}}$
direction. However, if other components are also important, differences
between $\grs$ and $\gcs$ can arise,
depending on the nonvanishing components of $\chi_{3}^{ijkl}$ and
their relative sizes.

The treatment of cross-phase modulation also follows that of the channel.
Considering ring fields at $\omega_{P}$ and $\omega_{S}$, the contribution
to the nonlinear Hamiltonian associated with cross-phase modulation
is 
\begin{align}
 \Hrx &=-\frac{1}{4\epsilon_{0}}\left(\frac{\hbar\omega_{P}}{2\mathcal{L}}\right)\left(\frac{\hbar\omega_{S}}{2\mathcal{L}}\right)4!\left(\int \dd\bm{r}_{\perp}d\zeta\Gamma_{3}^{ijkl}(\bm{r}_{\perp})\left(\mathsf{d}_{P}^{i}(\bm{r}_{\perp};\zeta)\mathsf{d}_{S}^{j}(\bm{r}_{\perp};\zeta)\right)^{*}\mathsf{d}_{P}^{k}(\bm{r}_{\perp};\zeta)\mathsf{d}_{S}^{l}(\bm{r}_{\perp};\zeta)\right)\\
 & \quad\times c_{P}^{\dagger}c_{S}^{\dagger}c_{P}c_{S}.\nonumber
\end{align}
As in the channel calculation, we introduce a nonlinear coefficient $\grx$ appropriate to describe cross-phase modulation
where there is a strong field at $\omega_{P}$ modifying the phase
of one at $\omega_{S}$, writing 
\begin{align}
 & \Hrx=-\frac{2\grx\hbar^{2}\omega_{P}v_{P}v_{S}}{\mathcal{L}}c_{P}^{\dagger}c_{S}^{\dagger}c_{P}c_{S},\label{eq:HringXPM}
\end{align}
where 
\begin{align}
 & \grx=\frac{3\omega_{S}\epsilon_{0}}{4v_{P}v_{S}}\frac{1}{\mathcal{L}}\int\chi_{3}^{ijkl}(\bm{r}_{\perp})\left(\mathsf{e}_{P}^{i}(\bm{r}_{\perp};\zeta)\mathsf{e}_{S}^{j}(\bm{r}_{\perp};\zeta)\right)^{*}\mathsf{e}_{P}^{k}(\bm{r}_{\perp};\zeta)\mathsf{e}_{S}^{l}(\bm{r}_{\perp};\zeta)\dd\bm{r}_{\perp}\dd\zeta,
\end{align}
where as for $\grx$ we have written the expression
in terms of the $\mathsf{e}_{P,S}(\bm{r_{\perp}};\zeta)$. Again
using the normalization condition (\ref{eq:ring_norm_use}), here
for both $\omega_{P}$ and $\omega_{S}$, we can write 
\begin{align}
 & \grx=\frac{3\omega_{S}\overline{\chi}_{3}}{4\epsilon_{0}\overline{n}_{P}\overline{n}_{S}c^{2}}\frac{1}{\Arx},
\end{align}
where 
\begin{align}
 & \frac{1}{\Arx}=\frac{\mathcal{L}^{-1}\int\left(\chi_{3}^{ijkl}(\bm{r}_{\perp})/\overline{\chi}_{3}\right)\left(\mathsf{e}_{P}^{i}(\bm{r}_{\perp};\zeta)\mathsf{e}_{S}^{j}(\bm{r}_{\perp};\zeta)\right)^{*}\mathsf{e}_{P}^{k}(\bm{r}_{\perp};\zeta)\mathsf{e}_{S}^{l}(\bm{r}_{\perp};\zeta)\dd\bm{r}_{\perp}\dd\zeta}{\left(\int\frac{n(\bm{r}_{\perp};\omega_{P})/\overline{n}_{P}}{v_{g}(\bm{r}_{\perp};\omega_{P})/v_{P}}\mathsf{e}_{P}^{*}(\bm{r}_{\perp};0)\cdot\mathsf{e}_{P}(\bm{r}_{\perp};0)\dd\bm{r}_{\perp}\right)\left(\int\frac{n(\bm{r}_{\perp};\omega_{S})/\overline{n}_{S}}{v_{g}(\bm{r}_{\perp};\omega_{S})/v_{S}}\mathsf{e}_{S}^{*}(\bm{r}_{\perp};0)\cdot\mathsf{e}_{S}(\bm{r}_{\perp};0)\dd\bm{r}_{\perp}\right)}.
\end{align}
 As in the channel calculation, here we have $\grs\approx\grx$
as long as $\omega_{J}$ and $\omega_{K}$ are close.

We close this section by deriving the promised correct expression
for the self-phase modulation term arising in Sec.~\ref{sec:TroubleWithE}.
In the special case considered there we can identify $\varepsilon_{3}(\bm{r})$
with $\chi_{3}^{zzzz}(\bm{r}),$ and from (\ref{eq:gamma_SPMring_work1})
we find 
\begin{align}
 & \grs=\frac{3\omega_{J}}{4v_{J}^{2}\mathcal{L}}K,
\end{align}
where $K$ is given by (\ref{eq:Kdef}). Using this in (\ref{eq:HringSPM})
we have 
\begin{align}
 & \Hrs=-\frac{3}{2}\left(\frac{\hbar\omega_{J}}{2\mathcal{L}}\right)^{2}Kc_{J}^{\dagger}c_{J}^{\dagger}c_{J}c_{J},
\end{align}
which leads to the expression (\ref{eq:shift_right}).

\section{Obtaining the ket from the equations of motion}\label{sec:Ket}
In this section we develop mathematical tools to describe the quantum states that are typically generated in quantum nonlinear optical processes. The Hamiltonians describing these processes are polynomials with degree strictly greater than two in the electromagnetic fields, such as those developed in the previous section. However, it is often the case that some of the fields can be described classically -- that is, replaced by their classical expectation values -- leading to Hamiltonians that, while still nonlinear in the fields, are only quadratic in quantum operators. The validity and usefulness of this approximation will be made apparent in subsequent sections, and thus for the moment we will focus on the mathematical tools necessary to obtain the quantum state generated by these interactions. We stress that while there is a considerable reduction in complexity working with Hamiltonians at most quadratic in the operators, there are still significant complications stemming from the fact that these Hamiltonians are still time-dependent and have non-vanishing commutators at different times.

Before heading on directly to the most general problem one can tackle with the tools from Gaussian quantum optics and the symplectic formalism \cite{serafini2017quantum,simon1988gaussian,dutta1995real,adesso2014continuous,weedbrook2012gaussian} we will consider simple examples that will help build intuition towards the most general results in Sec.~\ref{sec:sq_ham} . We will assume the reader is familiar with basic quantum mechanics, including the Schr\"{o}dinger and Heisenberg pictures \cite{shankar2012principles,sakurai2011modern}. In Sec.~\ref{sec:multimode} we introduce the Dyson, Magnus, and Trotter-Suzuki expansions as methods for the solution of linear differential equations, including the Schr\"{o}dinger and the Heisenberg equation. In  Sec.~\ref{sec:low-gain} we use the first order Magnus expansion to solve the low-gain multimode squeezing problem and to introduce the concept of Schmidt modes. In Sec.~\ref{sec:bogoliubov} we use the fact that the solutions to the Heisenberg equations must preserve equal-time commutation relations to derive important properties of the Heisenberg propagator. In Sec.~\ref{sec:theket} we use introduce the characteristic function and the Heisenberg propagator to derive the form of the quantum state generated by a general quadratic Hamiltonian in the spontaneous regime when the initial state is vacuum. We specialize some of the results derived to the case of non-degenerate squeezing in Sec.~\ref{sec:twin-beams}. 
Then we generalize the results related to the spontaneous problem, when the input state is vacuum, to arbitrary input states in Sec.~\ref{sec:estim}. In the last two subsections we study the effect of loss in Sec.~\ref{sec:loss} and finally show how to calculate the statistics of homodyne and photon-number resolved measurements in Sec.~\ref{sec:pnr_homodyne}.

\subsection{Squeezing Hamiltonians}\label{sec:sq_ham}
To gain some insight into the Gaussian formalism we study a very simple and well-known problem in quantum optics. We will solve, both in the Schr\"{o}dinger and Heisenberg pictures, the dynamics generated by the Hamiltonian
\begin{align}\label{eq:squeezing_hamiltonian}
H = i \frac{\hbar g}{2}(e^{i \phi} a^{\dagger 2} - e^{-i \phi}  a^2 ),
\end{align}
where $a,a^\dagger$ are the annihilation and creation operators of a harmonic oscillator that satisfy the usual commutation relations $[a,a^\dagger] = 1$. The evolution of any quantum mechanical system is dictated by its Hamiltonian $H(t)$ via the Schr\"odinger equation,
\begin{align}\label{eq:schrodinger}
i \hbar  \frac{\dd}{\dd t} \mathcal{U}(t,t_{\text{i}}) = H(t) \mathcal{U}(t,t_{\text{i}}),
\end{align}
where $\mathcal{U}(t,t_{\text{i}})$ is the time evolution operator that satisfies the boundary condition $\mathcal{U}(t_{\text{i}},t_{\text{i}}) = \mathbb{I}$ and  $\mathbb{I} $ is the Hilbert space identity. 

For the very simple time independent Hamiltonian of~\eqref{eq:squeezing_hamiltonian}, we can immediately write 
\begin{align}
\mathcal{U}(t_{\text{f}},t_{\text{i}}) = \exp\left( - i \frac{H}{\hbar}  (t_{\text{f}}-t_{\text{i}}) \right)  = \exp\left(\tfrac{r}{2}\left[e^{i \phi}  a^{\dagger 2} - e^{-i \phi} a^2 \right] \right), \quad r = g (t_{\text{f}}-t_{\text{i}}).
\end{align}
The unitary operator above corresponds to the so-called squeezing operator that when acted on vacuum yields a single-mode squeezed vacuum state, which can be written as~\cite{gerry2005introductory}
\begin{align}
\ket{\Psi(t_{\text{f}})} = \exp\left(\tfrac{r}{2}\left[ e^{i \phi} a^{\dagger 2} - e^{-i \phi} a^2 \right] \right)|0 \rangle = \frac{1}{\sqrt{\cosh r}}\sum_{n=0}^\infty e^{i n\phi} \tanh^n r \frac{\sqrt{2n!}}{2^n n!} |2n \rangle, \label{ket_easy}
\end{align}
where in the last equation we introduced the Fock states 
\begin{align}
\ket{n} = \frac{a^{\dagger n} }{ \sqrt{n!}} \ket{0}
\end{align}
and used the well-known disentangling formula (cf.~Appendix 5 of Barnett and Radmore~\cite{barnett2002methods})
\begin{align}
\exp\left(\tfrac{r}{2}\left[ e^{i \phi} a^{\dagger 2} - e^{-i \phi} a^2 \right] \right) &= \exp\left( \tfrac{1}{2} e^{i \phi } \tanh r \ a^{\dagger 2}\right) \\
&\quad\times \exp\left( - \left[ a^\dagger a+\tfrac{1}{2} \right] \ln \cosh r  \right)  \nonumber \\
&\quad\times \exp\left( -\tfrac{1}{2} e^{-i \phi } \tanh r \ a^{ 2}\right). \nonumber
\end{align}
One can easily verify the following expectation values
\begin{align} \label{eq:single_mode_moments}
\braket{a}_{t_{\text{f}}} &= \braket{\Psi(t_{\text{f}})|a|\Psi(t_{\text{f}})}  = 0, \\
n_{t_{\text{f}}} = \braket{a^\dagger a}_{t_{\text{f}}} &= \braket{\Psi(t_{\text{f}})|a^\dagger a|\Psi(t_{\text{f}})} = \sinh^2 r,\nonumber\\
m_{t_{\text{f}}} =\braket{a^2}_{t_{\text{f}}} &= \braket{\Psi(t_{\text{f}})|a^2|\Psi(t_{\text{f}})} =  \tfrac{1}{2}e^{i \phi} \sinh 2r,\nonumber\\
\braket{a^\dagger a a^\dagger a}_{t_\text{f}} &= 3 n_{t_\text{f}}^2+ 2n_{t_\text{f}}.\nonumber
\end{align}
as well as, introducing the Hermitian quadrature operators
\begin{align}
q = \sqrt{\frac{\hbar}{2}} \left(a + a^\dagger \right), \quad  p = -i  \sqrt{\frac{\hbar}{2}} \left(a - a^\dagger \right),
\end{align}
the moments
\begin{align}
\braket{q}_{t_{\text{f}}} &= \braket{p}_{t_{\text{f}}} = 0, \\
\braket{q^2}_{t_{\text{f}}} &= \tfrac{\hbar}{2}\left(2n_{t_{\text{f}}}+1+ m_{t_{\text{f}}}+m_{t_{\text{f}}}^*  \right) = \frac{\hbar}{2} \left(\sinh (2 r) \cos (\phi )+2 \sinh ^2(r)+1 \right),\nonumber\\
\braket{p^2}_{t_{\text{f}}} &= \tfrac{\hbar}{2}\left(2n_{t_{\text{f}}}+1- m_{t_{\text{f}}}-m_{t_{\text{f}}}^*  \right)=\frac{\hbar}{2} \left(-\sinh (2 r) \cos (\phi )+2 \sinh ^2(r)+1 \right),\nonumber\\
\frac{\braket{qp+pq}_{t_{\text{f}}}}{2} &= i \tfrac{\hbar}{2} (m^* - m) =\frac{\hbar}{2} \sin(\phi) \sinh(2r).\nonumber
\end{align}
These moments can be arranged into a covariance matrix
\begin{align}
V_{t_{\text{f}}} =\begin{pmatrix}
\braket{q}_{t_{\text{f}}}^2 & \frac{\braket{qp+pq}_{t_{\text{f}}}}{2} \\
\frac{\braket{qp+pq}_{t_{\text{f}}}}{2} & \braket{p^2}_{t_{\text{f}}}
\end{pmatrix} = \frac{\hbar}{2} R(\phi/2) \begin{pmatrix}
e^{2r} & 0 \\
0 & e^{-2r}
\end{pmatrix} R(\phi/2)^T, \quad  R(\phi) = \begin{pmatrix} 
\cos \phi & - \sin \phi \\
\sin \phi & \cos \phi 
\end{pmatrix}.
\end{align}
We see that the quadrature along the angle $\phi/2$ has increased fluctuations by amount $e^{2r}$ relative the vacuum level, while the quadrature in the direction perpendicular to $\phi/2$ has decreased (squeezed) fluctuations by amount  $e^{-2r}$. 

One can also solve the dynamics of this problem in the Heisenberg picture. The operators are now dynamical quantities defined by $O(t) = \mathcal{U}^\dagger(t,t_{\text{i}})O(t_{\text{i}}) \mathcal{U}(t,t_{\text{i}})$ and their dynamics is determined by the Heisenberg equation of motion,
\begin{align}
\ddt{} O(t) = \frac{i}{\hbar} [H(t),O(t)],
\end{align}
with the boundary condition $O(t_{\text{i}}) \equiv O$ where $O$ is simply the Schr\"{o}dinger picture operator. For the Hamiltonian of~\eqref{eq:squeezing_hamiltonian} and for $O \in  \{a,a^\dagger \}$ we find
\begin{align}
\ddt{} a &=  \frac{i}{\hbar} [H,O] = g e^{i \phi} a^\dagger, \quad \ddt{} a^\dagger =  \frac{i}{\hbar} [H,O] = g e^{-i \phi} a,
\end{align}
or, more compactly,
\begin{align}
\ddt{} \begin{pmatrix}
a \\
a^\dagger
\end{pmatrix} &= -i \underbrace{\begin{pmatrix}
0 & i g e^{i \phi}\\
i g e^{-i \phi} & 0
\end{pmatrix} }_{\equiv \bm{A}}\begin{pmatrix}
a \\
a^\dagger
\end{pmatrix}.
\end{align}
Note that the matrix $\bm{A}$ defined in the last equation is \emph{not} Hermitian. It is instead an element of the Lie algebra $\mathfrak{su}(1,1)$ that generates a Heisenberg propagator which is an element of the Lie group $SU(1,1)$ (cf.~Appendix 11.1.4 of Klimov  and  Chumakov~\cite{klimov2009group}).

We can immediately find the solution of the last differential equation by exponentiation,
\begin{align}
\begin{pmatrix}
a(t_{\text{f}}) \\
a^\dagger(t_{\text{f}})
\end{pmatrix} = \exp(-i \bm{A}(t_{\text{f}}-t_{\text{i}})) \begin{pmatrix}
a(t_{\text{i}}) \\
a^\dagger(t_{\text{i}})
\end{pmatrix} = 
\left(
\begin{array}{cc}
\cosh g (t_{\text{f}} - t_{\text{i}}) & e^{i \phi } \sinh g (t_{\text{f}} - t_{\text{i}}) \\
e^{-i \phi } \sinh g(t_{\text{f}} - t_{\text{i}}) & \cosh g (t_{\text{f}} - t_{\text{i}}) \\
\end{array}
\right) \begin{pmatrix}
a(t_{\text{i}}) \\
a^\dagger(t_{\text{i}})
\end{pmatrix},
\end{align}
from which we can recover the moments derived in the Schr\"odinger picture  by the usual rule $\braket{O}_t = \braket{\Psi(t_{\text{i}})|O(t)|\Psi(t_{\text{i}})} = \braket{\Psi(t)|O(t_{\text{i}})|\Psi(t)} $.

At this point it might seem like overkill to have derived the same expectation value using two different methods. However, as we shall see in a moment, it is useful to have two ways of looking at the problem because there are situations where solving the Schr\"{o}dinger equation for the ket $\ket{\Psi(t)}$ is hopeless while solving the Heisenberg equations is practical. Moreover, once the Heisenberg equations are solved one can then write the sought after time-evolved Schr\"{o}dinger picture ket.

\subsection{Multimode Squeezing: Dyson, Magnus and Suzuki-Trotter}\label{sec:multimode}
Let us consider a slightly more complicated Hamiltonian given by
\begin{align}\label{eq:quad_hamil}
H(t) = \hbar \left\{ \sum_{i,j =1}^\ell \Delta_{ij}(t) a_i^\dagger a_j + i \frac12 \sum_{k,l=1}^\ell \left[ \zeta_{kl}(t) a_k^{\dagger } a_{l}^{\dagger} -\text{H.c.} \right] \right\},
\end{align}
where we have introduced $\ell$ bosonic modes satisfying the usual canonical commutation relations
\begin{align} \label{equal-time}
[a_i,a_j] = [a_i^\dagger ,a_j^\dagger] = 0,  \quad [a_i,a_j^\dagger] = \delta_{ij},
\end{align}
we demand that $\Delta_{ij} = \Delta_{ji}^*$ for the Hamiltonian to be Hermitian, and assume without loss of generality that $\zeta_{kl} = \zeta_{lk}$, since any anti-symmetric contribution to $\zeta_{kl}$ will vanish upon contraction with the permutation symmetric term $a_k^\dagger a_l^\dagger$. 

In this section we do not attach any particular meaning to the mode labels $i$ and $j$ in the last set of equations. In subsequent sections these labels will correspond to, for example, discretized wavevectors near a reference or central wavevector used to obtain a simplified dispersion relation. In this case the term $\Delta_{ij}$ will correspond to a detunning between the different wavevectors (and possibly cross-phase modulation of a classical pump on the quantum modes) while $\zeta_{kl}$ will correspond to wave-mixing induced by a nonlinear process.

As before, we would like to find the ket obtained by evolving the vacuum under this Hamiltonian, the so-called ``spontaneous problem." Note that the Hamiltonian describing the dynamics is now time-dependent and has three different terms that do not commute with each other, making a more complex dynamics than that of the single-mode problem analyzed in Sec.~\ref{sec:sq_ham}. In what follows we use boldface to refer to matrices $\bm{M}$ while referring to their entries as $M_{ij}$. In matrix notation, the constraints described in the paragraph above are simply, $\bm{\Delta} = \bm{\Delta}^\dagger$ and $\bm{\zeta} = \bm{\zeta}^T$ where $\dagger$ and $T$ indicate the conjugate transpose and the transpose respectively.

We can formally solve the dynamics by writing the time-evolution operator associated with this Hamiltonian as
\begin{align}\label{eq:time-ordered}
\mathcal{U}(t,t_{\text{i}}) = \mathcal{T} \exp\left(-\frac{i}{\hbar} \int_{t_{\text{i}}}^t dt' H(t') \right),
\end{align}
where $\mathcal{T}$ is the so-called time-ordering operator. When applied to a product of Hamiltonians at diferent times $t_1,\ldots, t_n$ it simply orders them chronologically 
\begin{align}
\mathcal{T}\left[H(t_1) H(t_2) \ldots H(t_n) \right] = H(t_{i_1}) H(t_{i_2}) \ldots H(t_{i_n}),
\end{align}
where $i_1,i_2\ldots,i_n$ is a permutation of the set $[1,2,\ldots,n]$ such that $t_{i_1} \geq t_{i_2}\geq \ldots \geq t_{i_n}$~\cite{sakurai2011modern,shankar2012principles}.

In practice, the expression~\eqref{eq:time-ordered} can be interpreted as a power series of the Dyson~\cite{dyson1949radiation} or Magnus~\cite{blanes2009magnus,magnus1954exponential} type or can be approximated using Trotterization~\cite{suzuki1976generalized,trotter1959product}.
In the Dyson series the unitary evolution operator is written as an infinite power series
\begin{align}
\mathcal{U}(t_{\text{f}},t_{\text{i}})&= \mathbb{I}+(-i)\int_{t_{\text{i}}}^{t_{\text{f}}} \dd t' \frac{ H(t')}{\hbar}+(-i)^2\int_{t_{\text{i}}}^{t_{\text{f}}} \dd t' \int_{t_{\text{i}}}^{t'} \dd t'' \frac{H(t') H(t'')}{\hbar^2}+\ldots,
\end{align}
while in the Magnus series one writes the solution as the exponential of a series of nested commutators at different times
\begin{align}
\mathcal{U}(t_{\text{f}},t_{\text{i}})&=\exp\big( \Omega_1(t_{\text{f}},t_{\text{i}})+ \Omega_2(t_{\text{f}},t_{\text{i}})+ \Omega_3(t_{\text{f}},t_{\text{i}})+\ldots\big), \\
\Omega_1(t_{\text{f}},t_{\text{i}})&=-\frac{i}{\hbar} \int_{t_{\text{i}}}^{t_{\text{f}}} \dd t  H(t),\\
\label {ch4:mag2} \Omega_2(t_{\text{f}},t_{\text{i}})&=\frac{(-i)^2}{2 \hbar^2}\int_{t_{\text{i}}}^{t_{\text{f}}} \dd t \int_{t_{\text{i}}} ^{t} \dd t' \left[ H(t), H(t') \right],\\
\label{ch4:mag3a} \Omega_3(t_{\text{f}},t_{\text{i}})&=\frac{(-i)^3}{6 \hbar^3}\int_{t_{\text{i}}}^{t_{\text{f}}} \dd t \int_{t_{\text{i}}} ^{t} \dd t' \int_{t_{\text{i}}} ^{t'} \dd t'' \Big( \left[ H(t),\left[ H(t'), H(t'') \right] \right]\\
&\quad \quad\quad \quad\quad \quad \quad \quad\quad \quad\quad \quad +\left[\left[ H(t), H(t') \right], H(t'') \right] \Big). \nonumber
\end{align}
For the quadratic Hamiltonians we investigate here, the unitary evolution operator resulting from the Magnus expansion at any order respects the photon statistics that the exact operator is known to have, while that resulting from the Dyson expansion does not. Furthermore, in the Magnus expansion all terms beyond $\Omega_1$ are clearly time-ordering corrections, related to the fact that $\left[ H(t), H(t') \right] \neq 0$ whereas in the Dyson series each successive term contains a part associated with time-ordering corrections as well as a part associated with the solution that would be obtained by ignoring time-ordering corrections and naively writing $\mathcal{U}(t,t_{\text{i}}) = \exp\left(-\frac{i}{\hbar} \int_{t_{\text{i}}}^t \dd t' H(t') \right)$.

Finally, a third approach is to use the Trotter-Suzuki expansion to approximate the time evolution operator as a product of time evolution operators $\mathcal{U}(t_i,t_i+\Delta t)$ where each Hamiltonian is assumed to be roughly constant over small intervals of duration $\Delta t$ thus writing
\begin{align}
\mathcal{U}(t_{\text{f}},t_{\text{i}}) = \exp\left( -i  \frac{H(t_{N})}{\hbar} \Delta t \right)  \exp\left( -i  \frac{H(t_{N-1})}{\hbar} \Delta t \right) \ldots  \exp\left( -i  \frac{H(t_{\text{i}})}{\hbar} \Delta t \right),
\end{align}
where $t_j = t_{\text{i}} + j \Delta t |_{j=0}^N$ and $\Delta t = (t_{\text{f}}-t_{\text{i}})/N$.
Regardless of which strategy is chosen, one will have to deal with infinite dimensional creation and destruction operators acting on a Hilbert space. Even if these operators are truncated at a finite Fock cutoff $c$, the dimensionality of the operators appearing in any of the equations above will be $c^{2\ell}$, which will easily fit into computer memory only for very modest $c$ and $\ell$. 

\subsection{Low-gain solutions}\label{sec:low-gain}
The three strategies in the previous section have different strengths and weaknesses. If the Hamiltonian satisfies $[H(t),H(t')]=0$ for all the times $t,t'$ between $t_{\text{i}}$ and $t_{\text{f}}$ then the Magnus expansion at the first level immediately gives the exact solution.
If the Hamiltonian $H(t)$ is ``weak'' we can treat the Magnus or Dyson series perturbatively and keep only the first few terms of either of them. 

The simplest approximation is to keep only the first term of either of them. In this perturbative or low-gain limit one can obtain a simple solution to the spontaneous problem
\begin{align}\label{eq:ket_low_gain}
\ket{\overline{\Psi}(t_\text{f})}&\approx \overline{\mathcal{U}}(t_\text{f}, t_\text{i}) \ket{\text{vac}}, \\
\overline{\mathcal{U}}(t_\text{f}, t_\text{i})&= \exp\left(-\frac{i}{\hbar} \int_{t_{\text{i}}}^{t_\text{f}} \dd t' H(t') \right) =   \exp\left[ \tfrac{1}{2}\sum_{k,l} \overline{J}_{kl} a_l^\dagger a_k^\dagger  - \text{H.c.} \right] ,
\end{align}
where for simplicity we have assumed that $\Delta_{ij}(t) = 0$ \footnote{In the case where $\Delta_{ij}$ is not time-dependent this can always be assumed without loss of generality as shown in Appendix~\ref{app:interaction_picture}.} and introduced $\overline{\bm{J}} = \int_{t_{\text{i}}}^{t_{\text{f}}} \dd t  \ \bm{\zeta}(t)$, the so-called joint amplitude of the squeezed state above. As detailed below, it is often useful to rewrite this state by making use of the Takagi-Autonne decomposition of the symmetric matrix $\overline{\bm{J}} = \overline{\bm{J}}^T =  \overline{\bm{F}}  [\oplus_{\lambda=1}^\ell \overline{r}_\lambda] \overline{\bm{F}}^T$. Here $\overline{\bm{F}} $ is unitary $\overline{\bm{F}} \overline{\bm{F}}^\dagger = \mathbb{I}_\ell$, the low gain assumption is reflected in the fact that $\overline{r}_\lambda \ll 1$, we have used the direct-sum notation $\oplus_{\lambda=1}^\ell \overline{r}_\lambda$ to indicate a diagonal matrix square with entries $\{\overline{r}_\lambda\}$, and have used overbars to indicate that quantities are associated with a low-gain solution. Note that the symmetry of $\bm{J}$ simply follows from the symmetry of $\bm{\zeta}$. The Takagi-Autonne decomposition (cf. Corollary 4.4.4 of Horn et al. \cite{horn2012matrix}) is a singular value decomposition where it is made explicit that the matrix being decomposed is symmetric. Numerical routines to perform this decomposition can be found in the Python packages Strawberry Fields~\cite{killoran2019strawberryfields} and The Walrus~\cite{gupt2019walrus}.

Using the Takagi-Autonne decomposition lets us introduce the Schmidt (or broadband, or supermode) operators
\begin{align}
\overline{A}_\lambda^\dagger = \sum_{k=1}^\ell	a_k^\dagger \overline{F}_{k\lambda} \Longleftrightarrow a_k^\dagger = \sum_{i=1}^\ell \overline{F}^*_{k\lambda} \overline{A}_{\lambda}^\dagger,
\end{align}
which, due to the unitarity of $\overline{\bm{F}}$, satisfy bosonic canonical commutation relations
\begin{align}
[\overline{A}_{\lambda}, \overline{A}_{\lambda'}] = 0, \quad [\overline{A}_{\lambda}, \overline{A}_{\lambda'}^\dagger] = \delta_{\lambda,\lambda'}
\end{align}
that allows us to finally write~\cite{lvovsky2007decomposing,wasilewski2006pulsed}
\begin{align}
\ket{\overline{\Psi}(t_\text{f})} &= \exp\left[\tfrac12 \sum_{k,l=1}^\ell \overline{J}_{kl} a^\dagger_{k} a^\dagger_l  - \text{H.c.}\right] \ket{\text{vac}} = \exp \left[ \tfrac12 \sum_{\lambda=1}^\ell \overline{r}_\lambda  \overline{A}_\lambda^{\dagger 2}
-\text{H.c.} \right] \ket{\text{vac}} \\
&= \bigotimes_{\lambda=1}^\ell \exp \left[ \tfrac12 \overline{r}_\lambda  \overline{A}_\lambda^{\dagger 2}
-\text{H.c.} \right] \ket{\text{vac}}. \nonumber
\end{align}
The diagonal form of the exponential argument allows us to identify the state generated in the spontaneous problem as a manifold of squeezed states over the modes defined by $\overline{\bm{F}}$, i.e., the Schmidt modes of the system for which we can easily write their photon number expansion as
\begin{align}
\ket{\overline{\Psi}(t_\text{f})} = \bigotimes_{\lambda=1}^\ell \left[ \frac{1}{\sqrt{\cosh \overline{r}_\lambda}}\sum_{n=0}^\infty  \tanh^n \overline{r}_\lambda \frac{\sqrt{(2n)!}}{2^n n!} |2n \rangle_\lambda \right],
\end{align}
where $\ket{n}_\lambda \equiv \frac{\overline{A}_\lambda^{\dagger n}}{\sqrt{n!}} \ket{0}$ is a Fock state in the Schmidt mode labelled by $\lambda$.

We can also use the form of the low-gain time-evolution operator to transform the operators. Using the definitions of the Schmidt modes one can easily find
\begin{align}
a_i(t_{\text{f}}) &= \overline{\mathcal{U}}(t_\text{f}, t_\text{i})^\dagger a_i( t_\text{i}) \overline{\mathcal{U}}(t_\text{f}, t_\text{i}) = \sum_{j=1}^\ell \overline{V}_{ij} a_{j}(t_{\text{i}}) + \sum_{j=1}^\ell \overline{W}_{ij} a^\dagger _{j} (t_{\text{i}}), \\
a_i^\dagger(t_{\text{f}}) &= \overline{\mathcal{U}}(t_\text{f}, t_\text{i})^\dagger a_i^\dagger( t_\text{i}) \overline{\mathcal{U}}(t_\text{f}, t_\text{i}) = \sum_{j=1}^\ell \overline{V}^*_{ij} a^\dagger_{j}(t_{\text{i}}) + \sum_{j=1}^\ell \overline{W}^*_{ij} a _{j}(t_{\text{i}}),
\end{align}
where now 
\begin{align}
\overline{\bm{V}} = \overline{\bm{F}} \left[ \oplus_{\lambda=1}^\ell \cosh \overline{r}_{\lambda} \right] \overline{\bm{F}}^\dagger, \quad 
\overline{\bm{W}} = \overline{\bm{F}} \left[ \oplus_{\lambda=1}^\ell \sinh \overline{r}_{\lambda} \right] \overline{\bm{F}}^T.
\end{align}
To lowest non-vanishing order in the squeezing parameters $\overline{r}_\lambda \ll 1$ one has $\overline{\bm{V}} \approx \mathbb{I}_\ell$ and $\overline{\bm{W}} \approx \overline{\bm{J}}$.

\subsection{Properties of the solution to the Heisenberg equations of motion: Bogoliubov transformations}\label{sec:bogoliubov}

Let us now analyze this same problem in the Heisenberg picture for arbitrary gain. We now find
\begin{align}\label{eq:matrix_Heisenberg}
\ddt{} \begin{pmatrix}
\bm{a} \\
\bm{a}^\dagger
\end{pmatrix} &= -i \underbrace{\begin{pmatrix}
- \bm{\Delta}(t)  & i \bm{\zeta}(t) \\
i \bm{\zeta}^*(t)  &  \bm{\Delta}(t) 
\end{pmatrix} }_{\equiv \bm{A}(t)}\begin{pmatrix}
\bm{a} \\
\bm{a}^\dagger
\end{pmatrix},
\end{align}
where we have introduced
\begin{align}
\bm{a} = \begin{pmatrix} a_1 \\
\vdots \\
a_\ell \end{pmatrix}, \quad \bm{a}^\dagger = \begin{pmatrix} a^\dagger_1 \\
\vdots \\
a^\dagger_\ell \end{pmatrix}.
\end{align}
Just like the Schr\"{o}dinger equation for the time evolution operator of~\eqref{eq:schrodinger}, for quadratic bosonic Hamiltonians, the Heisenberg equations of motion are linear and dictated by the matrix $\bm{A}$. 
We can formally solve these equations using a time ordered exponential like before,
\begin{align}\label{eq:symplectic}
\begin{pmatrix}
\bm{a}(t_{\text{f}}) \\
\bm{a}^\dagger(t_{\text{f}})
\end{pmatrix} =  \mathcal{U}^\dagger(t_{\text{f}},t_{\text{i}})
\begin{pmatrix}
\bm{a}(t_{\text{i}}) \\
\bm{a}^\dagger(t_{\text{i}})
\end{pmatrix}  \mathcal{U}(t_{\text{f}},t_{\text{i}})=  \mathcal{T} \exp\left( - i \int_{t_{\text{i}}}^{_{\text{f}}} \dd t'\bm{A}(t') \right) \begin{pmatrix}
\bm{a}(t_{\text{i}}) \\
\bm{a}^\dagger(t_{\text{i}})
\end{pmatrix} =\bm{K}(t_{\text{f}},t_{\text{i}})\begin{pmatrix}
\bm{a}(t_{\text{i}}) \\
\bm{a}^\dagger(t_{\text{i}})
\end{pmatrix}.
\end{align}
yet while we still need need to consider a time-ordered exponential here, this problem is actually much simpler than the one associated with~\eqref{eq:time-ordered}. In particular, here $\bm{A}(t)$ is simply a $2 \ell \times 2 \ell$ matrix. It easily fits in the memory of a computer, and using Trotter-Suzuki we can approximate \begin{align}
\bm{K}(t_{\text{f}},t_{\text{i}}) \approx \exp\left( -i \Delta t \bm{A}(t_{N}) \right) \exp\left(-i \Delta t \bm{A}(t_{N-1}) \right) \ldots \exp\left(-i \Delta t \bm{A}(t_2) \right) \exp\left( -i \Delta t \bm{A}(t_{\text{i}}) \right),
\end{align}
where $t_j = t_{\text{i}} + j \Delta t |_{j=0}^N$ and $\Delta t = (t_{\text{f}}-t_{\text{i}})/N$ and we have discretized the time evolution into $N+1$ slices. The solution of the problem now boils down to the multiplication of matrices with size proportional to the number of modes, and there is no need to even introduce a cutoff in Fock space. Note that an alternative approach to approximate the Heisenberg propagator is to employ the Magnus expansion to a finite level as done in Lipfert et al. \cite{lipfert2018bloch}.

From the fact that the transformations connecting initial- and final-time Heisenberg operators are linear and the fact that these solutions must respect the canonical commutation relation at each time, one can obtain a number of useful properties.
To simplify notation, from now on we drop initial and final times from matrices, writing for example, $\bm{K}$ in place of $\bm{K}(t_{\text{f}},t_{\text{i}})$. 
Let us start by writing the Heisenberg propagator in block form 
\begin{align}
\bm{K} = \begin{pmatrix}
\bm{V} & \bm{W} \\
\bm{W}^* & \bm{V}^*
\end{pmatrix},
\end{align}
along with
\begin{align}\label{eq:in-out}
a_i(t_{\text{f}}) &= \mathcal{U}^\dagger a_i \mathcal{U} = \sum_{j=1}^\ell V_{ij} a_{j}(t_{\text{i}}) + \sum_{j=1}^\ell W_{ij} a^\dagger _{j} (t_{\text{i}}), \\
a_i^\dagger(t_{\text{f}}) &= \mathcal{U}^\dagger a_i^\dagger \mathcal{U} = \sum_{j=1}^\ell V^*_{ij} a^\dagger_{j}(t_{\text{i}}) + \sum_{j=1}^\ell W^*_{ij} a _{j}(t_{\text{i}}).
\end{align}
Since $\mathcal{U}$ is a unitary operator in Hilbert space, the bosonic operators at the final time must satisfy the same equations as their initial time versions, i.e.,~\eqref{equal-time} hold if we replace $a_i \to a_i(t)$.
From this alone we infer that 
\begin{align}
[a_i(t_{\text{f}}), a_j(t_{\text{f}})] = 0 &\Leftrightarrow \bm{W} \bm{V}^T = \bm{V}  \bm{W}^T, \\
[a_i(t_{\text{f}}), a^\dagger_j(t_{\text{f}})] = \delta_{ij} &\Leftrightarrow \bm{V}^* \bm{V}^T - \bm{W}^* \bm{W}^T = \mathbb{I}_M.
\end{align}
The constraints above imply that one can write the following \emph{joint} singular value decompositions~\cite{ekert1991relationship, braunstein2005squeezing,christ2013theory}
\begin{align}\label{eq:singular-values}
\bm{V} = \bm{F} \left[ \oplus_{\lambda=1}^\ell \cosh r_{\lambda} \right] \bm{G}, \quad 
\bm{W} = \bm{F} \left[ \oplus_{\lambda=1}^\ell \sinh r_{\lambda} \right] \bm{G}^*,
\end{align}
where $\bm{F}$ and $\bm{G}$ are unitary matrices. As we will see later, the matrix $\bm{F}$ will define the (output) Schmidt modes of the problem, the matrix $\bm{G}$ will define the input Schmidt modes of the problem and the quantities $r_\lambda$ will be the squeezing parameters.

It is useful to note that the entries of the matrix 
\begin{align}
\bm{M} = \bm{M}^T = \bm{W} \bm{V}^T
\end{align}
coincide with the following vacuum expectation value
\begin{align}\label{eq:Mdef}
M_{ij} = \braket{a_i(t_{\text{f}})a_j(t_{\text{f}})}_{\ket{\text{vac}}} = \bra{\text{vac}}\mathcal{U}^\dagger a_i a_j \mathcal{U} \ket{\text{vac}} = \left( \bm{F}  \left[ \oplus_{\lambda=1}^\ell \tfrac{1}{2}\sinh 2r_\lambda \right]  \bm{F}^T \right)_{ij}.
\end{align}
Similarly we find that the entries of the Hermitian matrix
\begin{align}
\bm{N} = \bm{N}^\dagger = \bm{W}^* \bm{W}^T
\end{align} 
coincide with
\begin{align}\label{eq:Ndef}
N_{ij} = \braket{a_i^\dagger(t_{\text{f}})a_j(t_{\text{f}})}_{\ket{\text{vac}}} = \bra{\text{vac}}\mathcal{U}^\dagger a_i^\dagger a_j \mathcal{U} \ket{\text{vac}} = \left( \bm{F}^*  \left[ \oplus_{\lambda=1}^\ell \sinh^2r_\lambda \right]  \bm{F}^T \right)_{ij}.
\end{align}
Finally, based on the decompositions of~\eqref{eq:singular-values} we can easily write
\begin{align}
\bm{K} = \begin{pmatrix}
\bm{F} & \bm{0} \\
\bm{0} & \bm{F}^*
\end{pmatrix}
\begin{pmatrix}
\oplus_{\lambda=1}^\ell \cosh r_{\lambda} & \oplus_{\lambda=1}^\ell \sinh r_{\lambda} \\
\oplus_{\lambda=1}^\ell \sinh r_{\lambda} & \oplus_{\lambda=1}^\ell \cosh r_{\lambda}
\end{pmatrix}
\begin{pmatrix}
\bm{G} & \bm{0} \\
\bm{0} & \bm{G}^*
\end{pmatrix},
\end{align}
and the inverse
\begin{align}\label{eq:backheisenberg}
\bm{K}^{-1} = \begin{pmatrix}
\bm{G}^\dagger & \bm{0} \\
\bm{0} & \bm{G}^T
\end{pmatrix}
\begin{pmatrix}
\oplus_{\lambda=1}^\ell \cosh r_{\lambda} & -\oplus_{\lambda=1}^\ell \sinh r_{\lambda} \\
-\oplus_{\lambda=1}^\ell \sinh r_{\lambda} & \oplus_{\lambda=1}^\ell \cosh r_{\lambda}
\end{pmatrix}
\begin{pmatrix}
\bm{F}^\dagger & \bm{0} \\
\bm{0} & \bm{F}^T
\end{pmatrix}=  \begin{pmatrix}
\tilde{\bm{V}} & \tilde{\bm{W}} \\
\tilde{\bm{W}}^* & \tilde{\bm{V}}^*
\end{pmatrix},
\end{align}
where we have introduced
\begin{align}
\tilde{\bm{V}} = \bm{G}^\dagger \left[ \oplus_{\lambda=1}^\ell \cosh r_\lambda \right] \bm{F}^\dagger = \bm{V}^\dagger, \quad \tilde{\bm{W}} = -\bm{G}^\dagger \left[ \oplus_{\lambda=1}^\ell \sinh r_\lambda \right] \bm{F}^T =  - \bm{W}^T,
\end{align}
for the blocks of the matrix $\bm{K}^{-1}$. These matrices are quite useful when later we need to write the backwards-evolved Heisenberg operators
\begin{align}
\mathcal{U}(t_{\text{f}},t_{\text{i}})
\begin{pmatrix}
\bm{a}(t_{\text{i}}) \\
\bm{a}^\dagger(t_{\text{i}})
\end{pmatrix}  \mathcal{U}^\dagger(t_{\text{f}},t_{\text{i}}) = \bm{K}^{-1} \begin{pmatrix}
\bm{a}(t_{\text{i}}) \\
\bm{a}^\dagger(t_{\text{i}})
\end{pmatrix} = \begin{pmatrix}
\tilde{\bm{V}} & \tilde{\bm{W}} \\
\tilde{\bm{W}}^* & \tilde{\bm{V}}^*
\end{pmatrix} \begin{pmatrix}
\bm{a}(t_{\text{i}}) \\
\bm{a}^\dagger(t_{\text{i}})
\end{pmatrix}.
\end{align}
Finally, we note that these transformations, that preserve the commutation relations, are called symplectic transformations. They form a group and have been extensively studied \cite{serafini2017quantum,simon1988gaussian,dutta1995real,adesso2014continuous,weedbrook2012gaussian}; we provide some details of their properties in Appendix~\ref{app:symplectic}.

\subsection{Obtaining the ket from the Bogoliubov transformations}\label{sec:theket}
In the previous section we showed how to solve the Heisenberg equations of motion, and, based on the fact that they correspond to unitary operations in Hilbert space, we derived some of their properties. In this section we show how one can obtain the state that evolved under $\mathcal{U}$. Before doing this we introduce a useful object that uniquely describes a quantum state, its characteristic function~\cite{barnett2002methods}
\begin{align}
\chi_{\rho}(\bm{\alpha}) = \text{tr}\left( \mathcal{D}(\bm{\alpha}) \rho \right),
\end{align}
where $\rho$ is the density matrix of quantum system. For a pure state $\rho = \ket{\Psi} \bra{\Psi}$ and we write its characteristic function as $\chi_{\ket{\Psi}}(\bm{\alpha})$. The displacement operator is defined as~\cite{cahill1969ordered}
\begin{align}
\mathcal{D}(\bm{\alpha}) &= \exp\left[ \bm{\alpha}^T \bm{a}^\dagger - \bm{\alpha}^\dagger \bm{a} \right] = \exp\left[ \bm{\alpha}^T \bm{a}^\dagger \right] \exp\left[ - \bm{\alpha}^\dagger \bm{a} \right] \exp[ -\tfrac12 ||\bm{\alpha} ||^2], \\
\bm{\alpha} &= (\alpha_1,\ldots,\alpha_M)^T \in \mathbb{C}^\ell,
\end{align}
and in the last line we used a well-known disentangling formula (cf. Appendix 5 of Barnett and Radmore~\cite{barnett2002methods}).

\begin{figure}
    \centering
    \includegraphics[width=1.0\textwidth]{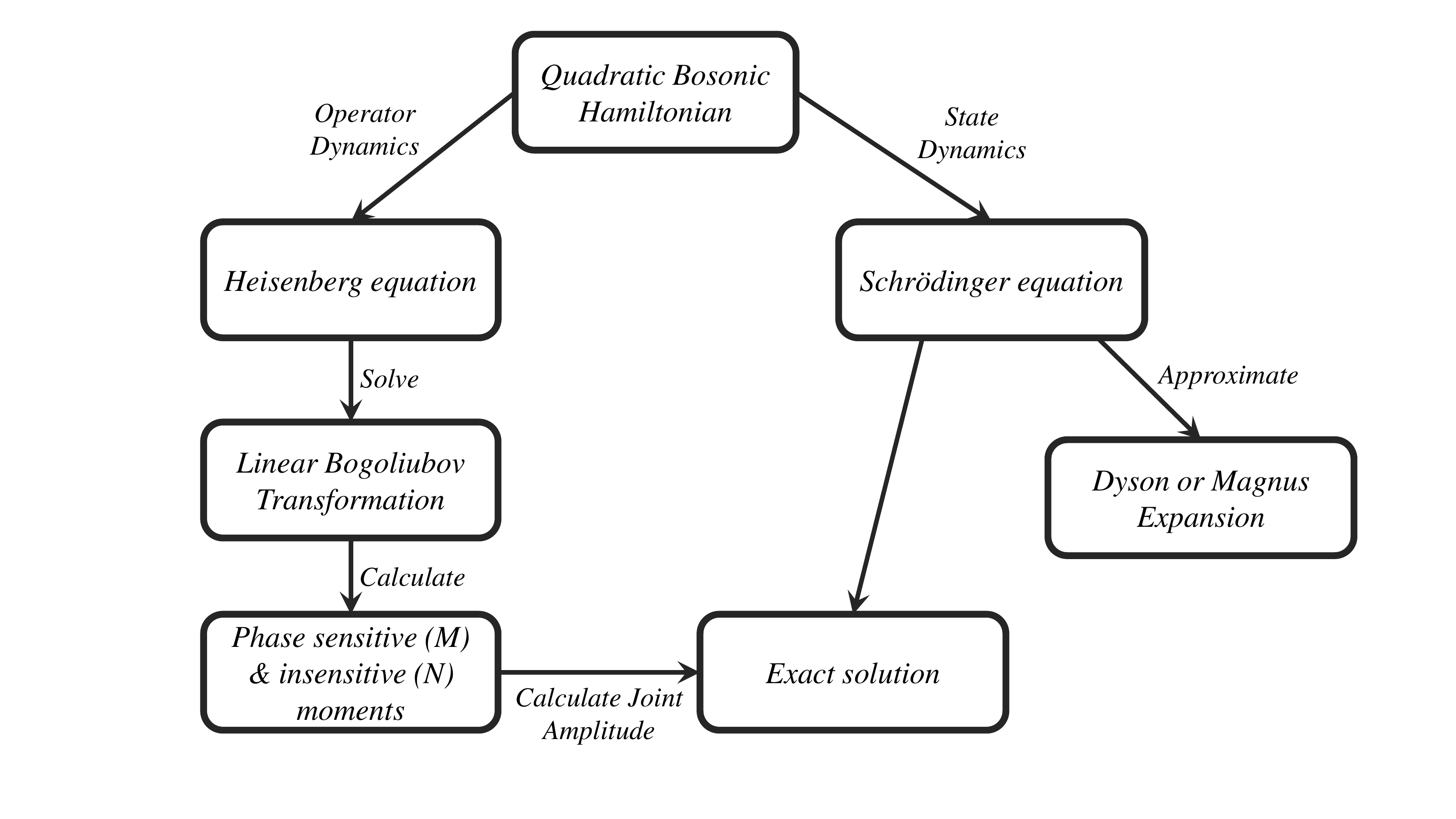}
    \caption{Block diagram describing how to calculate the state generated by a quadratic bosonic Hamiltonian. }
    \label{fig:flow}
\end{figure}

For the vacuum state the characteristic function is easily calculated, and found to be a Gaussian in the complex amplitudes $\bm{\alpha}$
\begin{align}
\chi_{\ket{\text{vac}}}(\bm{\alpha}) =  \exp[ -\tfrac12 ||\bm{\alpha} ||^2],
\end{align}
where we used the disentangling formula for the displacement operator and the cyclic property of the trace $\text{tr}(A B C) = \text{tr}(BCA)$ together with $ \bra{\text{vac}} \exp\left[ \bm{\alpha} \bm{a}^\dagger \right] = \bra{\text{vac}}$ and $ \exp\left[ - \bm{\alpha}^* \bm{a} \right] \ket{\text{vac}} = \ket{\text{vac}}$. 
The vacuum state is the simplest state that has a Gaussian characteristic function. More generally, any state that has a Gaussian characteristic function is called a Gaussian state, and we will show now that the state $\mathcal{U} \ket{\text{vac}}$ is a member of this set. To this end consider
\begin{align}\label{eq:charfunc}
\chi_{\mathcal{U}\ket{\text{vac}}}(\bm{\alpha}) &=  \text{tr}\left( \mathcal{D}(\bm{\alpha}) \mathcal{U}\ket{\text{vac}} \bra{\text{vac}} \mathcal{U}^\dagger  \right) \\
&= \text{tr}\left(\mathcal{U}^\dagger \mathcal{D}(\bm{\alpha}) \mathcal{U}\ket{\text{vac}} \bra{\text{vac}}   \right) \nonumber\\
& = \text{tr}\left(\mathcal{U}^\dagger \exp\left[ \bm{\alpha}^T \bm{a}^\dagger - \bm{\alpha}^\dagger \bm{a} \right]\mathcal{U}\ket{\text{vac}} \bra{\text{vac}}   \right) \nonumber \\
& = \text{tr}\left( \exp\left[ \bm{\alpha}^T \mathcal{U}^\dagger \bm{a}^\dagger \mathcal{U} - \bm{\alpha}^\dagger \mathcal{U}^\dagger \bm{a} \mathcal{U}\right]\ket{\text{vac}} \bra{\text{vac}}   \right) \nonumber \\
& = \text{tr}\left( \exp\left[ \bm{\alpha}^T \left\{ \bm{V}^*\bm{a}^\dagger + \bm{W}^* \bm{a} \right\}   - \bm{\alpha}^\dagger \left\{ \bm{V}\bm{a} + \bm{W} \bm{a}^\dagger \right\}\right]\ket{\text{vac}} \bra{\text{vac}}   \right) \nonumber \\
& = \text{tr}\left( \exp\left[ \left\{ \bm{\alpha}^T  \bm{V}^* - \bm{\alpha}^\dagger \bm{W}   \right\} \bm{a}^\dagger   -  \left\{   \bm{\alpha}^\dagger\bm{V} -  \bm{\alpha}^T \bm{W}^* \right\}  \bm{a}  \right]\ket{\text{vac}} \bra{\text{vac}}   \right) \nonumber \\
& = \text{tr}\left( \mathcal{D}( \bm{V}^\dagger \bm{\alpha}   -  \bm{W}^T \bm{\alpha}^* ) \ket{\text{vac}} \bra{\text{vac}} \right)\nonumber \\
& = \exp[ -\tfrac12 || \bm{\xi} ||^2], \quad \bm{\xi} = \bm{V}^\dagger \bm{\alpha}   -  \bm{W}^T \bm{\alpha}^*. \nonumber
\end{align}
We can now write
\begin{align}
|| \bm{\xi}||^2 =  \bm{\xi}^\dagger \bm{\xi} = \bm{\alpha}^\dagger \bm{V} \bm{V}^\dagger \bm{\alpha} - \bm{\alpha}^\dagger \bm{V} \bm{W}^T \bm{\alpha}^* - \bm{\alpha}^T \bm{W}^* \bm{V}^\dagger \bm{\alpha} + \bm{\alpha}^T \bm{W}^* \bm{W}^T \bm{\alpha}^*,
\end{align}
and using the singular value decompositions \eqref{eq:singular-values} we find
\begin{align}
\bm{V} \bm{V}^\dagger  &= \bm{F} \left[  \oplus_{\lambda=1}^\ell \cosh^2 r_{\lambda}  \right] \bm{F}^\dagger = \mathbb{I}_{M}+ \bm{N}^*, \\
 \bm{V} \bm{W}^T & = \bm{F} \left[  \oplus_{\lambda=1}^\ell \tfrac12 \sinh 2r_{\lambda}  \right] \bm{F}^T = \left( \bm{W}^* \bm{V}^\dagger \right)^\dagger = \bm{M},\\
  \bm{W}^* \bm{W}^T  & = \bm{F}^* \left[  \oplus_{\lambda=1}^\ell \sinh^2 r_{\lambda}  \right] \bm{F}^T = \bm{N},
\end{align}
from which we conclude that $\mathcal{U}\ket{\text{vac}}$ only depends on the left singular vectors $\bm{F}$ and the singular values $r_{\lambda}$, or equivalently, on the moments $\bm{M}$ and $\bm{N}$ of the state $\mathcal{U}\ket{\text{vac}}$ introduced in~\eqref{eq:Mdef} and~\eqref{eq:Ndef}, respectively.
Explicitly we have
\begin{align}
\chi_{\mathcal{U}\ket{\text{vac}}}(\bm{\alpha}) = \exp\left[- \tfrac12 \left\{ ||\bm{\alpha}||^2 + \bm{\alpha}^\dagger \bm{N}^* \bm{\alpha} - \bm{\alpha}^\dagger \bm{M} \bm{\alpha}^* - \bm{\alpha}^T \bm{M}^* \bm{\alpha} + \bm{\alpha}^T \bm{N} \bm{\alpha}^*  \right\}\right].
\end{align}
Based on this observation we propose that\footnote{Note that equation above does not imply that $\mathcal{U} = \exp\left[ \tfrac12 \sum_{k,l=1}^\ell J_{kl} a^\dagger_{k} a^\dagger_l  - \text{H.c.}\right] $, in particular we can postmultiply the exponential by any unitary operator that satisfies $\mathcal{U}_{0} \ket{\text{vac}} = \ket{\text{vac}}$ and still satisfy~\eqref{eq:final_ket}.
 	The form of the operator $\mathcal{U}_{0} $ is described in Appendix~\ref{app:form_of_U_0}.}
\begin{align}\label{eq:final_ket}
\ket{\Psi(t_{\text{f}})}& \equiv \mathcal{U}\ket{\text{vac}} = \exp\left[ \frac12 \sum_{k,l=1}^\ell J_{kl} a^\dagger_{k} a^\dagger_l  - \text{H.c.}\right]\ket{\text{vac}},\\
\bm{J} &= \bm{F} \left[ \oplus_{\lambda=1}^\ell  r_{\lambda}  \right] \bm{F}^T = \bm{J}^T,
\end{align}
where $\bm{J}$ is the (arbitrary gain) joint-amplitude of the squeezed state. To show that the equation above holds we need to show that the $\bm{M}$ and $\bm{N}$ moments when calculated with the right-hand side of the equation above coincide with the ones obtained for the state $\mathcal{U}\ket{\text{vac}}$. Since these uniquely determine the characteristic function, this is straightforward to do by reusing the results from Sec.~\ref{sec:low-gain} but now using the joint amplitude defined in the last equation. In fact, noting that the solution above has exactly the same \emph{form} as the low gain solution found at the end of Sec.~\ref{sec:low-gain}, we can write exactly the same equation as in that section but now removing the overbars to indicate an arbitrary gain solution. Additionally, just like in the low gain regime, we can introduce Schmidt operators and squeezing parameters. The fundamental difference is that here the joint amplitude is determined by the Takagi-Autonne singular values and vectors of the matrix moment $\bm{M}$ as defined in~\eqref{eq:Mdef} which is obtained by solving the Heisenberg equations of motion. The general procedure to solve for the ket is schematically represented in Fig.~\ref{fig:flow}.

\subsection{Non-degenerate squeezing}\label{sec:twin-beams}
As we will see later, it is often useful to split our $\ell$-modes into two separate groups, called the signal and idler modes, for which we write ladder operators $b_i$ and $c_j$ respectively satisfying canonical commutation relations
\begin{align}
[b_i, b_j] = [c_i,c_j] = [b_i,c_j] = [b_i,c_j^\dagger] = 0, \quad [b_i,b_j^\dagger] = [c_i,c_j^\dagger] = \delta_{ij}.
\end{align}
The signal and idlers have respectively $\ell_b$ and $\ell_c$ modes ($\ell_b+\ell_c = \ell$), and we write their non-degenerate squeezing Hamiltonian as
\begin{align}\label{eq:twin_hamil}
H(t) = \hbar \left\{\sum_{i,j=1}^{\ell} \Delta^{b}_{ij}(t) b_i^\dagger b_j + \sum_{i,j=1}^{\ell} \Delta^{c}_{ij}(t) c_i^\dagger c_j +  i \left[  \sum_{i=1}^{\ell_b} \sum_{j=1}^{\ell_c} \zeta^{b,c}_{ij}(t) b_i^\dagger c_j^\dagger - \text{H.c.} \right] \right\}.
\end{align}
The separation between the $b$ and $c$ modes is physically motivated by observing that there might be situations where two different polarizations, or spectral regions that are widely separated and have disjoint support, interact. It is then convenient to assign different operators to the two sets of modes participating in the Hamiltonian. Whenever this is the case, we call the process described by the Hamiltonian above non-degenerate squeezing; this is to contrast it with the degenerate squeezing discussed in the previous section.

The Hamiltonian in~\eqref{eq:twin_hamil} looks superficially different to the one in~\eqref{eq:quad_hamil} but it is easy to show that the former is just a special case of the latter by identifying
\begin{align}
\bm{a} = \begin{pmatrix}
\bm{b} \\
\bm{c}
\end{pmatrix}, \quad  
\bm{b} = \begin{pmatrix}
b_1 \\
\vdots \\
b_{\ell_b}
\end{pmatrix}, \quad 
\bm{c} = \begin{pmatrix}
c_1 \\
\vdots \\
c_{\ell_c}
\end{pmatrix}, \quad 
\bm{\Delta}(t) = \begin{pmatrix}
\bm{\Delta}_b(t) & \bm{0} \\
\bm{0} & \bm{\Delta}_c(t)
\end{pmatrix}, \quad \bm{\zeta}(t)  = \begin{pmatrix}
\bm{0} & \bm{\zeta}^{b,c}(t)\\
[\bm{\zeta}^{b,c}(t)]^T & \bm{0}
\end{pmatrix}.
\end{align}
Having written this problem in the general framework of the previous sections, we could simply use the techniques developed before to solve it. However, it is often more useful to study this problem separately from the general problem discussed before. 
Thus we start by looking at perturbative solutions, where we will assume for simplicity that $\bm{\Delta}^{b,c}(t)   = 0 $. Note that if these terms are time independent then one can without loss of generality transform to a rotating frame where the only term appearing in the Hamiltonian is the one corresponding to $\bm{\zeta}^{b,c}$. Under these assumptions we can use the Magnus series at the first order to write the perturbative solution to the spontaneous problem as~\cite{christ2013theory,branczyk2011time,grice1997spectral}
\begin{align}\label{eq:low-gain-twin}
\ket{\overline{\Psi}(t_\text{f})}&\approx \overline{\mathcal{U}}(t_\text{f}, t_\text{i}) \ket{\text{vac}} \\
\overline{\mathcal{U}}(t_\text{f}, t_\text{i})&= \exp\left(-\frac{i}{\hbar} \int_{t_{\text{i}}}^{t_\text{f}} \dd t' H(t') \right) =   \exp\left[  \sum_{i=1}^{\ell_b} \sum_{j=1}^{\ell_c} \overline{J}^{b,c}_{ij} b_i^\dagger c_j^\dagger  - \text{H.c.}\right],
\end{align}
where $\overline{\bm{J}}^{b,c} = \int_{t_{\text{i}}}^{t_{\text{f}}} dt\  \bm{\zeta}^{b,c}(t)$ is the joint amplitude of the non-degenerate squeezed state. As before,  it is often useful to rewrite this state by making use of a decomposition of $\bm{J}$, here the singular value decomposition (SVD)
\begin{align}
\overline{\bm{J}}^{b,c} = \overline{\bm{F}}^{b} \ \overline{\bm{D}} \  [\overline{\bm{F}}^{c} ]^T,
\end{align}
where $\overline{\bm{F}}^{b/c}$ are $\ell_{b/c} \times \ell_{b/c}$ unitary matrices and $\bm{D}$ is an $\ell_b \times \ell_{c}$ diagonal matrix with entries $\{\overline{r}_1,\ldots, \overline{r}_{\ell_{\min} } \}$ with $\ell_{\min} = \min(\ell_b, \ell_c)$.
We can use this SVD to introduce Schmidt modes for the signal and idler operators 
\begin{align}
\overline{B}_\lambda^\dagger = \sum_{k=1}^{\ell_b}	b_k^\dagger \overline{F}^b_{k,\lambda} \Longleftrightarrow b_k^\dagger = \sum_{i=1}^{\ell_b} [\overline{F}^b]^*_{k,\lambda} \overline{B}_{\lambda}^\dagger ,\\
\overline{C}_\lambda^\dagger = \sum_{k=1}^{\ell_c}	c_k^\dagger \overline{F}^c_{k,\lambda} \Longleftrightarrow c_k^\dagger = \sum_{i=1}^{\ell_c} [\overline{F}^c]^*_{k,\lambda} \overline{C}_{\lambda}^\dagger.
\end{align}
The Schmidt operators satisfy bona fide bosonic commutation relations and we can finally write the output ket as
\begin{align}
\ket{\overline{\Psi}(t_{\text{f}})} &= \exp\left[  \sum_{\lambda=1}^{\ell_{\min}} \overline{r}_{\lambda} \overline{B}_{\lambda}^\dagger \overline{C}_{\lambda}^\dagger  - \text{H.c.}\right] \ket{\text{vac}} = \bigotimes_{\lambda=1}^{\ell_{\min}} \exp\left[   \overline{r}_{\lambda} \overline{B}_{\lambda}^\dagger \overline{C}_{\lambda}^\dagger  - \text{H.c.}\right] \ket{\text{vac}} \\
&= \bigotimes_{\lambda=1}^{\ell_{\min}} \left[ \frac{1}{\cosh \overline{r}_\lambda} \sum_{n=0}^\infty \tanh^n \overline{r}_\lambda \ket{n,n}_{\lambda}  \right],
\end{align}
where in the last line we have used a well-known disentangling identity (cf. Appendix 5 of Barnett and Radmore~\cite{barnett2002methods})
\begin{align}
\exp(r e^{i \phi} b^\dagger c^\dagger -\text{H.c.}) =&  \exp\left( e^{i \phi}   \tanh r   \ c^{\dagger } b^\dagger\right) \\
& \times \exp\left( - \left[ c^\dagger c+ b^\dagger b +1\right] \ln \cosh r  \right) \nonumber \\
& \times \exp\left( - e^{-i \phi} \tanh r \ c b\right), \nonumber 
\end{align}
and introduced
\begin{align}
\ket{n,n}_{\lambda} = \frac{\overline{B}_\lambda^{\dagger n} \overline{C}_\lambda^{\dagger n}}{n!} \ket{\text{vac}}.
\end{align}

We can also use the low-gain perturbative solution in~\eqref{eq:low-gain-twin} to transform the operators, for which we find
\begin{align}\label{eq:twin-low-gain-fwd-heisen}
b_i(t_{\text{f}}) &= \overline{\mathcal{U}}^\dagger b_i(t_{\text{i}}) \overline{\mathcal{U}} = \sum_{j=1}^\ell \overline{V}^{b,b}_{ij} b_{j}(t_{\text{i}}) + \sum_{j=1}^\ell \overline{W}^{b,c}_{ij} c^\dagger _{j} (t_{\text{i}}), \\
c_i^\dagger(t_{\text{f}}) &= \overline{\mathcal{U}}^\dagger c_i^\dagger(t_{\text{i}}) \overline{\mathcal{U}} = \sum_{j=1}^\ell \left[\overline{V}^{(c,c)}_{ij}\right]^* c^\dagger_{j}(t_{\text{i}}) + \sum_{j=1}^\ell \left[\overline{W}^{c,b}_{ij}\right]^* b _{j}(t_{\text{i}}).
\end{align}
where now 
\begin{align}\label{eq:twin-beam-fwd-low-gain}
\overline{\bm{V}}^{b,b} &= \overline{\bm{F}}^{b} \left[ \oplus_{\lambda=1}^\ell \cosh \overline{r}_{\lambda} \right] \left[\overline{\bm{F}}^{b} \right]^\dagger, 
\quad 
\overline{\bm{V}}^{(c,c)} = \overline{\bm{F}}^{c} \left[ \oplus_{\lambda=1}^\ell \cosh \overline{r}_{\lambda} \right] \left[\overline{\bm{F}}^{c} \right]^\dagger, \\
\overline{\bm{W}}^{b,c} &= \overline{\bm{F}}^{b} \left[ \oplus_{\lambda=1}^\ell \sinh \overline{r}_{\lambda} \right] \left[\overline{\bm{F}}^{c} \right]^T \quad  \overline{\bm{W}}^{c,b} = \overline{\bm{F}}^{c} \left[ \oplus_{\lambda=1}^\ell \sinh \overline{r}_{\lambda} \right] \left[\overline{\bm{F}}^{b} \right]^T = \left[ \overline{\bm{W}}^{b,c} \right]^T.
\end{align}
To lowest non-vanishing order in the squeezing parameters $\overline{r}_\lambda \ll 1$ one has 
\begin{align}\label{eq:nondegen_lowgain_regime}
\overline{\bm{V}}^{b,b} \approx \mathbb{I}_{\ell_b}, \overline{\bm{V}}^{c,c} \approx \mathbb{I}_{\ell_c} \text{ and } \overline{\bm{W}}^{b,c} = \left[\overline{\bm{W}}^{c,b} \right]^T \approx \overline{\bm{J}}^{b,c}.
\end{align}
Since the inverse of the unitary operator $\overline{\mathcal{U}}$ can be obtained by reversing the sign of $\overline{\bm{J}}^{b,c}$ inside the exponential, it is straightforward to show the backwards-Heisenberg evolved operators $\overline{\mathcal{U}} b_i(t_{\text{i}}) \overline{\mathcal{U}} ^\dagger$ can be obtained from~\eqref{eq:twin-low-gain-fwd-heisen} by negating the sign of the terms $\overline{\bm{W}}^{b,c}$ and $\overline{\bm{W}}^{c,b}$.

\begin{figure}
    \centering
    \includegraphics[width=1.0\textwidth]{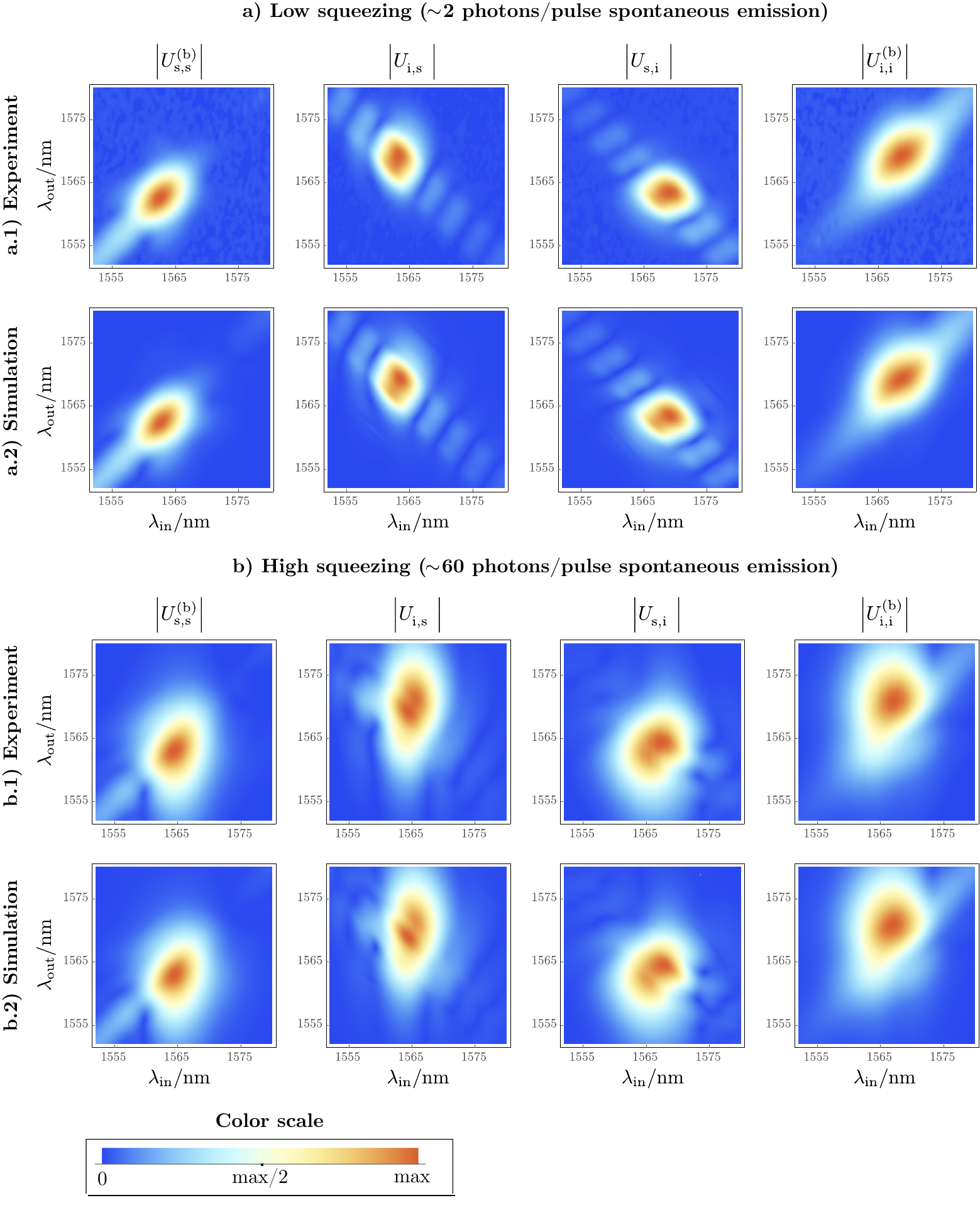}
    \caption{Experimental and theoretical prediction for the blocks of the transfer function $\bm{K}^{b,c}$ for non-degenerate squeezing in a periodically poled potassium titanyl phosphate spontaneous parametric down-conversion source. The labels $U_{s,s}, U_{i,s},  U_{s,i}, U_{i,i}$ in the figure correspond with $V_{b,b}, W_{c,b},  W_{b,c}, W_{c,c}$ in the notation used in this manuscript. The figure is reproduced with attribution to Triginer et al.~\cite{triginer2019understanding} under the terms of the Creative Commons Attribution 4.0 International license.}
    \label{fig:transferfunctions_high_gain}
\end{figure}

We can now move to the high gain-regime, where we will find that the solution to the spontaneous problem has exactly the same form as in the low gain-regime, but now the joint-amplitude of the non-degenerate squeezed state will be inferred from the SVD of a second order moment of the operators. Like before, we write the equations of motion for $\bm{b}$ and $\bm{c}^\dagger$ as
\begin{align}
\ddt{} \begin{pmatrix}
\bm{b} \\
\bm{c}^\dagger
\end{pmatrix} = -i \underbrace{\begin{pmatrix}
	- \bm{\Delta}_b(t)  & i \bm{\zeta}^{b,c}(t) \\
	i [\bm{\zeta}^{b,c}(t)]^\dagger  &  \bm{\Delta}_c(t) 
	\end{pmatrix} }_{\equiv \bm{A}^{b,c}(t)}  \begin{pmatrix}
\bm{b} \\
\bm{c}^\dagger
\end{pmatrix}.
\end{align}
The solution to this equation can be found with the same techniques as before, yielding
\begin{align}
\begin{pmatrix}
\bm{b}(t_{\text{f}}) \\
\bm{c}^\dagger(t_{\text{f}})
\end{pmatrix} =& \mathcal{T}\exp\left[ - i \int_{t_{\text{i}}}^{t_\text{f}} \dd t \bm{A}^{b,c}(t) \right] \begin{pmatrix}
\bm{b}(t_{\text{i}}) \\
\bm{c}^\dagger(t_{\text{i}})
\end{pmatrix} =
\begin{pmatrix}
\bm{V}^{b,b} & \bm{W}^{b,c} \\
(\bm{W}^{c,b})^* & (\bm{V}^{c,c})^*
\end{pmatrix}
\begin{pmatrix}
\bm{b}(t_{\text{i}}) \\
\bm{c}^\dagger(t_{\text{i}})
\end{pmatrix} \\
=& 
\bm{K}^{b,c}(t_{\text{f}},t_{\text{i}}) \begin{pmatrix}
\bm{b}(t_{\text{i}}) \\
\bm{c}^\dagger(t_{\text{i}})
\end{pmatrix}. \nonumber
\end{align}
These transfer functions can be accessed using classical intensity measurements as shown in Fig.~\ref{fig:transferfunctions_high_gain}. By correctly modelling the different terms in the equations of motion one can obtain excellent agreement between theory and simulation.

Using this input-output relation we can now calculate the (non-zero) second order moments
\begin{align}
N^b_{ij} = \bra{\text{vac}} b_i^\dagger(t_{\text{f}}) b_j(t_{\text{f}}) \ket{\text{vac}} = \bra{\text{vac}} \mathcal{U}^\dagger b_i^\dagger b_j \mathcal{U} \ket{\text{vac}},\\
N^c_{ij} = \bra{\text{vac}} c_i^\dagger(t_{\text{f}}) c_j(t_{\text{f}}) \ket{\text{vac}} = \bra{\text{vac}} \mathcal{U}^\dagger c_i^\dagger c_j \mathcal{U} \ket{\text{vac}},\\
M^{b,c}_{ij} = \bra{\text{vac}} b_i(t_{\text{f}}) c_j(t_{\text{f}}) \ket{\text{vac}} = \bra{\text{vac}} \mathcal{U}^\dagger b_i c_j \mathcal{U} \ket{\text{vac}},
\end{align}
and verify that the following moments are identically zero
\begin{align}
N^{b,c}_{ij} =  \bra{\text{vac}} b_i^\dagger(t_{\text{f}}) c_j(t_{\text{f}}) \ket{\text{vac}} = \bra{\text{vac}} \mathcal{U}^\dagger b_i^\dagger c_j \mathcal{U} \ket{\text{vac}} = 0,\\
M^{b}_{ij} = \bra{\text{vac}} b_i(t_{\text{f}}) b_j(t_{\text{f}}) \ket{\text{vac}} = \bra{\text{vac}} \mathcal{U}^\dagger b_i b_j \mathcal{U} \ket{\text{vac}} =0,\\
M^{c}_{ij} = \bra{\text{vac}} c_i(t_{\text{f}}) c_j(t_{\text{f}}) \ket{\text{vac}} = \bra{\text{vac}} \mathcal{U}^\dagger c_i c_j \mathcal{U} \ket{\text{vac}} =0.
\end{align}
Note that if we identify again the $\bm{b}$ and $\bm{c}$ modes as a subset of the larger set of $\bm{a}$ modes we can write the moments of the latter as
\begin{align}\label{eq:twin-into-degenerate}
\bm{N} = \begin{pmatrix}
\bm{N}^b & \bm{0} \\
\bm{0} & \bm{N}^c
\end{pmatrix}, \quad \bm{M} = \begin{pmatrix}
\bm{0} & \bm{M}^{b,c} \\
[\bm{M}^{b,c}]^T &  \bm{0}
\end{pmatrix}. 
\end{align}

Finally, after having the moments we simply need to find the SVD~\cite{triginer2019understanding}
\begin{align}\label{eq:SVDM}
\bm{M}^{b,c} = {\bm{F}}^{b} \ {\bm{D}} \  [{\bm{F}}^{c} ]^T,
\end{align}
where $\bm{F}^{b/c}$ are $\ell_{b/c} \times \ell_{b/c}$ unitary matrices and $\bm{D}$ is an $\ell_b \times \ell_{c}$ diagonal matrix with entries $\{\tfrac12 \sinh 2{r}_1,\ldots, \tfrac12 \sinh 2{r}_{\ell_{\min} } \}$ where $\ell_{\min} = \min(\ell_b, \ell_c)$.

Note that the output of the spontaneous problem is also a Gaussian state, characterized by its second order moments and that we can write the output ket as
\begin{align}
\ket{\Psi(t_{\text{f}})} = \mathcal{U} \ket{\text{vac}} = \exp\left[  \sum_{i=1}^{\ell_b} \sum_{j=1}^{\ell_c} {J}^{b,c}_{ij} b_i^\dagger c_j^\dagger  - \text{H.c.}\right] \ket{\text{vac}},
\end{align}
where now $\bm{J}^{b,c} =  {\bm{F}}^{b} \ {\bm{R}} \  [{\bm{F}}^{c} ]^T$ where $\bm{R}$ is an $\ell_b \times \ell_{c}$ diagonal matrix with entries $\{{r}_1,\ldots, {r}_{\ell_{\min} } \}$ that is directly related to the matrix $\bm{D}$ in the SVD of $\bm{M}^{b,c}$ of~\eqref{eq:SVDM}. This form is functionally the same as the ket in the low-gain regime, and thus arbitrary-gain Schmidt modes can be introduced in exactly the same way as before.

\subsection{Solution to the stimulated problem}\label{sec:estim}
Having solved the spontaneous problem associated with a general quadratic Hamiltonian, we now solve the associated stimulated problem, i.e., we study the evolution of other states different from vacuum. One can write an arbitrary input-state as
\begin{align}
\ket{\Psi(t_{\text{i}})} = f(a_i^\dagger) \ket{\text{vac}},
\end{align}
where $f$ is either a polynomial or a power series of the \emph{creation} operators. 
The parametrization above is the so-called Bargmann-Segal or stellar representation of a bosonic quantum state ~\cite{bargmann1961hilbert, segal1963mathematical,chabaud2020stellar}.
For a single photon and a coherent state we can write respectively
\begin{align}
\ket{1_{\bm{\alpha}}}  = \sum_{i=1}^\ell \alpha_i a_i^\dagger \ket{\text{vac}}, \quad \ket{\bm{\alpha}} = \mathcal{D}(\bm{\alpha})\ket{\text{vac}} =  e^{-\tfrac12 || \bm{\alpha}||^2} \exp\left[ \sum_{i=1}^\ell \alpha_i a_i^\dagger \right] \ket{\text{vac}}.
\end{align}
We can then write the output ket as
\begin{align}\label{eq:inputs1}
\ket{\Psi(t_{\text{f}})} = \mathcal{U}\ket{\Psi(t_{\text{i}})} = \mathcal{U} f(a_i^\dagger)\mathcal{U}^\dagger \mathcal{U}\ket{\text{vac}},
\end{align}
the part $\mathcal{U}\ket{\text{vac}}$ corresponds to the spontaneous problem solved in the previous section. For the second part,  $\mathcal{U} f(a_i^\dagger)\mathcal{U}^\dagger$ we can write
\begin{align}
\mathcal{U} f(a_i^\dagger)\mathcal{U}^\dagger =  f(\mathcal{U} a_i^\dagger \mathcal{U}^\dagger) = f\left(\sum_{k=1}^\ell \tilde{V}^*_{ik} a_k^\dagger + \sum_{k=1}^\ell \tilde{W}^*_{ik} a_k \right),
\end{align}
where in the last line we used the fact that $\mathcal{U} O^m \mathcal{U}^\dagger = \left[ \mathcal{U} O \mathcal{U}^\dagger \right]^m$
and the backwards Heisenberg transformation of~\eqref{eq:backheisenberg}. Going back to the initial states considered before in~\eqref{eq:inputs1} we find that an input single photon becomes a superposition of a photon added and a photon subtracted squeezed state
\begin{align}
\mathcal{U}\ket{1_{\bm{\alpha}}} = \left(\sum_{k=1}^\ell \tilde{V}^*_{ik} a_k^\dagger + \sum_{k=1}^\ell \tilde{W}^*_{ik} a_k \right) \mathcal{U} \ket{\text{vac}},
\end{align}
while for an input coherent state, the reader can confirm that
\begin{align}
\mathcal{U}\ket{\bm{\alpha}} = \mathcal{D}\left( \bm{V}\bm{\alpha} +  \bm{W} \bm{\alpha}^* \right) \mathcal{U} \ket{\text{vac}}
\end{align}
and thus the output is a displaced-squeezed state, where the input coherent displacement is transformed precisely like the bosonic operators are [cf.~\eqref{eq:in-out}]. Note also that the information about the joint spectral amplitude of the squeezed part of the state $\mathcal{U} \ket{\text{vac}}$ is imprinted in the displacement since it depends on $\bm{V}$ and $\bm{W}$ which uniquely determine $\bm{J}$. 
For the case of non-degenerate squeezing this connection can be made more explicit in the low-gain regime discussed at the beginning of Sec.~\ref{sec:twin-beams}. Consider the following coherent state input in the idler modes:
\begin{align}
\ket{\bm{\nu}} = \mathcal{D}_c(\bm{\nu}) \ket{\text{vac}} = \exp\left[ \bm{\nu}^T \bm{c}^\dagger - \bm{\nu}^\dagger \bm{c} \right] \ket{\text{vac}}.
\end{align}
Using the results from that section, and to lowest non-vanishing order in the squeezing parameters we find
\begin{align}
\overline{\mathcal{U}} \ket{\bm{\nu}} &= \overline{\mathcal{U}} \exp\left[ \bm{\nu}^T \bm{c}^\dagger - \bm{\nu}^\dagger \bm{c} \right] \overline{\mathcal{U}}^\dagger \overline{\mathcal{U}} \ket{\text{vac}} \\
&=\exp\left[ \bm{\nu}^T \overline{\mathcal{U}}  \bm{c}^\dagger \overline{\mathcal{U}}^\dagger - \bm{\nu}^\dagger \overline{\mathcal{U}}\bm{c}\overline{\mathcal{U}}^\dagger \right]  \overline{\mathcal{U}} \ket{\text{vac}}.\nonumber
\end{align}
Recalling that the backwards-Heisenberg operator $ \overline{\mathcal{U}}  \bm{c}^\dagger \overline{\mathcal{U}}^\dagger$ can be obtained from the forward-Heisenberg operators in the low-gain regime~\eqref{eq:twin-low-gain-fwd-heisen}, \eqref{eq:nondegen_lowgain_regime} by letting $r_\lambda \to -r_\lambda$ (equivalently letting $\bm{J}^{b,c} \to  - \bm{J}^{b,c}$) we obtain
\begin{align}
\overline{\mathcal{U}} \ket{\bm{\nu}} &=\exp\left[ \bm{\nu}^T \left(\bm{c}^\dagger - [\overline{\bm{J}}^{c,b}]^* \bm{b} \right) - \bm{\nu}^\dagger \left(\bm{c} - \overline{\bm{J}}^{c,b} \bm{b}^\dagger \right) \right]  \overline{\mathcal{U}} \ket{\text{vac}}  \\
&=\exp\left[ \bm{\nu}^T \bm{c}^\dagger - \bm{\nu}^\dagger \bm{c}  \right]
\exp\left[ \bm{\nu}^\dagger  \overline{\bm{J}}^{c,b}   \bm{b}^\dagger - \bm{\nu}^T [\overline{\bm{J}}^{c,b}]^* \bm{b}  \right] \overline{\mathcal{U}} \ket{\text{vac}}  \nonumber\\
&= \mathcal{D}_{\bm{c}}(\bm{\nu}) \mathcal{D}_{\bm{b}}(    [\overline{\bm{J}}^{c,b}]^T \bm{\nu}^*) \overline{\mathcal{U}} \ket{\text{vac}} \nonumber\\
& =  \mathcal{D}_{\bm{c}}(\bm{\nu}) \mathcal{D}_{\bm{b}}(    \overline{\bm{J}}^{b,c} \bm{\nu}^*) \exp\left[  \sum_{i=1}^{\ell_b} \sum_{j=1}^{\ell_c} \overline{J}^{b,c}_{ij} b_i^\dagger c_j^\dagger  - \text{H.c.}\right] \ket{\text{vac}}. \nonumber
\end{align}
Note that now the joint amplitude of the non-degenerate squeezed state not only determines the squeezed-part of the state but it also appears as a response function telling us the displacement imparted to the signal modes ($\bm{b}$) when the idler modes ($\bm{c}$) are seeded by a coherent displacement amplitude $\bm{\nu}$.

\subsection{Passive operations and Loss}\label{sec:loss}
An important set of processes in the evolution of bosonic modes are given by passive operations, i.e., linear operations where the energy is not increased. In these process, the modes of an environment prepared in the vacuum couple to the modes of interest via a beamsplitter-like interaction, which in the Heisenberg picture we write as~\cite{garcia2019simulating,brod2020classical,thomas2021general}
\begin{align}
\bm{a}(t_\text{f}) = \mathcal{U}_{\text{BS}}^\dagger \bm{a}(t_{\text{i}}) \mathcal{U}_{\text{BS}} = \bm{L} \bm{a}(t_{\text{i}}) + \sqrt{\mathbb{I}_\ell - \bm{L} \bm{L}^\dagger } \ \bm{e},
\end{align}
where $\bm{L}$ is a loss matrix, that in order to represent a physical process satisfies that all its singular values are bounded by 1 and $\bm{e}$ are bosonic destruction operators of the environment.  Note in particular that if the losses are uniform, then $\bm{L} = \sqrt{\eta} \mathbb{I}_\ell$ with $0 \leq \eta \leq 1$ being the net energy transmission. 
The case of non-lossy operation is obtained when the matrix $\bm{L}$ is unitary, in which case $\bm{L} = \bm{U}$ and thus
\begin{align}
\bm{a}(t_\text{f}) = \mathcal{U}_{\text{BS}}^\dagger \bm{a}(t_{\text{i}}) \mathcal{U}_{\text{BS}} = \bm{U} \bm{a}(t_{\text{i}}).
\end{align}

Since the transformations above are linear it is direct to show that if the input state at time $t_{\text{i}}$ is Gaussian then the output state is also Gaussian. This implies that the output state is uniquely characterized by the $\bm{M}$ and $\bm{N}$ moments. We can see how they transform by writing for example
\begin{align}
M_{ij} (t_{\text{f}}) &= \text{tr}\left( a_i a_j \mathcal{U}_{\text{BS}} \left[ \rho_{S} \otimes \ket{\text{vac}_e}  \bra{\text{vac}_e} \right] \mathcal{U}_{\text{BS}}^\dagger  \right) \\
&=\text{tr}\left( \mathcal{U}_{\text{BS}}^\dagger  a_i a_j \mathcal{U}_{\text{BS}} \left[ \rho_{S} \otimes \ket{\text{vac}_e}  \bra{\text{vac}_e} \right]  \right) \nonumber\\
&=\text{tr}\left( \left[\bm{L} \bm{a}(t_{\text{i}}) + \sqrt{\mathbb{I}_\ell - \bm{L} \bm{L}^\dagger } \ \bm{e}\right]_i \left[\bm{L} \bm{a}(t_{\text{i}}) + \sqrt{\mathbb{I}_\ell - \bm{L} \bm{L}^\dagger } \ \bm{e}\right]_j \left[ \rho_{S} \otimes \ket{\text{vac}_e}  \bra{\text{vac}_e} \right]  \right) \nonumber\\
&=\text{tr}\left( \left[\bm{L} \bm{a}(t_{\text{i}}) \right]_i \left[\bm{L} \bm{a}(t_{\text{i}}) \right]_j \left[ \rho_{S} \otimes \ket{\text{vac}_e}  \bra{\text{vac}_e} \right]  \right) \nonumber\\
&=\text{tr}\left( \sum_{k=1}^\ell L_{ik} {a}_k(t_{\text{i}}) \sum_{l=1}^\ell L_{jl} {a}_l(t_{\text{i}})  \rho_{S}  \right)  = \left[ \bm{L} \bm{M}(t_{\text{i}}) \bm{L}^T\nonumber \right]_{ij}.
\end{align}
where we used the cyclic property of the trace and the fact that $\bm{e} \ket{\text{vac}_e} = 0$, and wrote $\rho_S$ for the density operator of the input modes. 
Similarly, by using the cyclic property of the trace, one finds that
\begin{align}
\bm{N}(t_{\text{f}}) = \bm{L}^* \bm{N}(t_{\text{i}}) \bm{L}^T.
\end{align}
Note that in the special case where the loss is uniform we have
\begin{align}
\bm{N}(t_{\text{f}}) =\eta  \bm{N}(t_{\text{i}}), \quad  \bm{M}(t_{\text{f}}) = \eta \bm{M}(t_{\text{i}}),
\end{align}
which gives the identity operation when $\eta=1$ (no loss) and the complete loss channel when $\eta=0$. 

\subsection{Photon-number and homodyne statistics}\label{sec:pnr_homodyne}
In this section we study properties of the photon number and quadrature statistics of the Gaussian states described in the previous sections. We first look at pure lossless states of the form given in~\eqref{eq:final_ket} and ask what can be said about the statistical moments of the \emph{total} photon number observable
\begin{align}\label{eq:totalNa}
N_{\bm{a}} = \sum_{i=1}^\ell a_i^\dagger a_i.
\end{align}
While the mean photon number is simply the sum of the means of the different ``bare'' modes labelled by $i$, the variance of the total photon number is not the sum of the variances of the ``bare'' modes since these modes will be in general correlated. We can rewrite the total photon number in terms of the Schmidt operators finding
\begin{align}
N_{\bm{a}} = \sum_{\lambda=1}^\ell A_\lambda^\dagger A_\lambda,
\end{align}
and thus, since the Schmidt modes are not correlated, we can use~\eqref{eq:single_mode_moments} to obtain
\begin{align}
\braket{N_{\bm{a}}} = \sum_{\lambda=1}^{\ell} n_\lambda, \quad  \text{Var}(N_{\bm{a}}) = \sum_{\lambda=1}^{\ell} \text{Var}( A^\dagger_\lambda A_\lambda) = \sum_{\lambda=1}^{\ell} 2 n_\lambda (1+n_\lambda), \quad n_\lambda  = \braket{A^\dagger_\lambda A_\lambda} = \sinh^2 r_\lambda,
\end{align}
where we introduced the variance as $\text{Var}(O) = \braket{O^2} - \braket{O}^2$.
Similarly, for non-degenerate squeezed states we can easily use their Schmidt decomposition to write
\begin{align}
\braket{N_{\bm{b}}} &= \braket{N_{\bm{c}}} =  \sum_{\lambda=1}^{\min\{\ell_b, \ell_c\}} n_\lambda,\\
\text{Var}(N_{\bm{b}}) &= \text{Var}(N_{\bm{c}}) =  \sum_{\lambda=1}^{\min\{\ell_b, \ell_c\}} \text{Var}( B^\dagger_\lambda B_\lambda) = \sum_{\lambda=1}^{\min\{\ell_b, \ell_c\}}  n_\lambda (1+n_\lambda), \quad n_\lambda  = \braket{A^\dagger_\lambda A_\lambda} = \sinh^2 r_\lambda
\end{align}

More generally, assume that we have a state $\rho$ and that we wish to evaluate some normal-ordered moment of the form
\begin{align}
\left\langle \prod_{i=1}^\ell a_i^{\dagger m_i } \prod_{i=1}^\ell a_i^{n_i } \right\rangle.
\end{align}
These moments can be obtained by taking derivatives of the normal-ordered characteristic function $\chi^{N}(\bm{\alpha})$ which is related to the (symmetric-)ordered characteristic function introduced in~\eqref{eq:charfunc} as 
\begin{align}
\chi^{N}(\bm{\alpha}) = \chi(\bm{\alpha}) e^{\tfrac12 ||\bm{\alpha}||^2} = \left\langle  \exp\left[ \bm{\alpha}^T \bm{a}^\dagger \right] \exp\left[ - \bm{\alpha}^\dagger \bm{a} \right]  \right\rangle,
\end{align}
from which we easily confirm that
\begin{align}
\left\langle \prod_{i=1}^\ell a_i^{\dagger m_i } \prod_{i=1}^\ell a_i^{n_i } \right\rangle = \left. (-1)^{\sum_{i=1}^\ell n_i}\frac{\partial^{m_1}}{\partial \alpha_1^{m_1}} \ldots \frac{\partial^{m_\ell}}{\partial \alpha_\ell^{m_\ell}} \frac{\partial^{n_1}}{\partial \alpha_1^{*n_1}} \ldots \frac{\partial^{n_\ell}}{\partial \alpha_\ell^{* n_\ell}} \chi^{N}(\bm{\alpha}) \right|_{\bm{\alpha}=\bm{\alpha}^* = 0}.
\end{align}
For a Gaussian state (having a Gaussian characteristic function) one easily finds
\begin{align}\label{eq:moments_funcs}
\braket{n_i} &= N_{i,i},	\\
\text{Var}(n_i) &= \braket{n_i^2} - \braket{n_i}^2 = 	|M_{i,i}|^2 + N_{i,i}(N_{i,i}+1),\\
\text{Cov}(n_i,n_j) &= |M_{ij}|^2 + |N_{ij}|^2,\\
\text{Var}(n_i-n_j)&= \text{Var}(n_i) + \text{Var}(n_j) - 2 \  \text{Cov}(n_i,n_j)  \\
&= |M_{i,i}|^2 + N_{i,i}(N_{i,i}+1) + |M_{j,j}|^2 + N_{j,j}(N_{j,j}+1) - 2 \left( |M_{ij}|^2 + |N_{ij}|^2 \right),\nonumber\\
\text{Var}(N) &= \sum_{i=1}^\ell \text{Var}(n_i) + 2 \sum_{i<j}^\ell \text{Cov}(n_i,n_j), 
\end{align}
with $n_i = a_i^\dagger a_i$ and $n_i^2 = a^\dagger_i a_i a^\dagger_i a_i = a_i^{\dagger 2}a_i^2 + a^\dagger_i a_i$. 

Let us apply these results to study so-called coherence functions~\cite{laiho2021measuring}. For a beam with destruction operators $b_i$ the second-order coherence function is defined as
\begin{align}\label{eq:g2def}
g^{(2)}_{\bm{b}} = \frac{\braket{N_{\bm{b}}^2 - N_{\bm{b}}}}{\braket{N_{\bm{b}}}^2} = \frac{\sum_{i=1}^{\ell_b} \sum_{j=1}^{\ell_b} \braket{b_i^\dagger b_j^\dagger b_i b_j} }{\left( \sum_{i=1}^{\ell_b} \braket{b_i^\dagger b_i}\right)^2} ,
\end{align}
where in analogy to \eqref{eq:totalNa} we define the total photon number of the signal modes $\bm{b}$ as $N_{\bm{b}} = \sum_{i=1}^{\ell_b} b_i^\dagger b_i$. Note that a similar quantity can be introduced for the idler modes by letting $b \to c$ in the last equation.
One can also introduce the cross-beam second order coherence function as
\begin{align}\label{eq:g11def}
g^{(1,1)}_{\bm{b},\bm{c}} = \frac{\braket{N_{\bm{b}} N_{\bm{c}}}}{\braket{N_{\bm{b}}} \braket{N_{\bm{c}}}} = \frac{\sum_{i=1}^{\ell_b} \sum_{j=1}^{\ell_c} \braket{b_i^\dagger c_j^\dagger b_i c_j} }{\left( \sum_{i=1}^{\ell_b} \braket{b_i^\dagger b_i}\right) \left( \sum_{i=1}^{\ell_c} \braket{c_i^\dagger c_i}\right)} .
\end{align}
The first interesting property of these quantities is that they are not affected by a loss. 
Assume for example that a beam is prepared in certain quantum state $\rho_{\bm{b}}$ and then undergoes uniform loss by energy transmission $\eta_{\bm{b}}$ as described in Sec.~\ref{sec:loss}. It is straighforward to see that since the numerator and denominator of ~\eqref{eq:g2def} are normal ordered in the creation and annihilation operators of the beam $\bm{b}$ then each of them will pick up a factor of $\eta_{\bm{b}}^2$ that will cancel out. Similarly, for the case of the cross-beam second order correlation function, if beams $\bm{b}$ and $\bm{c}$ undergo uniform loss by energy transmissions $\eta_{\bm{b}}$ and $\eta_{\bm{c}}$ respectively, then both the numerator and denominator in the right-hand side of ~\eqref{eq:g11def} will pick up a factor $\eta_{\bm{b}} \eta_{\bm{c}}$ that will cancel out. We emphasize that these results are true irrespective of the state prepared before loss. We will now use these results to investigate the correlation functions of degenerate and non-degenerate squeezed light. 
For non-degenerate squeezed states it is easy to show~\cite{christ2011probing}
\begin{align}
g^{(2)}_{\bm{b}}  = g^{(2)}_{\bm{c}} = 1 + \frac{1}{K}, \text{ where } K = \frac{ \left[ \sum_{\lambda=1}^{\min(\ell_c,\ell_b)} \braket{A_\lambda^\dagger A_\lambda} \right]^2 }{\sum_{\lambda=1}^{\min(\ell_c,\ell_b)} \braket{A_\lambda^\dagger A_\lambda}^2}
\end{align}
The quantity $K$ is the so called Schmidt number of the of non-degenerate squeezed state. Note that this quantity is equal to one only if one of the Schmidt modes is occupied $\braket{A_1^\dagger A_1} >0$ and all the rest ($\lambda >1$) have zero photons $\braket{A_\lambda^\dagger A_\lambda} =0$.  If more than one Schmidt is occupied then $K>1$ and $g^{(2)}_{\bm{b}/\bm{c}} <2$.
Similarly, we find that the cross correlation function is given by
\begin{align}
g^{(1,1)}_{\bm{b},\bm{c}} = 1 + \frac{1}{K} + \frac{1}{\braket{N}}
\end{align}
where we wrote the total photon number of either the signal or idler beam as $\braket{N} = \sum_{\lambda=1}^{\min(\ell_c,\ell_b)} \braket{A_\lambda^\dagger A_\lambda}$.
Note that this is the total photon number before any propagation loss occurs, thus, by subtracting the second-order correlation function of either of the beams from the cross-correlation coherence function one has access to the absolute number of photons generated by the source. These can be compared with the total average photon numbers measured in the detectors allowing to obtain the inefficiencies, or energy transmission factors,  $\eta_{\bm{b}},\eta_{\bm{c}}$; if we write the measured mean photon numbers after transmission $\eta_{\bm{b}/\bm{c}}$ as
\begin{align}
\braket{N_{\bm{b}/\bm{c}}} = \eta_{\bm{b}/\bm{c}} \braket{N},
\end{align}
then we can obtain an absolute determination of the transmission by forming
\begin{align}
\eta_{\bm{b}/\bm{c}} = \braket{N_{\bm{b}/\bm{c}}} \left( g^{(2)}_{\bm{b}/\bm{c}} - g^{(1,1)}_{\bm{b},\bm{c}} \right),
\end{align}
which is a photon-number resolved extension of the elegant Klyshko method ~\cite{klyshko2018photons}.

The question of calculating individual photon-number probabilities of a multimode Gaussian state has led to the introduction of the Gaussian Boson Sampling problem~\cite{hamilton2017gaussian,kruse2019detailed,quesada2018gaussian_boson}. In general the evaluation of this probabilitites is computationally hard. Moreover, even the computational task of generating samples that distribute according to this probability distribution is believed to be hard (modulo reasonable complexity-theoretic assumptions~\cite{deshpande2021quantum}). The study of this problem is beyond the scope of this tutorial, however we point the reader to a number of recent theoretical~\cite{thomas2021general,bromley2020applications,quesada2020exact} and experimental advances~\cite{zhong2019experimental,zhong2020quantum}. 

One can also study homodyne measurements where now one is interested in statistics of the quadrature observable 
\begin{align} 
X_{\phi} = e^{i \phi} \sum_{i=1}^\ell \alpha_i^* a_i + \text{H.c.}
\end{align}
It is direct to see that this observable has $\braket{X_\phi} = 0$ and variance
\begin{align}
V_{\phi} = ||\bm{\alpha}||^2 + 2\bm{\alpha}^T \bm{N} \bm{\alpha}^* + \left(  e^{2i\phi} \bm{\alpha}^\dagger \bm{M} \bm{\alpha}^*  +\text{c.c} \right).
\end{align}
Note that only the phase-sensitive moment $\bm{M}$ couples to the local oscillator phase $\phi$. The phase insensitive  moment $\bm{N}$ only adds noise to the variance. Moreover, if we select $\alpha_k = F_{k,\lambda}$ for a fixed Schmidt mode $\lambda$ then the maximum and minimum variance as $\phi$ is varied are given by
\begin{align}
V_{\phi}^{\max / \min} =  e^{\pm 2 r_\lambda}
\end{align}
where $r_\lambda$ is the squeezing parameter of the Schmidt mode $\lambda$ and we have used the fact that the Schmidt mode functions have unit norm. This is exactly like the single mode case described at the end of Sec.~\ref{sec:sq_ham}.

Considering now non-degenerate squeezing, assuming the local oscillators for the signal and idler modes are given by the amplitudes $\bm{\mu}$ and $\bm{\nu}$ respectively and defining the quadrature observable
\begin{align}
X_{\phi} = e^{i \phi} \left[ \sum_{i=1}^{\ell_b} \mu_i^* b_i +  \sum_{j=1}^{\ell_c} \nu_j^* c_j\right] +  \text{H.c.},
\end{align}
we find that the variance is given by
\begin{align}
V_{\phi} = ||\bm{\nu}||^2 + ||\bm{\mu}||^2 + 2\bm{\mu}^T \bm{N}^{b} \bm{\mu}^* + 2 \bm{\nu}^T \bm{N}^{c} \bm{\nu}^* +\left[ 2 e^{2i \phi}  \bm{\mu}^\dagger \bm{M}^{b,c} \bm{\nu}^* + \text{c.c.} \right].
\end{align}
To obtain this result we use~\eqref{eq:twin-into-degenerate} mapping a non-degenerate state into a degenerate squeezer and write the equivalent degenerate squeezer local-oscillator as amplitude $\bm{\alpha} = \left( \begin{smallmatrix} \bm{\mu} \\ \bm{\nu} \end{smallmatrix} \right)$.

\section{Heisenberg equations for waveguides}\label{sec:Waveguides}
In this section, following~\cite{helt2020degenerate} and~\cite{quesada2020theory}, we present a general formalism for considering degenerate as well as non-degenerate squeezing in waveguides where either a second- or third-order nonlinearity is dominant. The six nonlinear optical processes that this includes are spontaneous parametric downconversion (SPDC), single-pump spontaneous four-wave mixing (SFWM), and dual-pump SFWM for the creation of degenerate squeezed vacuum (DSV) states such as those in Section~\ref{sec:theket} as well as SPDC, single-pump SFWM, and dual-pump SFWM for the creation of non-degenerate squeezed vacuum (NDSV) states such as those in Section~\ref{sec:twin-beams}. Beginning with Hamiltonians of the form presented in Section~\ref{sec:Quantization}, suited to integrated photonics structures, we find that the Heisenberg equations of motion, in the limit of strong classical pumps and the undepleted pump approximation, lead to the expected classical coupled mode equations for the pump fields, as well as coupled operator equations for the generated fields. We then discretize coupled operator equations to arrive at compact equations of the form of~\eqref{eq:matrix_Heisenberg}, which can subsequently be solved according to the methods of Section~\ref{sec:bogoliubov}. Perhaps somewhat surprisingly, despite having different physical origins, all of the coupled operator equations can be cast in the same general form.

All of the processes have the same linear Hamiltonian, which we write as~\eqref{eq:channel_linear_Hamiltonian_J} plus the group velocity dispersion (GVD) term, i.e.
\begin{align}\label{eq:waveguide_linear_Hamiltonian}
\Hl&=\sum_{J}\int\dd x\,\hbar\left\{ \omega_{J}\psi_{J}^{\dagger}\left(x\right)\psi_{J}\left(x\right)+\frac{i}{2}v_{J}\left[\frac{\partial\psi_{J}^{\dagger}\left(x\right)}{\partial x}\psi_{J}\left(x\right)-\psi_{J}^{\dagger}\left(x\right)\frac{\partial\psi_{J}\left(x\right)}{\partial x}\right]\right.\\
&\left.\quad+\frac{v_{J}^{\prime}}{2}\frac{\partial\psi_{J}^{\dagger}\left(x\right)}{\partial x}\frac{\partial\psi_{J}\left(x\right)}{\partial x}+\cdots\right\},\nonumber
\end{align}
Nonlinear Hamiltonians can be constructed from~\eqref{eq:HNL_general} as in Section~\ref{sec:LinearQO}.

As a concrete example we present the relevant Hamiltonian and resulting mode and coupled operator equations for the process of single-pump (SP) SFWM to produce a NDSV state here, but note that the remaining five are presented in Appendix~\ref{sec:WG_Hamiltonians}. Assuming three relevant modes of interest $J=\left\{P,S,I\right\}$ for pump, signal, and idler, respectively, satisfying $k_{S}+k_{I}=2k_{P}$, we find that the Hamiltonian consists of a SP-SFWM term, an SPM term for the pump, an XPM term for the pump's effect on the signal, and an XPM term for the pump's effect on the idler. Putting it all together we have
\begin{align}\label{eq:H_SP-SFWM}
H_{\text{NDSV}}^{\text{SP-SFWM}}&=-\gamma_{\text{chan}}^{SIPP}\hbar^{2}\overline{\omega}_{SIPP}\overline{v}_{SIPP}^{2}\int\dd{x}\,\psi_{S}^{\dagger}\left(x\right)\psi_{I}^{\dagger}\left(x\right)\psi_{P}\left(x\right)\psi_{P}\left(x\right)+\hc\\
&\quad-\frac{\gamma_{\text{chan}}^{PPPP}}{2}\hbar^{2}\omega_{P}v_{P}^{2}\int\dd{x}\,\psi_{P}^{\dagger}\left(x\right)\psi_{P}^{\dagger}\left(x\right)\psi_{P}\left(x\right)\psi_{P}\left(x\right)\nonumber\\
&\quad-2\gamma_{\text{chan}}^{PSPS}\hbar^{2}\sqrt{\omega_{P}\omega_{S}}v_{P}v_{S}\int\dd{x}\,\psi_{P}^{\dagger}\left(x\right)\psi_{S}^{\dagger}\left(x\right)\psi_{P}\left(x\right)\psi_{S}\left(x\right)\nonumber\\
&\quad-2\gamma_{\text{chan}}^{PIPI}\hbar^{2}\sqrt{\omega_{P}\omega_{I}}v_{P}v_{I}\int\dd{x}\,\psi_{P}^{\dagger}\left(x\right)\psi_{I}^{\dagger}\left(x\right)\psi_{P}\left(x\right)\psi_{I}\left(x\right),\nonumber
\end{align}
where it is convenient to introduce a general form for the coupling coefficients, 
\begin{equation}\label{eq:gamma_chan_general}
\gamma_{\text{chan}}^{J_{1}J_{2}J_{3}J_{4}} = \frac{3\overline{\omega}_{J_{1}J_{2}J_{3}J_{4}}\overline{\chi}_{3}}{4\epsilon_{0}\left(\overline{n}_{J_{1}}\overline{n}_{J_{2}}\overline{n}_{J_{3}}\overline{n}_{J_{4}}\right)^{1/2}c^{2}}\frac{1}{A_{\text{chan}}^{J_{1}J_{2}J_{3}J_{4}}},
\end{equation}
with effective areas $A_{\text{chan}}^{J_{1}J_{2}J_{3}J_{4}}$ given by
\begin{equation}
\frac{1}{A_{\text{chan}}^{J_{1}J_{2}J_{3}J_{4}}}=\frac{\int\dd{y}\dd{z}\frac{\chi^{ijkl}_{3}\left(y,z\right)}{\overline{\chi}_{3}}\left[e_{J_{1}}^{i}\left(y,z\right)e_{J_{2}}^{j}\left(y,z\right)\right]^{*}e_{J_{3}}^{k}\left(y,z\right)e_{J_{4}}^{l}\left(y,z\right)}{\mathcal{N}_{J_{1}}\mathcal{N}_{J_{2}}\mathcal{N}_{J_{3}}\mathcal{N}_{J_{4}}},
\end{equation}
where
\begin{equation}\label{eq:v_bar_and_omega_bar}
\overline{\omega}_{J_{1}J_{2}J_{3}J_{4}}=\left(\omega_{J_{1}}\omega_{J_{2}}\omega_{J_{3}}\omega_{J_{4}}\right)^{1/4}, \quad\overline{v}_{J_{1}J_{2}J_{3}J_{4}}=\left(v_{J_{1}}v_{J_{2}}v_{J_{3}}v_{J_{4}}\right)^{1/4},
\end{equation}
and
\begin{equation}\label{eq:N_norm}
\mathcal{N}_{J}=\sqrt{\int\text{d}y\text{d}z\frac{n\left(y,z;\omega_{J}\right)/\overline{n}_{J}}{v_{g}\left(y,z;\omega\right)/v_{J}}\bm{e}_{J}^{*}\left(y,z\right)\cdot\bm{e}_{J}\left(y,z\right)}.
\end{equation}
One can readily identify $\gamma_{\text{chan}}^{SIPP}$ as characterizing the strength of SP-SFWM, $\gamma_{\text{chan}}^{PPPP}$ as characterizing the strength of self-phase modulation of the pump, and $\gamma_{\text{chan}}^{PSPS}$ and $\gamma_{\text{chan}}^{PIPI}$ as characterizing the strength of the cross-phase modulations of the signal and the idler, respectively, by the pump. Comparing with our earlier expressions ~\eqref{eq:gamma_SPM_chan_result} and \eqref{eq:gamma_XPM_chan_result}, we see that special cases of the coefficients $\gamma_{\text{chan}}^{J_1J_2J_3J_4}$ are related to the standard coefficient describing self-phase modulation of the pump $P$,
\begin{equation}
\gamma_{\text{chan}}^{\text{SPM}}=\gamma_{\text{chan}}^{PPPP},
\end{equation}
and to the standard coefficient describing cross-phase modulation of the signal $S$ by the pump $P$,
\begin{equation}
\gamma_{\text{chan}}^{\text{XPM}}=\sqrt{\frac{\omega_S}{\omega_P}}\gamma_{\text{chan}}^{PSPS}.    
\end{equation}

To simplify the equations resulting from the sum of the Hamiltonians $H^{\text{L}}$ and $H_{\text{NDSV}}^{\text{SP-SFWM}}$ above, we put
\begin{equation}\label{eq:psi_overline}
\overline{\psi}_{J}\left(x,t\right)=e^{i\omega_{J}t}\psi_{J}\left(x,t\right),
\end{equation}
such that $\overline{\psi}$ is slowly varying in both time and space. With $\omega_{S}+\omega_{I}=2\omega_{P}$, we find that the Heisenberg equations of motion yield
\begin{equation}\label{eq:NDSV_SP_SFWM_pump}
\left(\frac{\partial}{\partial t}+v_{P}\frac{\partial}{\partial x}-i\frac{v_{P}^{\prime}}{2}\frac{\partial^{2}}{\partial x^{2}}\right)\left\langle \overline{\psi}_{P}\left(x,t\right)\right\rangle =i\gamma_{\text{chan}}^{PPPP}\hbar\omega_{P}v_{P}^{2}\left|\left\langle \overline{\psi}_{P}\left(x,t\right)\right\rangle \right|^{2}\left\langle \overline{\psi}_{P}\left(x,t\right)\right\rangle, 
\end{equation}
\begin{align}\label{eq:NDSV_SP_SFWM_gen}
\left(\frac{\partial}{\partial t}+v_{S}\frac{\partial}{\partial x}-i\frac{v_{S}^{\prime}}{2}\frac{\partial^{2}}{\partial x^{2}}\right)\overline{\psi}_{S}\left(x,t\right)&=i\gamma_{\text{chan}}^{SIPP}\hbar\overline{\omega}_{SIPP}\overline{v}_{SIPP}^{2}\left\langle \overline{\psi}_{P}\left(x,t\right)\right\rangle ^{2}\overline{\psi}_{I}^{\dagger}\left(x,t\right)\\
&\quad+2i\gamma_{\text{chan}}^{PSPS}\hbar\sqrt{\omega_{P}\omega_{S}}v_{P}v_{S}\left|\left\langle \overline{\psi}_{P}\left(x,t\right)\right\rangle \right|^{2}\overline{\psi}_{S}\left(x,t\right),\nonumber\\
\left(\frac{\partial}{\partial t}+v_{I}\frac{\partial}{\partial x}-i\frac{v_{I}^{\prime}}{2}\frac{\partial^{2}}{\partial x^{2}}\right)\overline{\psi}_{I}\left(x,t\right)&=i\gamma_{\text{chan}}^{SIPP}\hbar\overline{\omega}_{SIPP}\overline{v}_{SIPP}^{2}\left\langle \overline{\psi}_{P}\left(x,t\right)\right\rangle ^{2}\overline{\psi}_{S}^{\dagger}\left(x,t\right)\nonumber\\
&\quad+2i\gamma_{\text{chan}}^{PIPI}\hbar\sqrt{\omega_{P}\omega_{I}}v_{P}v_{I}\left|\left\langle \overline{\psi}_{P}\left(x,t\right)\right\rangle \right|^{2}\overline{\psi}_{I}\left(x,t\right),\nonumber
\end{align}
where, under the assumption of strong classical pumps, we have made the approximation
\begin{equation}
\psi_{P}\left(x,t\right)\rightarrow\left\langle\psi_{P}\left(x,t\right)\right\rangle.
\end{equation}

With this example in front of us, we note that its coupled operator equations~\eqref{eq:NDSV_SP_SFWM_gen} are of the form
\begin{align}\label{eq:NDSV_COEs}
\left(\frac{\partial}{\partial t}+v_{S}\frac{\partial}{\partial x}-i\frac{v_{S}^{\prime}}{2}\frac{\partial^{2}}{\partial x^{2}}\right)\overline{\psi}_{S}\left(x,t\right)&=i\mathcal{\tilde{S}}\left(x,t\right)\overline{\psi}_{I}^{\dagger}\left(x,t\right)+2i\mathcal{\tilde{M}_{S}}\left(x,t\right)\overline{\psi}_{S}\left(x,t\right),\\
\left(\frac{\partial}{\partial t}+v_{I}\frac{\partial}{\partial x}-i\frac{v_{I}^{\prime}}{2}\frac{\partial^{2}}{\partial x^{2}}\right)\overline{\psi}_{I}\left(x,t\right)&=i\mathcal{\tilde{S}}\left(x,t\right)\overline{\psi}_{S}^{\dagger}\left(x,t\right)+2i\mathcal{\tilde{M}_{I}}\left(x,t\right)\overline{\psi}_{I}\left(x,t\right).\nonumber
\end{align}
In fact, as we show in Appendix~\ref{sec:WG_equations}, whether due to waveguide SPDC, single-pump SFWM, or dual-pump SFWM, coupled operator equations for generating NDSV states can be placed in this general form. Similarly, as shown by Helt et al. \cite{helt2020degenerate}, coupled operator equations for SPDC, single-pump SFWM, or dual-pump SFWM generating DSV states in waveguides can be written in the form
\begin{align}\label{eq:DSV_COEs}
\left(\frac{\partial}{\partial t}+v\frac{\partial}{\partial x}-i\frac{v^{\prime}}{2}\frac{\partial^{2}}{\partial x^{2}}\right)\overline{\psi}\left(x,t\right)=i\tilde{\mathcal{S}}\left(x,t\right)\overline{\psi}^{\dagger}\left(x,t\right)+2i\tilde{\mathcal{M}}\left(x,t\right)\overline{\psi}\left(x,t\right),
\end{align}
Equations for mean fields, e.g.~\eqref{eq:NDSV_SP_SFWM_pump}, can be solved using standard approaches, such as split-step Fourier or finite difference methods~\cite{agrawal2007nonlinear}. The operator equations, on the other hand, can be solved using methods first sketched in Vidrighin et al.~\cite{vidrighin2017quantum} and further elaborated by Helt et al.~\cite{helt2020degenerate} and Quesada et al.~\cite{quesada2020theory}, which we detail below.

\subsection{Solving the coupled operator equations}
Defining
\begin{align}
\omega_{J}\left(\kappa\right)&=v_{J}\kappa+\frac{v_{J}^\prime}{2}\kappa^{2},\\
\tilde{\mathcal{S}}\left(x,t\right)&=\int\dd{\kappa}\frac{e^{i\kappa x}}{\sqrt{2\pi}}\mathcal{S}\left(\kappa,t\right),\nonumber\\
\tilde{\mathcal{M}_{J}}\left(x,t\right)&=\int\dd{\kappa}\frac{e^{i\kappa x}}{\sqrt{2\pi}}\mathcal{M}_{J}\left(\kappa,t\right),\nonumber
\end{align}
where $\mathcal{M}_{J}\left(\kappa,t\right)=\mathcal{M}_{J}^{*}\left(-\kappa,t\right)$ and using
\begin{equation}
\overline{\psi}_{J}\left(x,t\right)=\int\dd{\kappa}\frac{e^{i\kappa x}}{\sqrt{2\pi}}b_{J}\left(\kappa,t\right),
\end{equation}
where we have put $b_{J}\left(\kappa,t\right)=a_{J\left(k_{J}+\kappa\right)}\left(t\right)$, we can rewrite \eqref{eq:DSV_COEs} and \eqref{eq:NDSV_COEs} as
\begin{align}\label{eq:NDSV_QFields}
\left[\frac{\partial}{\partial t}+i\omega_{S}\left(\kappa\right)\right]b_{S}\left(\kappa,t\right)&=i\int\frac{\dd{\kappa}^{\prime}}{\sqrt{2\pi}}\mathcal{S}\left(\kappa+\kappa^{\prime},t\right)b_{I}^{\dagger}\left(\kappa^{\prime},t\right)+2i\int\frac{\dd{\kappa}^{\prime}}{\sqrt{2\pi}}\mathcal{M}_{S}\left(\kappa-\kappa^{\prime},t\right)b_{S}\left(\kappa^{\prime},t\right),\\
\left[\frac{\partial}{\partial t}+i\omega_{I}\left(\kappa\right)\right]b_{I}\left(\kappa,t\right)&=i\int\frac{\dd{\kappa}^{\prime}}{\sqrt{2\pi}}\mathcal{S}\left(\kappa+\kappa^{\prime},t\right)b_{S}^{\dagger}\left(\kappa^{\prime},t\right)+2i\int\frac{\dd{\kappa}^{\prime}}{\sqrt{2\pi}}\mathcal{M}_{I}\left(\kappa-\kappa^{\prime},t\right)b_{I}\left(\kappa^{\prime},t\right),\nonumber
\end{align}
and
\begin{equation}\label{eq:DSV_QFields}
\left[\frac{\partial}{\partial t}+i\omega\left(\kappa\right)\right]b\left(\kappa,t\right)=i\int\frac{\dd{\kappa}^{\prime}}{\sqrt{2\pi}}\mathcal{S}\left(\kappa+\kappa^{\prime},t\right)b^{\dagger}\left(\kappa^{\prime},t\right)+2i\int\frac{\dd{\kappa}^{\prime}}{\sqrt{2\pi}}\mathcal{M}\left(\kappa-\kappa^{\prime},t\right)b\left(\kappa^{\prime},t\right),
\end{equation}
respectively, forms highly amenable to numeric evaluation. We note that higher-order dispersion can be considered by including more terms in the linear Hamiltonian \eqref{eq:waveguide_linear_Hamiltonian}, thus resulting in more terms in $\omega_{J}\left(\kappa\right)$ above.

To solve these equations, we discretize $\kappa_{j}=j\Delta\kappa$ and write
\begin{align}
\int\frac{\dd{\kappa}^{\prime}}{\sqrt{2\pi}}\mathcal{S}\left(\kappa+\kappa^{\prime},t\right)b_{J}^{\dagger}\left(\kappa^{\prime},t\right)
&\approx\sum_{j^{\prime}}\sqrt{\frac{\Delta\kappa}{2\pi}}\mathcal{S}\left(\kappa_{j}+\kappa_{j^{\prime}},t\right)a_{J;j^{\prime}}^{\dagger}\left(t\right),\\
\int\frac{\dd{\kappa}^{\prime}}{\sqrt{2\pi}}\mathcal{M}_{J}\left(\kappa-\kappa^{\prime},t\right)b_{J}\left(\kappa^{\prime},t\right)
&\approx\sum_{j^{\prime}}\sqrt{\frac{\Delta\kappa}{2\pi}}\mathcal{M}_{J}\left(\kappa_{j}-\kappa_{j^{\prime}},t\right)a_{J;j^{\prime}}\left(t\right),\nonumber
\end{align}
where $a_{J;j}\left(t\right)=\sqrt{\Delta\kappa}b_{J}\left(\kappa_{j},t\right)$, such that $\left[a_{J;j}\left(t\right),a^\dagger_{J^\prime;j^\prime}\left(t\right)\right]=\delta_{JJ^\prime}\delta_{jj^\prime}$. Introducing
\begin{equation}
\mathbf{A}^{J_{1}J_{2}}\left(t\right)=\left(\begin{array}{cc}
\bm{\Delta}_{J_{1}}\left(t\right) & \bm{\zeta}\left(t\right)\\
 -\bm{\zeta}^{*}\left(t\right)  & -\bm{\Delta}_{J_{2}}^{*}\left(t\right)
\end{array}\right),
\end{equation}
where
\begin{align}
\Delta_{J;jj^{\prime}}\left(t\right)&=-\frac{\omega_{J}\left(\kappa_{j}\right)}{\Delta\kappa}\delta_{jj^{\prime}}+2\sqrt{\frac{\Delta\kappa}{2\pi}}\mathcal{M}_{J}\left(\kappa_{j}-\kappa_{j^{\prime}},t\right),\nonumber\\
\zeta_{jj^{\prime}}\left(t\right)&=\sqrt{\frac{\Delta\kappa}{2\pi}}\mathcal{S}\left(\kappa_{j}+\kappa_{j^{\prime}},t\right),
\end{align}
as well as the vector $\mathbf{a}_{J}(t)$ with entries $a_{J;j}(t)$, 
then allows us to rewrite~\eqref{eq:NDSV_QFields} and~\eqref{eq:DSV_QFields} compactly as
\begin{equation}
\frac{\partial}{\partial t}\left(\begin{array}{c}
\mathbf{a}_{S}\left(t\right)\\
\mathbf{a}_{I}^{\dagger}\left(t\right)
\end{array}\right)=i\mathbf{A}^{SI}\left(t\right)\left(\begin{array}{c}
\mathbf{a}_{S}\left(t\right)\\
\mathbf{a}_{I}^{\dagger}\left(t\right)
\end{array}\right),
\end{equation}
and
\begin{equation}
\frac{\partial}{\partial t}\left(\begin{array}{c}
\mathbf{a}\left(t\right)\\
\mathbf{a}^{\dagger}\left(t\right)
\end{array}\right)=i\mathbf{A}\left(t\right)\left(\begin{array}{c}
\mathbf{a}\left(t\right)\\
\mathbf{a}^{\dagger}\left(t\right)
\end{array}\right),
\end{equation}
respectively.
Note that the matrix $\bm{\zeta}=\bm{\zeta}^{T}$ is symmetric and $\bm{\Delta}_{J}=\bm{\Delta}_{J}^{\dagger}$ is Hermitian. For a small enough propagation forward in time $\Delta t$ these have solution [recall~\eqref{eq:symplectic}]
\begin{equation}\label{eq:FwdFields}
\left(\begin{array}{c}
\mathbf{a}_{J_{1}}\left(t+\Delta t\right)\\
\mathbf{a}_{J_{2}}^{\dagger}\left(t+\Delta t\right)
\end{array}\right)=\mathbf{K}^{J_{1}J_{2}}(t)\left(\begin{array}{c}
\mathbf{a}_{J_{1}}\left(t\right)\\
\mathbf{a}_{J_{2}}^{\dagger}\left(t\right)
\end{array}\right),
\end{equation}
where the (single-time) infinitesimal propagator is defined as
\begin{equation}
\mathbf{K}^{J_{1}J_{2}}(t)=\exp\left[i\Delta t\mathbf{A}^{J_{1}J_{2}}\left(t\right)\right]=\left(\begin{array}{cc}
\mathbf{V}_{J_{1}}(t) & \mathbf{W}_{J_{1}J_{2}}(t)\\
 \mathbf{W}_{J_{2}J_{1}}^{*}(t) & \mathbf{V}_{J_{2}}^{*}(t) 
\end{array}\right).
\end{equation}
Propagation over a finite interval can then be calculated by concatenating infinitesimal propagators $\mathbf{K}^{J_{1}J_{2}}(t)$ to form the (two-argument) Heisenberg picture propagator~\cite{quesada2020theory} 
\begin{align}\label{eq:cron}
\mathbf{K}^{J_{1}J_{2}}(t,t_0) = \prod_{j=1}^\ell \mathbf{K}^{J_{1}J_{2}}(t_j) = \left(\begin{array}{cc}
\mathbf{V}_{J_{1}}(t,t_0) & \mathbf{W}_{J_{1}J_{2}}(t,t_0)\\
\mathbf{W}_{J_{2}J_{1}}^{*}(t,t_0) & \mathbf{V}_{J_{2}}^{*}(t,t_0)
\end{array}\right),
\end{align}
where $t_j = t_0+j \Delta t$ and $\Delta t = (t-t_0)/\ell$,
as
\begin{align}
\left(\begin{array}{c}
\mathbf{a}_{J_{1}}\left(t\right)\\
\mathbf{a}_{J_{2}}^{\dagger}\left(t\right)
\end{array}\right) = \mathbf{K}^{J_{1}J_{2}}(t,t_0) \left(\begin{array}{c}
\mathbf{a}_{J_{1}}\left(t_{0}\right)\\
\mathbf{a}_{J_{2}}^{\dagger}\left(t_{0}\right)
\end{array}\right),
\end{align}
and the product in~\eqref{eq:cron} is taken in chronological order.
Reverting back to continuous notation, we note that we may also write the solutions of~\eqref{eq:NDSV_QFields} and~\eqref{eq:DSV_QFields} as
\begin{align}
b_{S}(\kappa,t) &= \int \dd \kappa'\, \mathcal{V}_{S}(\kappa,\kappa';t,t_0) b_{S}(\kappa',t_0)+ \int \dd \kappa'\, \mathcal{W}_{SI}(\kappa,\kappa';t,t_0) b_{I}^\dagger(\kappa',t_0)\\
b_{I}^\dagger(\kappa,t) &= \int \dd \kappa'\, \mathcal{W}_{IS}^{*}(\kappa,\kappa';t,t_0) b_{S}(\kappa',t_0) + \int \dd \kappa'\, \mathcal{V}_{I}^{*}(\kappa,\kappa';t,t_0) b_{I}^\dagger(\kappa',t_0)\nonumber
\end{align}
and
\begin{align}
b(\kappa,t) = \int \dd \kappa'\, \mathcal{V}(\kappa,\kappa';t,t_0) b(\kappa',t_0)+ \int \dd \kappa'\, \mathcal{W}(\kappa,\kappa';t,t_0) b^\dagger(\kappa',t_0),
\end{align}
where 
\begin{align}\label{eq:conttodisc}
\mathcal{V}_{J_{1}J_{2}}(\kappa_j,\kappa_{j'};t,t_0) &= V_{J_{1}J_{2};j,j'}(t,t_0)/\Delta \kappa,\\
\mathcal{W}_{J_{1}J_{2}}(\kappa_j,\kappa_{j'};t,t_0) &= W_{J_{1}J_{2};j,j'}(t,t_0)/\Delta \kappa.\nonumber
\end{align}
We note that similar input-output relations were also derived in~\cite{mckinstrie2013quadrature, christ2013theory,horoshko2017bloch,kolobov1999spatial}, albeit with different labels. 
\subsection{Including loss}
\label{ssec:loss}
With these expressions known, we may now form the phase sensitive and phase insensitive moments (cf. Sections~\ref{sec:theket} and~\ref{sec:twin-beams})
\begin{align}\label{eq:MomentsDef}
N_{J;i,j}\left(t\right)&=\left\langle a_{J;i}^{\dagger}\left(t\right)a_{J;j}\left(t\right)\right\rangle,\\ 
M_{J_{1}J_{2};i,j}\left(t\right)&=\left\langle a_{J_{1};i}\left(t\right)a_{J_{2};j}\left(t\right)\right\rangle.\nonumber
\end{align}
We note that, because the generated state is Gaussian, it is completely characterized by these two moments. Following Section~\ref{sec:loss}, including loss is then simply a matter of properly updating the moments. In particular, for uniform waveguide loss 
\begin{align}\label{eq:FieldsUpdateLoss}
\mathbf{N}_{J}(t+\Delta t) = \eta_{J} \mathbf{N}_{J}(t), \quad \mathbf{M}_{J_{1}J_{2}}(t+\Delta t) = \sqrt{\eta_{J_{1}}\eta_{J_{2}}} \mathbf{M}_{J_{1}J_{2}}(t),
\end{align}
where
\begin{align}\label{eq:expdecay}
\eta_{J} = \exp(-\rho_{J} \Delta t),
\end{align}
and $\rho_{J}$ can be easily obtained from the standard attenuation constant $\alpha_{J}$~\cite{agrawal2007nonlinear}. 

\subsection{Example: ``Separable'' joint spectral amplitudes}
The formalism presented above is quite general in that it allows for the calculation of squeezed vacuum states in waveguides where either a second- or third-order nonlinearity is dominant, correctly accounting for dispersion to any desired order, self- and cross-phase modulation, if present, and loss, in either the so-called low- or high-gain regime. However it is also something of a black box, in that it relies on numerics. Thus, to help build intuition as one moves from the low- to the high-gain regime, here we provide an illustrative calculation of a lossless waveguide SP-SFWM process engineered to produce a separable joint spectral amplitude (JSA) in the low-gain regime ignoring group velocity dispersion, self- and cross-phase modulation under three scenarios: i) ignoring time-ordering corrections to arrive at an idealized analytic result, ii) considering leading-order time-ordering corrections via the Magnus expansion, and iii) using the full formalism presented above. We note that we neglect group velocity dispersion, self- and cross-phase modulation in iii) not because the numerical formalism cannot include them, but because doing so simplifies comparisons against the analytic results.

For completeness, using~\eqref{eq:channel_Hamiltonian} and~\eqref{eq:H_SP-SFWM} we write the linear and nonlinear Hamiltonians as
\begin{align}
\Hl &= \sum_{J=P,S,I}\int\dd{k}\,\hbar\omega_{Jk}a_{J}^{\dagger}\left(k\right)a_{J}\left(k\right)\\
\Hnl &= -\frac{\gamma_{\text{chan}}^{SIPP}\hbar^{2}\overline{\omega}_{SIPP}\overline{v}_{SIPP}^{2}}{4\pi^{2}}\\
&\quad\times\int\dd{k}_{1}\dd{k}_{2}\dd{k}_{3}\dd{k}_{4}\dd{x}\,e^{-i\left(k_{1}+k_{2}-k_{3}-k_{4}\right)x}a_{S}^{\dagger}\left(k_{1}\right)a_{I}^{\dagger}\left(k_{2}\right) a_{P}\left(k_{3}\right)a_{P}\left(k_{4}\right)+\hc,\nonumber
\end{align}
respectively. Switching to a description in terms of frequency\cite{yang2008spontaneous}, and moving to the interaction picture, we write the corresponding interaction Hamiltonian as
\begin{align}
H_{I}\left(t\right)&=-\frac{\gamma_{\text{chan}}^{SIPP}\hbar^{2}\overline{\omega}_{SIPP}}{4\pi^{2}}\int\dd{\omega}_{1}\dd{\omega}_{2}\dd{\omega}_{3}\dd{\omega}_{4}\int\dd{x}\,e^{i\left(\omega_{1}+\omega_{2}-\omega_{3}-\omega_{4}\right)t}\\
&\quad \times e^{-i\left[k_{S}\left(\omega_{1}\right)+k_{I}\left(\omega_{2}\right)-k_{P}\left(\omega_{3}\right)-k_{P}\left(\omega_{4}\right)\right]x}a_{S}^{\dagger}\left(\omega_{1}\right)a_{I}^{\dagger}\left(\omega_{2}\right)a_{P}\left(\omega_{3}\right)a_{P}\left(\omega_{4}\right)+\hc\nonumber
\end{align}
Expanding dispersion relations
\begin{equation}
k_{J}\left(\omega\right)=k_{J}+\frac{\omega-\omega_{J}}{v_{J}},
\end{equation}
and approximating the pump operators as classical waveforms
\begin{equation}
a_{P}\left(\omega\right)\rightarrow\left\langle a_{P}\left(\omega\right)\right\rangle =\sqrt{\frac{N_{P}\tau}{\left(\pi/2\right)^{1/2}}}e^{-\tau^{2}\left(\omega-\omega_{P}\right)^{2}},
\end{equation}
normalized such that $\int\dd{\omega}\left\langle a^{\dagger}_{P}\left(\omega\right)a_{P}\left(\omega\right)\right\rangle = N_{P}$, we first focus on the integration over $\omega_{3}$ and $\omega_{4}$. In particular, putting
\begin{equation}
\omega_{\pm}=\omega_{3}\pm\omega_{4},
\end{equation}
we find
\begin{align}
&\int\dd{\omega_{3}}\dd{\omega_{4}}\,e^{-i\left(\omega_{3}+\omega_{4}\right)t}e^{i\frac{\omega_{3}+\omega_{4}-2\omega_{P}}{v_{P}}x}e^{-\tau^{2}\left(\omega_{3}-\omega_{P}\right)^{2}}e^{-\tau^{2}\left(\omega_{4}-\omega_{P}\right)^{2}}\\
=&\int\dd{\omega_{+}}\,e^{-i\omega_{+}t}e^{i\frac{\omega_{+}-2\omega_{P}}{v_{P}}x}e^{-\frac{\tau^{2}}{2}\left(\omega_{+}-2\omega_{P}\right)^{2}}\int\frac{\dd{\omega_{-}}}{2}e^{-\frac{\tau^{2}}{2}\omega_{-}^{2}}\nonumber\\
=&\int\dd{\omega_{+}}\,e^{-i\omega_{+}t}e^{i\frac{\omega_{+}-2\omega_{P}}{v_{P}}x}e^{-\frac{\tau^{2}}{2}\left(\omega_{+}-2\omega_{P}\right)^{2}}\frac{\left(\pi/2\right)^{1/2}}{\tau},\nonumber
\end{align}
and thus can rewrite
\begin{align}\label{eq:wg_interaction_H}
H_{I}\left(t\right)	&=-\frac{\gamma_{\text{chan}}^{SIPP}\hbar^{2}\overline{\omega}_{SIPP}L N_{P}}{4\pi^{2}}\int\text{d}\omega_{1}\dd{\omega_{2}}\dd{\omega_{+}}e^{i\left(\omega_{1}+\omega_{2}-\omega_{+}\right)t}F\left(\omega_{1},\omega_{2},\omega_{+}\right)\\
&\quad\times e^{-\frac{\tau^{2}}{2}\left(\omega_{+}-2\omega_{P}\right)^{2}}a_{S}^{\dagger}\left(\omega_{1}\right)a_{I}^{\dagger}\left(\omega_{2}\right)+\hc,\nonumber
\end{align}
where the phase-matching function (PMF)
\begin{align}
F\left(\omega_{1},\omega_{2},\omega_{+}\right)&=\int_{-L/2}^{L/2}\frac{\dd{x}}{L}e^{-i\left(\frac{\omega_{1}-\omega_{S}}{v_{S}}+\frac{\omega_{2}-\omega_{I}}{v_{I}}-\frac{\omega_{+}-2\omega_{P}}{v_{P}}\right)x}\\
&=\text{sinc}\left[\left(\frac{\omega_{1}-\omega_{S}}{v_{S}}+\frac{\omega_{2}-\omega_{I}}{v_{I}}-\frac{\omega_{+}-2\omega_{P}}{v_{P}}\right)\frac{L}{2}\right],\nonumber
\end{align}
and we have taken the nonlinear region to extend from $-L/2$ to $L/2$. The associated time evolution operator for this Hamiltonian is
\begin{equation}
\mathcal{U}=\mathcal{T}\exp{\left(-\frac{i}{\hbar}\int_{-\infty}^{\infty} \dd{t^{\prime}}H_{I}\left(t^{\prime}\right)\right)},
\end{equation}
with $\mathcal{T}$ the time-ordering operator, which can be applied to a vacuum state to yield the state of generated photons. A similar Hamiltonian can be formed for SPDC generating NDSV states in waveguides~\cite{quesada2014effects}.

\subsubsection{The low-gain regime}
In the low-gain regime a common, and safe, approximation is to neglect time-ordering effects and instead, using $\int_{-\infty}^{\infty}\dd{t^{\prime}}\,e^{i\Delta t^{\prime}}=2\pi\delta\left(\Delta\right)$, write
\begin{align}\label{eq:naive_unitary}
\mathcal{U}&=\exp{\left(-\frac{i}{\hbar}\int_{-\infty}^{\infty} \dd{t^{\prime}}H_{I}\left(t^{\prime}\right)\right)}\\
&=\exp{\left(i\int\dd{\omega_{1}}\dd{\omega_{2}}\,J_{1;SI}\left(\omega_{1},\omega_{2}\right)a_{S}^{\dagger}\left(\omega_{1}\right)a_{I}^{\dagger}\left(\omega_{2}\right)+\hc\right)},\nonumber
\end{align}
where we have introduced the JSA neglecting time-ordering corrections
\begin{equation}
J_{1;SI}\left(\omega_{1},\omega_{2}\right)=\frac{\Phi\tau}{2\pi}F\left(\omega_{1},\omega_{2},\omega_{1}+\omega_{2}\right)e^{-\frac{\tau^{2}}{2}\left(\omega_{1}+\omega_{2}-2\omega_{P}\right)^{2}},
\end{equation}
and the dimensionless quantity
\begin{equation}
\Phi=\gamma_{\text{chan}}^{SIPP}\frac{\hbar\overline{\omega}_{SIPP}N_{P}}{\tau}L,
\end{equation}
similar to the so-called nonlinear phase shift~\cite{agrawal2007nonlinear}. Further approximating the phase-matching sinc function as a Gaussian via $\text{sinc}\left(x\right)\approx\exp\left(-sx^{2}\right)$, for $s\approx 0.193$, or in fact engineering the nonlinearity along the waveguide to produce a Gaussian phase matching function~\cite{branczyk2011engineered, bendixon2013spectral, tambasco2016domain, dosseva2016shaping, graffitti2017pure}, allows us to go further and calculate the Schmidt decomposition of $J_{1;SI}\left(\omega_{1},\omega_{2}\right)$ analytically.

Here we choose to write the Schmidt decomposition of such a JSA as
\begin{equation}
J_{SI}(\omega,\omega')=\sum_{\lambda} \xi_{\lambda} f_{\lambda;S}(\omega) f_{\lambda;I}(\omega'),    
\end{equation}
with Schmidt number
\begin{align}\label{eq:schmidt_number}
K=\frac{\left(\sum_{\lambda} \sinh^2\xi_\lambda\right)^2}{\sum_{\lambda} \sinh^4\xi_\lambda}.
\end{align}
Note that the Schmidt number of a separable JSA,
$J_{SI}(\omega,\omega')=\xi_0 f_{0;S}(\omega) f_{0;I}(\omega')$, is 1, whereas JSAs that are not separable have $K>1$. Further note that we have chosen to work with unnormalized JSAs. While this marks a departure from more common usage we find it useful here for comparing the results of this subsection against those that move beyond the approximations used in deriving $J_{1;SI}\left(\omega_{1},\omega_{2}\right)$. In the low-gain regime, where $K\approx\left(\sum_{\lambda} \xi^{2}_{\lambda}\right)^2/\sum_{\lambda} \xi^{4}_{\lambda}$, the Schmidt number is insensitive to a scaling transformation of $J_{SI}$. That is, $J_{SI}$ and $l J_{SI}$ have the same Schmidt number for any $l \neq 0$, though this does not hold in general.

Putting
\begin{align}
\Omega_{1} &= \omega_{1} - \omega_{S},\\
\Omega_{2} &= \omega_{2} - \omega_{I},\nonumber
\end{align}
we write
\begin{align}
J_{1;SI}\left(\Omega_{1},\Omega_{2}\right)&=\frac{\Phi\tau}{2\pi}e^{-s\left(\Omega_{1}v_{S}^{-1}+\Omega_{2}v_{I}^{-1}-\left(\Omega_{1}+\Omega_{2}\right)v_{P}^{-1}\right)^{2}\frac{L^{2}}{4}}e^{-\frac{\tau^{2}}{2}\left(\Omega_{1}+\Omega_{2}\right)^{2}}\\
&= \frac{\Phi\tau}{2\pi}e^{-\Omega_{1}^{2}\left(s\left(v_{S}^{-1}-v_{P}^{-1}\right)^{2}\frac{L^{2}}{4}+\frac{\tau^{2}}{2}\right)}e^{-\Omega_{2}^{2}\left(s\left(v_{I}^{-1}-v_{P}^{-1}\right)^{2}\frac{L^{2}}{4}+\frac{\tau^{2}}{2}\right)}\nonumber\\
&\quad\times e^{-2\Omega_{1}\Omega_{2}\left(s\left(v_{S}^{-1}-v_{P}^{-1}\right)\left(v_{I}^{-1}-v_{P}^{-1}\right)\frac{L^{2}}{4}+\frac{\tau^{2}}{2}\right)},\nonumber
\end{align}
where we have used the fact that $\omega_{S}+\omega_{I}=2\omega_{P}$, and see that the JSA is separable when
\begin{equation}\label{eq:separable_JSA_condition}
s\left(v_{S}^{-1}-v_{P}^{-1}\right)\left(v_{I}^{-1}-v_{P}^{-1}\right)\frac{L^{2}}{4}+\frac{\tau^{2}}{2}=0.
\end{equation}
We note, in passing, that the only square integrable function that is separable and has elliptical (or circular) contours is precisely a two dimensional Gaussian, and that moreover, only a combination of a Gaussian PMF and a Gaussian pump can give a separable JSA in the low gain regime\cite{uren2006generation,quesada2018gaussian_functions}.
Under this condition~\eqref{eq:separable_JSA_condition}, and introducing
\begin{equation}
\tau_{J}^{2}=s\left(v_{J}^{-1}-v_{P}^{-1}\right)^{2}\frac{L^{2}}{4}+\frac{\tau^{2}}{2},
\end{equation}
we write the final form of the JSA, neglecting GVD, SPM, XPM, and time-ordering corrections, that has been engineered to be separable in the low-gain regime as
\begin{equation}\label{eq:zeroth_order_Magnus}
J_{1;SI}\left(\Omega_{1},\Omega_{2}\right)=\xi_{0}f_{0;S}\left(\Omega_{1}\right)f_{0;I}\left(\Omega_{2}\right),
\end{equation}
with
\begin{equation}\label{eq:zeroth_schmidt_function}
f_{0;J}\left(\Omega\right)=\sqrt{\frac{\tau_{J}}{\left(\pi/2\right)^{1/2}}}e^{-\Omega^{2}\tau_{J}^{2}}\equiv \phi_{0;J}\left(\Omega\right),
\end{equation}
the Schmidt functions corresponding to the (only nonzero) Schmidt value
\begin{equation}\label{eq:zeorth_schmidt_value}
\xi_{0}=\frac{\Phi\tau}{2\left(2\pi\tau_{S}\tau_{I}\right)^{1/2}}\equiv \overline{\xi}.
\end{equation}

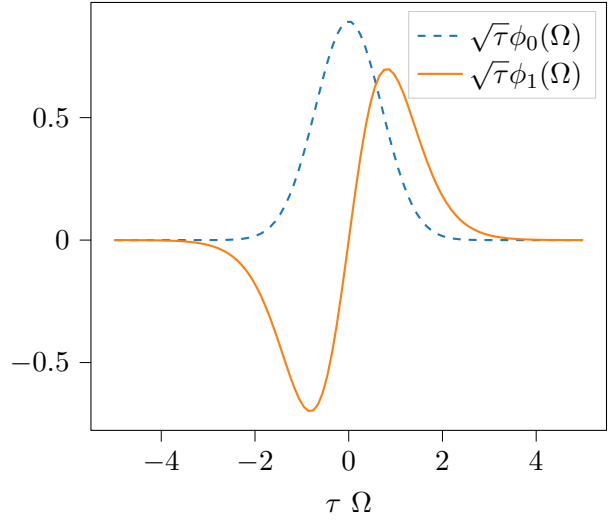
\begin{figure}
    \centering
\begin{tikzpicture}

\definecolor{color0}{rgb}{0.12156862745098,0.466666666666667,0.705882352941177}
\definecolor{color1}{rgb}{1,0.498039215686275,0.0549019607843137}

\begin{axis}[
legend cell align={left},
legend style={fill opacity=0.8, draw opacity=1, text opacity=1, draw=white!80!black},
tick align=outside,
tick pos=left,
x grid style={white!69.0196078431373!black},
xlabel={\(\displaystyle  \tau \ \Omega\)},
xmin=-5.5, xmax=5.5,
xtick style={color=black},
y grid style={white!69.0196078431373!black},
ymin=-0.776880110042291, ymax=0.970389646976529,
ytick style={color=black}
]
\addplot [thick, color0, dashed]
table {%
-5 1.24053203317822e-11
-4.8989898989899 3.37177164263223e-11
-4.7979797979798 8.97937431763067e-11
-4.6969696969697 2.34299709430886e-10
-4.5959595959596 5.99011521155585e-10
-4.49494949494949 1.50050121927535e-09
-4.39393939393939 3.68277580621953e-09
-4.29292929292929 8.85629270562056e-09
-4.19191919191919 2.08673064402455e-08
-4.09090909090909 4.81746510331727e-08
-3.98989898989899 1.08970391860995e-07
-3.88888888888889 2.4151060145671e-07
-3.78787878787879 5.2444703925439e-07
-3.68686868686869 1.11584745709875e-06
-3.58585858585859 2.32619313526221e-06
-3.48484848484848 4.7514313965417e-06
-3.38383838383838 9.50913270041528e-06
-3.28282828282828 1.86464046455274e-05
-3.18181818181818 3.58250704240988e-05
-3.08080808080808 6.74398743712624e-05
-2.97979797979798 0.000124389638027109
-2.87878787878788 0.000224796428533288
-2.77777777777778 0.000398045173032664
-2.67676767676768 0.000690578523570201
-2.57575757575758 0.00117390114418315
-2.47474747474747 0.00195518445406254
-2.37373737373737 0.00319066858682369
-2.27272727272727 0.00510168229172916
-2.17171717171717 0.00799250421026374
-2.07070707070707 0.0122684610858828
-1.96969696969697 0.0184516435944039
-1.86868686868687 0.0271905364860888
-1.76767676767677 0.0392589108966653
-1.66666666666667 0.0555387971834304
-1.56565656565657 0.0769825768116533
-1.46464646464646 0.104550513381881
-1.36363636363636 0.139122587709257
-1.26262626262626 0.181387299887797
-1.16161616161616 0.231714844172308
-1.06060606060606 0.290027105793321
-0.959595959595959 0.355681332654005
-0.858585858585859 0.427387000062137
-0.757575757575758 0.503175277791033
-0.656565656565657 0.580436926378312
-0.555555555555555 0.656037287054678
-0.454545454545455 0.726506937673316
-0.353535353535354 0.78829497727942
-0.252525252525253 0.838060756376721
-0.151515151515151 0.872971371029396
-0.0505050505050502 0.890968294384764
0.0505050505050502 0.890968294384764
0.151515151515151 0.872971371029396
0.252525252525253 0.838060756376721
0.353535353535354 0.78829497727942
0.454545454545454 0.726506937673317
0.555555555555555 0.656037287054678
0.656565656565657 0.580436926378312
0.757575757575758 0.503175277791033
0.858585858585858 0.427387000062138
0.959595959595959 0.355681332654005
1.06060606060606 0.290027105793321
1.16161616161616 0.231714844172308
1.26262626262626 0.181387299887797
1.36363636363636 0.139122587709257
1.46464646464646 0.104550513381881
1.56565656565657 0.0769825768116533
1.66666666666667 0.0555387971834303
1.76767676767677 0.0392589108966654
1.86868686868687 0.0271905364860889
1.96969696969697 0.0184516435944039
2.07070707070707 0.0122684610858828
2.17171717171717 0.00799250421026376
2.27272727272727 0.00510168229172917
2.37373737373737 0.00319066858682369
2.47474747474747 0.00195518445406254
2.57575757575758 0.00117390114418315
2.67676767676768 0.000690578523570203
2.77777777777778 0.000398045173032664
2.87878787878788 0.000224796428533288
2.97979797979798 0.000124389638027109
3.08080808080808 6.74398743712622e-05
3.18181818181818 3.58250704240988e-05
3.28282828282828 1.86464046455275e-05
3.38383838383838 9.50913270041528e-06
3.48484848484848 4.75143139654172e-06
3.58585858585859 2.32619313526219e-06
3.68686868686869 1.11584745709875e-06
3.78787878787879 5.24447039254394e-07
3.88888888888889 2.41510601456709e-07
3.98989898989899 1.08970391860995e-07
4.09090909090909 4.81746510331731e-08
4.19191919191919 2.08673064402455e-08
4.29292929292929 8.85629270562063e-09
4.39393939393939 3.68277580621949e-09
4.49494949494949 1.50050121927535e-09
4.5959595959596 5.99011521155589e-10
4.6969696969697 2.34299709430886e-10
4.7979797979798 8.97937431763067e-11
4.8989898989899 3.3717716426322e-11
5 1.24053203317822e-11
};
\addlegendentry{$\sqrt{\tau} \phi_{0}(\Omega)$}
\addplot [thick, color1]
table {%
-5 -5.31014403553053e-05
-4.8989898989899 -7.57484622730622e-05
-4.7979797979798 -0.000107379359169284
-4.6969696969697 -0.000151272691508948
-4.5959595959596 -0.00021179106711295
-4.49494949494949 -0.000294698788530154
-4.39393939393939 -0.000407559255783046
-4.29292929292929 -0.000560228430346864
-4.19191919191919 -0.000765463255662532
-4.09090909090909 -0.0010396669400388
-3.98989898989899 -0.00140379660690544
-3.88888888888889 -0.00188446323740288
-3.78787878787879 -0.00251525934978575
-3.68686868686869 -0.0033383567339516
-3.58585858585859 -0.00440642485804942
-3.48484848484848 -0.00578492986992589
-3.38383838383838 -0.00755488301998743
-3.28282828282828 -0.00981611268870503
-3.18181818181818 -0.0126911301545684
-3.08080808080808 -0.0163296361278567
-2.97979797979798 -0.0209136585261614
-2.87878787878788 -0.026663202671186
-2.77777777777778 -0.0338421100655314
-2.67676767676768 -0.0427635382107993
-2.57575757575758 -0.0537940756583989
-2.47474747474747 -0.0673549956063815
-2.37373737373737 -0.0839185616588612
-2.27272727272727 -0.103996711291945
-2.17171717171717 -0.128118994496213
-2.07070707070707 -0.156796534603559
-1.96969696969697 -0.190469246236177
-1.86868686868687 -0.22943483590131
-1.76767676767677 -0.27376040651284
-1.66666666666667 -0.323180823840815
-1.56565656565657 -0.376992180909835
-1.46464646464646 -0.43395321417423
-1.36363636363636 -0.492211567560965
-1.26262626262626 -0.549274310122438
-1.16161616161616 -0.60204196286259
-1.06060606060606 -0.646921543569931
-0.959595959595959 -0.680026355193264
-0.858585858585859 -0.697458757450527
-0.757575757575758 -0.695658238272371
-0.656565656565657 -0.671782892230572
-0.555555555555555 -0.624080660748818
-0.454545454545455 -0.552200209104318
-0.353535353535354 -0.457392341078288
-0.252525252525253 -0.342562400814786
-0.151515151515151 -0.212151550676304
-0.0505050505050502 -0.0718477787442739
0.0505050505050502 0.0718477787442739
0.151515151515151 0.212151550676304
0.252525252525253 0.342562400814786
0.353535353535354 0.457392341078288
0.454545454545454 0.552200209104317
0.555555555555555 0.624080660748818
0.656565656565657 0.671782892230572
0.757575757575758 0.695658238272371
0.858585858585858 0.697458757450527
0.959595959595959 0.680026355193264
1.06060606060606 0.646921543569931
1.16161616161616 0.60204196286259
1.26262626262626 0.549274310122438
1.36363636363636 0.492211567560965
1.46464646464646 0.43395321417423
1.56565656565657 0.376992180909835
1.66666666666667 0.323180823840815
1.76767676767677 0.27376040651284
1.86868686868687 0.22943483590131
1.96969696969697 0.190469246236177
2.07070707070707 0.156796534603559
2.17171717171717 0.128118994496213
2.27272727272727 0.103996711291945
2.37373737373737 0.0839185616588612
2.47474747474747 0.0673549956063815
2.57575757575758 0.0537940756583989
2.67676767676768 0.0427635382107993
2.77777777777778 0.0338421100655314
2.87878787878788 0.026663202671186
2.97979797979798 0.0209136585261614
3.08080808080808 0.0163296361278566
3.18181818181818 0.0126911301545684
3.28282828282828 0.00981611268870505
3.38383838383838 0.00755488301998743
3.48484848484848 0.0057849298699259
3.58585858585859 0.00440642485804941
3.68686868686869 0.0033383567339516
3.78787878787879 0.00251525934978576
3.88888888888889 0.00188446323740288
3.98989898989899 0.00140379660690544
4.09090909090909 0.0010396669400388
4.19191919191919 0.000765463255662532
4.29292929292929 0.000560228430346865
4.39393939393939 0.000407559255783044
4.49494949494949 0.000294698788530154
4.5959595959596 0.000211791067112951
4.6969696969697 0.000151272691508948
4.7979797979798 0.000107379359169284
4.8989898989899 7.57484622730619e-05
5 5.31014403553053e-05
};
\addlegendentry{$\sqrt{\tau} \phi_{1}(\Omega)$}
\end{axis}

\end{tikzpicture}

    \caption{Amplitudes of the orthonormal functions $\phi_{0}(\Omega)$ and $\phi_{1}(\Omega)$ needed to expand the time-ordering corrected joint spectral amplitude of~\eqref{eq:magnus3_JSA}.}
    \label{fig:schmidt_functs_waveguides}
\end{figure}

We note that this JSA is a two-dimensional Gaussian with elliptical contours, becoming a two-dimensional Gaussian with circular contours for $\tau_{S}=\tau_{I}=\tau$. Using ~\eqref{eq:separable_JSA_condition} to show that
\begin{equation}
\tau_{S}^{2}+\tau_{I}^{2} = s\left(v_{S}^{-1}-v_{I}^{-1}\right)^{2}\frac{L^{2}}{4},
\end{equation}
then allows interpretation of the factor
\begin{equation}
\sqrt{\frac{\tau^{2}}{\tau_{S}\tau_{I}}}=\sqrt{\frac{2\tau_{S}\tau_{I}}{\tau_{S}^{2}+\tau_{I}^{2}}}=\sqrt{\frac{2R}{1+R^{2}}}\le 1,
\end{equation}
where the aspect ratio
\begin{equation}
R = \frac{\tau_{S}}{\tau_{I}},
\end{equation}
as measuring the departure from circular contours.
\subsubsection{The Magnus expansion}
Such an analysis describes the state of generated photons well for small enough $\Phi$. However, as $\Phi$ increases, whether through an increase in $\gamma_{\text{chan}}^{SIPP}$, waveguide length $L$, or pump pulse energy via $N_{P}$, time-ordering corrections must be considered for a correct description. Using the Magnus expansion, discussed in Section~\ref{sec:multimode}, one can show that the leading-order corrections rewrite the JSA as~\cite{quesada2014effects, quesada2015time}
\begin{equation}
J_{SI}\left(\Omega_{1},\Omega_{2}\right)\approx J_{1;SI}\left(\Omega_{1},\Omega_{2}\right) + J_{3;SI}\left(\Omega_{1},\Omega_{2}\right)- i K_{3;SI}\left(\Omega_{1},\Omega_{2}\right),
\end{equation}
where the terms with subscript 3 are proportional to $\Phi^{3}$ and the next-order corrections are proportional to $\Phi^{5}$.

The quantities $J_{3;SI}$ and $K_{3;SI}$ have been calculated previously~\cite{quesada2014effects, quesada2015time, quesada2015very} and we can write them in closed form when \eqref{eq:separable_JSA_condition} holds as
\begin{align}\label{ch5:j3elip}
J_{3;SI}(\Omega_1,\Omega_2)-i K_{3;SI}(\Omega_1,\Omega_2)&=\overline{\xi}^3 \Bigg(\frac{\phi_{0;S}\left(\Omega_{1}\right)\phi_{0;I}\left(\Omega_{2}\right)-\phi_{1;S}\left(\Omega_{1}\right)\phi_{1;I}\left(\Omega_{2}\right)}{12}\nonumber\\
&\quad  -i\frac{\phi_{0;S}\left(\Omega_{1}\right)\phi_{1;I}\left(\Omega_{2}\right)-\phi_{1;S}\left(\Omega_{1}\right)\phi_{0;I}\left(\Omega_{2}\right)}{4\sqrt{3}}\Bigg),\nonumber
\end{align}
where we have introduced (recall~\eqref{eq:zeroth_schmidt_function} and see Fig.~\ref{fig:schmidt_functs_waveguides})
\begin{equation}
\phi_{1;J}\left(\Omega\right)=\sqrt{3}\phi_{0;J}\left(\Omega\right)\text{erfi}\left(\sqrt{\frac{2}{3}}\tau_{J}\Omega\right),
\end{equation}
and the imaginary error function $\text{erfi}\left(z\right)=-i\text{erf}\left(iz\right)$ with $\text{erf}$ the error function, satisfying
\begin{equation}
\int\dd{\Omega} \ \phi_{i;J}^*(\Omega) \phi_{j;J}(\Omega)=\delta_{i,j}.
\end{equation}
This allows us to write the JSA up to these corrections as
\begin{equation}\label{eq:magnus3_JSA}
J_{SI}\left(\Omega_{1},\Omega_{2}\right)\approx \mathbf{u}_{S}\left(\Omega_{1}\right)\mathbf{L}\left[\mathbf{u}_{I}\left(\Omega_{2}\right)\right]^{\dagger},
\end{equation}
where
\begin{align}
\mathbf{L}&=\left(\begin{array}{cc}
\overline{\xi}+\frac{\overline{\xi}^{3}}{12} & -i\frac{\overline{\xi}^{3}}{4\sqrt{3}}\\
i\frac{\overline{\xi}^{3}}{4\sqrt{3}} & -\frac{\overline{\xi}^{3}}{12}
\end{array}\right)\\
&=\frac{\overline{\xi}}{2}\mathbb{I}_{2}+\frac{\overline{\xi}^{3}}{4\sqrt{3}}\sigma_{2}+\left(\frac{\overline{\xi}}{2}+\frac{\overline{\xi}^{3}}{12}\right)\sigma_{3}\nonumber\\
&=\frac{\overline{\xi}}{2}\left(\mathbb{I}_{2}+\sqrt{1+\frac{\overline{\xi}^{2}}{3}+\frac{\overline{\xi}^{4}}{9}}\mathbf{n}\cdot\mathbf{\bm{\sigma}}\right),
\end{align}
with $\mathbb{I}_{2}$ the 2x2 identity matrix, $\bm{\sigma}=\left(\sigma_{1}, \sigma_{2}, \sigma_{3}\right)$, $\sigma_{i}$ the Pauli matrices, $\mathbf{n}=\left(0,\sin\theta,\cos\theta\right)$,
and
\begin{equation}
\mathbf{u}_{J}\left(\Omega\right)=\left(\phi_{0;J}\left(\Omega\right),\phi_{1;J}\left(\Omega\right)\right).
\end{equation}
Note that when $\Phi \ll 1$ the matrix $\mathbf{L}\approx\left(
\begin{array}{cc}
 \overline{\xi} & 0\\
 0 & 0\\
\end{array}
\right)$
and we recover the results obtained in~\eqref{eq:zeroth_order_Magnus}. More
interestingly, past this limit we can obtain the new time-ordering
corrected Schmidt values and functions and verify that they are no
longer linear functions of $\Phi$ and independent of
$\Phi$, respectively. 

Since we have written the JSA as quadratic form (the matrix $\mathbf{L}$) in the orthonormal basis $\{\phi_{i;J}\}$, obtaining the Schmidt decomposition of $J_{SI}$ is equivalent to obtaining the singular value decomposition (SVD) of the matrix $\mathbf{L}$. In particular,  because $\mathbf{L}$ is Hermitian, its SVD is related to its eigendecomposition and we can rewrite~\eqref{eq:magnus3_JSA} as
\begin{equation}
J_{SI}\left(\Omega_{1},\Omega_{2}\right)\approx \mathbf{u}^\prime_{S}\left(\Omega_{1}\right)\mathbf{L}^\prime\left[\mathbf{u}^\prime_{I}\left(\Omega_{2}\right)\right]^{\dagger},
\end{equation}
where the diagonal matrix $\mathbf{L}^\prime=\mathbf{R}^{\dagger}\left(\theta\right)\mathbf{L}\mathbf{R}\left(\theta\right)=\left(
\begin{array}{cc}
 \xi_{+} & 0\\
 0 & \xi_{-}\\
\end{array}
\right)$ with
\begin{equation}
\xi_{\pm}=\frac{\overline{\xi}}{2}\left(1\pm\sqrt{1+\frac{\overline{\xi}^{2}}{3}+\frac{\overline{\xi}^{4}}{9}}\right),
\end{equation}
and $\mathbf{u}^{\prime}_{J}\left(\Omega\right)= \mathbf{u}_{J}\left(\Omega\right) \mathbf{R}\left(\theta\right)$ with
\begin{equation}
\mathbf{R}\left(\theta\right)=\left(\begin{array}{cc}
\cos\left(\frac{\theta}{2}\right) & i\sin\left(\frac{\theta}{2}\right)\\
i\sin\left(\frac{\theta}{2}\right) & \cos\left(\frac{\theta}{2}\right)
\end{array}\right),\,\tan(\theta)=\frac{\overline{\xi}^2}{2\sqrt{3}(1+\overline{\xi}^2/6)}.
\end{equation}
Written in this form, it is easy to see that now the Schmidt values are simply the absolute values of the eigenvalues of $\mathbf{L}$ [cf.~\eqref{eq:zeorth_schmidt_value}], $\xi_0=|\xi_+|, \  \xi_1=|\xi_-|$ and the Schmidt functions are the eigenfunctions of $\mathbf{L}$
\begin{align}
f_{0;J}\left(\Omega\right)	&=\cos\left(\tfrac{\theta}{2}\right)\phi_{0;J}\left(\Omega\right)+i\sin\left(\tfrac{\theta}{2}\right)\phi_{1;J}\left(\Omega\right),\\
f_{1;J}\left(\Omega\right)	&=\cos\left(\tfrac{\theta}{2}\right)\phi_{1;J}\left(\Omega\right)+i\sin\left(\tfrac{\theta}{2}\right)\phi_{0;J}\left(\Omega\right).\nonumber
\end{align}
Thus the JSA that started as separable in the low-gain regime does not remain so as $\Phi$ increases sufficiently.

\begin{figure}
    \centering
    \includegraphics[width=0.5\textwidth]{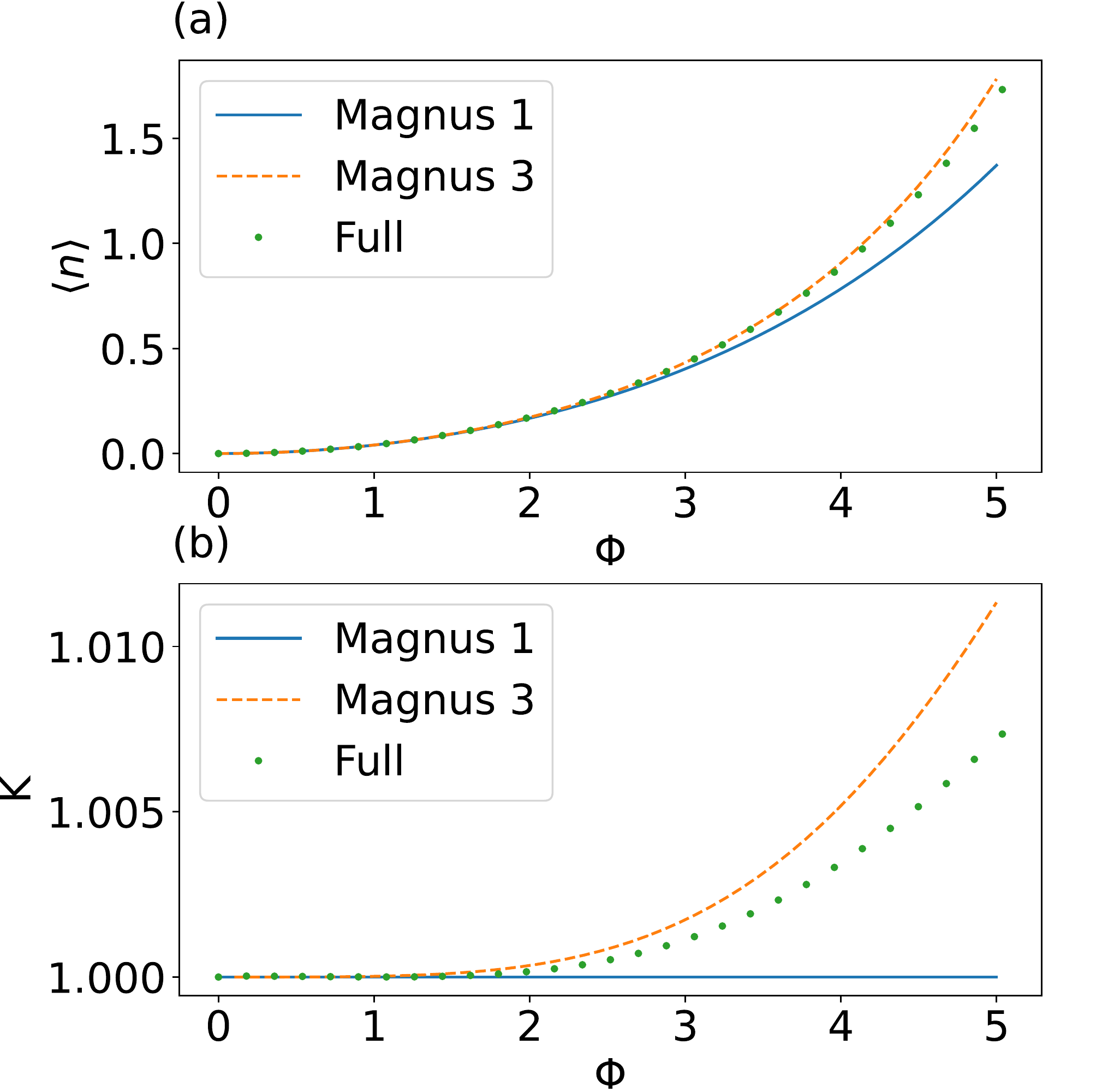}
    \caption{Comparison of the generated mean photon number per pump pulse $\langle n \rangle$ and Schmidt number $K$ for no time-ordering corrections (Magnus 1), time-ordering corrections up to $\Phi^{3}$ (Magnus 3), and a full numerical solution.}
    \label{fig:n_K_waveguides}
\end{figure}

\subsubsection{Comparisons}
\begin{figure}
    \centering
    \includegraphics[width=0.5\textwidth]{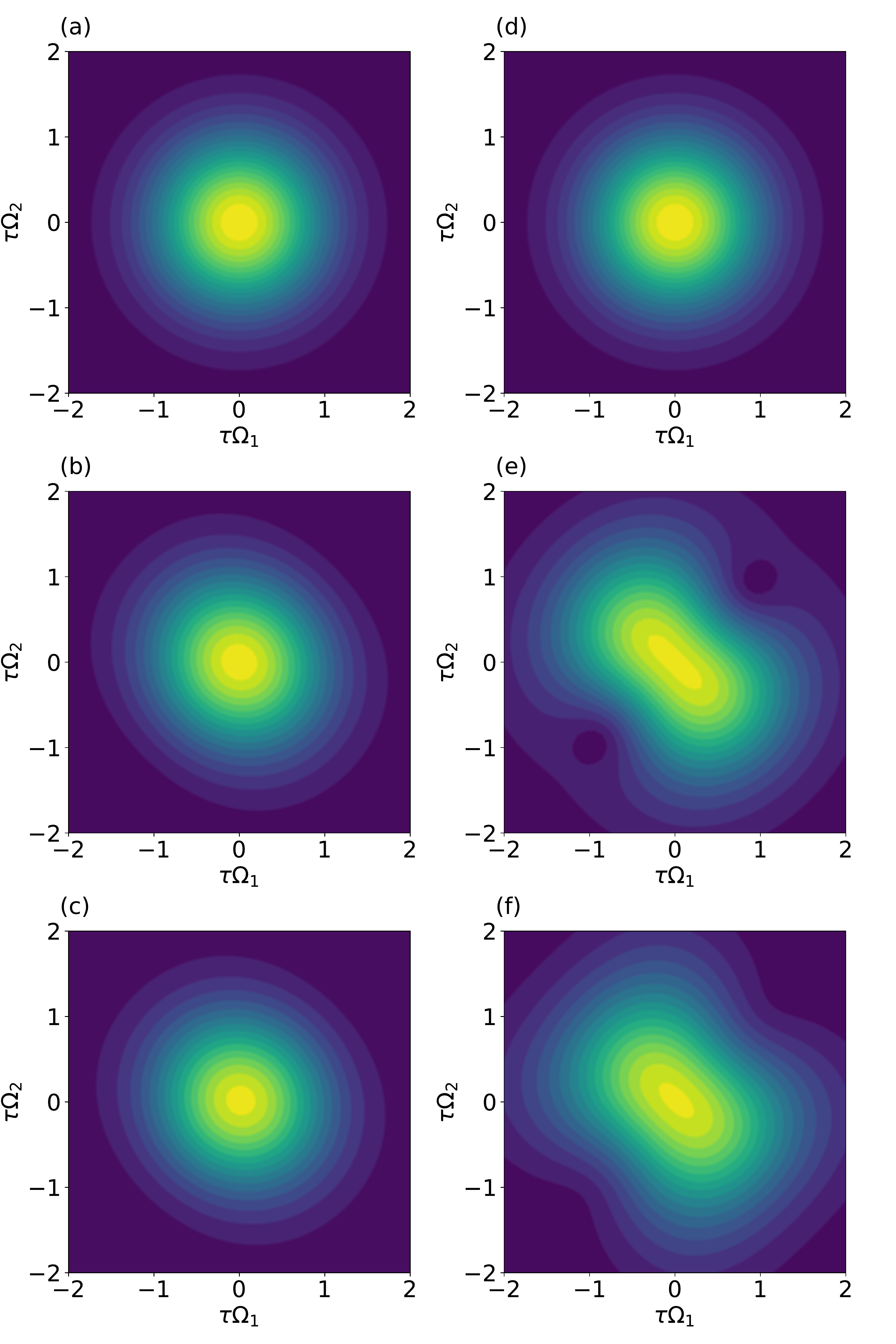}
    \caption{Comparison of the absolute values of the generated JSAs $\left\vert J_{SI}(\Omega_{1},\Omega_{2})\right\vert$ with $\Phi=4$ for no time-ordering corrections (a), time-ordering corrections up to $\Phi^{3}$ (b), and a full numerical solution (c) as well as $\Phi=11$ for no time-ordering corrections (d), time-ordering corrections up to $\Phi^{3}$ (e), and a full numerical solution (f).}
    \label{fig:JSAs_waveguides}
\end{figure}

We are now in a position to compare the average number of generated pairs per pump pulse
\begin{align}
\left\langle n_{\text{pairs}}\right\rangle&=\bvac\mathcal{U}^{\dagger}\int\text{d}\Omega\,a_{S}^{\dagger}\left(\Omega\right)a_{S}\left(\Omega\right)\mathcal{U}\vac\\
&=\bvac\mathcal{U}^{\dagger}\int\text{d}\Omega\,a_{I}^{\dagger}\left(\Omega\right)a_{I}\left(\Omega\right)\mathcal{U}\vac\nonumber\\
&=\sum_{\lambda}\sinh^{2}\xi_{\lambda},\nonumber
\end{align}
and Schmidt number~\eqref{eq:schmidt_number}, for the approximations in 4.3.1, and 4.3.2 as well as the full numerical apparatus as $\Phi$ increases. In particular, neglecting time-ordering corrections we have
\begin{equation}
\left\langle n_{\text{pairs}}\right\rangle=\sinh^{2}\overline{\xi},
\end{equation}
and
\begin{equation}
K=1,
\end{equation}
while including time-ordering corrections up to $\Phi^3$ we have
\begin{equation}
\left\langle n_{\text{pairs}}\right\rangle=\sinh^{2}\left[\frac{\overline{\xi}}{2}\left(1+\sqrt{1+\frac{\overline{\xi}^{2}}{3}+\frac{\overline{\xi}^{4}}{9}}\right)\right]+\sinh^{2}\left[\frac{\overline{\xi}}{2}\left(1-\sqrt{1+\frac{\overline{\xi}^{2}}{3}+\frac{\overline{\xi}^{4}}{9}}\right)\right],
\end{equation}
and
\begin{equation}
K=1+2\frac{\sinh^{2}\left[\frac{\overline{\xi}}{2}\left(1+\sqrt{1+\frac{\overline{\xi}^{2}}{3}+\frac{\overline{\xi}^{4}}{9}}\right)\right]\sinh^{2}\left[\frac{\overline{\xi}}{2}\left(1-\sqrt{1+\frac{\overline{\xi}^{2}}{3}+\frac{\overline{\xi}^{4}}{9}}\right)\right]}{\sinh^{4}\left[\frac{\overline{\xi}}{2}\left(1+\sqrt{1+\frac{\overline{\xi}^{2}}{3}+\frac{\overline{\xi}^{4}}{9}}\right)\right]+\sinh^{4}\left[\frac{\overline{\xi}}{2}\left(1-\sqrt{1+\frac{\overline{\xi}^{2}}{3}+\frac{\overline{\xi}^{4}}{9}}\right)\right]}.
\end{equation}
We plot these, as well as a full numerical solution, as functions of $\Phi$, setting $\tau_{S}=\tau_{I}=\tau$, so that
\begin{equation}
\overline{\xi} = \frac{\Phi}{2\left(2\pi\right)^{1/2}},
\end{equation}
for convenience, in Figs.~\ref{fig:n_K_waveguides} (a) and (b) respectively. Note how the time-uncorrected approximations only remain valid up until $\Phi\approx 2$ or so. We also plot the absolute values of the corresponding JSAs at $\Phi=4$ in Figs.~\ref{fig:JSAs_waveguides} (a), (b), and (c), as well as $\Phi=11$ in Figs.~\ref{fig:JSAs_waveguides} (d), (e), and (f).

\section{Heisenberg equations for rings}\label{sec:Rings}
In this section we build on top of Sec.~\ref{sec:Quantization} to continue the analysis
of ring resonators in the presence of second- or third-order nonlinearities,
leveraging input-output (IO) theory. Integrated optical resonators are some of
the most efficient and successful sources of quantum states of light,
thanks to the large field enhancements and the possibility to tailor
independently the density of states of each mode involved in the nonlinear
process, thereby controlling the statistical properties of the generated
quantum light \cite{rabus2007integrated, vernon2017microresonators}. A wide range of integrated source geometries have been proposed and analyzed, including Fabry-Perot
cavities, ring resonators, photonic crystal cavities, and multi-resonator
designs, each with a variety of coupling and pumping schemes.

\begin{figure}[ht]\centering
    \includegraphics[width=1.0\textwidth]{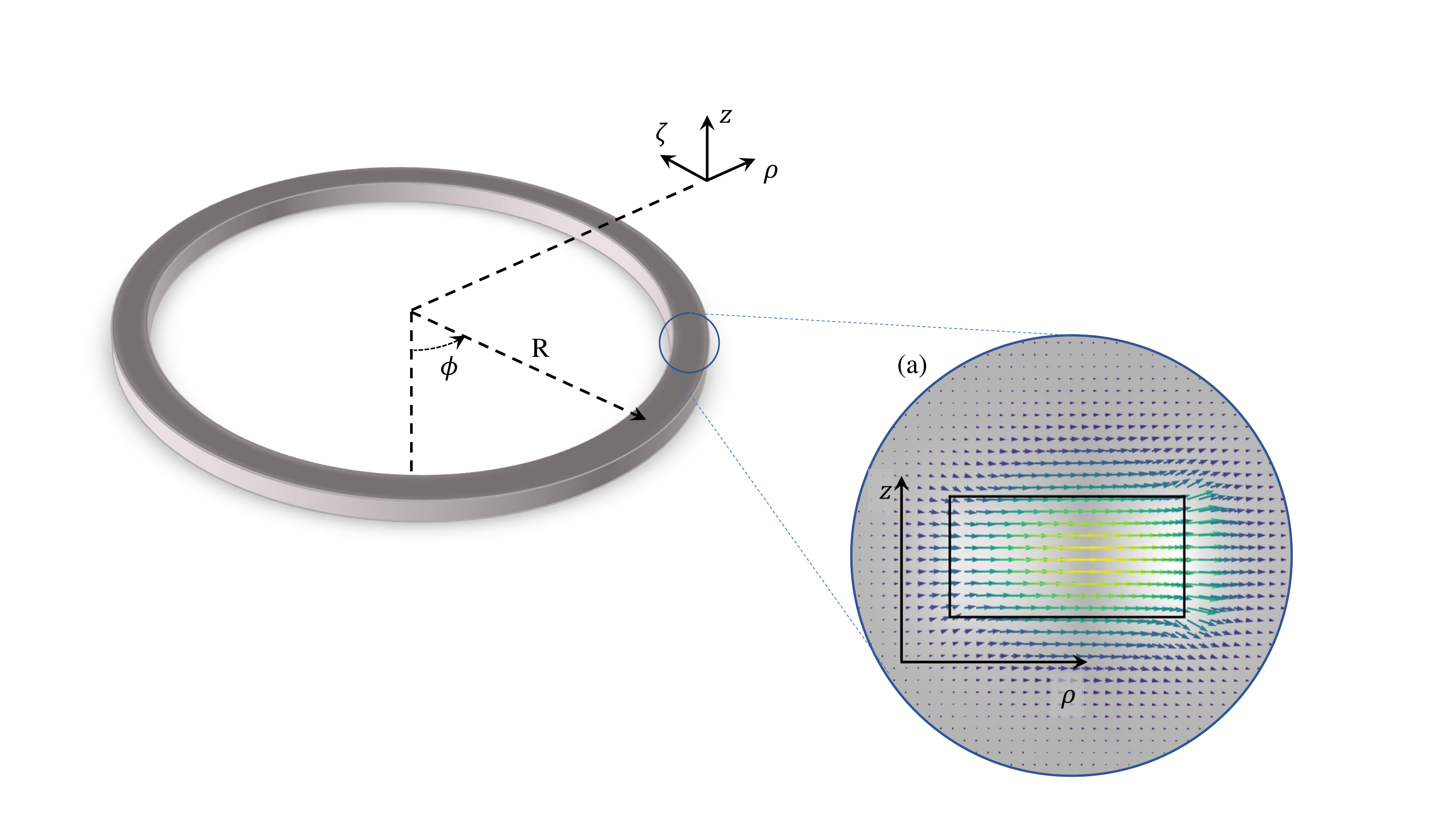}
    \caption{\label{fig:ring_resonator} Schematic representation of an integrated ring resonator of radius R point-coupled to a bus waveguide. The interaction is regulated by the self-coupling and cross-coupling coefficients $\sigma$ and $\kappa$, linked by $\sigma^2+\kappa^2=1$ \cite{yariv2007photonics}.}
\end{figure}

Here, in order to provide a simple and intuitive example, we restrict
the discussion to a simple ring resonator point-coupled to a channel
waveguide following the prescriptions of Sec.~\ref{sec:Quantization}, and sketched in Fig.~\ref{fig:ring_resonator}. Similar derivations can be carried out with
more complex resonator or coupler designs \cite{vernon2017truly, christensen2018engineering}, resulting in analogous conclusions.

When this idealized structure is pumped by bright classical fields,
it proves an efficient, stable and compact source of non-degenerate squeezed vacuum (NDSV) \cite{vaidya2019broadband} or degenerate squeezed vacuum (DSV) states \cite{vernon2019scalable, zhao2020near}. In the following we discuss the generation of such states using second- and third-order nonlinearities in the most common pumping schemes, and present some simplified cases.

\subsubsection{The SFWM Hamiltonian}

Considering a ring with a third-order nonlinearity, first we can extend Sec.~\ref{sec:SXPM_rings} to include
spontaneous four-wave mixing, the driver of squeezing in such a ring.
Starting from the form \eqref{eq:field_expansions_ring} of the $\bm{D}(\bm{r})$ operators, we restrict our attention to three resonant modes, with equally spaced wave vectors, conventionally labeled Pump, Signal, and Idler, with frequencies $\omega_{\text{P}}$, $\omega_{\text{S}}$ and $\omega_{\text{I}}$, respectively. Here we have chosen to reduce the scope of this discussion to a three-resonance system, since it greatly simplifies the description of its dynamics. Indeed, with an unconstrained set of resonances, other processes, usually referred to as \emph{spurious}, and that involve resonant modes beyond the three of interest here can lead to nonlinear frequency shifts and undesired parametric noise being generated completely or partly in the Signal mode, affecting the quantum properties of the photons generated by SFWM \cite{zhang2021squeezed, Seifoory2021}.

The third-order SFWM contribution to the nonlinear Hamiltonian \eqref{eq:Hfull} is
\begin{align}
H_{\text{ring}}^{\text{SFWM}}  =&-\frac{1}{4\varepsilon_{0}}\left(\frac{\hbar\omega_{\text{P}}}{2\mathcal{L}}\right)\sqrt{\frac{\hbar\omega_{\text{S}}}{2\mathcal{L}}\frac{\hbar\omega_{\text{I}}}{2\mathcal{L}}}\left(\frac{4!}{2!1!1!}\right)\label{eq:H_ring_SFWM}\\
&\times  \int \dd\bm{r}_{\perp}d\zeta\ \Gamma_{3}^{ijkl}(\bm{r}_{\perp},\zeta)\left(\mathsf{d}_{\text{S}}^{k}(\bm{r}_{\perp},\zeta)\mathsf{d}_{\text{I}}^{l}(\bm{r}_{\perp},\zeta)\right)^{*}\mathsf{d}_{\text{P}}^{i}(\bm{r}_{\perp},\zeta)\mathsf{d}_{\text{P}}^{j}(\bm{r}_{\perp},\zeta)e^{i\Delta\kappa\zeta}c_{\text{S}}^{\dagger}c_{\text{I}}^{\dagger}c_{\text{P}}c_{\text{P}},\nonumber 
\end{align}
with $\Delta\kappa=2\kappa_{\text{P}}-\kappa_{\text{S}}-\kappa_{\text{I}}$ and we assume
$\Delta\kappa=0$ for the resonances are equally spaced in $\kappa$.
Similarly to the general case of SPM and XPM, cf. (\ref{eq:HringSPM}, \ref{eq:HringXPM}), and for nonlinear processes in channel waveguides, cf. (\ref{eq:H_SP-SFWM_pre}, \ref{eq:H_SP-SFWM}, \ref{eq:H_DP-SFWM_pre}), we can introduce a general form of the nonlinear coefficient $\gamma_{\text{ring}}^{J_1J_2J_3J_4}$ such that

\begin{equation}
H_{\text{ring}}^{\text{SFWM}}=-\frac{\hbar^{2}\left(\omega_{\text{P}}^{2}\omega_{\text{S}}\omega_{\text{I}}\right)^{\frac{1}{4}}v_{\text{P}}\sqrt{v_{\text{S}}v_{\text{I}}}\gamma_{\text{ring}}^{\text{SIPP}}}{\mathcal{L}}c_{\text{S}}^{\dagger}c_{\text{I}}^{\dagger}c_{\text{P}}c_{\text{P}},
\end{equation}
where
\begin{align}\label{eq:gamma_ring_general}
\gamma_{\text{ring}}^{J_1J_2J_3J_4}=&\frac{3\left(\omega_{J_1}\omega_{J_2}\omega_{J_3}\omega_{J_4}\right)^{\frac{1}{4}}\varepsilon_{0}}{4\sqrt{v_{J_1}v_{J_2}v_{J_3}v_{J_4}}}\frac{1}{\mathcal{L}}\\
&\times\int \dd\bm{r}_{\perp}\dd\zeta\ \chi_{3}^{ijkl}(\bm{r}_{\perp},\zeta)\left(\mathsf{e}_{J_3}^{k}(\bm{r}_{\perp},\zeta)\mathsf{e}_{J_4}^{l}(\bm{r}_{\perp},\zeta)\right)^{*}\mathsf{e}_{J_1}^{i}(\bm{r}_{\perp},\zeta)\mathsf{e}_{J_2}^{j}(\bm{r}_{\perp},\zeta).\nonumber
\end{align}
The expression \eqref{eq:gamma_ring_general}, which will be soon useful to describe also SPM and XPM in different pumping regimes, considers that $\mathsf{d}_{J}(\bm{r}_{\perp},\zeta)=\epsilon_{0}n^{2}(\bm{r}_{\perp};\omega_{J})\mathsf{e}_{J}(\bm{r}_{\perp},\zeta)$,
and uses $n_{J}(\bm{r}_{\perp};\omega)=\sqrt{\varepsilon_{1}(\bm{r}_{\perp};\omega_{J})}$.
Using the normalization condition \eqref{eq:ring_norm_use} and, as usual, introducing characteristic group velocities $\overline{n}_{J}$ and $\overline{\chi}_{3}$ for the indices $n(\bm{r}_{\perp};\omega_{J})$ and the nonvanishing components of $\chi_{3}^{ijkl}(\bm{r}_{\perp},\zeta)$, the general nonlinear
coefficient $\gamma_{\text{ring}}^{J_1J_2J_3J_4}$ takes the same form of \eqref{eq:gamma_chan_general}
\begin{equation}\label{eq:gamma_ring_general_Aeff}
\gamma_{\text{ring}}^{J_1J_2J_3J_4}=\frac{3\left(\omega_{J_1}\omega_{J_2}\omega_{J_3}\omega_{J_4}\right)^{\frac{1}{4}}\overline{\chi}_{3}}{4\varepsilon_{0}\sqrt{\overline{n}_{J_1}\overline{n}_{J_2}\overline{n}_{J_3}\overline{n}_{J_4}}c^{2}}\frac{1}{A_{\text{ring }}^{J_1J_2J_3J_4}},
\end{equation}
with the effective area $A_{\text{ring}}^{J_1J_2J_3J_4}$ defined as
\begin{align}\label{eq:Aeff_ring_general}
\frac{1}{A_{\text{ring}}^{J_1J_2J_3J_4}} =&\frac{\mathcal{L}^{-1}\int \dd\bm{r}_{\perp}d\zeta\ \left(\chi_{3}^{ijkl}(\bm{r}_{\perp},\zeta)\slash\overline{\chi}_{3}\right)\left(\mathsf{e}_{J_3}^{k}(\bm{r}_{\perp},\zeta)\mathsf{e}_{J_4}^{l}(\bm{r}_{\perp},\zeta)\right)^{*}\mathsf{e}_{J_1}^{i}(\bm{r}_{\perp},\zeta)\mathsf{e}_{J_2}^{j}(\bm{r}_{\perp},\zeta)}{\mathcal{N}_{J_{1}}\mathcal{N}_{J_{2}}\mathcal{N}_{J_{3}}\mathcal{N}_{J_{4}}},
\end{align}
with
\begin{equation}
\mathcal{N}_{J} = \sqrt{\int\dd\bm{r}_{\perp}\frac{n(\bm{r}_{\perp};\omega_{J})\slash\overline{n}_{J}}{v_{g}(\bm{r}_{\perp};\omega_{J})/v_{J}}\mathsf{e}_{J}^{*}(\bm{r}_{\perp},0)\cdot\mathsf{e}_{J}(\bm{r}_{\perp},0)}.
\end{equation}

In the expression of the effective area \eqref{eq:Aeff_ring_general}, as well as throughout the rest of the section, we assume that dispersion can be neglected within the frequency range of each ring resonance. Therefore, in the field normalization we can truncate the dispersion to the first derivative of the wave vectors. However, considering nonlinear processes that involve multiple resonances, we will not make any further assumption on the impact of material and waveguide dispersion on the central frequencies of the resonant modes.

\subsubsection{The SPDC Hamiltonian}

When the ring is characterized by a second-order nonlinear response,
and we now consider the three relevant modes conventionally labeled
Second Harmonic (SH), and Fundamentals (F1 and F2) with frequencies
$\omega_{\text{SH}}$, $\omega_{\text{F1}}$ and $\omega_{\text{F2}}$, respectively, the
SPDC contribution to the nonlinear Hamiltonian \eqref{eq:HNL_general} is 

\begin{align}
H_{\text{ring}}^{\text{SPDC}}  =&-\frac{2}{3\varepsilon_{0}^{2}}\Big(\frac{3!}{1!1!1!}\Big)\sqrt{\frac{\hbar\omega_{\text{SH}}}{2\mathcal{L}}}\sqrt{\frac{\hbar\omega_{\text{F1}}}{2\mathcal{L}}\frac{\hbar\omega_{\text{F2}}}{2\mathcal{L}}}\\
&\times  \int\dd\bm{r}_{\perp}\dd\zeta\ \Gamma_{ijk}^{(2)}(\bm{r}_{\perp},\zeta)\mathsf{d}_{\text{SH}}^{i}(\bm{r}_{\perp},\zeta)\left[\mathsf{d}_{\text{F1}}^{j}(\bm{r}_{\perp},\zeta)\right]^{*}\left[\mathsf{d}_{\text{F2}}^{k}(\bm{r}_{\perp},\zeta)\right]^{*}e^{i\Delta\kappa\zeta}c_{\text{SH}}c_{\text{F1}}^{\dagger}c_{\text{F2}}^{\dagger}\nonumber \\
  \equiv&-\hbar\Lambda_{\text{ring}}^{\text{SPDC}}c_{\text{SH}}c_{\text{F1}}^{\dagger}c_{\text{F2}}^{\dagger}+\text{H.c.},\nonumber 
\end{align}
where $\Delta\kappa=\kappa_{\text{SH}}-\kappa_{\text{F1}}-\kappa_{\text{F2}}=0$ in a ring
with equally spaced resonances, and we immediately jump to the definition
of
\begin{align}
\Lambda_{\text{ring}}^{\text{SPDC}}  =&-\frac{1}{3\varepsilon_{0}}\Big(\frac{3!}{1!1!1!}\Big)\sqrt{\frac{\hbar\omega_{\text{SH}}}{2\mathcal{L}}}\sqrt{\frac{\hbar\omega_{\text{F1}}}{2\mathcal{L}}}\label{eq: Lambda_SPDC}\\
&\times  \sqrt{\frac{\hbar\omega_{\text{F2}}}{2\mathcal{L}}}\int \dd\bm{r}_{\perp}\dd\zeta\ \Gamma_{ijk}^{(2)}(\bm{r}_{\perp},\zeta)\mathsf{d}_{\text{SH}}^{i}(\bm{r}_{\perp},\zeta)\left[\mathsf{d}_{\text{F1}}^{j}(\bm{r}_{\perp},\zeta)\right]^{*}\left[\mathsf{d}_{\text{F2}}^{k}(\bm{r}_{\perp},\zeta)\right]^{*}\nonumber 
\end{align}
as the characteristic coefficient of the SPDC process in the ring.

\subsubsection{General expression of the dynamic equation}

Before diving into particular scenarios, it is useful to write down the expression of the Heisenberg equation

\begin{equation}
i\hbar\ddt{\mathcal{O}(t)}=\left[\mathcal{O}(t),H_{\text{ring}}\right],\label{eq: general_Heisenberg}
\end{equation}
with $H_{\text{ring}}$ being the nonlinear Hamiltonian for a ring resonator point-coupled to a channel waveguide and $\mathcal{O}(t)$ any of the ring Heisenberg
operators. Following (\ref{eq:ring+channel_H}, \ref{eq:H_phantom}) the full Hamiltonian is
\begin{align}
H_{\text{ring}}  =&H_{\text{ring}}^{\text{L}}+H_{\text{ring}}^{\text{NL}}\\
  =&\sum_{J}\left(\hbar\omega_{J}\int\psi_{J}^{\dagger}(x)\psi_{J}(x)\dd x-\frac{1}{2}i\hbar v_{J}\int\left(\psi_{J}^{\dagger}(x)\frac{\partial\psi_{J}(x)}{\partial x}-\frac{\partial\psi_{J}^{\dagger}(x)}{\partial x}\psi_{J}(x)\right)\dd x\right)\nonumber \\
  &+\sum_{J}\hbar\omega_{J}c_{J}^{\dagger}c_{J}+\sum_{J}(\hbar\gamma_{J}c_{J}^{\dagger}\psi_{J}(0)+\mathrm{H.c.})\nonumber \\
 & +\sum_{J}\left(\hbar\omega_{J}\int\phi_{J}^{\dagger}(x)\phi_{J}(x)\dd x-\frac{1}{2}i\hbar v_{J\text{ph}}\int\left(\phi_{J}^{\dagger}(x)\frac{\partial\phi_{J}(x)}{\partial x}-\frac{\partial\phi_{J}^{\dagger}(x)}{\partial x}\phi_{J}(x)\right)\dd x\right)\nonumber \\
 & +\sum_{J}(\hbar\gamma_{J\text{ph}}c_{J}^{\dagger}\phi_{J}(0)+\mathrm{H.c.})+H_{\text{ring}}^{\text{NL}},\nonumber 
\end{align}
where of course the expression of $H_{\text{ring}}^{\text{NL}}$ is case-specific. Using this in \eqref{eq: general_Heisenberg} we obtain
\begin{align}
i\hbar\frac{d\mathcal{O}(t)}{dt}  =&\sum_{J}\hbar\omega_{J}\left[\mathcal{O}(t),c_{J}^{\dagger}c_{J}\right]+\sum_{J}(\hbar\gamma_{J}\left[\mathcal{O}(t),c_{J}^{\dagger}\right]\psi_{J}(0)+\mathrm{H.c.})\label{eq: general_equation_motion}\\
 & +\sum_{J}\hbar\omega_{J}\left[\mathcal{O}(t),c_{J}^{\dagger}c_{J}\right]+\sum_{J}(\hbar\gamma_{J\text{ph}}\left[\mathcal{O}(t),c_{J}^{\dagger}\right]\phi_{J}(0)+\mathrm{H.c.})\nonumber \\
 & +\left[\mathcal{O}(t),H_{\text{ring}}^{\text{NL}}\left(\left\{ c_{J}(t)\right\} ,\left\{ c_{J}^{\dagger}(t)\right\};t\right)\right],\nonumber 
\end{align}
where the otherwise implicit dependence of the nonlinear Hamiltonian on the ring operators and time is made explicit. In the limiting case of vanishing nonlinearity, the Heisenberg equation is just \eqref{eq:Heisenberg_O} and after solving the commutators in \eqref{eq: general_equation_motion} we recover \eqref{eq:gen_lin_dyn}.

\subsection{Single-pump SFWM}
\begin{figure}[ht]\centering
    \includegraphics[width=1.0\textwidth]{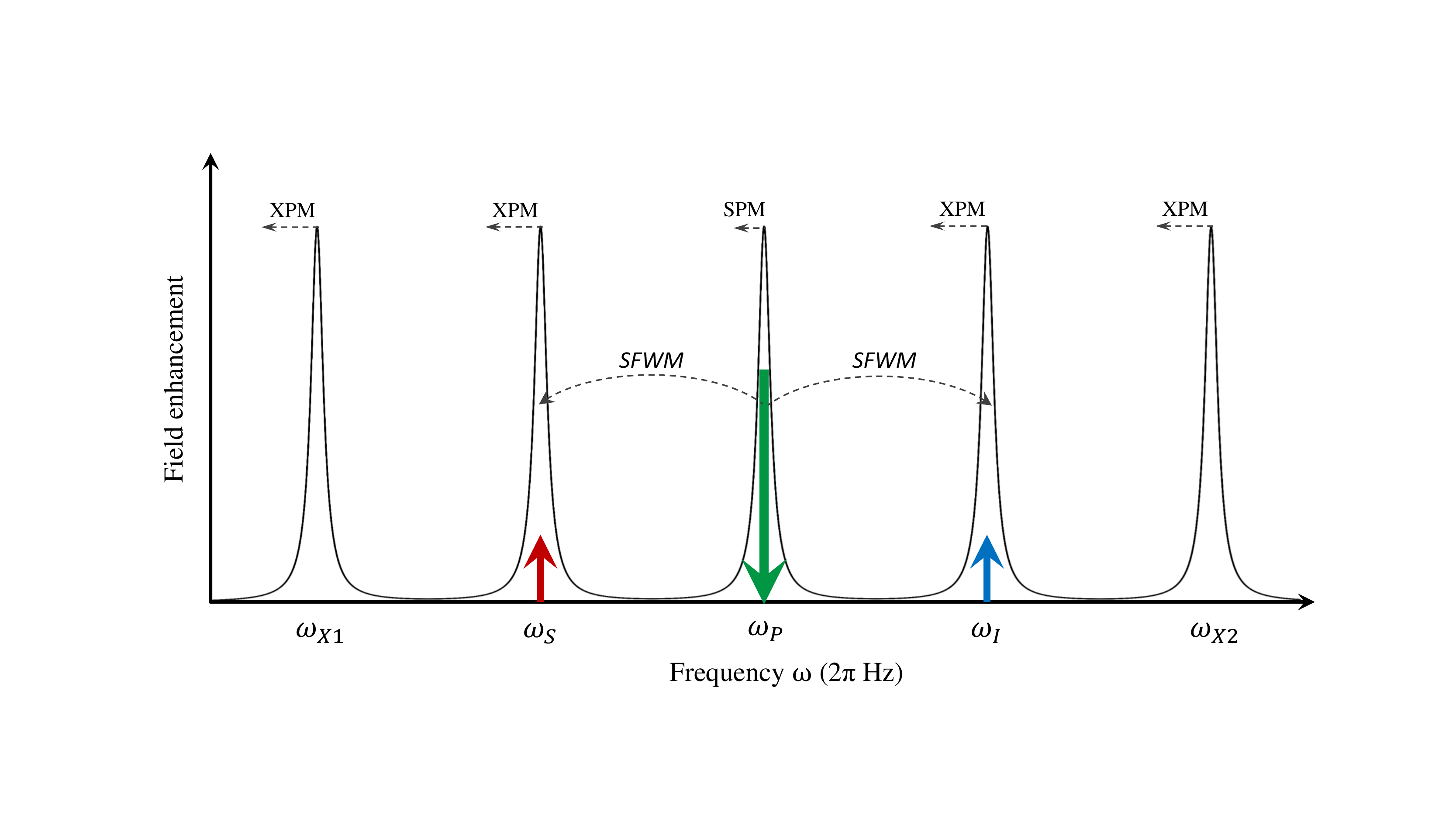}
    \caption{Ring resonances involved in a SP-SFWM process, and the associated SPM and XPM wavelength shifts.}
    \label{fig:SP-SFWM}
\end{figure}

The resonant wavelengths of the idealized three-mode system introduced above are solutions of $\kappa_{m}\mathcal{L}=\frac{2\pi}{\lambda_{m}}n_{\text{P}}(\omega_{m})=2\pi m$
with $m\in\mathbb{Z}$ and are equally spaced in $\kappa$. Solutions
are solely affected by the ring geometry, the material and the waveguide
dispersion. Let the pump be the central resonance, surrounded by the signal and idler resonances, as sketched in Fig.~\ref{fig:SP-SFWM}. While here we consider the interaction between nearest-neighbor resonances, any triple can be considered, as long as equally spaced in $\kappa$.

In Sec.~\ref{sec:ring_channel_coupling} we have already introduced the channel
field operators \eqref{eq:channel_field}, the IO formalism for the channel-ring coupling (\ref{eq:ring+channel_H} - \ref{eq:cdynamics}), and the related derivation of the full coupling coefficients $\overline{\Gamma}_{J}$ including the scattering loss modeled by a phantom
channel (\ref{eq:Gamma_J}, \ref{eq:Gammabar_J}, \ref{eq:Gamma_Jph}). It is worth recalling at this point that such an approach is only suitable for high finesse systems \eqref{eq:Finesse}, but this
is typically verified for integrated photonic structures at frequencies
below the bandgap in most common material platforms such as silicon on insulator (SOI), silicon nitride ($\text{Si}_3\text{N}_4$), lithium niobate on insulator (LNOI),
aluminum gallium arsenide (AlGaAs), aluminum nitride (AlN), chalcogenide glasses, etc.

We illuminate the ring with a single, bright light source, where it is implicit that its bandwidth cannot exceed that of the central resonance. We refer to this case as single-pump SFWM (SP-SFWM). From the full Hamiltonian of the system by composition of the linear Hamiltonian \eqref{eq:ring+channel_H} and the nonlinear response due to SPM \eqref{eq:HSPMwork-1}, XPM \eqref{eq:HringXPM}, and SFWM
\eqref{eq:H_ring_SFWM}, we obtain
\begin{align}
H_{\text{ring}}^{\text{SP-SFWM}}  =&H_{\text{ring}}^{\text{L}}+H_{\text{ring}}^{\text{SP-SFWM,NL}}\label{eq:H_ring_full_SP-SFWM}\\
= & \sum_{J\in\{\text{P,S,I}\}}\left(\hbar\omega_{J}\int\psi_{J}^{\dagger}(x)\psi_{J}(x)\dd x-\frac{1}{2}i\hbar v_{J}\int\left(\psi_{J}^{\dagger}(x)\frac{\partial\psi_{J}(x)}{\partial x}-\frac{\partial\psi_{J}^{\dagger}(x)}{\partial x}\psi_{J}(x)\right)\dd x\right)\nonumber \\
&+  \sum_{J\in\{\text{P,S,I}\}}\hbar\omega_{J}c_{J}^{\dagger}c_{J}+\sum_{J}(\hbar\gamma_{J}c_{J}^{\dagger}\psi_{J}(0)+\text{H.c.})\nonumber \\
&+  \sum_{J\in\{\text{P,S,I}\}}\left(\hbar\omega_{J}\int\phi_{J}^{\dagger}(x)\phi_{J}(x)\dd x-\frac{1}{2}i\hbar v_{J\text{ph}}\int\left(\phi_{J}^{\dagger}(x)\frac{\partial\phi_{J}(x)}{\partial x}-\frac{\partial\phi_{J}^{\dagger}(x)}{\partial x}\phi_{J}(x)\right)\dd x\right)\nonumber \\
 & +\sum_{J}(\hbar\gamma_{J\text{ph}}c_{J}^{\dagger}\phi_{J}(0)+\text{H.c.})\nonumber \\
&+  \left(-\hbar\Lambda_{\text{ring}}^{\text{SFWM}}c_{\text{P}}c_{\text{P}}c_{\text{S}}^{\dagger}c_{\text{I}}^{\dagger}+\text{H.c.}\right)-\hbar\eta_{\text{ring}}^{\text{SPM}}c_{\text{P}}^{\dagger}c_{\text{P}}^{\dagger}c_{\text{P}}c_{\text{P}}-\hbar\zeta_{\text{ring}}^{\text{XPM}}\left(c_{\text{S}}^{\dagger}c_{\text{P}}^{\dagger}c_{\text{S}}c_{\text{P}}+c_{\text{I}}^{\dagger}c_{\text{P}}^{\dagger}c_{\text{I}}c_{\text{P}}\right),\nonumber 
\end{align}
where using the general definition of the nonlinear coefficient \eqref{eq:gamma_ring_general}
\begin{align}
\Lambda_{\text{ring}}^{\text{SFWM}} & =\frac{\hbar\left(\omega_{\text{P}}^{2}\omega_{\text{S}}\omega_{\text{I}}\right)^{\frac{1}{4}}v_{\text{P}}\sqrt{v_{\text{S}}v_{\text{I}}}\gamma_{\text{ring}}^{\text{SIPP}}}{\mathcal{L}}\\
\eta_{\text{ring}}^{\text{SPM}} & =\frac{\hbar\omega_{\text{P}}v_{\text{P}}^{2}\gamma_{\text{ring}}^{\text{PPPP}}}{2\mathcal{L}}\nonumber \\
\zeta_{\text{ring}}^{\text{XPM}} & =\frac{2\hbar\omega_{\text{P}}v_{\text{P}}v_{S(I)}\gamma_{\text{ring}}^{\text{SS(II)PP}}}{\mathcal{L}},\nonumber 
\end{align}
and for the last expression we allow $\omega_{\text{S}}\simeq\omega_{\text{I}}$.
It is also worth pointing out that, in most practical cases, $\omega_{\text{P}}\simeq\omega_{\text{S}}\simeq\omega_{\text{I}}$
and therefore $\eta_{\text{ring}}^{\text{SPM}}\simeq\frac{1}{2}\Lambda_{\text{ring}}^{\text{SFWM}}$
and $\zeta_{\text{ring}}^{\text{XPM}}\simeq2\Lambda_{\text{ring}}^{\text{SFWM}}$.
From the comparison of the nonlinear Hamiltonian in \eqref{eq:H_ring_full_SP-SFWM}
and the expression \eqref{eq:quad_hamil} it becomes apparent that once the pump operators are replaced by their expectation values, such a process generates a NDSV state in the signal and idler modes.

In the interest of notation, throughout the rest of this section we will drop the superscripts and subscripts unless necessary. Given the full Hamiltonian \eqref{eq:H_ring_full_SP-SFWM} we can now calculate the Heisenberg equations of motion for the ring operators following \eqref{eq: general_equation_motion}, and obtain 
\begin{align}
\left(\ddt{}+\overline{\Gamma}_{\text{P}}+i\omega_{\text{P}}-2i\eta c_{\text{P}}^{\dagger}(t)c_{\text{P}}(t)\right)c_{\text{P}}(t) & =-i\gamma_{\text{P}}^{*}\psi_{\text{\text{P}<}}(0,t)-i\gamma_{\text{Pph}}^{*}\phi_{\text{P}<}(0,t)+2i\Lambda^{*}c_{\text{P}}^{\dagger}(t)c_{\text{S}}(t)c_{\text{I}}(t)\label{eq:Equations_of_motion_ring}\\
\left(\ddt{}+\overline{\Gamma}_{\text{S}}+i\omega_{\text{S}}-i\zeta c_{\text{P}}^{\dagger}(t)c_{\text{P}}(t)\right)c_{\text{S}}(t) & =-i\gamma_{\text{S}}^{*}\psi_{\text{S<}}(0,t)-i\gamma_{\text{Sph}}^{*}\phi_{\text{S<}}(0,t)+i\Lambda c_{\text{P}}(t)c_{\text{P}}(t)c_{\text{I}}^{\dagger}(t)\nonumber \\
\left(\ddt{}+\overline{\Gamma}_{\text{I}}+i\omega_{\text{I}}-i\zeta c_{\text{P}}^{\dagger}(t)c_{\text{P}}(t)\right)c_{\text{I}}(t) & =-i\gamma_{\text{I}}^{*}\psi_{\text{I<}}(0,t)-i\gamma_{\text{Iph}}^{*}\phi_{\text{I<}}(0,t)+i\Lambda c_{\text{P}}(t)c_{\text{P}}(t)c_{\text{S}}^{\dagger}(t),\nonumber 
\end{align}
with damping rates $\overline{\Gamma}_{J}=\Gamma_{J}+\Gamma_{J\text{ph}}=\frac{|\gamma_{J}|^{2}}{2v_{J}}+\frac{|\gamma_{J\text{ph}}|^{2}}{2v_{J\text{ph}}}$.
Note that within the approximations introduced, for each resonance
the coupling constant $\gamma_{J}$ can be related to the channel-ring self-coupling
coefficient $\sigma$ using \eqref{eq:gamsig}. That self-coupling coefficient, which can be different for different resonances $J$, can be computed numerically in a straightforward way, e.g. via FDTD simulations \cite{hagness1997fdtd}.

Let us then remove the rapidly oscillating part of the fields, by
defining $\overline{c}_{J}(t)=e^{i\omega_{J}t}c_{J}(t)$, so that the equations
of motion become

\begin{align}
\left(\ddt{}+\overline{\Gamma}_{\text{P}}-2i\eta\overline{c}_{\text{P}}^{\dagger}(t)\overline{c}_{\text{P}}(t)\right)\overline{c}_{\text{P}}(t) & =-i\gamma_{\text{P}}^{*}\overline{\psi}_{\text{P}<}(0,t)-i\gamma_{\text{Pph}}^{*}\overline{\phi}_{\text{P}<}(0,t)+2i\Lambda^{*}\overline{c}_{\text{P}}^{\dagger}(t)\overline{c}_{\text{S}}(t)\overline{c}_{\text{I}}(t)e^{i\Delta_{\text{ring}}t}\label{eq:Equations_of_motion_ring_rot}\\
\left(\ddt{}+\overline{\Gamma}_{\text{S}}-i\zeta\overline{c}_{\text{P}}^{\dagger}(t)\overline{c}_{\text{P}}(t)\right)\overline{c}_{\text{S}}(t) & =-i\gamma_{\text{S}}^{*}\overline{\psi}_{\text{S<}}(0,t)-i\gamma_{\text{Sph}}^{*}\overline{\phi}_{\text{S<}}(0,t)+i\Lambda\overline{c}_{\text{P}}(t)\overline{c}_{\text{P}}(t)\overline{c}_{\text{I}}^{\dagger}(t)e^{-i\Delta_{\text{ring}}t}\nonumber \\
\left(\ddt{}+\overline{\Gamma}_{\text{I}}-i\zeta\overline{c}_{\text{P}}^{\dagger}(t)\overline{c}_{\text{P}}(t)\right)\overline{c}_{\text{I}}(t) & =-i\gamma_{\text{I}}^{*}\overline{\psi}_{\text{I<}}(0,t)-i\gamma_{\text{Iph}}^{*}\overline{\phi}_{\text{I<}}(0,t)+i\Lambda\overline{c}_{\text{P}}(t)\overline{c}_{\text{P}}(t)\overline{c}_{\text{S}}^{\dagger}(t)e^{-i\Delta_{\text{ring}}t}\nonumber 
\end{align}
where $\Delta_{\text{ring}}=2\omega_{\text{P}}-\omega_{\text{S}}-\omega_{\text{I}}$. Note
that commonly $\Delta_{\text{ring}}=0$ since, neglecting the waveguide
group velocity dispersion, the ring resonances are equally spaced
in wavelength.

In order to solve equations \eqref{eq:Equations_of_motion_ring_rot}
let us restrict the discussion to the case of a coherent laser pump,
thus treated classically. In this instance, the pump mode operator
is replaced by its expectation value
\begin{equation}
\overline{c}_{\text{P}}(t)\rightarrow\left\langle \overline{c}_{\text{P}}(t)\right\rangle e^{i\Delta_{\text{P}}t}\equiv\tilde{\beta}_{\text{P}}(t)e^{i\Delta_{\text{P}}t},
\end{equation}
where we allow for a possible pump detuning $\Delta_{\text{P}}$ from $\omega_{\text{P}}$.
Furthermore, if we assume that the pump is undepleted, then we can
safely neglect the term $\overline{c}_{\text{P}}^{\dagger}(t)\overline{c}_{\text{S}}(t)\overline{c}_{\text{I}}(t)e^{i\Delta_{\text{ring}}t}$
in the pump operator equation. The pump equation is now fully self-contained,
and reads
\begin{equation}
\left(\ddt{}+\overline{\Gamma}_{\text{P}}+i\Delta_{\text{P}}-2i\eta|\tilde{\beta}_{\text{P}}(t)|^{2}\right)\tilde{\beta}_{\text{P}}(t)=-i\gamma_{\text{P}}^{*}\tilde{\psi}_{\text{P}<}(0,t),\label{eq: Exact_equation_pump}
\end{equation}
where we defined $\tilde{\psi}_{\text{P}<}(0,t)=\overline{\psi}_{\text{P}<}(0,t)e^{-i\Delta_{\text{P}}t}$
and we have additionally ignored any input contribution form the phantom
channel. Within these limits and assumptions, the equation for the
pump can be solved independently. Once such solution is found, analytically
or numerically, the signal and idler equations can be written 

\begin{align}
\left(\ddt{}+i\Delta_{\text{P}}+\overline{\Gamma}_{\text{S}}-i\zeta\left|\tilde{\beta}_{\text{P}}(t)\right|^{2}\right)\tilde{c}_{\text{S}}(t) & =-i\gamma_{\text{S}}^{*}\tilde{\psi}_{\text{S<}}(0,t)-i\gamma_{\text{Sph}}^{*}\tilde{\phi}_{\text{S<}}(0,t)+i\Lambda\left[\tilde{\beta}_{\text{P}}(t)\right]^{2}\tilde{c}_{\text{I}}^{\dagger}(t)e^{-i\Delta_{\text{ring}}t}\label{eq: SP_SFWM_Signal_Idler_undeplated_pump}\\
\left(\ddt{}+i\Delta_{\text{P}}+\overline{\Gamma}_{\text{I}}-i\zeta\left|\tilde{\beta}_{\text{P}}(t)\right|^{2}\right)\tilde{c}_{\text{I}}(t) & =-i\gamma_{\text{I}}^{*}\tilde{\psi}_{\text{I<}}(0,t)-i\gamma_{\text{Iph}}^{*}\tilde{\phi}_{\text{I<}}(0,t)+i\Lambda\left[\tilde{\beta}_{\text{P}}(t)\right]^{2}\tilde{c}_{\text{S}}^{\dagger}(t)e^{-i\Delta_{\text{ring}}t},\nonumber 
\end{align}
where we have defined $\tilde{c}_{\text{S,I}}(t)=e^{-i\Delta_{\text{P}}t}\overline{c}_{\text{S,I}}(t)$,
$\tilde{\psi}_{\text{S,I}<}(0,t)=e^{-i\Delta_{\text{P}}t}\overline{\psi}_{\text{S<}}(0,t)$
and $\tilde{\phi}_{\text{S,I<}}(0,t)=e^{-i\Delta_{\text{P}}t}\overline{\phi}_{\text{S<}}(0,t)$.

We can reformulate the Heisenberg equations of motion \eqref{eq: SP_SFWM_Signal_Idler_undeplated_pump} for a ring under SP-SFWM in matrix form as

\begin{equation}
\ddt{}\begin{pmatrix}\tilde{c}_{\text{S}}(t)\\
\tilde{c}_{\text{I}}^{\dagger}(t)
\end{pmatrix}=M_{\text{SP-SFWM}}(t)\begin{pmatrix}\tilde{c}_{\text{S}}(t)\\
\tilde{c}_{\text{I}}^{\dagger}(t)
\end{pmatrix}+D_{\text{SP-SFWM}}(t),\label{eq: SP_Exact_equation_signal_and_idler}
\end{equation}
where 
\begin{equation}
M_{\text{SP-SFWM}}(t)=\begin{pmatrix}-\overline{\Gamma}_{\text{S}}-i\Delta_{\text{P}}+i\zeta\left|\tilde{\beta}_{\text{P}}(t)\right|^{2} & i\Lambda\left[\tilde{\beta}_{\text{P}}(t)\right]^{2}e^{-i\Delta_{\text{ring}}t}\\
-i\Lambda^{*}\left[\tilde{\beta}_{\text{P}}^{*}(t)\right]^{2}e^{i\Delta_{\text{ring}}t} & -\overline{\Gamma}_{\text{I}}+i\Delta_{\text{P}}-i\zeta^{*}\left|\tilde{\beta}_{\text{P}}(t)\right|^{2}
\end{pmatrix},\label{eq: M_Exact}
\end{equation}
and 
\begin{equation}
D_{\text{SP-SFWM}}(t)=\begin{pmatrix}-i\gamma_{\text{S}}^{*}\tilde{\psi}_{\text{S<}}(0,t)-i\gamma_{\text{Sph}}^{*}\tilde{\phi}_{\text{S<}}(0,t)\\
i\gamma_{\text{I}}\tilde{\psi}_{\text{I<}}^{\dagger}(0,t)+i\gamma_{\text{Iph}}\tilde{\phi}_{\text{I<}}^{\dagger}(0,t)
\end{pmatrix}.\label{eq: D_Exact}
\end{equation}
The solution of \eqref{eq: SP_Exact_equation_signal_and_idler} is exact within the assumptions introduced, and can in general be found numerically. One common approach to solve \eqref{eq: SP_Exact_equation_signal_and_idler} is to use the Green function formalism, hence identifying a matrix $G(t,t')$ that satisfies $G(t_{0},t_{0})=\mathbb{I}$ and
\begin{equation}
\ddt{}G(t,t')=M_{\text{SP-SFWM}}(t)G(t,t').\label{eq: Green_function}
\end{equation}
The full solution can be eventually written as
\begin{equation}\label{eq: Green solution SP}
\begin{pmatrix}\tilde{c}_{\text{S}}(t)\\
\tilde{c}_{\text{I}}^{\dagger}(t)
\end{pmatrix}=G(t_{0},t)\begin{pmatrix}\tilde{c}_{\text{S}}(t)\\
\tilde{c}_{\text{I}}^{\dagger}(t)
\end{pmatrix}+\int_{-\infty}^{t}\dd t^{\prime}\ \theta(t-t')G\left(t,t^{\prime}\right)D\left(t^{\prime}\right).
\end{equation}
Once the Green function has been calculated, it can be used to derive the $N$ and $M$ moments of the converted photon distribution, defined as
\begin{align}\label{eq: N_M_def_SPSFWM}
    N_{\text{S(I)}}(t_{1},t_{2})&\equiv v_{\text{S(I)}}\left\langle \tilde{\psi}_{\text{S(I)>}}^{\dagger}(0,t_{1})\tilde{\psi}_{\text{S(I)>}}(0,t_{2})\right\rangle \\
    M_{\text{SI(IS)}}(t_{1},t_{2})&\equiv\sqrt{v_{\text{S(I)}}v_{\text{I(S)}}}\left\langle\tilde{\psi}_{\text{S(I)>}}(0,t_{1})\tilde{\psi}_{\text{I(S)>}}(0,t_{2})\right\rangle\nonumber,
\end{align}
where the connection between the $\tilde{\psi}_{J>}(0,t)$ and $\tilde{\psi}_{J<}(0,t)$ operators is provided by
\begin{equation}\label{eq:phi_out_cj}
    \tilde{\psi}_{J>}(0,t)=\tilde{\psi}_{J<}(0,t)-\frac{i\gamma_{J}}{v_{J}}\tilde{c}_{J}(t),
\end{equation}
and thanks to \eqref{eq: Green solution SP} the expression of the $\tilde{c}_{J}(t)$ is now known. The explicit calculation of the N moment for the signal mode yields
\begin{align}
    N_{\text{S}}(t_{1},t_{2}) =& v_{\text{S}}\left\langle \left[\tilde{\psi}_{\text{S<}}^{\dagger}(0,t_{1})+\frac{i\gamma_{\text{S}}^{*}}{v_{\text{S}}}\tilde{c}_{\text{S}}^{\dagger}(t_{1})\right]\left[\tilde{\psi}_{\text{S<}}(0,t_{2})-\frac{i\gamma_{\text{S}}}{v_{\text{S}}}\tilde{c}_{\text{S}}(t_{2})\right]\right\rangle \\
    =&v_{\text{S}}\frac{|\gamma_{\text{S}}|^{2}}{v_{\text{S}}^{2}}\int_{-\infty}^{t_{1}}\dd t'\ \int_{-\infty}^{t_{2}}\dd t''\ G_{12}^{*}\left(t_{1},t'\right)G_{12}(t_{2},t'')\nonumber\\
    &\times\left\langle \left(-i\gamma_{\text{I}}^{*}\tilde{\psi}_{\text{I<}}(0,t')-i\mu_{\text{I}}^{*}\tilde{\phi}_{\text{I<}}(0,t')\right)\left(i\gamma_{\text{I}}\tilde{\psi}_{\text{I<}}^{\dagger}(0,t'')+i\mu_{\text{I}}\tilde{\phi}_{\text{I<}}^{\dagger}(0,t'')\right)\right\rangle \nonumber \\
    =&2\Gamma_{\text{S}}2\overline{\Gamma}_{\text{I}}\int_{-\infty}^{t_{1}}\dd t'\ \int_{-\infty}^{t_{2}}\dd t''\ G_{12}^{*}\left(t_{1},t'\right)G_{12}(t_{2},t'')\delta(t'-t'')\nonumber\\
    =&4\Gamma_{\text{S}}\overline{\Gamma}_{\text{I}}\int \dd\tau\ G_{12}^{*}\left(t_{1},\tau\right)\theta(t_{1}-\tau)G_{12}(t_{2},\tau)\theta(t_{2}-\tau),\nonumber
\end{align}
and a similar expression holds for the idler mode. The $M_{\text{SI}}$ moment is
\begin{align}\label{eq: MSI_SPSFWM}
    M_{\text{SI}}(t_{1},t_{2})=&\sqrt{v_{\text{S}}v_{\text{I}}}\left\langle \tilde{\psi}_{\text{S>}}(0,t_{1})\tilde{\psi}_{\text{I>}}(0,t_{2})\right\rangle \\
    =&-2\sqrt{\eta_{\text{S}}\eta_{\text{I}}\overline{\Gamma}_{\text{S}}\overline{\Gamma}_{\text{I}}}\nonumber\\
    &\times\left(\overline{\Gamma}_{\text{S}}+\overline{\Gamma}_{\text{I}}\right)\int \dd\tau\ \left[G_{21}^{*}(t_{1},\tau)G_{22}^{*}(t_{2},\tau)-G_{22}^{*}(t_{1},\tau)G_{21}^{*}(t_{2},\tau)\right]\theta(t_{2}-t_{1})\theta(t_{1}-\tau)\nonumber\\
    &+2\overline{\Gamma}_{\text{S}}\int \dd\tau\ G_{11}(t_{1},\tau)\theta(t_{1}-\tau)G_{21}^{*}(t_{2},\tau)\theta(t_{2}-\tau).\nonumber
\end{align}
As discussed in Sec.~\ref{sec:pnr_homodyne} the $N$ and $M$ moments are instrumental in calculating key quantities such as the average photon number of the squeezed state, its variance and covariance matrix \eqref{eq:moments_funcs}.

\subsection*{}

\subsection{Dual-pump SFWM}

\begin{figure}[ht]\centering
    \includegraphics[width=1.0\textwidth]{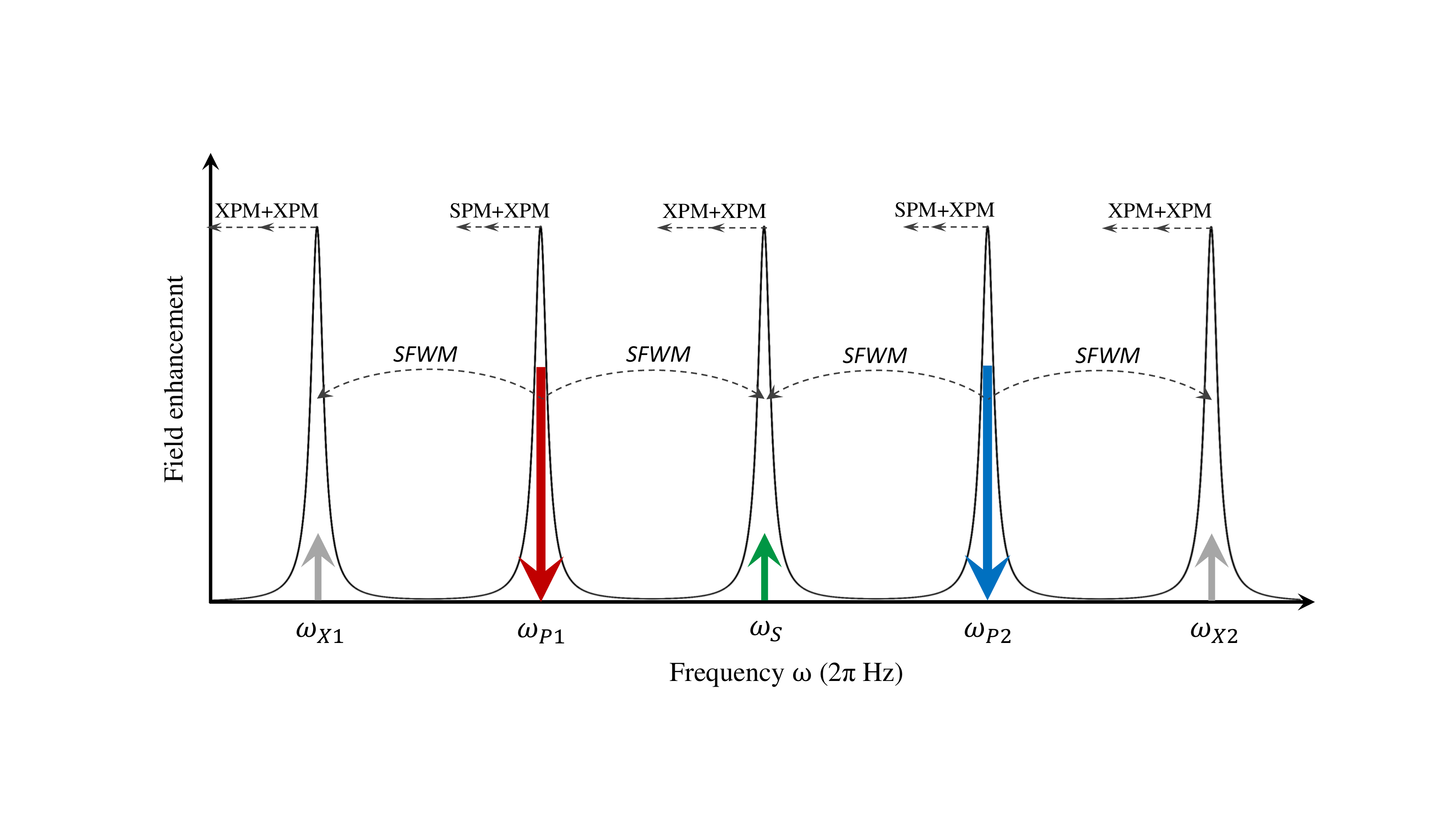}
    \caption{Ring resonances involved in a DP-SFWM process, and the associated SPM and XPM wavelength shifts.}
    \label{fig:DP-SFWM}
\end{figure}

In a variety of applications, such as quantum sensing or photonic
quantum computing, it is preferable to generate DSV rather than NDSV states; yet, the single pump approach described in the previous section (where one neglects the bright squeezed state generated in the P mode), is not suitable for the task.
Nonetheless, with the same high-finesse resonator and still restricting our discussion
to three isolated modes, one can invert the pumping scheme and readily obtain a DSV state. Let us illuminate the ring with two individual pumps, named $\text{P1}$ and $\text{P2}$, addressing the side resonances as prescribed in the previous section and represented in Fig.~\ref{fig:DP-SFWM}.

Incidentally, we note that
a two-lobed pump where each lobe is in correspondence of a ring resonance
is a completely equivalent view of the pumping scheme. We can conveniently
rename the three ring resonances as P1, S, and P2, respectively, where
it is assumed that the DSV state is generated in the S resonance
by SFWM. We refer to this pumping scheme as dual-pump SFWM (DP-SFWM).

The full Hamiltonian of the process is 
\begin{align}
H_{\text{ring}}^{\text{DP-SFWM}} & =H_{\text{ring}}^{\text{L}}+H_{\text{ring}}^{\text{DP-SFWM},\text{NL}}\label{eq:H_ring_full_DP}\\
= & \sum_{J\in\{\text{S,P1,P2}\}}\left(\hbar\omega_{J}\int\psi_{J}^{\dagger}(x)\psi_{J}(x)\dd x-\frac{1}{2}i\hbar v_{J}\int\left(\psi_{J}^{\dagger}(x)\frac{\partial\psi_{J}(x)}{\partial x}-\frac{\partial\psi_{J}^{\dagger}(x)}{\partial x}\psi_{J}(x)\right)\dd x\right)\nonumber \\
 & +\sum_{J\in\{\text{S,P1,P2}\}}\hbar\omega_{J}c_{J}^{\dagger}c_{J}+\sum_{J\in\{\text{S,P1,P2}\}}(\hbar\gamma_{J}c_{J}^{\dagger}\psi_{J}(0)+\text{H.c.})\nonumber \\
&+  \sum_{J\in\{\text{S,P1,P2}\}}\left(\hbar\omega_{J}\int\phi_{J}^{\dagger}(x)\phi_{J}(x)\dd x-\frac{1}{2}i\hbar v_{J\text{ph}}\int\left(\phi_{J}^{\dagger}(x)\frac{\partial\phi_{J}(x)}{\partial x}-\frac{\partial\phi_{J}^{\dagger}(x)}{\partial x}\phi_{J}(x)\right)\dd x\right)\nonumber \\
 & +\sum_{J\in\{\text{S,P1,P2}\}}(\hbar\gamma_{J\text{ph}}c_{J}^{\dagger}\phi_{J}(0)+\text{H.c.})\nonumber \\
 & -\left(\hbar\Lambda_{\text{ring}}^{\text{DP-SFWM}}c_{\text{P1}}c_{\text{P2}}c_{\text{S}}^{\dagger}c_{\text{S}}^{\dagger}+\text{H.c.}\right)-\hbar\eta_{\text{ring}}^{\text{SPM}}(c_{\text{P1}}^{\dagger}c_{\text{P1}}^{\dagger}c_{\text{P1}}c_{\text{P1}}+c_{\text{P2}}^{\dagger}c_{\text{P2}}^{\dagger}c_{\text{P2}}c_{\text{P2}})\nonumber \\
 & -\hbar\zeta_{\text{ring}}^{\text{XPM}}\left(c_{\text{S}}^{\dagger}c_{\text{P1}}^{\dagger}c_{\text{S}}c_{\text{P1}}+c_{\text{S}}^{\dagger}c_{\text{P2}}^{\dagger}c_{\text{S}}c_{\text{P2}}+c_{\text{P1}}^{\dagger}c_{\text{P2}}^{\dagger}c_{\text{P1}}c_{\text{P2}}\right)\nonumber 
\end{align}
where now, together with the SFWM term, we have two SPM terms -- one for each pump -- and three XPM terms -- between each pump and
the signal mode, and between the two pumps. As one can now appreciate
from the comparison of \eqref{eq:H_ring_full_DP} and \eqref{eq:squeezing_hamiltonian}, DP-SFWM leads to the generation of a DSV state in the signal resonance.

As before, in the following we will drop the unnecessary subscripts
for the sake of clarity. We can proceed with the same reasoning as
in the SP-SFWM case and work out the Heisenberg equations of motion,
to obtain
\begin{align}
&\left(\ddt{}+\overline{\Gamma}_{\text{P1}}-2i\eta\overline{c}_{\text{P1}}^{\dagger}(t)\overline{c}_{\text{P1}}(t)-i\zeta\overline{c}_{\text{P2}}^{\dagger}(t)\overline{c}_{\text{P2}}(t)\right)\overline{c}_{\text{P1}}(t)  =\label{eq: DP_Equations_of_motion}\\
&\quad \quad-i\gamma_{\text{P1}}^{*}\overline{\psi}_{\text{P1<}}(0,t)-i\gamma_{\text{P1ph}}^{*}\overline{\phi}_{\text{P1<}}(0,t)+i\Lambda^{*}\overline{c}_{\text{P2}}^{\dagger}(t)\overline{c}_{\text{S}}(t)\overline{c}_{\text{S}}(t)e^{i\Delta_{\text{ring}}t}\nonumber \\
&\left(\ddt{}+\overline{\Gamma}_{\text{P2}}-2i\eta\overline{c}_{\text{P2}}^{\dagger}(t)\overline{c}_{\text{P2}}(t)-i\zeta\overline{c}_{\text{P1}}^{\dagger}(t)\overline{c}_{\text{P1}}(t)\right)\overline{c}_{\text{P2}}(t)  =\nonumber \\
&\quad \quad-i\gamma_{\text{P2}}^{*}\overline{\psi}_{\text{P2<}}(0,t)-i\gamma_{\text{P2ph}}^{*}\overline{\phi}_{\text{P2<}}(0,t)+i\Lambda^{*}\overline{c}_{\text{P1}}^{\dagger}(t)\overline{c}_{\text{S}}(t)\overline{c}_{\text{S}}(t)e^{i\Delta_{\text{ring}}t}\nonumber \\
&\left(\ddt{}+\overline{\Gamma}_{\text{S}}-i\zeta\overline{c}_{\text{P1}}^{\dagger}(t)\overline{c}_{\text{P1}}(t)-i\zeta\overline{c}_{\text{P2}}^{\dagger}(t)\overline{c}_{\text{P2}}(t)\right)\overline{c}_{\text{S}}(t)  =\nonumber \\
&\quad \quad-i\gamma_{\text{S}}^{*}\overline{\psi}_{\text{S<}}(0,t)-i\gamma_{S\text{ph}}^{*}\overline{\phi}_{\text{S<}}(0,t)+2i\Lambda\overline{c}_{\text{P1}}(t)\overline{c}_{\text{P2}}(t)\overline{c}_{\text{S}}^{\dagger}(t)e^{-i\Delta_{\text{ring}}t}\nonumber 
\end{align}
where now $\Delta_{\text{ring}}=\omega_{\text{P1}}+\omega_{\text{P2}}-2\omega_{\text{S}}$.

As usual, we assume the pumps to be coherent lasers, thus treated
classically. Each pump mode operator is then replaced by its expectation
value

\begin{equation}
\overline{c}_{\text{P1(P2)}}(t)\rightarrow\left\langle \overline{c}_{\text{P1(P2)}}(t)\right\rangle e^{i\Delta_{\text{P1(P2)}}t}\equiv\tilde{\beta}_{\text{P1(P2)}}(t)e^{i\Delta_{\text{P1(P2)}}t},
\end{equation}
once more allowing for optional pump detunings $\Delta_{\text{P1(P2)}}$.
With undepleted pumps, neglecting the $\overline{c}_{\text{P1(P2)}}^{\dagger}(t)\overline{c}_{\text{S}}(t)\overline{c}_{\text{S}}(t)e^{i\Delta_{\text{ring}}t}$
terms, we have
\begin{align}
\left(\ddt{}+i\Delta_{\text{P1}}+\overline{\Gamma}_{\text{P1}}-2i\eta\left|\tilde{\beta}_{\text{P1}}(t)\right|^{2}-i\zeta\left|\tilde{\beta}_{\text{P2}}(t)\right|^{2}\right)\tilde{\beta}_{\text{P1}}(t) & =-i\gamma_{\text{P1}}^{*}\tilde{\psi}_{\text{P1<}}(0,t)\\
\left(\ddt{}+i\Delta_{\text{P2}}+\overline{\Gamma}_{\text{P2}}-2i\eta\left|\tilde{\beta}_{\text{P2}}(t)\right|^{2}-i\zeta\left|\tilde{\beta}_{\text{P1}}(t)\right|^{2}\right)\overline{\beta}_{\text{P2}}(t) & =-i\gamma_{\text{P2}}^{*}\tilde{\psi}_{\text{P2<}}(0,t),\nonumber 
\end{align}
where $\tilde{\psi}_{\text{P1(P2)<}}(0,t)=\overline{\psi}_{\text{P1(P2)<}}(0,t)e^{-i\Delta_{\text{P1(P2)}}t}$,
and we ignored the input contribution of the phantom channel. The dynamic equation of the signal mode is now
\begin{align}
&\left(\ddt{}+i\frac{\Delta_{\text{P1}}+\Delta_{\text{P2}}}{2}+\overline{\Gamma}_{\text{S}}-i\zeta\left|\tilde{\beta}_{\text{P1}}(t)\right|^{2}-i\zeta\left|\tilde{\beta}_{\text{P2}}(t)\right|^{2}\right)\tilde{c}_{\text{S}}(t)  =\label{eq: DP_SFWM_Signal_undeplated_pumps}\\
&\quad \quad -i\gamma_{\text{S}}^{*}\tilde{\psi}_{\text{S<}}(0,t)-i\gamma_{\text{Sph}}^{*}\tilde{\phi}_{\text{S<}}(0,t)+2i\Lambda\tilde{\beta}_{\text{P1}}(t)\tilde{\beta}_{\text{P2}}(t)\tilde{b}_{\text{S}}^{\dagger}(t)e^{-i\Delta_{\text{ring}}t}. \nonumber 
\end{align}
where, as for the pump equations, we defined $\tilde{c}_{\text{S}}(t)=e^{-i\frac{\Delta_{\text{P1}}+\Delta_{\text{P2}}}{2}t}\overline{c}_{\text{S}}(t)$,
and $\tilde{\psi}_{\text{S<}}(0,t)=\overline{\psi}_{\text{S<}}(0,t)e^{-i\frac{\Delta_{\text{P1}}+\Delta_{\text{P2}}}{2}t}$
and $\tilde{\phi}_{\text{S<}}(0,t)=\overline{\phi}_{\text{S<}}(0,t)e^{-i\frac{\Delta_{\text{P1}}+\Delta_{\text{P2}}}{2}t}$.

The analogue of the Heisenberg equations of motion (\ref{eq: SP_Exact_equation_signal_and_idler})
for DP-SFWM is
\begin{equation}
\ddt{}\begin{pmatrix}\tilde{c}_{\text{S}}(t)\\
\tilde{c}_{\text{S}}^{\dagger}(t)
\end{pmatrix} = M_{\text{DP-SFWM}}(t)\begin{pmatrix}\tilde{c}_{\text{S}}(t)\\
\tilde{c}_{\text{S}}^{\dagger}(t)
\end{pmatrix}+D_{\text{DP-SFWM}}(t),\label{eq: DP_Exact_equation_signal}
\end{equation}
but now 
\begin{align}\label{eq: DPSFWM_M_matrix}
M_{\text{DP-SFWM}}(t)=&-\overline{\Gamma}_{\text{S}}\mathbb{I} \\
&+ 
\begin{pmatrix}
-i\frac{\Delta_{\text{P1}}+\Delta_{\text{P2}}}{2}+i\zeta\left[\left|\tilde{\beta}_{\text{P1}}(t)\right|^{2}+\left|\tilde{\beta}_{\text{P2}}(t)\right|^{2}\right] & 2i\Lambda\tilde{\beta}_{\text{P1}}(t)\tilde{\beta}_{\text{P2}}(t)e^{-i\Delta_{\text{ring}}t}\\
-2i\Lambda^{*}\tilde{\beta}_{\text{P1}}^{*}(t)\tilde{\beta}_{\text{P2}}^{*}(t)e^{i\Delta_{\text{ring}}t} & i\frac{\Delta_{\text{P1}}+\Delta_{\text{P2}}}{2}-i\zeta^{*}\left[\left|\tilde{\beta}_{\text{P1}}(t)\right|^{2}+\left|\tilde{\beta}_{\text{P2}}(t)\right|^{2}\right]
\end{pmatrix},\nonumber
\end{align}
and
\begin{equation}
D_{\text{DP-SFWM}}(t)=\begin{pmatrix}-i\gamma_{\text{S}}^{*}\tilde{\psi}_{\text{S<}}(0,t)-i\gamma_{\text{Sph}}^{*}\tilde{\phi}_{\text{S<}}(0,t)\\
i\gamma_{\text{S}}\tilde{\psi}_{\text{S<}}^{\dagger}(0,t)+i\gamma_{\text{Sph}}\tilde{\phi}_{\text{S<}}^{\dagger}(0,t)
\end{pmatrix}.
\end{equation}

We can finally calculate the $N$ and $M$ moments of the photon distribution as
\begin{align}
    N_{\text{S}}(t_{1},t_{2})&\equiv v_{\text{S}}\left\langle\tilde{\psi}_{\text{S>}}^{\dagger}(0,t_{1})\tilde{\psi}_{\text{S>}}(0,t_{2})\right\rangle \\
    M_{\text{SS}}(t_{1},t_{2})&\equiv\sqrt{v_{\text{S}}v_{\text{S}}}\left\langle\tilde{\psi}_{\text{S>}}(0,t_{1})\tilde{\psi}_{\text{S>}}(0,t_{2})\right\rangle, \nonumber
\end{align}
and the output fields are connected to the ring operators by \eqref{eq:phi_out_cj}. Having solved the Green function equation, the moments can be expressed as
\begin{align}\label{eq:DP_SFWM_N}
    N_{\text{S}}(t_{1},t_{2})&=2\Gamma_{\text{S}}2\overline{\Gamma}_{\text{S}}\int_{-\infty}^{t_{1}}\dd t'\ \int_{-\infty}^{t_{2}}\dd t''\ G_{21}\left(t_{1},t'\right)G_{12}(t_{2},t'')\delta(t'-t'')\\
    &=4\Gamma_{\text{S}}\overline{\Gamma}_{\text{S}}\int \dd\tau\ G_{21}\left(t_{1},\tau\right)\theta(t_{1}-\tau)G_{12}(t_{2},\tau)\theta(t_{2}-\tau)\nonumber
\end{align}
and
\begin{align}\label{eq:DP_SFWM_M}
M_{\text{SS}}(t_{1},t_{2})=&v_{\text{S}}\left[\frac{\gamma_S^2}{v_S^2}\int_{-\infty}^{t_2}\dd t''\ G_{12}(t_2,t'')\delta(t_1-t'')\right. \\
&\quad \quad -\left.\frac{\gamma_{\text{S}}^{2}}{v_{\text{S}}^{2}}\int_{-\infty}^{t_{1}}\dd t'\ \int_{-\infty}^{t_{2}}\dd t''\ G_{11}(t_{1},t')G_{12}(t_{2},t'')\delta(t'-t'')\left[\frac{|\gamma_{\text{S}}|^{2}}{v_{\text{S}}}+\frac{|\mu_{\text{S}}|^{2}}{v_{\text{S}}}\right]\right]\nonumber\\
=&-4\frac{\gamma_{\text{S}}^{2}}{2v_{\text{S}}}\overline{\Gamma}_{\text{S}}\int \dd\tau\ \left[G_{12}(t_{1},\tau)G_{11}(t_{2},\tau)\theta(t_{1}-\tau)\theta(t_{2}-t_{1})\theta(t_{2}-\tau)\right.\nonumber\\
&+\left.G_{11}(t_{1},\tau)G_{12}(t_{2},\tau)\theta(t_{2}-\tau)\theta(t_{1}-t_{2})\theta(t_{1}-\tau)\right].\nonumber
\end{align}

\subsubsection{Dynamics of the cavity and field operators for DP-SFWM}

So far our main goal was to express the Heisenberg equations of motion
in an analytic form, with little insight on how one could solve them
and obtain the full dynamics of the cavity. That, in turn, allows one
to deduce relevant physical quantities such as the squeezing level,
the average photon numbers in the converted modes, the Schmidt number
or the distribution of the Schmidt modes. Unsurprisingly, a general
analytical solution of those equations cannot be provided. The pumping
scheme adopted, the amplitude and phase profile of the pump(s), the
strength of the nonlinear terms involved and the resonances detunings
are some of the key parameters that affect the solution in a non-trivial
way.

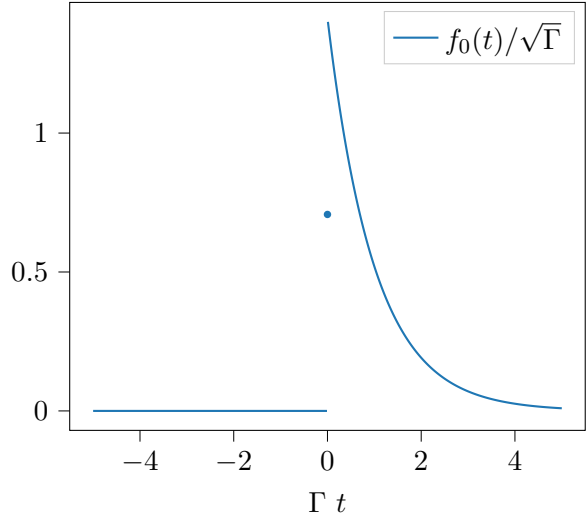
\begin{figure}
    \centering
\begin{tikzpicture}

\definecolor{color0}{rgb}{0.12156862745098,0.466666666666667,0.705882352941177}

\begin{axis}[
legend cell align={left},
legend style={fill opacity=0.8, draw opacity=1, text opacity=1, draw=white!80!black},
tick align=outside,
tick pos=left,
x grid style={white!69.0196078431373!black},
xlabel={\(\displaystyle  \Gamma  \ t\)},
xmin=-5.5, xmax=5.5,
xtick style={color=black},
y grid style={white!69.0196078431373!black},
ymin=-0.0700070951156651, ymax=1.47014899742897,
ytick style={color=black}
]
\addplot [thick, color0]
table {%
-5 0
-4.99 0
-4.98 0
-4.97 0
-4.96 0
-4.95 0
-4.94 0
-4.93 0
-4.92 0
-4.91 0
-4.9 0
-4.89 0
-4.88 0
-4.87 0
-4.86 0
-4.85 0
-4.84 0
-4.83 0
-4.82 0
-4.81 0
-4.8 0
-4.79 0
-4.78 0
-4.77 0
-4.76 0
-4.75 0
-4.74 0
-4.73 0
-4.72 0
-4.71 0
-4.7 0
-4.69 0
-4.68 0
-4.67 0
-4.66 0
-4.65 0
-4.64 0
-4.63 0
-4.62 0
-4.61 0
-4.6 0
-4.59 0
-4.58 0
-4.57 0
-4.56 0
-4.55 0
-4.54 0
-4.53 0
-4.52 0
-4.51 0
-4.5 0
-4.49 0
-4.48 0
-4.47 0
-4.46 0
-4.45 0
-4.44 0
-4.43 0
-4.42 0
-4.41 0
-4.4 0
-4.39 0
-4.38 0
-4.37 0
-4.36 0
-4.35 0
-4.34 0
-4.33 0
-4.32 0
-4.31 0
-4.3 0
-4.29 0
-4.28 0
-4.27 0
-4.26 0
-4.25 0
-4.24 0
-4.23 0
-4.22 0
-4.21 0
-4.2 0
-4.19 0
-4.18 0
-4.17 0
-4.16 0
-4.15 0
-4.14 0
-4.13 0
-4.12 0
-4.11 0
-4.1 0
-4.09 0
-4.08 0
-4.07 0
-4.06 0
-4.05 0
-4.04 0
-4.03 0
-4.02 0
-4.01 0
-4 0
-3.99 0
-3.98 0
-3.97 0
-3.96 0
-3.95 0
-3.94 0
-3.93 0
-3.92 0
-3.91 0
-3.9 0
-3.89 0
-3.88 0
-3.87 0
-3.86 0
-3.85 0
-3.84 0
-3.83 0
-3.82 0
-3.81 0
-3.8 0
-3.79 0
-3.78 0
-3.77 0
-3.76 0
-3.75 0
-3.74 0
-3.73 0
-3.72 0
-3.71 0
-3.7 0
-3.69 0
-3.68 0
-3.67 0
-3.66 0
-3.65 0
-3.64 0
-3.63 0
-3.62 0
-3.61 0
-3.6 0
-3.59 0
-3.58 0
-3.57 0
-3.56 0
-3.55 0
-3.54 0
-3.53 0
-3.52 0
-3.51 0
-3.5 0
-3.49 0
-3.48 0
-3.47 0
-3.46 0
-3.45 0
-3.44 0
-3.43 0
-3.42 0
-3.41 0
-3.4 0
-3.39 0
-3.38 0
-3.37 0
-3.36 0
-3.35 0
-3.34 0
-3.33 0
-3.32 0
-3.31 0
-3.3 0
-3.29 0
-3.28 0
-3.27 0
-3.26 0
-3.25 0
-3.24 0
-3.23 0
-3.22 0
-3.21 0
-3.2 0
-3.19 0
-3.18 0
-3.17 0
-3.16 0
-3.15 0
-3.14 0
-3.13 0
-3.12 0
-3.11 0
-3.1 0
-3.09 0
-3.08 0
-3.07 0
-3.06 0
-3.05 0
-3.04 0
-3.03 0
-3.02 0
-3.01 0
-3 0
-2.99 0
-2.98 0
-2.97 0
-2.96 0
-2.95 0
-2.94 0
-2.93 0
-2.92 0
-2.91 0
-2.9 0
-2.89 0
-2.88 0
-2.87 0
-2.86 0
-2.85 0
-2.84 0
-2.83 0
-2.82 0
-2.81 0
-2.8 0
-2.79 0
-2.78 0
-2.77 0
-2.76 0
-2.75 0
-2.74 0
-2.73 0
-2.72 0
-2.71 0
-2.7 0
-2.69 0
-2.68 0
-2.67 0
-2.66 0
-2.65 0
-2.64 0
-2.63 0
-2.62 0
-2.61 0
-2.6 0
-2.59 0
-2.58 0
-2.57 0
-2.56 0
-2.55 0
-2.54 0
-2.53 0
-2.52 0
-2.51 0
-2.5 0
-2.49 0
-2.48 0
-2.47 0
-2.46 0
-2.45 0
-2.44 0
-2.43 0
-2.42 0
-2.41 0
-2.4 0
-2.39 0
-2.38 0
-2.37 0
-2.36 0
-2.35 0
-2.34 0
-2.33 0
-2.32 0
-2.31 0
-2.3 0
-2.29 0
-2.28 0
-2.27 0
-2.26 0
-2.25 0
-2.24 0
-2.23 0
-2.22 0
-2.21 0
-2.2 0
-2.19 0
-2.18 0
-2.17 0
-2.16 0
-2.15 0
-2.14 0
-2.13 0
-2.12 0
-2.11 0
-2.1 0
-2.09 0
-2.08 0
-2.07 0
-2.06 0
-2.05 0
-2.04 0
-2.03 0
-2.02 0
-2.01 0
-2 0
-1.99 0
-1.98 0
-1.97 0
-1.96 0
-1.95 0
-1.94 0
-1.93 0
-1.92 0
-1.91 0
-1.9 0
-1.89 0
-1.88 0
-1.87 0
-1.86 0
-1.85 0
-1.84 0
-1.83 0
-1.82 0
-1.81 0
-1.8 0
-1.79 0
-1.78 0
-1.77 0
-1.76 0
-1.75 0
-1.74 0
-1.73 0
-1.72 0
-1.71 0
-1.7 0
-1.69 0
-1.68 0
-1.67 0
-1.66 0
-1.65 0
-1.64 0
-1.63 0
-1.62 0
-1.61 0
-1.6 0
-1.59 0
-1.58 0
-1.57 0
-1.56 0
-1.55 0
-1.54 0
-1.53 0
-1.52 0
-1.51 0
-1.5 0
-1.49 0
-1.48 0
-1.47 0
-1.46 0
-1.45 0
-1.44 0
-1.43 0
-1.42 0
-1.41 0
-1.4 0
-1.39 0
-1.38 0
-1.37 0
-1.36 0
-1.35 0
-1.34 0
-1.33 0
-1.32 0
-1.31 0
-1.3 0
-1.29 0
-1.28 0
-1.27 0
-1.26 0
-1.25 0
-1.24 0
-1.23 0
-1.22 0
-1.21 0
-1.2 0
-1.19 0
-1.18 0
-1.17 0
-1.16 0
-1.15 0
-1.14 0
-1.13 0
-1.12 0
-1.11 0
-1.1 0
-1.09 0
-1.08 0
-1.07 0
-1.06 0
-1.05 0
-1.04 0
-1.03 0
-1.02 0
-1.01 0
-1 0
-0.99 0
-0.98 0
-0.97 0
-0.96 0
-0.95 0
-0.94 0
-0.93 0
-0.92 0
-0.91 0
-0.9 0
-0.89 0
-0.88 0
-0.87 0
-0.86 0
-0.85 0
-0.84 0
-0.83 0
-0.82 0
-0.81 0
-0.8 0
-0.79 0
-0.78 0
-0.77 0
-0.76 0
-0.75 0
-0.74 0
-0.73 0
-0.72 0
-0.71 0
-0.7 0
-0.69 0
-0.68 0
-0.67 0
-0.66 0
-0.649999999999999 0
-0.64 0
-0.63 0
-0.62 0
-0.61 0
-0.6 0
-0.59 0
-0.58 0
-0.57 0
-0.56 0
-0.55 0
-0.54 0
-0.53 0
-0.52 0
-0.51 0
-0.5 0
-0.49 0
-0.48 0
-0.47 0
-0.46 0
-0.45 0
-0.44 0
-0.43 0
-0.42 0
-0.41 0
-0.399999999999999 0
-0.39 0
-0.38 0
-0.37 0
-0.36 0
-0.35 0
-0.34 0
-0.33 0
-0.32 0
-0.31 0
-0.3 0
-0.29 0
-0.28 0
-0.27 0
-0.26 0
-0.25 0
-0.24 0
-0.23 0
-0.22 0
-0.21 0
-0.2 0
-0.19 0
-0.18 0
-0.17 0
-0.16 0
-0.149999999999999 0
-0.14 0
-0.13 0
-0.12 0
-0.11 0
-0.0999999999999996 0
-0.0899999999999999 0
-0.0800000000000001 0
-0.0700000000000003 0
-0.0599999999999996 0
-0.0499999999999998 0
-0.04 0
-0.0300000000000002 0
-0.0199999999999996 0
-0.00999999999999979 0
};
\addlegendentry{$f_0(t) / \sqrt{\Gamma}$}
\addplot [thick, color0, mark=*, mark size=1, mark options={solid}, only marks, forget plot]
table {%
0 0.707106781186548
};
\addplot [thick, color0, forget plot]
table {%
0.00999999999999979 1.4001419023133
0.0200000000000005 1.38621025761053
0.0300000000000002 1.37241723508869
0.04 1.35876145543406
0.0499999999999998 1.34524155305726
0.0600000000000005 1.33185617595682
0.0700000000000003 1.31860398558385
0.0800000000000001 1.30548365670828
0.0899999999999999 1.29249387728629
0.100000000000001 1.27963334832911
0.11 1.26690078377312
0.12 1.25429491035127
0.13 1.24181446746571
0.14 1.22945820706173
0.15 1.21722489350302
0.16 1.20511330344801
0.17 1.19312222572762
0.18 1.18125046122406
0.19 1.16949682275101
0.2 1.15786013493482
0.21 1.14633923409701
0.22 1.13493296813789
0.23 1.12364019642137
0.24 1.11245978966086
0.25 1.10139062980637
0.26 1.09043160993269
0.27 1.07958163412869
0.28 1.06883961738777
0.29 1.05820448549928
0.3 1.04767517494119
0.31 1.03725063277366
0.32 1.02692981653378
0.33 1.01671169413133
0.34 1.00659524374556
0.350000000000001 0.996579453722993
0.36 0.986663322476286
0.37 0.976845858384049
0.38 0.967126079691691
0.39 0.957503014413243
0.4 0.947975700234158
0.41 0.93854318441508
0.42 0.929204523696565
0.43 0.919958784204759
0.44 0.910805041358008
0.45 0.9017423797744
0.46 0.892769893180224
0.47 0.883886684319343
0.48 0.875091864863469
0.49 0.866384555323327
0.5 0.857763884960707
0.51 0.849228991701389
0.52 0.840779022048933
0.53 0.832413130999334
0.54 0.824130481956515
0.55 0.81593024664867
0.56 0.807811605045433
0.57 0.79977374527588
0.58 0.791815863547334
0.59 0.783937164064992
0.600000000000001 0.77613685895234
0.61 0.768414168172365
0.62 0.760768319449556
0.63 0.753198548192667
0.64 0.745704097418265
0.65 0.738284217675026
0.66 0.730938166968794
0.67 0.723665210688376
0.68 0.716464621532084
0.69 0.709335679435
0.7 0.702277671496975
0.71 0.695289891911333
0.72 0.688371641894292
0.73 0.681522229615085
0.74 0.674740970126778
0.75 0.668027185297769
0.76 0.661380203743981
0.77 0.654799360761719
0.78 0.648283998261201
0.79 0.641833464700749
0.8 0.635447115021629
0.81 0.629124310583552
0.82 0.622864419100805
0.83 0.616666814579025
0.84 0.610530877252592
0.850000000000001 0.604455993522662
0.86 0.5984415558958
0.87 0.592486962923231
0.88 0.586591619140694
0.89 0.5807549350089
0.9 0.57497632685457
0.91 0.569255216812074
0.92 0.563591032765641
0.93 0.557983208292145
0.94 0.552431182604465
0.95 0.546934400495407
0.96 0.541492312282179
0.97 0.536104373751424
0.98 0.530770046104798
0.99 0.525488795905094
1 0.520260095022889
1.01 0.515083420583738
1.02 0.509958254915882
1.03 0.504884085498485
1.04 0.499860404910376
1.05 0.49488671077931
1.06 0.489962505731728
1.07 0.485087297343023
1.08 0.480260598088293
1.09 0.475481925293591
1.1 0.470750801087654
1.11 0.466066752354119
1.12 0.461429310684211
1.13 0.456838012329896
1.14 0.452292398157514
1.15 0.447792013601859
1.16 0.443336408620726
1.17 0.438925137649903
1.18 0.434557759558617
1.19 0.430233837605419
1.2 0.425952939394512
1.21 0.421714636832506
1.22 0.417518506085614
1.23 0.413364127537263
1.24 0.409251085746137
1.25 0.405178969404629
1.26 0.401147371297713
1.27 0.397155888262216
1.28 0.39320412114651
1.29 0.38929167477059
1.3 0.385418157886558
1.31 0.381583183139497
1.32 0.377786367028738
1.33 0.374027329869504
1.34 0.370305695754948
1.35 0.366621092518556
1.36 0.362973151696935
1.37 0.359361508492962
1.38 0.355785801739307
1.39 0.352245673862315
1.4 0.348740770846249
1.41 0.345270742197886
1.42 0.341835240911468
1.43 0.338433923434006
1.44 0.335066449630916
1.45 0.331732482752013
1.46 0.328431689397829
1.47 0.325163739486279
1.48 0.321928306219648
1.49 0.318725066051914
1.5 0.31555369865639
1.51 0.312413886893694
1.52 0.309305316780033
1.53 0.306227677455806
1.54 0.303180661154514
1.55 0.30016396317199
1.56 0.29717728183592
1.57 0.294220318475682
1.58 0.291292777392476
1.59 0.288394365829755
1.6 0.285524793943945
1.61 0.282683774775468
1.62 0.27987102422004
1.63 0.27708626100026
1.64 0.274329206637486
1.65 0.271599585423984
1.66 0.268897124395358
1.67 0.266221553303254
1.68 0.263572604588332
1.69 0.260950013353513
1.7 0.258353517337489
1.71 0.255782856888494
1.72 0.25323777493834
1.73 0.250718016976713
1.74 0.248223331025716
1.75 0.245753467614675
1.76 0.24330817975519
1.77 0.240887222916439
1.78 0.238490355000719
1.79 0.236117336319242
1.8 0.233767929568162
1.81 0.231441899804846
1.82 0.229139014424379
1.83 0.226859043136305
1.84 0.224601757941594
1.85 0.222366933109846
1.86 0.220154345156715
1.87 0.217963772821563
1.88 0.21579499704533
1.89 0.213647800948631
1.9 0.211521969810068
1.91 0.209417291044754
1.92 0.20733355418306
1.93 0.205270550849563
1.94 0.20322807474221
1.95 0.201205921611688
1.96 0.199203889241
1.97 0.19722177742524
1.98 0.195259387951574
1.99 0.19331652457942
2 0.191392993020822
2.01 0.18948860092102
2.02 0.187603157839219
2.03 0.185736475229537
2.04 0.18388836642216
2.05 0.182058646604666
2.06 0.180247132803548
2.07 0.178453643865917
2.08 0.176678000441384
2.09 0.174920024964128
2.1 0.173179541635135
2.11 0.171456376404622
2.12 0.169750356954631
2.13 0.168061312681795
2.14 0.166389074680278
2.15 0.164733475724887
2.16 0.163094350254347
2.17 0.161471534354745
2.18 0.159864865743138
2.19 0.158274183751327
2.2 0.156699329309786
2.21 0.15514014493176
2.22 0.153596474697511
2.23 0.152068164238729
2.24 0.150555060723095
2.25 0.149057012838996
2.26 0.147573870780396
2.27 0.146105486231852
2.28 0.144651712353686
2.29 0.1432124037673
2.3 0.141787416540634
2.31 0.140376608173778
2.32 0.138979837584721
2.33 0.13759696509524
2.34 0.136227852416932
2.35 0.13487236263739
2.36 0.133530360206505
2.37 0.132201710922917
2.38 0.13088628192059
2.39 0.129583941655527
2.4 0.128294559892616
2.41 0.127018007692607
2.42 0.125754157399216
2.43 0.124502882626361
2.44 0.123264058245521
2.45 0.122037560373226
2.46 0.120823266358666
2.47 0.119621054771429
2.48 0.118430805389353
2.49 0.117252399186509
2.5 0.116085718321295
2.51 0.11493064612465
2.52 0.113787067088395
2.53 0.11265486685367
2.54 0.111533932199511
2.55 0.110424151031517
2.56 0.109325412370646
2.57 0.108237606342118
2.58 0.107160624164422
2.59 0.106094358138443
2.6 0.10503870163669
2.61 0.103993549092633
2.62 0.102958795990147
2.63 0.10193433885306
2.64 0.100920075234803
2.65 0.0999159037081706
2.66 0.0989217238551722
2.67 0.0979374362569944
2.68 0.0969629424840571
2.69 0.0959981450861708
2.7 0.0950429475827919
2.71 0.0940972544533741
2.72 0.0931609711278162
2.73 0.0922340039770054
2.74 0.0913162603034543
2.75 0.0904076483320307
2.76 0.0895080772007803
2.77 0.0886174569518401
2.78 0.0877356985224434
2.79 0.0868627137360122
2.8 0.0859984152933405
2.81 0.0851427167638636
2.82 0.0842955325770157
2.83 0.083456778013672
2.84 0.0826263691976772
2.85 0.0818042230874577
2.86 0.0809902574677173
2.87 0.0801843909412159
2.88 0.0793865429206291
2.89 0.07859663362049
2.9 0.0778145840492104
2.91 0.0770403160011814
2.92 0.0762737520489531
2.93 0.0755148155354912
2.94 0.0747634305665121
2.95 0.0740195220028928
2.96 0.0732830154531569
2.97 0.0725538372660358
2.98 0.0718319145231029
2.99 0.0711171750314825
3 0.0704095473166297
3.01 0.0697089606151834
3.02 0.0690153448678897
3.03 0.0683286307125957
3.04 0.0676487494773136
3.05 0.0669756331733534
3.06 0.0663092144885238
3.07 0.065649426780401
3.08 0.0649962040696642
3.09 0.064349481033498
3.1 0.06370919299906
3.11 0.063075275937013
3.12 0.0624476664551225
3.13 0.0618263017919176
3.14 0.0612111198104139
3.15 0.0606020589919007
3.16 0.0599990584297886
3.17 0.0594020578235188
3.18 0.0588109974725333
3.19 0.0582258182703043
3.2 0.0576464616984241
3.21 0.0570728698207525
3.22 0.0565049852776239
3.23 0.0559427512801107
3.24 0.0553861116043447
3.25 0.0548350105858944
3.26 0.0542893931141986
3.27 0.0537492046270555
3.28 0.0532143911051664
3.29 0.0526848990667331
3.3 0.0521606755621108
3.31 0.0516416681685121
3.32 0.0511278249847651
3.33 0.0506190946261233
3.34 0.0501154262191268
3.35 0.0496167693965152
3.36 0.0491230742921907
3.37 0.0486342915362313
3.38 0.0481503722499544
3.39 0.0476712680410279
3.4 0.0471969309986317
3.41 0.0467273136886663
3.42 0.0462623691490093
3.43 0.0458020508848193
3.44 0.0453463128638863
3.45 0.0448951095120284
3.46 0.0444483957085344
3.47 0.0440061267816517
3.48 0.0435682585041191
3.49 0.0431347470887439
3.5 0.0427055491840233
3.51 0.0422806218698092
3.52 0.041859922653016
3.53 0.0414434094633716
3.54 0.0410310406492097
3.55 0.0406227749733054
3.56 0.0402185716087509
3.57 0.0398183901348728
3.58 0.0394221905331904
3.59 0.0390299331834131
3.6 0.0386415788594793
3.61 0.0382570887256329
3.62 0.03787642433254
3.63 0.0374995476134441
3.64 0.0371264208803594
3.65 0.0367570068203015
3.66 0.0363912684915565
3.67 0.0360291693199869
3.68 0.0356706730953736
3.69 0.0353157439677956
3.7 0.0349643464440441
3.71 0.0346164453840742
3.72 0.0342720059974899
3.73 0.0339309938400653
3.74 0.0335933748103007
3.75 0.0332591151460117
3.76 0.0329281814209534
3.77 0.0326005405414773
3.78 0.0322761597432226
3.79 0.0319550065878391
3.8 0.0316370489597437
3.81 0.0313222550629085
3.82 0.0310105934176817
3.83 0.0307020328576388
3.84 0.0303965425264668
3.85 0.0300940918748779
3.86 0.0297946506575551
3.87 0.0294981889301269
3.88 0.0292046770461737
3.89 0.0289140856542624
3.9 0.0286263856950117
3.91 0.0283415483981859
3.92 0.028059545279818
3.93 0.0277803481393611
3.94 0.0275039290568684
3.95 0.0272302603902015
3.96 0.0269593147722655
3.97 0.026691065108273
3.98 0.0264254845730339
3.99 0.0261625466082734
4 0.025902224919976
4.01 0.0256444934757558
4.02 0.0253893265022536
4.03 0.0251366984825596
4.04 0.024886584153661
4.05 0.0246389585039168
4.06 0.0243937967705555
4.07 0.0241510744371994
4.08 0.0239107672314131
4.09 0.0236728511222756
4.1 0.0234373023179777
4.11 0.0232040972634429
4.12 0.0229732126379711
4.13 0.0227446253529076
4.14 0.0225183125493333
4.15 0.0222942515957792
4.16 0.0220724200859633
4.17 0.0218527958365498
4.18 0.0216353568849306
4.19 0.0214200814870295
4.2 0.0212069481151272
4.21 0.0209959354557089
4.22 0.0207870224073329
4.23 0.0205801880785201
4.24 0.0203754117856655
4.25 0.0201726730509689
4.26 0.0199719516003881
4.27 0.0197732273616107
4.28 0.0195764804620472
4.29 0.0193816912268436
4.3 0.0191888401769142
4.31 0.0189979080269933
4.32 0.0188088756837067
4.33 0.0186217242436626
4.34 0.018436434991561
4.35 0.0182529893983223
4.36 0.0180713691192343
4.37 0.0178915559921178
4.38 0.0177135320355102
4.39 0.0175372794468674
4.4 0.0173627806007838
4.41 0.0171900180472292
4.42 0.0170189745098045
4.43 0.0168496328840133
4.44 0.0166819762355519
4.45 0.0165159877986157
4.46 0.0163516509742228
4.47 0.0161889493285537
4.48 0.0160278665913083
4.49 0.0158683866540786
4.5 0.015710493568738
4.51 0.0155541715458465
4.52 0.0153994049530713
4.53 0.0152461783136244
4.54 0.015094476304714
4.55 0.0149442837560128
4.56 0.0147955856481409
4.57 0.0146483671111635
4.58 0.0145026134231042
4.59 0.0143583100084728
4.6 0.0142154424368075
4.61 0.0140739964212321
4.62 0.0139339578170272
4.63 0.0137953126202157
4.64 0.0136580469661624
4.65 0.0135221471281874
4.66 0.0133875995161937
4.67 0.013254390675308
4.68 0.0131225072845353
4.69 0.0129919361554264
4.7 0.0128626642307597
4.71 0.012734678583235
4.72 0.0126079664141809
4.73 0.0124825150522749
4.74 0.0123583119522762
4.75 0.0122353446937715
4.76 0.0121136009799322
4.77 0.0119930686362856
4.78 0.011873735609497
4.79 0.011755589966164
4.8 0.0116386198916241
4.81 0.0115228136887722
4.82 0.0114081597768915
4.83 0.0112946466904954
4.84 0.0111822630781806
4.85 0.0110709977014921
4.86 0.0109608394337997
4.87 0.0108517772591847
4.88 0.0107438002713388
4.89 0.0106368976724733
4.9 0.0105310587722391
4.91 0.010426272986658
4.92 0.0103225298370642
4.93 0.0102198189490563
4.94 0.0101181300514598
4.95 0.0100174529753002
4.96 0.00991777765278605
4.97 0.00981909411630205
4.98 0.00972139249741228
4.99 0.00962466302587344
5 0.00952889602865776
};
\end{axis}

\end{tikzpicture}
    \caption{Amplitude of the ``ringdown'' function $f_{0}(t)$ in \eqref{eq:ringdown}. Note that we make explicit the fact that the Heaviside function takes the value 1/2 when evaluated at 0.}
    \label{fig:ringdown_amplitude}
\end{figure}

Yet, there exist some particular simplified scenarios where an analytical
solution can indeed be written in a closed form. For simplicity, here
we focus on the DP-SFWM case, but one can get to similar conclusions
for SP-SFWM and non-degenerate SPDC. Furthermore, we imagine that
XPM from the pump modes to the signal mode can be neglected and that
the pumps are on-resonance (hence $\Delta_{P1(P2)}=0$). Under these
conditions we can rephrase \eqref{eq: DPSFWM_M_matrix} as 
\begin{align}
M_{\text{DP-SFWM}}(t) & =-\overline{\Gamma}_{\text{S}}\mathcal{\mathbb{I}}_{2}+B(t),\\
B(t)= & \left[\begin{array}{lr} 0 & f(t)\\
f^{*}(t) & 0
\end{array}\right]
\end{align}
where $f(t)=2i\Lambda\tilde{\beta}_{\text{P1}}(t)\tilde{\beta}_{\text{P2}}(t)e^{-i\Delta_{\text{ring}}t}$
. Ignoring any additional contribution to and from the phantom channel
(hence, working with a lossless system), we can express the solution
using the Green function formalism as
\begin{equation}
\begin{pmatrix}\tilde{c}_{\text{S}}(t)\\
\tilde{c}_{\text{S}}^{\dagger}(t)
\end{pmatrix}=G(t_{0},t)\begin{pmatrix}\tilde{c}_{\text{S}}(t)\\
\tilde{c}_{\text{S}}^{\dagger}(t)
\end{pmatrix}+\int_{-\infty}^{t}\dd t^{\prime}\ \theta(t-t')G\left(t,t^{\prime}\right)\begin{pmatrix}-i\gamma_{\text{S}}^{*}\tilde{\psi}_{\text{S<}}(0,t^{\prime})\\
i\gamma_{\text{S}}\tilde{\psi}_{\text{S<}}^{\dagger}(0,t^{\prime})
\end{pmatrix}\label{eq: DPSFWM_euqations_green}
\end{equation}
with $G(t_{0},t_{0})=\mathbb{I}_{2}$. The general solution of \eqref{eq: DPSFWM_euqations_green}
is
\begin{align}
G\left(t,t^{\prime}\right) & =e^{-\overline{\Gamma}_{\text{S}}\left(t-t^{\prime}\right)}S\left(t,t^{\prime}\right)\\
S\left(t,t^{\prime}\right) & =\mathcal{T}\exp\left(\int_{t^{\prime}}^{t}\dd t^{\prime\prime}B\left(t^{\prime\prime}\right)\right)\in SU(1,1)\nonumber 
\end{align}
where $\mathcal{T}$ is the time-ordering operator. We can now make
an assumption on the shape of the pump, and take it as real and infinitely
short, such that $f(t)=f^{*}(t)=\mu\delta(t)$. In these conditions
the time-ordering correction is superfluous and the solution becomes 

\begin{equation}
S\left(t,t^{\prime}\right)=\begin{pmatrix}\cosh\left(\mu\left[\theta(t)-\theta\left(t^{\prime}\right)\right]\right) & \sinh\left(\mu\left[\theta(t)-\theta\left(t^{\prime}\right)\right]\right)\\
\sinh\left(\mu\left[\theta(t)-\theta\left(t^{\prime}\right)\right]\right) & \cosh\left(\mu\left[\theta(t)-\theta\left(t^{\prime}\right)\right]\right)
\end{pmatrix}.
\end{equation}
Finally, using \eqref{eq:across_jump} we obtain
\begin{align}
\psi_{>}(t)  =&\psi_{<}(t)-2\overline{\Gamma}_{\text{S}}\int_{-\infty}^{\infty}\dd t^{\prime}\theta\left(t-t^{\prime}\right)e^{-\overline{\Gamma}_{\text{S}}\left(t-t^{\prime}\right)}S_{1,1}\left(t,t^{\prime}\right)\psi_{\text{in }}\left(t^{\prime}\right)\\
 & +2\overline{\Gamma}_{\text{S}}\int_{-\infty}^{\infty}\dd t^{\prime}\theta\left(t-t^{\prime}\right)e^{-\overline{\Gamma}_{\text{S}}\left(t-t^{\prime}\right)}S_{1,2}\left(t,t^{\prime}\right)\psi_{\text{in }}^{\dagger}\left(t^{\prime}\right),\nonumber \\
  =&\int \dd t^{\prime}\left[\delta\left(t-t^{\prime}\right)-\left(2\overline{\Gamma}_{\text{S}}\theta\left(t-t^{\prime}\right)e^{-\overline{\Gamma}_{\text{S}}\left(t-t^{\prime}\right)}+[\cosh\mu-1]f_{0}(t)f_{0}\left(-t^{\prime}\right)\right)\right]\psi_{\text{in }}\left(t^{\prime}\right)\nonumber \\
 & +\int \dd t^{\prime}\sinh\mu f_{0}(t)f_{0}(-t)\psi_{\text{in }}^{\dagger}\left(t^{\prime}\right),\nonumber 
\end{align}
where (see Fig.~\ref{fig:ringdown_amplitude})
\begin{equation}\label{eq:ringdown}
f_{0}(t)=\sqrt{2\overline{\Gamma}_{\text{S}}}e^{-\overline{\Gamma}_{\text{S}}t}\theta(t).
\end{equation}
So, in the idealized case of infinitely short pump pulses and neglecting
both loss and XPM we can analytically express the input-output relation
for the squeezed mode and assess the single-Schmidt mode nature of
the generated state \cite{christensen2018engineering, vernon2017truly}.

\subsubsection{Example: Squeezing spectrum of a nanophotonic molecule}
To illustrate the practical applications of the formalism presented in this section, we now consider the real-case scenario of \cite{zhang2021squeezed}. The integrated resonator the authors propose is a ``photonic molecule" made of two coupled rings, where the additional cavity serves the purpose of suppressing spurious nonlinear processes, allowing for an effective three-mode modelling of the squeezer, precisely as we have so far assumed in all the previous derivations.

To illustrate the practical applications of the formalism presented in this section, we now consider the real-case scenario of \cite{zhang2021squeezed}. The integrated resonator the authors propose is a ``photonic molecule" made of two coupled rings, where the additional cavity serves the purpose of suppressing spurious nonlinear processes, allowing for an effective three-mode modelling of the squeezer, precisely as we have so far assumed in all the previous derivations.
In this particular instance, the rings are fabricated in silicon nitride and have a $1.5\times 0.8\ \mu m$ cross section. From the field distribution calculated by eigenmode simulation and the other geometrical parameters reported in \cite{zhang2021squeezed} one can extract a nonlinear parameter $\Lambda\sim 5\ Hz$.
The Hamiltonian of the system is the same as \eqref{eq:H_ring_full_DP}, and therefore the Heisenberg equation of motion for the signal mode is \eqref{eq: DP_SFWM_Signal_undeplated_pumps}.

\begin{figure}[ht]\centering
    \includegraphics[width=0.75\textwidth]{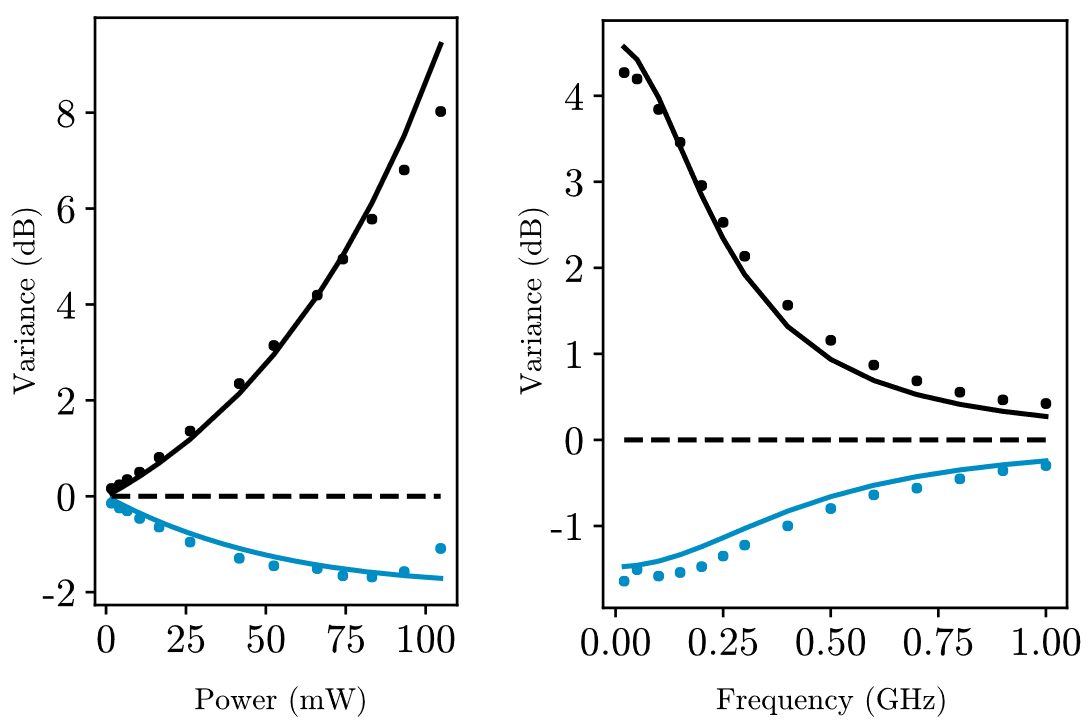}
    \caption{Squeezing and anti-squeezing spectra (blue and black, respectively) as a function of the pump power (left) and sideband frequency (right), as reported in \eqref{eq:sqz_spectrum}.}
    \label{fig:Squeezing_spectrum}
\end{figure}

In \cite{zhang2021squeezed}, the cavity is pumped with two CW lasers of equal intensity, and the squeezing spectrum is measured by balanced homodyne detection \cite{yuen1983noise, lvovsky2015squeezed}, that consists of interfering the signal with a local oscillator (LO) on a 50-50 splitter and recording the difference photocurrent while varying the phase of the LO.
Once the N and M moments are calculated as in \eqref{eq:DP_SFWM_N} and \eqref{eq:DP_SFWM_M}, it is more conveinet to work in the frequency domain by taking their Fourier transform $N_{\text{S}}(\Omega, \Omega')$ and $M_{\text{SS}}(\Omega, \Omega')$ (given the CW nature of the pumps). Then, the squeezing spectrum can be expressed as

\begin{equation}\label{eq:sqz_spectrum}
	S(\Omega) = 1 + N_{\text{S}}(\Omega, \Omega) + N_{\text{S}}(-\Omega, -\Omega) + 2\Re\big[e^{-2i\phi_{\text{LO}}} M_{\text{SS}}(\Omega, -\Omega)\big],
\end{equation}
with $\phi_{\text{LO}}$ the LO phase. This expression then allows to predict the squeezing spectra as a function of pumping (and cavity) parameters, such as the pump power and sideband frequency. We report in Fig.~\ref{fig:Squeezing_spectrum} the same conclusions from \cite{zhang2021squeezed}, where the experimental values are compared to the fitting following \eqref{eq:sqz_spectrum}. 

\subsection{Parametric Down-Conversion}

\begin{figure}[ht]\centering
    \includegraphics[width=1.0\textwidth]{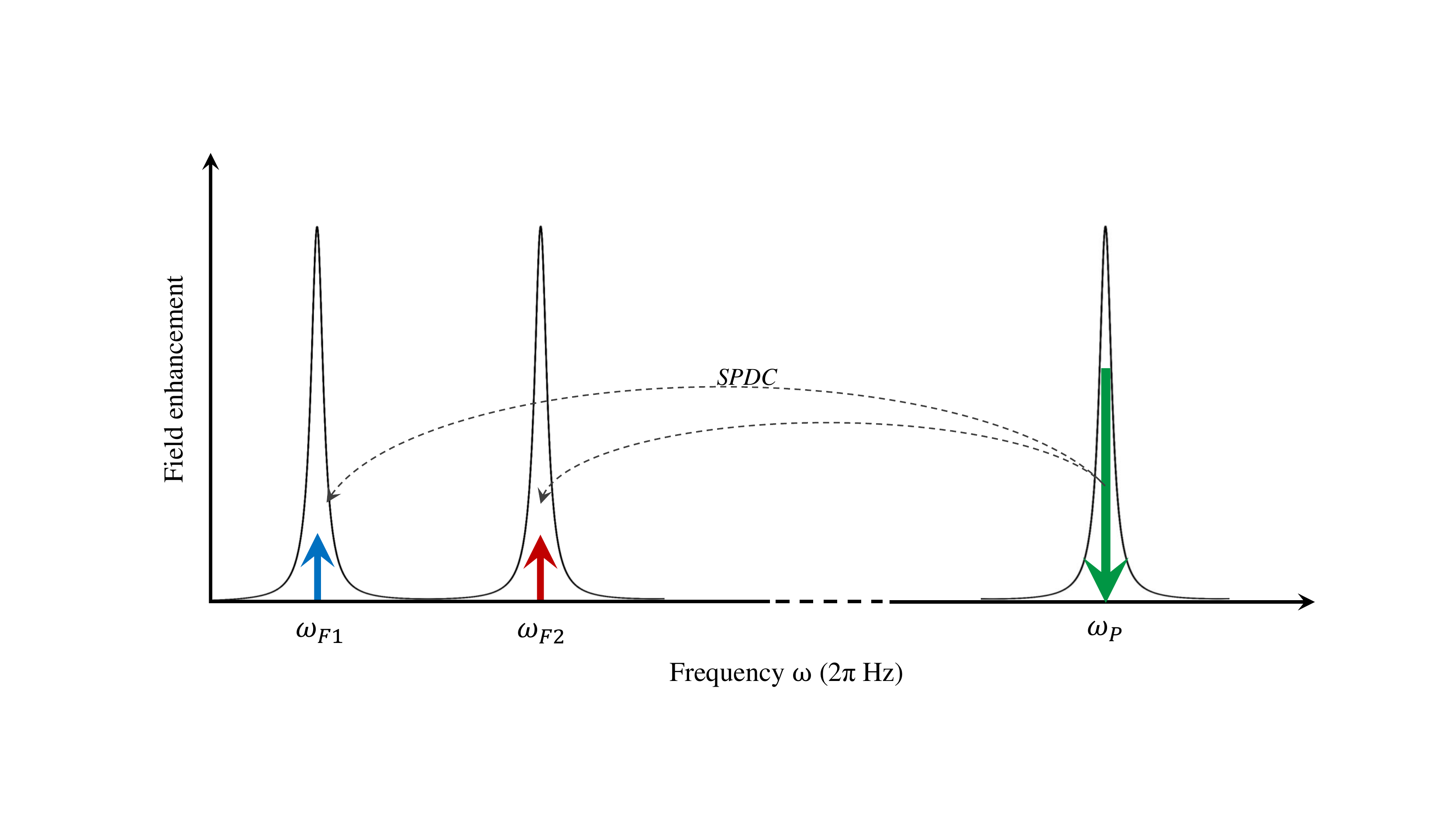}
    \caption{Ring resonances involved in a non-degenerate SPDC process.}
    \label{fig:SPDC}
\end{figure}

We can consider one last time the ring introduced above, and focus
on the second-order nonlinear response, specifically on SPDC. To this
end, we can select three resonances conventionally labeled second
harmonic (SH) and two fundamentals (F1 and F2), where the energy of
the former approximately equals the sum of the energies of the latter, as depicted in Fig.~\ref{fig:SPDC}.

Pumping the SH resonance with a bright field, the Hamiltonian governing the system becomes

\begin{align}
H_{\text{ring}}^{\text{SPDC}}  =&H_{\text{ring}}^{\text{L}}+H_{\text{ring}}^{\text{SPDC},\text{NL}}\label{eq: ND_SPDC_Hamiltonian}\\
  =&\sum_{J\in\{\text{F1,F2,SH}\}}\left(\hbar\omega_{J}\int\psi_{J}^{\dagger}(x)\psi_{J}(x)\dd x-\frac{1}{2}i\hbar v_{J}\int\left(\psi_{J}^{\dagger}(x)\frac{\partial\psi_{J}(x)}{\partial x}-\frac{\partial\psi_{J}^{\dagger}(x)}{\partial x}\psi_{J}(x)\right)\dd x\right)\nonumber \\
 & +\sum_{J\in\{\text{F1,F2,SH}\}}\hbar\omega_{J}c_{J}^{\dagger}c_{J}+\sum_{J\in\{\text{F1,F2,SH}\}}(\hbar\gamma_{J}c_{J}^{\dagger}\psi_{J}(0)+\text{H.c.})\nonumber \\
 & +\sum_{J\in\{\text{F1,F2,SH}\}}\left(\hbar\omega_{J}\int\phi_{J}^{\dagger}(x)\phi_{J}(x)\dd x-\frac{1}{2}i\hbar v_{J\text{ph}}\int\left(\phi_{J}^{\dagger}(x)\frac{\partial\phi_{J}(x)}{\partial x}-\frac{\partial\phi_{J}^{\dagger}(x)}{\partial x}\phi_{J}(x)\right)\dd x\right)\nonumber \\
 & +\sum_{J\in\{\text{F1,F2,SH}\}}(\hbar\gamma_{J\text{ph}}c_{J}^{\dagger}\phi_{J}(0)+\text{H.c.})\nonumber \\
 & -\hbar\Lambda_{\text{ring}}^{\text{SPDC}}c_{\text{SH}}c_{\text{F1}}^{\dagger}c_{\text{F2}}^{\dagger}+\text{H.c.,}\nonumber 
\end{align}
where $\Lambda_{\text{ring}}^{\text{SPDC}}$ is given by \eqref{eq: Lambda_SPDC}
and the only nonlinear term is responsible for the destruction of
a photon in the pump resonance and the correspondent creation of two
photons in the fundamental resonances. As for the SFWM cases, we can
drop the unnecessary notation, and express the equations of motion
as

\begin{align}
\Big(\ddt{}+\overline{\Gamma}_{\text{SH}}+i\omega_{\text{SH}}\Big)c_{\text{SH}}(t) & =i\Lambda^{*}c_{\text{F1}}(t)c_{\text{F2}}(t)-i\gamma_{\text{SH}}^{*}\psi_{SH<}\left(0,t\right)-i\gamma_{\text{SHph}}^{*}\phi_{SH<}(0,t)\\
\Big(\ddt{}+\overline{\Gamma}_{\text{F1}}+i\omega_{\text{F1}}\Big)c_{\text{F1}}(t) & =i\Lambda c_{\text{SH}}(t)c_{\text{F2}}^{\dagger}(t)-i\gamma_{\text{F1}}^{*}\psi_{F1<}(0,t)-i\gamma_{\text{F1ph}}^{*}\phi_{F1<}(0,t)\nonumber \\
\Big(\ddt{}+\overline{\Gamma}_{\text{F2}}+i\omega_{\text{F2}}\Big)c_{\text{F2}}(t) & =i\Lambda c_{\text{SH}}(t)c_{\text{F1}}^{\dagger}(t)-i\gamma_{\text{F2}}^{*}\psi_{F2<}(0,t)-i\gamma_{\text{F2ph}}^{*}\phi_{F2<}(0,t)\nonumber 
\end{align}
and in terms of the slowly-varying operators $\overline{c}_{J}(t)=e^{i\omega_{J}t}c_{J}(t)$,
\begin{align}
\Big(\ddt{}+\overline{\Gamma}_{\text{SH}}\Big)\overline{c}_{\text{SH}}(t) & =i\Lambda^{*}\overline{c}_{\text{F1}}(t)\overline{c}_{\text{F2}}(t)e^{i\Delta_{\text{ring}}t}-i\gamma_{\text{SH}}^{*}\overline{\psi}_{SH<}\left(0,t\right)-i\gamma_{\text{SHph}}^{*}\overline{\phi}_{SH<}(0,t)\label{eq: SPDC_Dynamic_Equations}\\
\Big(\ddt{}+\overline{\Gamma}_{\text{F1}}\Big)\overline{c}_{\text{F1}}(t) & =i\Lambda\overline{c}_{\text{SH}}(t)\overline{c}_{\text{F2}}^{\dagger}(t)e^{-i\Delta_{\text{ring}}t}-i\gamma_{\text{F1}}^{*}\overline{\psi}_{F1<}(0,t)-i\gamma_{\text{F1ph}}^{*}\overline{\phi}_{F1<}(0,t)\nonumber \\
\Big(\ddt{}+\overline{\Gamma}_{\text{F2}}\Big)\overline{c}_{\text{F2}}(t) & =i\Lambda\overline{c}_{\text{SH}}(t)\overline{c}_{\text{F1}}^{\dagger}(t)e^{-i\Delta_{\text{ring}}t}-i\gamma_{\text{F2}}^{*}\overline{\psi}_{F2<}(0,t)-i\gamma_{\text{F2ph}}^{*}\overline{\phi}_{F2<}(0,t)\nonumber 
\end{align}
with now $\Delta_{\text{ring}}=\omega_{\text{SH}}-\omega_{\text{F1}}-\omega_{\text{F2}}$.
Pumping the SH resonance with an intense pump we can make the substitutions

\begin{align}
\overline{c}_{\text{SH}}(t) & \rightarrow\left\langle \overline{c}_{\text{SH}}(t)\right\rangle e^{i\Delta_{\text{SH}}t}\equiv\beta_{\text{SH}}(t)e^{i\Delta_{\text{SH}}t}\\
\overline{\psi}_{\text{SH}<}(0,t) & \rightarrow\left\langle \overline{\psi}_{\text{SH}<}(0,t)\right\rangle e^{i\Delta_{\text{SH}}t}.\nonumber 
\end{align}
Once more assuming an undepleted pump  and ignoring all contributions
from the phantom channel, the pump equation becomes

\begin{equation}
\ddt{}\beta_{\text{SH}}(t)=\left(-\overline{\Gamma}_{\text{SH}}-i\Delta_{\text{SH}}\right)\beta_{\text{SH}}(t)-i\gamma_{\text{SH}}^{*}\left\langle \overline{\psi}_{SH<}(0,t)\right\rangle .\label{eq: SPDC_pump_equation}
\end{equation}
Finally, the down-converted photons Heisenberg equation can be expressed
in matrix form as 
\begin{equation}
\ddt{}\begin{pmatrix}\tilde{c}_{\text{F1}}(t)\\
\tilde{c}_{\text{F2}}^{\dagger}(t)
\end{pmatrix}=M_{\text{SPDC}}(t)\begin{pmatrix}\tilde{c}_{\text{F1}}(t)\\
\tilde{c}_{\text{F2}}^{\dagger}(t)
\end{pmatrix}+D_{\text{SPDC}}(t),\label{eq: Exact_equation_SPDC}
\end{equation}
where 
\begin{align}
M_{\text{SPDC}}(t) & =\begin{pmatrix}-\overline{\Gamma}_{\text{F1}}-i\frac{\Delta_{\text{SH}}-\Delta_{\text{ring}}}{2} & i\Lambda\beta_{\text{SH}}(t)\\
-i\Lambda^{*}\beta_{\text{SH}}^{*}(t) & -\overline{\Gamma}_{\text{F2}}+i\frac{\Delta_{\text{SH}}-\Delta_{\text{ring}}}{2}
\end{pmatrix},\\
D_{\text{SPDC}}(t) & =\begin{pmatrix}-i\gamma_{\text{F1}}^{*}\tilde{\psi}_{\text{F1}<}(0,t)-i\gamma_{\text{F1}ph}^{*}\tilde{\phi}_{\text{F1}<}(0,t)\\
i\gamma_{\text{F2}}\tilde{\psi}_{\text{F2}<}^{\dagger}(0,t)+i\gamma_{\text{F2}ph}\tilde{\phi}_{\text{F2}<}^{\dagger}(0,t)
\end{pmatrix}
\end{align}
and we have performed the usual transformation $\ensuremath{\tilde{c}_{\text{F1},\text{F2}}(t)=e^{-\frac{i}{2}(\Delta_{\text{SH}}-\Delta_{\text{ring}})t}\overline{c}_{\text{F1},\text{F2}}(t)}$.

The calculation of the $N$ and $M$ moments follow precisely the same path as for SP-SFWM (\ref{eq: N_M_def_SPSFWM}-\ref{eq: MSI_SPSFWM}) with the natural substitutions $\text{S}\rightarrow\text{F1}$ and $\text{I}\rightarrow\text{F2}$.

It is worth noting that the nonlinear Hamiltonian in \eqref{eq: ND_SPDC_Hamiltonian}
drives the creation of a NDSV state [cf. \eqref{eq:quad_hamil}] in modes F1 and F2.
However, in the limit $\text{F1}\rightarrow\text{F2}\equiv\text{F}$, the solution transitions to a DSV state in the F mode, and all of the apparatus for the calculation
of the Heisenberg equations is unchanged, provided that the SPDC nonlinear
coefficient is replaced by

\begin{equation}
\Lambda_{\text{ring}}^{\text{SPDC}}\rightarrow2\Lambda_{\text{ring}}^{\text{SPDC}}.
\end{equation}

\section{Connections between classical and quantum nonlinear optics}\label{sec:Connections}
In Sec.~\ref{sec:estim} we presented the evolution of a state different from vacuum in the presence of a quadratic Hamiltonian. We saw that even in the presence of a seed field, the joint amplitude appearing in the spontaneous case can act as a response function of the system. In other words, it seems as if there is a deep connection between spontaneous and stimulated nonlinear processes. Starting from these results, we now discuss this relation from a different point of view, with the aim of clarifying the connection between classical and quantum nonlinear optics, thus between those nonlinear phenomena that can be explained in the framework of a classical electromagnetic theory (e.g. second harmonic generation) and those that require the quantization of the electromagnetic field (e.g. spontaneous parametric down conversion). On the one hand, such a connection is interesting from a fundamental point of view, for it allows one to separate the features of a nonlinear phenomenon that are intrinsically quantum from those that are not. On the other hand, the link between classical and quantum nonlinear optics can also have important practical consequences. First, many systems that have been designed to enhance classical nonlinear processes can inspire new solutions to control and amplify quantum nonlinear interactions.  Second, as we shall see in this section, one can gain important information about the quantum correlations of nonclassical light generated by parametric fluorescence by studying the same system and  nonlinear interaction in a regime in which they can both be described classically, with advantages in terms of speed and accuracy. 

Here we restrict our analysis to the generation of squeezed light via spontaneous parametric down conversion or spontaneous four-wave mixing. As we have seen, both of these processes result in the generation of photon pairs and are associated with either the second- or third-order nonlinear optical response of the system. The generic name with which SPDC and SFWM are often referred to is \emph{parametric fluorescence}, which suggests a certain analogy with the spontaneous emission of light, for example occurring in an atomic system. This analogy is quite strong. Indeed, likewise the emission of a photon by an atomic system can be either spontaneous or stimulated by the radiation field, there exist also for SPDC and SFWM two \emph{corresponding} nonlinear processes, difference frequency generation (DFG) and stimulated four-wave mixing (FWM), in which the emission of photons pairs is stimulated by the presence of an additional field, usually labelled \emph{seed} \cite{helt2012does}. More interestingly, in the limit of a sufficiently weak seed field, thus neglecting SPM and XPM induced by it, both the stimulated and spontaneous emission of pairs are described by the very same Hamiltonian, and the probabilities of spontaneous and stimulated emission of photon pairs are related. Under these hypotheses, the study of the stimulated processes, i.e. DFG and FWM, allows one to gain information about the quantum properties and the generation rate of pairs that would be emitted spontaneously through SPDC or SFWM in the absence of the seed field \cite{liscidini2013stimulated}. 

\subsection{Low-gain regime: generation of two-photon states}
The connection between stimulated and spontaneous emission of photon pairs is manifest in the context of the generation of two-photon states, i.e. in the low-gain regime, for which one has a small probability of generating a photon pair per pump pulse. As an example, we consider the case of SPDC and DFG in a waveguide in the limit of a continuous wave pump. In this case the photon pair generation rates for spontaneous and stimulated emission can be written as \cite{helt2012does}:
\begin{equation}\label{eq:rate_spdc_sp}
    R_\mathrm{SPDC,wg}=\frac{P_P}{\mathcal{PA}}\frac{L^{3/2}}{\frac{3}{2}\sqrt{2\pi\left|\beta_2(\omega_P/2)\right|}}
\end{equation}
and 
\begin{equation}\label{eq:rate_spdc_st}
    R_\mathrm{DFG,wg}=\frac{P_P}{\mathcal{PA}}\frac{L^2}{\frac{1}{2}\hbar\omega_P}P_S.
\end{equation}
Here we assume phase matching between $\omega_P$ and $\omega_P/2$, with $P_P$ and $P_S$ the pump and seed powers, respectively. Finally, $\mathcal{P}$ is a characteristic power, which depends on the material nonlinearity, $\mathcal{A}$ is the nonlinear effective area (see \eqref{eq:chi2_A} in Appendix~\ref{sec:WG_Hamiltonians}), which we assume to be constant over the frequency range of interest here, $L$ is the length of the nonlinear interaction region, and $\beta_2$ is the group velocity dispersion.

\begin{figure}
    \centering
    \includegraphics[width=0.8\textwidth]{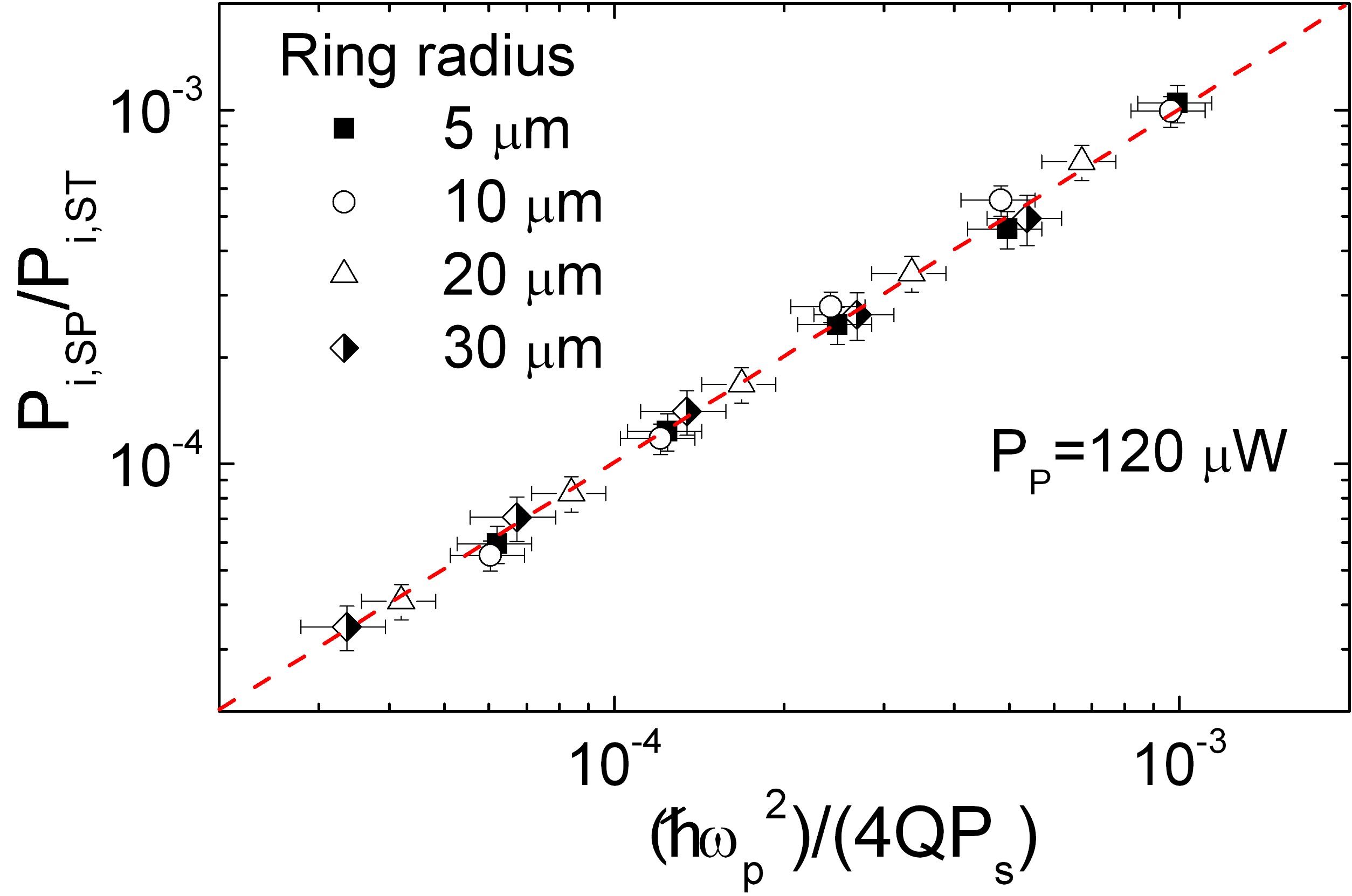}
    \caption{Ratio between the generated idler photons in spontaneous and stimulated FWM for several ring resonators. Data from Azzini et al.~\cite{azzini2012classical}}
    \label{fig:fwm_sfwm}
\end{figure}

By taking the ratio of  \eqref{eq:rate_spdc_st} and \eqref{eq:rate_spdc_sp}, one obtains a simple relation between the two rates 
\begin{equation}\label{eq:ratio_SPDC_DFG_wg}
    \frac{R_\mathrm{SPDC,wg}}{R_\mathrm{DFG,wg}}=\frac{\hbar\omega_P}{3\sqrt{2\pi\left|\beta_2(\omega_P/2)\right|L}}\frac{1}{P_S},
\end{equation}
in which SPDC appears as a process stimulated by a certain "vacuum power"
\begin{equation}
P_\mathrm{vac,wg}= \frac{\hbar\omega_P}{3\sqrt{2\pi\left|\beta_2(\omega_P/2)\right|L}},
\end{equation}
which is associated with the generation bandwidth, i.e. the frequency range over which the energy of the generated photons can fluctuate. 
Analogous  results can also be found for SPDC in resonant systems. For example, one can derive the generation rates for SPDC and DFG in the case of cw pumping in  ring resonators \cite{helt2012does}, with \begin{equation}\label{spdc_ring_cw}
R_\mathrm{SPDC,ring}=\frac{P_p}{\mathcal{PA}}\frac{4v_F^2v_P}{\pi\omega_p\omega_F}\frac{Q_PQ_F}{R},    
\end{equation}
and
\begin{equation}\label{dfg_ring_cw}
R_\mathrm{DFG,ring}=\frac{P_p}{\mathcal{PA}}\frac{4v_F^2v_P}{\hbar\pi\omega_p\omega_F^3}\frac{Q_PQ_F^2}{R},       
\end{equation}
where $\omega_P=2\omega_F$, $v_{F(P)}$ and $Q_{F(P)}$ are the group velocity and the quality factor at the fundamental (pump) frequency, respectively, and $R$ is the ring the radius. Please, note that in deriving these formula losses due to either scattering or absorption are neglected. By dividing \eqref{spdc_ring_cw} by \eqref{dfg_ring_cw}, one obtains
\begin{equation}\label{eq:ratio_SPDC_DFG_ring}
    \frac{R_\mathrm{SPDC,ring}}{R_\mathrm{DFG,ring}}=\frac{\hbar\omega_F^2}{8Q_F}\frac{1}{P_S},
\end{equation}
where one can identify
\begin{equation}
    P_\mathrm{vac,ring}= \frac{\hbar\omega_F^2}{8Q_F}
\end{equation}
as the power associated with the vacuum. Similar results can be obtained for photonic crystal structures \cite{introini2020spontaneous}, and for SFWM in either non-resonant or resonant systems \cite{helt2012does, azzini2012classical}. In Fig. \ref{fig:fwm_sfwm} Azzini et al., demonstrated that, for a given seed power, the ratio of the idler powers generated by stimulated and spontaneous FWM in ring resonators depends only the quality factor, which ultimately determined the power vacuum fluctuations in this system. This confirms that in the case of parametric nonlinear processes, spontaneous emission of photon pairs can be viewed as stimulated by vacuum power fluctuations. In addition,~\eqref{eq:ratio_SPDC_DFG_wg}, \eqref{eq:ratio_SPDC_DFG_ring} and similar relations, allow one to estimate the efficiency of the spontaneous process from that of the corresponding stimulated one. 

\begin{figure}
    \centering
    \includegraphics[width=0.8\textwidth]{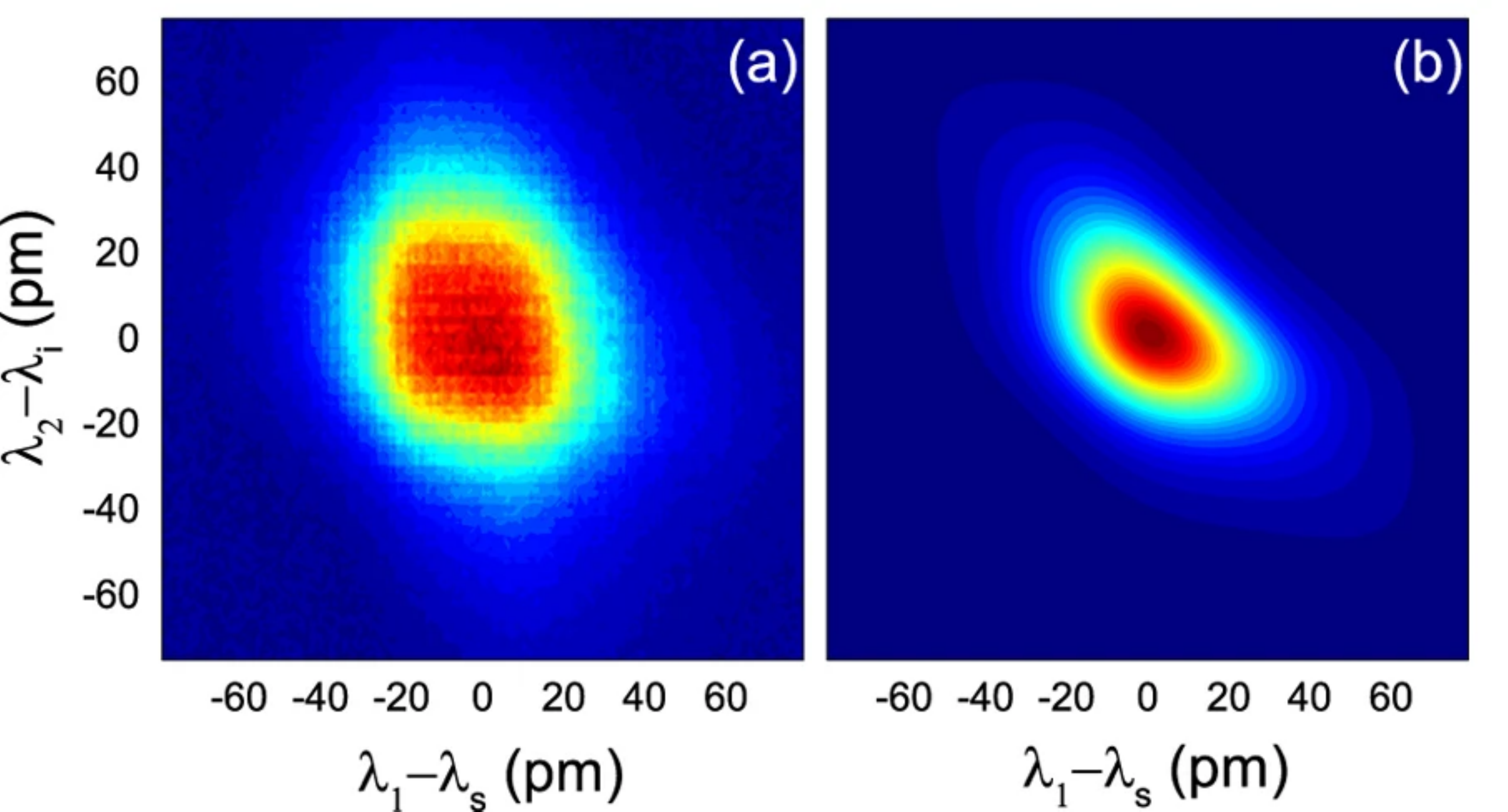}
    \caption{Joint spectral density measured by stimulated emission tomography (a) and calculated (b) in the case of photon pairs generated under pulsed pumping in a micro ring resonator. Image from Grassani et al.~\cite{grassani2016energy}}
    \label{fig:set_ring}
\end{figure}

In \cite{liscidini2013stimulated}, Liscidini and Sipe went even further and demonstrated that the biphoton wavefunction describing the spontaneously generated two-photon state acts as the response function characterizing how the seed pulse stimulates the emission of photon pairs in the corresponding classical process. Assuming the same pumping condition of the spontaneous process and a seed pulse sufficiently intense to be treated classically, but sufficiently weak to neglect any effect of SPM and XPM induced by the seed, it follows that
\begin{equation}\label{eq:stimulated_spontaneous_SET}
    \beta_{\sigma,\mathbf{k_2}}\approx\sqrt{2}\gamma\sum_{\sigma^{\prime}}\int\mathrm{d}\mathbf{k}_1\phi_{\sigma^{\prime},\sigma}(\mathbf{k_1},\mathbf{k_2})\beta_{\sigma^{\prime},\mathbf{k_1}}^{*},
\end{equation}
where $\beta_{\sigma,\mathbf{k_2}}$ and $\beta_{\sigma^{\prime},\mathbf{k_1}}^{*}$  are the amplitudes of the stimulated and seeded classical fields associated with light exiting the systems. Here, $\phi_{\sigma^{\prime},\sigma}(\mathbf{k_1},\mathbf{k_2})$ and $|\gamma|^2$ are the biphoton wavefunction and the pair generation rate, were they generated spontaneously. Finally $\sigma,\sigma^{\prime}$ and $\mathbf{k_1},\mathbf{k_2}$ label all the discrete (e.g. polarization) and continuous (e.g. momentum) degrees of freedom, respectively. Starting from the result of \eqref{eq:stimulated_spontaneous_SET}, one can imagine to characterize the two-photon state that would be generated by parametric fluorescence by performing a \emph{stimulated emission tomography} (SET), in which the seed is used to mimic one of the photon of the pairs and the stimulated beam is analysed to reconstruct the biphoton wavefunction. This has been experimentally demonstrated in the case of polarization entangled photon pairs \cite{rozema2015characterizing,fang2016multidimensional} and for two-photon states hyper-entangled in path and polarization \cite{ciampini2019stimulated}. Particularly interesting from a practical point of view is the study of energy correlations, which can be can particularly challenging for resonant systems, for it typically requires high spectral resolution. In this respect, stimulated emission has been used to reconstruct the joint spectral density (JSD) in several systems, with pairs generated either by SPDC \cite{eckstein2014high} or SFWM \cite{fang2014fast,jizan2015bi,grassani2016energy}. This is particularly useful in the case of high-Q resonant systems, in which high resolution is required. As an example, in Fig. \ref{fig:set_ring}, we show the comparison between the joint spectral density measured by SET for a silicon micro ring resonator and the expected theoretical result.
Finally, while most of the experiments have been focused on the JSD, there are also cases in which stimulated emission was used to determine the phase of the biphoton wavefunction \cite{jizan2016phase,borghi2020phase}.

\subsection{High-gain regime: characterization of squeezed light}
While the vast majority of the results regarding the connection between classical and quantum nonlinear processes have been obtained in the low-gain regime, when only a few photon pairs are generated per pump pulse, a similar link is expected to hold also in the high-gain regime, e.g. in the case of the generation bright squeezed light. Indeed, when effects associated with SPM and XPM induced by the seed field are negligible, stimulated and spontaneous processes are described by the same nonlinear Hamiltonian. This opens to the possibility of characterizing nonclassical light generated in a non-perturbative regime by exploiting the stimulated emission of photon pairs. 

Recently, Triginer et al. proposed in \cite{triginer2019understanding} that cascaded stimulated emission of photon pairs can be sued to measure the strength of squeezed light generated by SPDC. Unlike in SET, where the connection between the biphoton wavefunction and the stimulated idler is direct (see~\eqref{eq:stimulated_spontaneous_SET}), in this case the authors used two cascaded DFG processes to have a self-reference field, which is made to interfere with the generated squeezed light. From such an interference, it is possible to gain information about the squeezing strength and phase information, which depends on a non trivial interplay of SPM  and XPM associated with the pump field. 

\section{The world beyond Gaussian states: Tensor networks for the perplexed}\label{sec:Tensor}
\subsection{The curse of dimensionality in many-body systems and Matrix Product States}
In the sections~\ref{sec:Ket},~\ref{sec:Waveguides}, and~\ref{sec:Rings} we have introduced methods to describe the quantum states generated in quantum nonlinear optics whenever the Heisenberg equations of motion can be linearized. Under these circumstances we can use the methods from Sec.~\ref{sec:Ket}. In a nutshell, whenever we can linearize the equations of motion for the field operators we can always write the output of the spontaneous problem as a multimode squeezed state, which in the low-gain limit corresponds to pair production.
While these methods provide useful tools for many problems, they leave open the question of what can be done beyond linearized dynamics and Gaussian state generation.
A naive approach to this problem is to expand every possible time, position or frequency bin into an orthonormal basis, like for example the Fock basis. Thus a state over $\ell$ bins would be written as
\begin{align}
\ket{\Psi} = \sum_{i_0=0}^{c-1} \ldots \sum_{i_{\ell-1}=0}^{c-1} R_{i_0\ldots i_{\ell-1}} \ket{i_0,\ldots,i_{\ell-1}}
\end{align}
where $R_{i_0\ldots i_{\ell-1}}$ is an $\ell-$rank tensor $\ket{i_0 \ldots i_{\ell-1}} \equiv \ket{i_0} \otimes \ldots \otimes \ket{i_{\ell-1}}$ and $\ket{i_k} = \tfrac{a_k^{\dagger i}}{\sqrt{i!}} \ket{\text{vac}}$ is a Fock state with $i$ photons in the $k$-th bin. If we assume a minimal cutoff of one photon per mode ($c=2$), and consider for example $\ell = 100$ bins we arrive to the conclusion that we need to store in memory the $2^{100} \approx 10^{30}$ elements of the tensor $\bm{R}$. This is simply impractical and seems to put the simulation of these systems beyond any current or future classical (super)computer.

Now let us assume that the tensor $\bm{R}$ can be written in the following so-called Matrix Product State (MPS)~\cite{schollwock2011density,orus2014practical,montangero2018introduction} form
\begin{align}\label{eq:mps}
 R_{i_0,\ldots,i_{\ell-1}} =\sum_{\alpha_1=0}^{D_1-1} \ldots \sum_{\alpha_{\ell-1} =0}^{D_{\ell-1} - 1} (S^{[0]})_{i_0,\alpha_1} (S^{[1]})_{i_1,\alpha_2}^{\alpha_1} \ldots (S^{[\ell-2]})_{i_{\ell-2},\alpha_{\ell-1}}^{\alpha_{\ell-2}} (S^{[\ell-1]})_{i_{\ell-1}}^{\alpha_{\ell-1}}  ,
\end{align}
where for $n \in \{1,\ldots,\ell-2\}$ the  $S^{[n]}$ are rank-three tensors while $S^{[0]}$ and $S^{[\ell-1]}$ are rank-two tensors (matrices). For different notations used when describing MPSs see appendix ~\ref{app:MPS}.
The indices $i_0,\ldots,i_{\ell-1}$ are called physical indices, while the contracted indices $\alpha_k$ in the sum in the last equation are known as auxiliary indices.
In general, an arbitrary tensor with $\ell$ indices can always be taken into the matrix product form above~\ref{app:MPS}, however, it is not obvious a-priori that the expression of~\eqref{eq:mps} is useful. In particular, we do not know the possible values $\{D_k\}$ the auxiliary indices $\{\alpha_k\}$ can take. The maximum of these quantities over all possible auxiliary indices, $D = \max\{D_1,\ldots, D_{\ell-1}\}$, is called the bond dimension of the tensor, and in principle for an arbitrary tensor this quantity can be exponentially large in the number of physical indices $\ell$. However, if this is not the case and the bond dimension is constant or grows only polynomially in $\ell$, then we can obtain significant savings in representing a tensor similarly to the form of~\eqref{eq:mps}. Indeed, if the bond dimension of the matrix product expression in~\eqref{eq:mps} is a constant $D$ then each of the rank-three tensors will have a total number of elements bounded by $D c^2$ and if we have a total of $\ell-2$ rank-three tensors the memory required to represent the tensor $\bm{R}$ will not be larger than $\ell D c^2$. In practice the cutoff can be chosen self-consistently, by increasing it until there is convergence.

If a quantum state can be efficiently represented as an MPS, then one can easily calculate expectation values of local observables such as correlations functions. As recently shown by Yanagimoto et al.~\cite{yanagimoto2021efficient} one can also extract the density matrix of any Schmidt (or broadband) mode $A_\lambda^\dagger  = \sum_{k=0}^{\ell-1} F_{k,\lambda} a_k^\dagger$ defined as a linear combination of the bare modes $\{a_k \}$.

In the following section we show how to construct a constant bond-dimension MPS representation of a resonant system with an arbitrary Hamiltonian coupled to a waveguide with a linear dispersion relation. Note that very recently it was also shown that MPSs provide an efficient description of quantum states where the nonlinearity is distributed along a dispersive nonlinear waveguide~\cite{yanagimoto2021efficient}.

\subsection{Matrix Product States in Cavity QED}
We start by writing a Hamiltonian for an isolated resonance coupled to a waveguide \cite{helt2010spontaneous}
\begin{align}
\tilde H = \tilde H_{\text{res}}    + \tilde H_{\text{wg}} + \tilde V,
\end{align}
where
\begin{align}
\tilde H_{\text{res}} &= \hbar \omega_0 a^\dagger a    + \tilde H_{\text{NL}}(a^\dagger,a,t), \\
\tilde H_{\text{wg}} &= \hbar \omega_1 \int \dd x \ \psi^\dagger(x) \psi(x) + \frac{i}{2} \hbar v \int \dd x \left( \left[\frac{\partial}{\partial x} \psi^\dagger(x) \right] \psi(x) - \text{H.c.} \right), \\
\tilde V &= \hbar \gamma \left( \psi^\dagger(0)a + \text{H.c.} \right),
\end{align}
and $\omega_0/\omega_1$ are the frequencies of the resonator/waveguide, $v$ is the group velocity of the waveguide mode and $\gamma^2/(2v)$ is the decay rate of the resonator into the waveguide. We furthermore assume point coupling between the waveguide and the resonator at position $x=0$.
All the calculations in the section will be done in the interaction picture, meaning that we will move some trivial time evolution into the operators, while the nontrivial part will be carried by the state vector. Moreover, we will ignore dispersion within each field in the waveguide, as done in Sec.~\ref{sec:Rings}. 

The nonzero canonical commutation relations of the field and resonance are
\begin{align}
[\psi(x),\psi^\dagger(x')]=\delta(x-x') \text{ and } [a,a^\dagger] = 1.
\end{align}
Note that $H_{\text{NL}}$ is a nonlinear Hamiltonian that is written in terms of resonator operators and that is possibly time-dependent.
We leave this completely general but it can be taken to be, for example, the one of a classical field with frequency $\omega_c \approx 3\omega_0$ coupled to the resonance of interest; for this case we would write the Hamiltonian
\begin{align}
\tilde{H}_{NL}(a^\dagger, a , t) = \mu \beta(t) e^{i \omega_c t} a^3 + \mu^* \beta^* e^{-i \omega_c t} a^{\dagger 3},
\end{align}
where $\mu$ is proportional to the coupling between the modes inside the resonator and $\beta(t) $ is the slowly varying amplitude of the field in the classical resonance.

We can now go to an interaction picture rotating at frequency $\omega_1$ by transforming the Hamiltonian using the following unitary~\ref{app:interaction_picture}
\begin{align}
\mathcal{ U}_1 &= \exp\left(- \frac{i}{\hbar} H_1 (t-t_0)  \right),     \quad H_1 = \hbar \omega_1\left[a^\dagger a + \int \dd x \ \psi^\dagger(x) \psi(x) \right],
\end{align}
which now gives the Hamiltonian
\begin{align}
\tilde H = \tilde H_{\text{res}}    + \tilde H_{\text{wg}} + \tilde V,    
\end{align}
where
\begin{align}
\bar H_{\text{res}} &= \hbar \delta a^\dagger a    + H_{\text{NL}}(t), \quad \delta = \omega_0 - \omega_1, \\
\bar H_{\text{wg}} &=  \frac{i}{2} \hbar v \int \dd x \left( \left[ \frac{\partial}{\partial x} \psi^\dagger(x) \right] \psi(x) - \text{H.c.} \right), \\
H_{\text{NL}}(t) &= \tilde{H}_{\text{NL}}({e^{i \omega_1 t}a^\dagger},e^{-i \omega_1 t} a,t),\\
\bar V &= \tilde V.
\end{align}
Note that by going into the interaction with respect to $\omega_1$ the nonlinear Hamiltonian is now evaluated at the time evolving operators  $e^{i \omega_1 t}a^\dagger$ and $e^{-i \omega_1 t} a$.
Finally, we can go to a second interaction picture with the unitary operation
\begin{align}
\mathcal{U}_2 &= \exp\left( - \frac{i}{\hbar} (t-t_0) \bar{H}_{\text{wg}}  \right) \\
&= \exp\left(  (t-t_0)  \frac{v}{2}  \int \dd x \left\{ \left[ \frac{\partial}{\partial x} \psi^\dagger(x) \right] \psi(x) - \text{H.c.} \right\} \right),    \nonumber
\end{align}
to obtain the final form of the Hamiltonian
\begin{equation}\label{eq:HfinalMPS}
H =\hbar \delta  a^\dagger a    + H_{\text{NL}}(t)  +  \hbar \gamma \left( \psi^\dagger(-v (t-t_0))a + \text{H.c.} \right).
\end{equation}
Note that the last term in the RHS can be written as
\begin{equation}
\hbar \gamma  \left( \psi^\dagger(-v (t-t_0))a + \text{H.c.} \right)=  \hbar \gamma \left[ a \int \dd x \  \psi^\dagger (x)\delta(x+v(t-t_0))     +\text{H.c.}  \right],
\end{equation}
where it is made explicit that the Hamiltonian is time-dependent but it is only made of Schr\"odinger operators.

This (interaction-picture) Hamiltonian has a very nice physical interpretation. The evolution of the resonator is given by the first two-terms in~\eqref{eq:HfinalMPS}; then, at a given time $t$, the section of waveguide field located at $x=-v(t-t_0)$ gets to exchange some energy with the resonator by a beamsplitter (hopping) interaction mediated by the last term in~\eqref{eq:HfinalMPS}.

Now we are ready to discretize time and space. To this end we introduce $t_j = t_0 + j \Delta t$ and write,
\begin{align}
\psi_j &= \frac{1}{\sqrt{\Delta x}} \int_{-(j+1/2)\Delta x}^{-(j-1/2)\Delta x} \dd x \  \psi(x) \\
&\approx \sqrt{\Delta x} \ \psi(- j \Delta x) = \sqrt{\Delta x} \ \psi(- j v \Delta t).
\end{align}
This operator satisfies $[\psi_j, \psi_k^\dagger] = \delta_{j,k}$ and we can now write the time dependent field-resonance Hamiltonian as
\begin{equation}
H_j = \hbar \delta  a^\dagger a    + H_{\text{NL}}(t_0+j\Delta t)  +i \hbar \sqrt{\frac{\gamma v}{\Delta x }} \left( \psi_j^\dagger a -\text{H.c.} \right)    .
\end{equation}
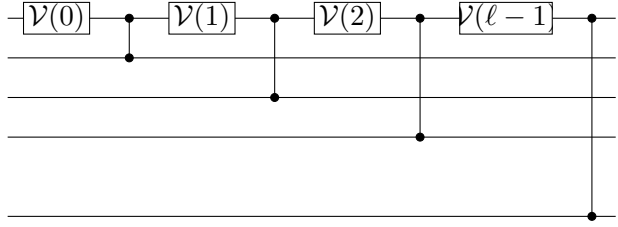
\begin{figure}[!t]\centering
    \begin{tikzpicture}[scale=1.000000,x=1pt,y=1pt]
\filldraw[color=white] (0.000000, -7.500000) rectangle (230.000000, 82.500000);
\draw[color=black] (0.000000,75.000000) -- (230.000000,75.000000);
\draw[color=black] (0.000000,60.000000) -- (230.000000,60.000000);
\draw[color=black] (0.000000,45.000000) -- (230.000000,45.000000);
\draw[color=black] (0.000000,30.000000) -- (230.000000,30.000000);
\draw[color=white] (0.000000,15.000000) -- (230.000000,15.000000);
\draw[color=white] (0.000000,15.000000) node[left] {$\vdots$};
\draw[color=black] (0.000000,0.000000) -- (230.000000,0.000000);
\begin{scope}
\draw[fill=white] (18.500000, 75.000000) +(-45.000000:17.677670pt and 8.485281pt) -- +(45.000000:17.677670pt and 8.485281pt) -- +(135.000000:17.677670pt and 8.485281pt) -- +(225.000000:17.677670pt and 8.485281pt) -- cycle;
\clip (18.500000, 75.000000) +(-45.000000:17.677670pt and 8.485281pt) -- +(45.000000:17.677670pt and 8.485281pt) -- +(135.000000:17.677670pt and 8.485281pt) -- +(225.000000:17.677670pt and 8.485281pt) -- cycle;
\draw (18.500000, 75.000000) node {$\mathcal{V}(0)$};
\end{scope}
\draw (46.000000,75.000000) -- (46.000000,60.000000);
\filldraw (46.000000, 75.000000) circle(1.500000pt);
\filldraw (46.000000, 60.000000) circle(1.500000pt);
\begin{scope}
\draw[fill=white] (73.500000, 75.000000) +(-45.000000:17.677670pt and 8.485281pt) -- +(45.000000:17.677670pt and 8.485281pt) -- +(135.000000:17.677670pt and 8.485281pt) -- +(225.000000:17.677670pt and 8.485281pt) -- cycle;
\clip (73.500000, 75.000000) +(-45.000000:17.677670pt and 8.485281pt) -- +(45.000000:17.677670pt and 8.485281pt) -- +(135.000000:17.677670pt and 8.485281pt) -- +(225.000000:17.677670pt and 8.485281pt) -- cycle;
\draw (73.500000, 75.000000) node {$\mathcal{V}(1)$};
\end{scope}
\draw (101.000000,75.000000) -- (101.000000,45.000000);
\filldraw (101.000000, 75.000000) circle(1.500000pt);
\filldraw (101.000000, 45.000000) circle(1.500000pt);
\begin{scope}
\draw[fill=white] (128.500000, 75.000000) +(-45.000000:17.677670pt and 8.485281pt) -- +(45.000000:17.677670pt and 8.485281pt) -- +(135.000000:17.677670pt and 8.485281pt) -- +(225.000000:17.677670pt and 8.485281pt) -- cycle;
\clip (128.500000, 75.000000) +(-45.000000:17.677670pt and 8.485281pt) -- +(45.000000:17.677670pt and 8.485281pt) -- +(135.000000:17.677670pt and 8.485281pt) -- +(225.000000:17.677670pt and 8.485281pt) -- cycle;
\draw (128.500000, 75.000000) node {$\mathcal{V}(2)$};
\end{scope}
\draw (156.000000,75.000000) -- (156.000000,30.000000);
\filldraw (156.000000, 75.000000) circle(1.500000pt);
\filldraw (156.000000, 30.000000) circle(1.500000pt);
\begin{scope}
\draw[fill=white] (188.500000, 75.000000) +(-45.000000:24.748737pt and 8.485281pt) -- +(45.000000:24.748737pt and 8.485281pt) -- +(135.000000:24.748737pt and 8.485281pt) -- +(225.000000:24.748737pt and 8.485281pt) -- cycle;
\clip (188.500000, 75.000000) +(-45.000000:24.748737pt and 8.485281pt) -- +(45.000000:24.748737pt and 8.485281pt) -- +(135.000000:24.748737pt and 8.485281pt) -- +(225.000000:24.748737pt and 8.485281pt) -- cycle;
\draw (188.500000, 75.000000) node {$\mathcal{V}(\ell-1)$};
\end{scope}
\draw (221.000000,75.000000) -- (221.000000,0.000000);
\filldraw (221.000000, 75.000000) circle(1.500000pt);
\filldraw (221.000000, 0.000000) circle(1.500000pt);
\draw[color=white] (230.000000,15.000000) node[right] {$\vdots$};
\end{tikzpicture}

    \caption{\label{fig:circuit1} Circuit representation of the unitary evolution operator in~\eqref{eq:evol} in a discretization using four modes. Note that in this figure the first wire represents the resonator mode while the rest of the wires represent the different discretized sections of the waveguide.
    We use vertical lines with dots at the ends to represent the beamsplitter operation $\mathcal{B}_{0,j}$ between modes 0 (the resonator) and the $j^{\text{th}}$ part of the waveguide.
    }
\end{figure}

\begin{figure*}[!ht] \centering
    \begin{tikzpicture}[scale=1.000000,x=1pt,y=1pt]
\filldraw[color=white] (0.000000, -7.500000) rectangle (302.000000, 82.500000);
\draw[color=black] (0.000000,75.000000) -- (302.000000,75.000000);
\draw[color=black] (0.000000,60.000000) -- (302.000000,60.000000);
\draw[color=black] (0.000000,45.000000) -- (302.000000,45.000000);
\draw[color=black] (0.000000,30.000000) -- (302.000000,30.000000);
\draw[color=white] (0.000000,15.000000) -- (302.000000,15.000000);
\draw[color=white] (0.000000,15.000000) node[left] {$\vdots$};
\draw[color=black] (0.000000,0.000000) -- (302.000000,0.000000);
\begin{scope}
\draw[fill=white] (18.500000, 75.000000) +(-45.000000:17.677670pt and 8.485281pt) -- +(45.000000:17.677670pt and 8.485281pt) -- +(135.000000:17.677670pt and 8.485281pt) -- +(225.000000:17.677670pt and 8.485281pt) -- cycle;
\clip (18.500000, 75.000000) +(-45.000000:17.677670pt and 8.485281pt) -- +(45.000000:17.677670pt and 8.485281pt) -- +(135.000000:17.677670pt and 8.485281pt) -- +(225.000000:17.677670pt and 8.485281pt) -- cycle;
\draw (18.500000, 75.000000) node {$\mathcal{V}(0)$};
\end{scope}
\draw (46.000000,75.000000) -- (46.000000,60.000000);
\filldraw (46.000000, 75.000000) circle(1.500000pt);
\filldraw (46.000000, 60.000000) circle(1.500000pt);
\draw (64.000000,75.000000) -- (64.000000,60.000000);
\begin{scope}
\draw (61.878680, 72.878680) -- (66.121320, 77.121320);
\draw (61.878680, 77.121320) -- (66.121320, 72.878680);
\end{scope}
\begin{scope}
\draw (61.878680, 57.878680) -- (66.121320, 62.121320);
\draw (61.878680, 62.121320) -- (66.121320, 57.878680);
\end{scope}
\begin{scope}
\draw[fill=white] (91.500000, 60.000000) +(-45.000000:17.677670pt and 8.485281pt) -- +(45.000000:17.677670pt and 8.485281pt) -- +(135.000000:17.677670pt and 8.485281pt) -- +(225.000000:17.677670pt and 8.485281pt) -- cycle;
\clip (91.500000, 60.000000) +(-45.000000:17.677670pt and 8.485281pt) -- +(45.000000:17.677670pt and 8.485281pt) -- +(135.000000:17.677670pt and 8.485281pt) -- +(225.000000:17.677670pt and 8.485281pt) -- cycle;
\draw (91.500000, 60.000000) node {$\mathcal{V}(1)$};
\end{scope}
\draw (119.000000,60.000000) -- (119.000000,45.000000);
\filldraw (119.000000, 60.000000) circle(1.500000pt);
\filldraw (119.000000, 45.000000) circle(1.500000pt);
\draw (137.000000,60.000000) -- (137.000000,45.000000);
\begin{scope}
\draw (134.878680, 57.878680) -- (139.121320, 62.121320);
\draw (134.878680, 62.121320) -- (139.121320, 57.878680);
\end{scope}
\begin{scope}
\draw (134.878680, 42.878680) -- (139.121320, 47.121320);
\draw (134.878680, 47.121320) -- (139.121320, 42.878680);
\end{scope}
\begin{scope}
\draw[fill=white] (164.500000, 45.000000) +(-45.000000:17.677670pt and 8.485281pt) -- +(45.000000:17.677670pt and 8.485281pt) -- +(135.000000:17.677670pt and 8.485281pt) -- +(225.000000:17.677670pt and 8.485281pt) -- cycle;
\clip (164.500000, 45.000000) +(-45.000000:17.677670pt and 8.485281pt) -- +(45.000000:17.677670pt and 8.485281pt) -- +(135.000000:17.677670pt and 8.485281pt) -- +(225.000000:17.677670pt and 8.485281pt) -- cycle;
\draw (164.500000, 45.000000) node {$\mathcal{V}(2)$};
\end{scope}
\draw (192.000000,45.000000) -- (192.000000,30.000000);
\filldraw (192.000000, 45.000000) circle(1.500000pt);
\filldraw (192.000000, 30.000000) circle(1.500000pt);
\draw (210.000000,45.000000) -- (210.000000,30.000000);
\begin{scope}
\draw (207.878680, 42.878680) -- (212.121320, 47.121320);
\draw (207.878680, 47.121320) -- (212.121320, 42.878680);
\end{scope}
\begin{scope}
\draw (207.878680, 27.878680) -- (212.121320, 32.121320);
\draw (207.878680, 32.121320) -- (212.121320, 27.878680);
\end{scope}
\begin{scope}
\draw[fill=white] (242.500000, 30.000000) +(-45.000000:24.748737pt and 8.485281pt) -- +(45.000000:24.748737pt and 8.485281pt) -- +(135.000000:24.748737pt and 8.485281pt) -- +(225.000000:24.748737pt and 8.485281pt) -- cycle;
\clip (242.500000, 30.000000) +(-45.000000:24.748737pt and 8.485281pt) -- +(45.000000:24.748737pt and 8.485281pt) -- +(135.000000:24.748737pt and 8.485281pt) -- +(225.000000:24.748737pt and 8.485281pt) -- cycle;
\draw (242.500000, 30.000000) node {$\mathcal{V}(\ell-1)$};
\end{scope}
\draw (275.000000,30.000000) -- (275.000000,0.000000);
\filldraw (275.000000, 30.000000) circle(1.500000pt);
\filldraw (275.000000, 0.000000) circle(1.500000pt);
\draw (293.000000,30.000000) -- (293.000000,0.000000);
\begin{scope}
\draw (290.878680, 27.878680) -- (295.121320, 32.121320);
\draw (290.878680, 32.121320) -- (295.121320, 27.878680);
\end{scope}
\begin{scope}
\draw (290.878680, -2.121320) -- (295.121320, 2.121320);
\draw (290.878680, 2.121320) -- (295.121320, -2.121320);
\end{scope}
\draw[color=white] (302.000000,15.000000) node[right] {$\vdots$};
\end{tikzpicture}

    \caption{\label{fig:circuit2} Circuit representation of the unitary evolution operator in~\eqref{eq:evol} and for a discretization using 4 modes. Note that in this figure we apply swap operations after each beamsplitter. Thus after the $i^\text{th}$ swap the resonator mode lies in the $i^\text{th}$ wire, and in particular at the end of the interaction the resonator mode is represented by the last wire. We use vertical lines with dots at the end to represent the beamsplitter operation $\mathcal{B}_{i,i+1}$ between wire $i$ and $i+1$ and similarly we use vertical lines with crosses to represent swap operations between wires  $i$ and $i+1$. }
\end{figure*}
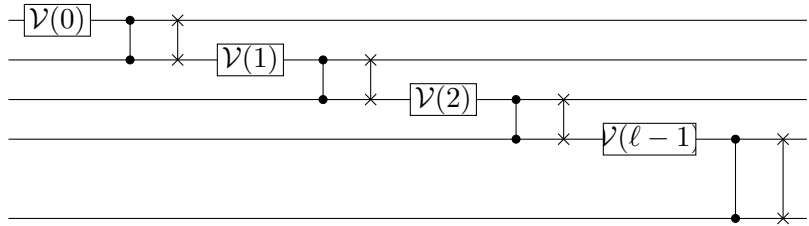
The Trotterized time-evolution operator after discretizing time in sliace of size  and space into $\ell$ slices can be written as
\begin{align}\label{eq:evol}
\mathcal{ U} &=    \left[ \mathcal{B}_{0,\ell-1} \mathcal{V}_0(\ell-1) \right]   \ldots    \left[ \mathcal{B}_{0,0} \mathcal{V}_0(0) \right] , \\
\mathcal{V}_0(j) &=  \exp\left(-\frac{i}{\hbar} \Delta t \left[     \hbar \delta  a^\dagger a    + H_{\text{NL}}(t_0+j \Delta t ) \right]     \right),  \\
 \mathcal{B}_{0,j}&= \exp\left( i \gamma \sqrt{ \frac{\Delta t} {v}} \left( \psi^\dagger_j a + \text{H.c.} \right)  \right), 
\end{align}
where for each time ``slice'' ($\Delta t = \Delta x/v$) two things happen:
\begin{enumerate}
    \item At time $j$ the local unitary $\mathcal{V}_0(j)$ is applied on the resonator mode. Note that we allow for an explicit dependence on time $j$.
    \item A beamsplitter by rotation angle $\theta = \gamma \sqrt{ \frac{\Delta t} {v}}$ is applied between the resonator mode and the waveguide section labelled by the index $j$.
\end{enumerate}
These two sets of operations can be represented by a quantum circuit diagram as shown in Fig.~\ref{fig:circuit1}.
Note that in this diagram the first wire refers to the resonance while the rest of the wires represent the different discretized sections of the waveguide. In this representation we are required to carry out beamsplitter interaction between non-nearest neighbour modes. Thus in this representation the interactions between the different wires are non-local. To simplify the evaluation of the circuit we will now permute the order of the wires, that is, after each beamsplitter operation we will \emph{swap} the resonator wire with the corresponding waveguide section~\cite{dhand2018proposal}. This will simplify the calculation since the interaction between the different modes will now only consist of nearest neighbours beamsplitters as shown in  Fig.~\ref{fig:circuit2}.
Physically this corresponds to a moving frame in which the field is no longer propagating at its group velocity but instead the resonance is moving in the opposite direction at the same group velocity.

Introducing the following two-wire unitary
\begin{align}
\mathcal{W}(i+1,i) = \text{SWAP}_{i+1,i} \mathcal{B}_{i+1,i} \mathcal{V}_i(i) ,
\end{align}
we can write the Trotterized evolution operator as
\begin{align}
\mathcal{U} = \mathcal{W}(\ell, \ell-1) \ldots  \mathcal{W}(1,0),
\end{align}
with the added convention that the resonator mode starts in the first wire but ends in the last one. We consider a total of $\ell+1$ modes where $\ell$ modes are used for the discretization of the waveguide and the remaining one for the resonance.
Now assume that the initial state has product form
\begin{align}\label{eq:initial_ketMPS}
| \Psi (t_0) \rangle = \bigotimes_{i=0}^\ell | 0_i \rangle.
\end{align}
where $| 0_i \rangle$ is the vacuum state in either the resonance or a section of the waveguide (this can be generalized to any product state such as a broadband coherent state).

We now want to argue that if one starts with a product state, then the time evolved state has matrix product state form. To this end we will calculate the tensor
\begin{align}
R_{i_0,i_1,\ldots,r_\ell} = \langle i_0,i_1,\ldots,i_\ell|\mathcal{U}|\Psi \rangle,
\end{align}
and will show that it has matrix-product state (MPS) form. We consider a discretization with $\ell=4$ modes and with Fock cutoff $c$ but everything follows in a straightforward manner for an arbitrary number of modes.
We thus want to find
\begin{align}
R_{i_0, i_1, i_2, i_3, i_4} &= \langle i_0, i_1, i_2, i_3, i_4|\mathcal{ U}|0_0,0_1,0_2,0_3,0_4 \rangle
\\
&= \langle i_0, i_1, i_2, i_3, i_4|W(4,3)W(3,2)W(2,1)W(1,0)|0_0,0_1,0_2,0_3,0_4 \rangle. \nonumber
\end{align}
We start by writing
\begin{align}
\langle i_0|W(0,1)|0_0,0_1 \rangle=\sum_{\alpha_1=0}^{c-1} (T^{[0]})^{i_0,\alpha_1}_{0_0,0_1}    | \alpha_1 \rangle \leftrightarrow \langle i_0,  \alpha_1|W(1,0)|0_0,0_1 \rangle= (T^{[0]})^{i_0,\alpha_1}_{0_0,0_1},
\end{align}
and then find
\begin{align}
R_{i_0, i_1, i_2, i_3, i_4} =     \sum_{\alpha_1=0}^{c-1} (T^{[0]})^{i_0,\alpha_1}_{0_0,0_1} \langle  i_1, i_2, i_3, i_4|W(4,3)W(3,2)W(2,1)|\alpha_1,0_2,0_3,0_4 \rangle.    
\end{align}
Next we introduce
\begin{align}\label{eq:W12}
\langle i_1|W(2,1)|\alpha_1,0_2 \rangle=\sum_{\alpha_2=0}^{c-1} (T^{[1]})^{i_1,\alpha_2}_{\alpha_1,0_2}    | \alpha_2 \rangle,
\end{align}
to find
\begin{align}
R_{i_0, i_1, i_2, i_3, i_4} = \sum_{\alpha_1,\alpha_2=0}^{c-1} (T^{[0]})^{i_0,\alpha_1}_{0_0,0_1} (T^{[1]})^{i_1,\alpha_2}_{\alpha_1,0_2} \langle  i_2, i_3, i_4|W(4,3)W(3,2)|\alpha_2,0_3,0_4 \rangle.\nonumber
\end{align}

We can proceed like this two more times, defining
\begin{align}
\langle i_2|W(3,2)|\alpha_2,0_3 \rangle&=\sum_{\alpha_3=0}^{c-1} (T^{[2]})^{i_2,\alpha_3}_{\alpha_2,0_3}    | \alpha_3 \rangle,    \\
\langle i_3|W(4,3)|\alpha_3,0_4 \rangle&=\sum_{\alpha_4=0}^{c-1} (T^{[3]})^{i_3,\alpha_4}_{\alpha_3,0_4}    | \alpha_4 \rangle,    
\end{align}
to finally write the coefficients of the time evolved ket in MPS form
\begin{align}
R_{i_0, i_1, i_2, i_3, i_4} =     \sum_{\alpha_1,\alpha_2,\alpha_3,\alpha_4=0}^{c-1} (T^{[0]})^{i_0,\alpha_1}_{0_0,0_1} (T^{[1]})^{i_1,\alpha_2}_{\alpha_1,0_2} (T^{[2]})^{i_2,\alpha_3}_{\alpha_2,0_3}  (T^{[3]})^{i_3,\alpha_4}_{\alpha_3,0_4}     \langle i_4|\alpha_4 \rangle.
\end{align}
Defining the simplified notation $(T^{[k]})^{x,y}_{w,0_p} \equiv (S^{[k]})_{x,y}^{w}$ for $0<k< \ell = 4$ (recall that $0_p$ is fixed from the initial ket~\eqref{eq:initial_ketMPS}) and also $(T^{[0]})^{x,y}_{0_0,0_1} \equiv (S^{[0]})_{i_0,\alpha_1}$ and $(S^{[4]})_{i_4}^{j_4} = \langle i_4|\alpha_4 \rangle $ we can finally write the MPS form
\begin{align}
R_{i_0, i_1, i_2, i_3, i_4} =     \sum_{\alpha_1,\alpha_2,\alpha_3,\alpha_4=0}^{c-1} (S^{[0]})_{i_0,\alpha_1} (S^{[1]})_{i_1,\alpha_2}^{\alpha_1} (S^{[2]})_{i_2,\alpha_3}^{\alpha_2}  (S^{[3]})_{i_3,i_4}^{\alpha_3} (S^{[4]})^{\alpha_4}_{i_4},
\end{align}
which is the MPS form of the ket. Note that for this ket the bond dimension is precisely the Fock cutoff $c$ and thus our $(\ell+1)$-mode time-evolved system can be represented in memory with only on the order $\ell c^3$ coefficients, as opposed to the $c^{\ell+1}$ coefficients a naive Fock representation would entail.

As stated in the previous subsection, once the MPS form of a state is obtained, most useful properties of the state can be obtained in a straightforward manner. The methods to obtain expectation values from MPS state are presented and derived elsewhere [cite relevant literature]. We would just like to point out that standard computer libraries already exist for the calculation of expectation values of local observables from MPSs~\cite{roberts2021tensornetwork,fishman2020itensor,jaschke2018open,bauer2011alps}.\cite{fishman2020itensor}

Finally, note that the result just derived is a qu\textbf{d}it extension of the qu\textbf{b}it results derived by Sch\"{o}n \emph{et al.} \cite{schon2005sequential,schon2007sequential}, where they consider ``the deterministic generation of entangled multiqubit states by the sequential coupling of an ancillary system to initially uncorrelated qubits'' and ``characterize all achievable states in terms of classes of matrix-product states''.

\section{Summary, conclusions, and future outlook}\label{sec:Conclusion}
Since the initial experiments on squeezing carried out more than three decades ago, there has been a constant drive to move quantum optics from  a collection of bulk optical elements spanning large tabletops to a single chip-scale device no larger than a fingernail. Such a transition not only reduces the system size and power requirements, but it also offers the promise of scalability and an unprecedented control over the device parameters, and the properties of the generated light. While advances in  the fabrication of integrated photonic device have accelerated this evolution, they have also ushered in an era of ``high-gain'' quantum optics, for example enabling single photon generation via post-selection involving photon-number resolving detection, efficient photon frequency conversion, and state generation beyond the single photon pair regime. For these and other high-gain applications, a first-order perturbation theory, the traditional tool of choice, is no longer sufficient to describe the nonlinear interaction. A key theoretical issue is that relevant Hamiltonian terms describing the nonlinear interaction do not commute with themselves at different times. In short, this means that weakly-transformed photon amplitudes at early times influence their own transformation at later times.

Drawing on earlier works, in this tutorial we have shown how to overcome this difficulty, starting from a rigorous first-principles quantization procedure and employing a Trotter-Suzuki expansion of linearized Heisenberg picture dynamics to reduce the problem to the multiplication of matrices with a size proportional to the number of bosonic modes involved. In particular, in Sec.~\ref{sec:Quantization} we detailed the quantization of nonlinear optics in integrated photonic structures, paying particular attention to channel waveguides and ring resonators. In Sec.~\ref{sec:Ket} we presented the tools necessary to calculate the output states of a few nonlinear optical processes, whether in the low-gain or high-gain regimes, and whether the initial quantum state is vacuum or not, resulting in a process that is spontaneous or stimulated, respectively. These tools correctly account for loss and material dispersion and, as we showed, they can be readily applied to calculate the results of homodyne and photon-number resolved measurements of the generated states. With these mathematical tools firmly established, we then turned to the particular structures singled out in Sec.~\ref{sec:Quantization}, going into further detail regarding their application to waveguides in Sec.~\ref{sec:Waveguides} and ring resonators in Sec.~\ref{sec:Rings} for various SPDC and SFWM processes. In Sec.~\ref{sec:Connections} we discussed the deep connection between spontaneous and stimulated processes, showing how the knowledge of the latter can be used to inform the former. Finally, in Sec.~\ref{sec:Tensor} we examined how tensor networks and matrix product states can be used to move beyond linearized equations of motion and Gaussian state generation. Throughout this tutorial, we have also emphasized connections to relevant theoretical and experimental works.

It is our hope that the tools provided here will benefit the entire integrated quantum photonics community, leading to devices with still greater complexity and enhanced functionalities than their present-day counterparts, especially as state generation moves beyond the single or few photon pairs regime. Indeed, integrated quantum photonics technologies promise ultimate information security, enhanced measurement precision, and greater computing power than traditional technologies, and their realization is now within reach.

\section*{Acknowledgments}
N.Q. thanks A.M. Bra\'{n}czyk, M.D Vidrighin, G. Triginer, A.L. Grimsmo and I. Dhand for useful discussions. M.L acknowledges the support from Ministero dell’Istruzione, dell’Università e della Ricerca (Dipartimenti di Eccellenza Program (2018–2022)). J.E.S. acknowledges support from the Natural Sciences and Engineering Research Council of Canada.



\bibliography{main}

\appendix
\section{Field normalization and dispersion}\label{sec:Normalization}
As a first example we look at a uniform medium in the dispersionless
limit, where $\varepsilon_{1}(\boldsymbol{r})=\varepsilon_{1}$. Then
the equations (\ref{eq:mode_equations}) for the $\boldsymbol{B}_{\alpha}(\boldsymbol{r})$
become simply 
\begin{align*}
 & \boldsymbol{\nabla}\times\left(\boldsymbol{\nabla\times}\boldsymbol{B}_{\alpha}(\boldsymbol{r})\right)=\frac{\omega_{\alpha}^{2}\varepsilon_{1}}{c^{2}}\boldsymbol{B}_{\alpha}(\boldsymbol{r}),\\
 & \boldsymbol{\nabla\cdot}\boldsymbol{B}_{\alpha}(\boldsymbol{r})=0.
\end{align*}
Solutions satisfying periodic boundary conditions are of the form
\textbf{$\boldsymbol{b}\exp(i\boldsymbol{k}\cdot\boldsymbol{r}),$}
where the vector $\boldsymbol{b}$ is perpendicular to $\boldsymbol{k}$
and
\begin{align}
 & \omega_{\alpha}=c\left|\boldsymbol{k}\right|/n,\label{eq:uniform_no_dispersion}
\end{align}
 where $n=\sqrt{\varepsilon_{1}}$. If the normalization volume $V$
is a cube with each side of length $L$, $V=L^{3}$, the allowed $\boldsymbol{k}$
are of the form 
\begin{align*}
 & \boldsymbol{k}=\frac{2\pi}{L}(m_{x}\boldsymbol{\hat{x}}+m_{y}\boldsymbol{\hat{y}}+m_{z}\boldsymbol{\hat{z}})
\end{align*}
where the $m_{i}$ are integers and $\boldsymbol{\hat{x}}$ is a unit
vector pointing in the $x$ direction, etc. There are two independent
vectors perpendicular to $\boldsymbol{k}$; call them $\boldsymbol{\hat{e}}_{1\boldsymbol{k}}$
and $\boldsymbol{\hat{e}}_{2\boldsymbol{k}}$, and define them such
that 
\begin{align*}
 & \boldsymbol{\hat{e}}_{1\boldsymbol{k}}\times\boldsymbol{\hat{e}}_{2\boldsymbol{k}}=\boldsymbol{\hat{k}}.
\end{align*}

In a uniform medium it is often useful to work with left- and right-handed
circularly polarized components of light, which can be identified
by the polarization vectors 
\begin{align*}
 & \boldsymbol{\hat{e}}_{L\boldsymbol{k}}=-\frac{1}{\sqrt{2}}\left(\boldsymbol{\hat{e}}_{1\boldsymbol{k}}+i\boldsymbol{\hat{e}}_{2\boldsymbol{k}}\right),\\
 & \boldsymbol{\hat{e}}_{R\boldsymbol{k}}=\frac{1}{\sqrt{2}}\left(\boldsymbol{\hat{e}}_{1\boldsymbol{k}}-i\boldsymbol{\hat{e}}_{2\boldsymbol{k}}\right)
\end{align*}
respectively. Using $I$ to denote either $L$ or $R$ (positive or
negative helicity respectively), and $\overline{I}$ to equal the opposite
helicity, we have 
\begin{align*}
 & \boldsymbol{\hat{e}}_{I\boldsymbol{k}}\cdot\boldsymbol{\hat{e}}_{I\boldsymbol{k}}^{*}=1,\\
 & \boldsymbol{\hat{e}}_{I\boldsymbol{k}}\cdot\boldsymbol{\hat{e}}_{\overline{I}\boldsymbol{k}}^{*}=0,
\end{align*}
and 
\begin{align*}
 & i\boldsymbol{\hat{k}}\times\boldsymbol{\hat{e}}_{I\boldsymbol{k}}=s_{I}\boldsymbol{\hat{e}}_{I\boldsymbol{k}},\\
 & \hat{\boldsymbol{e}}_{I\boldsymbol{k}}\times\boldsymbol{\hat{e}}_{I\boldsymbol{k}}^{*}=-is_{I}\boldsymbol{\hat{k}}
\end{align*}
where \textbf{$\boldsymbol{\hat{k}}\equiv\boldsymbol{k}/\left|\boldsymbol{k}\right|$}
and $s_{L}=1$, $s_{R}=-1$ identifies the helicity. It is convenient
to link the definitions of $\boldsymbol{\hat{e}}_{1\boldsymbol{k}}$
and $\boldsymbol{\hat{e}}_{2\boldsymbol{k}}$ to those of $\boldsymbol{\hat{e}}_{1(-\boldsymbol{k})}$
and $\boldsymbol{\hat{e}}_{2(-\boldsymbol{k})}$ according to 
\begin{align*}
 & \boldsymbol{\hat{e}}_{1(-\boldsymbol{k})}=\boldsymbol{\hat{e}}_{1\boldsymbol{k}},\\
 & \boldsymbol{\hat{e}}_{2(-\boldsymbol{k})}=-\boldsymbol{\hat{e}}_{2\boldsymbol{k}},
\end{align*}
for then 
\begin{align*}
 & \boldsymbol{\hat{e}}_{I\boldsymbol{k}}^{*}=\boldsymbol{\hat{e}}_{I(-\boldsymbol{k})},\\
 & \boldsymbol{\hat{e}}_{I\boldsymbol{k}}\cdot\boldsymbol{\hat{e}}_{I'(-\boldsymbol{k})}=\delta_{II'}.
\end{align*}
We can label the modes $\alpha$ by the pair of indices $I,\boldsymbol{k}$,
and seeking $\boldsymbol{b}$ vectors proportional to $s_{I}\boldsymbol{\hat{e}}_{I\boldsymbol{k}}$
the normalization condition (\ref{eq:norm_B}) means we can take 
\begin{align}
 & \boldsymbol{B}_{I\boldsymbol{k}}(\boldsymbol{r})=\sqrt{\frac{\mu_{0}}{V}}s_{I}\boldsymbol{\hat{e}}_{I\boldsymbol{k}}e^{i\boldsymbol{k}\cdot\boldsymbol{r}},\label{eq:B_mode_uniform}
\end{align}
while from the last of (\ref{eq:mode_equations}) we then find 
\begin{align}
 & \boldsymbol{D}_{I\boldsymbol{k}}(\boldsymbol{r})=in\sqrt{\frac{\epsilon_{0}}{V}}\boldsymbol{\hat{e}}_{I\boldsymbol{k}}e^{i\boldsymbol{k}\cdot\boldsymbol{r}}.\label{eq:D_mode_uniform}
\end{align}
and for the mode frequencies we write 
\begin{align*}
 & \omega_{I\boldsymbol{k}}=\frac{c}{n}\left|\boldsymbol{k}\right|,
\end{align*}
which of course only depend on $\left|\boldsymbol{k}\right|.$ Note
that the partner mode of a mode of polarization $I$ with propagation
wave vector $\boldsymbol{k}$ is the mode of polarization $I$ with
propagation wave vector $-\boldsymbol{k}$. Clearly other choices
of the modes besides (\ref{eq:B_mode_uniform},\ref{eq:D_mode_uniform})
are possible; one could multiply both by an overall phase factor or
the factor $s_{I}$, or one could work with the linearly polarized
basis $\boldsymbol{\hat{e}}_{1\boldsymbol{k}}$ and $\boldsymbol{\hat{e}}_{2\boldsymbol{k}}$
at each $\boldsymbol{k}.$ But choosing (\ref{eq:B_mode_uniform},\ref{eq:D_mode_uniform})
in (\ref{eq:field_expansions}) we find 
\begin{align}
 & \boldsymbol{D}(\boldsymbol{r},t)=i\sum_{I,\boldsymbol{k}}\sqrt{\frac{\hbar cn\left|\boldsymbol{k}\right|\epsilon_{0}}{2V}}a_{I\boldsymbol{k}}(t)\boldsymbol{\hat{e}}_{I\boldsymbol{k}}e^{i\boldsymbol{k}\cdot\boldsymbol{r}}+h.c.,\label{eq:field_expansions-1}\\
 & \boldsymbol{B}(\boldsymbol{r},t)=\sum_{I,\boldsymbol{k}}\sqrt{\frac{\hbar c\left|\boldsymbol{k}\right|\mu_{0}}{2nV}}a_{I\boldsymbol{k}}(t)s_{I}\boldsymbol{\hat{e}}_{I\boldsymbol{k}}e^{i\boldsymbol{k}\cdot\boldsymbol{r}}+h.c.\nonumber 
\end{align}

Finally, we consider the passage to an infinite normalization volume.
Here the analogues of (\ref{eq:1Ddtoc},\ref{eq:1Ddtoc_second}) in
three dimensions are 
\begin{align}
 & \sum_{\boldsymbol{k}}\rightarrow\frac{\int d\boldsymbol{k}}{\left(\frac{2\pi}{L}\right)^{3}}=\frac{V}{8\pi^{3}}\int d\boldsymbol{k}\label{eq:3Ddtoc}
\end{align}
and 
\begin{align}
 & a_{I\boldsymbol{k}}(t)\rightarrow\sqrt{\frac{8\pi^{3}}{V}}a_{I}(\boldsymbol{k},t)\label{eq:3Ddtoc_second}
\end{align}
We then have 
\begin{align}
 & \left[a_{I}(\boldsymbol{k},t),a_{I'}^{\dagger}(\boldsymbol{k'},t)\right]=\delta_{II'}\delta(\boldsymbol{k}-\boldsymbol{k'}),\label{eq:comm_continuous_first}\\
 & \left[a_{I}(\boldsymbol{k},t),a_{I'}(\boldsymbol{k'},t)\right]=0.\nonumber 
\end{align}
In terms of these variables $a_{I}(\boldsymbol{k},t)$ (\ref{eq:field_expansions-1})
become 
\begin{align}
 & \boldsymbol{D}(\boldsymbol{r},t)=i\sum_{I}\int\sqrt{\frac{\hbar cn\left|\boldsymbol{k}\right|\epsilon_{0}}{16\pi^{3}}}a_{I}(\boldsymbol{k},t)\boldsymbol{\hat{e}}_{I\boldsymbol{k}}e^{i\boldsymbol{k}\cdot\boldsymbol{r}}d\boldsymbol{k}+h.c.,\label{eq:field_expansions_infinite_no_disp}\\
 & \boldsymbol{B}(\boldsymbol{r},t)=\sum_{I}\int\sqrt{\frac{\hbar c\left|\boldsymbol{k}\right|\mu_{0}}{16\pi^{3}n}}a_{I}(\boldsymbol{k},t)s_{I}\boldsymbol{\hat{e}}_{I\boldsymbol{k}}e^{i\boldsymbol{k}\cdot\boldsymbol{r}}d\boldsymbol{k}+h.c.,\nonumber 
\end{align}
where we keep the same notation for $\boldsymbol{\hat{e}}_{I\boldsymbol{k}}$
and $\omega_{I\boldsymbol{k}}$ as in (\ref{eq:field_expansions-1}),
although $\boldsymbol{k}$ now ranges continuously. And the Hamiltonian
(\ref{eq:basic_Hamiltonian}) becomes 
\begin{align}
 & H=\sum_{I}\int\hbar\omega_{I\boldsymbol{k}}a_{I}(\boldsymbol{k},t)a_{I}(\boldsymbol{k,}t)d\boldsymbol{k}.\label{eq:Hamiltonian_infinite}
\end{align}

We now turn to the inclusion of dispersion. The consequences of this
new factor $v_{p}(\boldsymbol{r};\omega_{\alpha})/v_{g}(\boldsymbol{r};\omega_{\alpha})$
in the normalization condition (\ref{eq:norm_D-3}) can be easiest
seen by first considering periodic boundary conditions. Since in a
uniform medium the relative dielectric constant and the velocities
are independent of position, we can put $\varepsilon_{1}(\boldsymbol{r};\omega_{I\boldsymbol{k}})\rightarrow\varepsilon_{1}(\omega_{I\boldsymbol{k}})$,
$n(\boldsymbol{r},\omega_{I\boldsymbol{k}})\rightarrow n(\omega_{I\boldsymbol{k}})$,
$v_{p,g}(\boldsymbol{r};\omega_{I\boldsymbol{k}})\rightarrow v_{p,g}(\omega_{I\boldsymbol{k}})$,
and the modes (\ref{eq:B_mode_uniform},\ref{eq:D_mode_uniform})
in the dispersionless limit are replaced by 
\begin{align}
 & \boldsymbol{B}_{I\boldsymbol{k}}(\boldsymbol{r})=\sqrt{\frac{\mu_{0}v_{g}(\omega_{I\boldsymbol{k}})}{Vv_{p}(\omega_{I\boldsymbol{k}})}}s_{I}\boldsymbol{\hat{e}}_{I\boldsymbol{k}}e^{i\boldsymbol{k}\cdot\boldsymbol{r}},\label{eq:B_mode_uniform_dispersion}
\end{align}
and 
\begin{align}
 & \boldsymbol{D}_{I\boldsymbol{k}}(\boldsymbol{r})=in(\omega_{I\boldsymbol{k}})\sqrt{\frac{\epsilon_{0}v_{g}(\omega_{I\boldsymbol{k}})}{Vv_{p}(\omega_{I\boldsymbol{k}})}}\boldsymbol{\hat{e}}_{I\boldsymbol{k}}e^{i\boldsymbol{k}\cdot\boldsymbol{r}},\label{eq:D_mode_uniform_dispersion}
\end{align}
where now the mode frequency for a given $\boldsymbol{k}$ is determined
by self-consistently solving
\begin{align}
 & n(\omega_{I\boldsymbol{k}})\omega_{I\boldsymbol{k}}=c\left|\boldsymbol{k}\right|\label{eq:uniform_dispersion}
\end{align}
for $\omega_{I\boldsymbol{k}}$ (compare (\ref{eq:uniform_no_dispersion})),
and the field expansions (\ref{eq:field_expansions_infinite_no_disp})
become 
\begin{align}
 & \boldsymbol{D}(\boldsymbol{r},t)=i\sum_{I,\boldsymbol{k}}\sqrt{\frac{\hbar\left|\boldsymbol{k}\right|\epsilon_{0}\varepsilon_{1}(\omega_{I\boldsymbol{k}})v_{g}(\omega_{I\boldsymbol{k}})}{2V}}a_{I\boldsymbol{k}}(t)\boldsymbol{\hat{e}}_{I\boldsymbol{k}}e^{i\boldsymbol{k}\cdot\boldsymbol{r}}+h.c.,\label{eq:field_expansions-1-2}\\
 & \boldsymbol{B}(\boldsymbol{r},t)=\sum_{I,\boldsymbol{k}}\sqrt{\frac{\hbar\left|\boldsymbol{k}\right|\mu_{0}v_{g}(\omega_{I\boldsymbol{k}})}{2V}}a_{I\boldsymbol{k}}(t)s_{I}\boldsymbol{\hat{e}}_{I\boldsymbol{k}}e^{i\boldsymbol{k}\cdot\boldsymbol{r}}+h.c.\nonumber 
\end{align}
Again we form the Poynting vector $\boldsymbol{S}(\boldsymbol{r},t)=\boldsymbol{E}(\boldsymbol{r},t)\times\boldsymbol{H}(\boldsymbol{r},t)$;
using (\ref{eq:disp_fields},\ref{eq:disp_constitutive}) we have
\begin{align}
 & \boldsymbol{E}(\boldsymbol{r},t)=i\sum_{I,\boldsymbol{k}}\sqrt{\frac{\hbar\left|\boldsymbol{k}\right|v_{g}(\omega_{I\boldsymbol{k}})}{2V\epsilon_{0}\varepsilon_{1}(\omega_{I\left|\boldsymbol{k}\right|})}}a_{I\boldsymbol{k}}(t)\boldsymbol{\hat{e}}_{I\boldsymbol{k}}e^{i\boldsymbol{k}\cdot\boldsymbol{r}}+h.c.,\label{eq:field_expansions-1-2-1}\\
 & \boldsymbol{H}(\boldsymbol{r},t)=\sum_{I,\boldsymbol{k}}\sqrt{\frac{\hbar\left|\boldsymbol{k}\right|v_{g}(\omega_{I\boldsymbol{k}})}{2V\mu_{0}}}a_{I\boldsymbol{k}}(t)s_{I}\boldsymbol{\hat{e}}_{I\boldsymbol{k}}e^{i\boldsymbol{k}\cdot\boldsymbol{r}}+h.c.,\nonumber 
\end{align}
and considering a state $\left|\Psi\right\rangle $ with one photon
in mode $I,\boldsymbol{k}$,
\begin{align*}
 & \left|\Psi\right\rangle =a_{I\boldsymbol{k}}^{\dagger}(0)\left|vac\right\rangle ,
\end{align*}
where $\left|vac\right\rangle $ is the vacuum state, we find 
\begin{align*}
 & \int_{V}\left\langle \boldsymbol{S}(\boldsymbol{r},t)\right\rangle d\boldsymbol{r}=\hbar\omega_{I\boldsymbol{k}}v_{g}(\omega_{I\boldsymbol{k}}),
\end{align*}
with the group velocity governing the energy flow, as expected.

For reference, we give the field expansion for a uniform infinite
medium, the appropriate generalization of (\ref{eq:field_expansions_infinite_no_disp}):
\begin{align}
 & \boldsymbol{D}(\boldsymbol{r},t)=i\sum_{I}\int\sqrt{\frac{\hbar\left|\boldsymbol{k}\right|\epsilon_{0}\varepsilon_{1}(\omega_{I\boldsymbol{k}})v_{g}(\omega_{I\boldsymbol{k}})}{16\pi^{3}}}a_{I}(\boldsymbol{k},t)\boldsymbol{\hat{e}}_{I}(\boldsymbol{k})e^{i\boldsymbol{k}\cdot\boldsymbol{r}}d\boldsymbol{k}+h.c.,\label{eq:field_expansions_infinite_disp}\\
 & \boldsymbol{B}(\boldsymbol{r},t)=\sum_{I}\int\sqrt{\frac{\hbar\left|\boldsymbol{k}\right|\mu_{0}v_{g}(\omega_{I\boldsymbol{k}})}{16\pi^{3}}}a_{I}(\boldsymbol{k},t)s_{I}\boldsymbol{\hat{e}}_{I}(\boldsymbol{k})e^{i\boldsymbol{k}\cdot\boldsymbol{r}}d\boldsymbol{k}+h.c.,\nonumber 
\end{align}
where the dispersion relation is given by (\ref{eq:uniform_dispersion}),
although $\boldsymbol{k}$ now varies continuously, and the commutation
relations (\ref{eq:comm_continuous_first}) and the form of the Hamiltonian
(\ref{eq:Hamiltonian_infinite}) are as they were in the limit of
no dispersion.

\section{Group velocity}\label{sec:Group_velocity}
In this appendix we confirm the expression (\ref{eq:vIk}) for the
channel mode group velocity. Temporarily introducing 
\begin{align}
 & \mathcal{B}_{Ik}(\boldsymbol{r})=\boldsymbol{b}_{Ik}(y,z)e^{ikx},\label{eq:script_defs}\\
 & \mathcal{D}_{Ik}(\boldsymbol{r})=\boldsymbol{d}_{Ik}(y,z)e^{ikx},\nonumber 
\end{align}
we begin with the master equation (recall the first of (\ref{eq:mode_equations_dispersion}),
which we can write as
\begin{align}
 & \boldsymbol{\nabla}\times\left(\frac{\boldsymbol{\nabla}\times\mathcal{B}_{Ik}(\boldsymbol{r})}{\varepsilon_{1}(y,z;\omega_{Ik})}\right)=\frac{\omega_{Ik}^{2}}{c^{2}}\mathcal{B}_{Ik}(\boldsymbol{r}),\label{eq:masterwork}
\end{align}
and take its derivative with respect to $k$. Using 
\begin{align*}
 & \frac{\partial\mathcal{B}_{Ik}(\boldsymbol{r})}{\partial k}=\frac{\partial\boldsymbol{b}_{Ik}(y,z)}{\partial k}e^{ikx}+ix\mathcal{B}_{Ik}(\boldsymbol{r}),\\
 & \frac{\partial}{\partial k}\left(\frac{1}{\varepsilon_{1}(y,z;\omega_{Ik})}\right)=\frac{\partial\omega_{Ik}}{\partial k}\frac{\partial}{\partial\omega_{Ik}}\left(\frac{1}{\varepsilon_{1}(y,z;\omega_{Ik})}\right)
\end{align*}
and putting 
\begin{align*}
 & v_{Ik}\equiv\frac{\partial\omega_{Ik}}{\partial k},
\end{align*}
the group velocity of the mode at $k$ including both modal and material
dispersion effects, we find 
\begin{align}
 & \boldsymbol{\nabla}\times\left(\boldsymbol{C}_{1}(\boldsymbol{r})+\boldsymbol{C}_{2}(\boldsymbol{r})+v_{Ik}\boldsymbol{C}_{3}(\boldsymbol{r)}\right)\nonumber \\
 & =\frac{2\omega_{Ik}}{c^{2}}v_{Ik}\mathcal{B}_{Ik}(\boldsymbol{r})+\frac{\omega_{Ik}^{2}}{c^{2}}\left(\frac{\partial\boldsymbol{b}_{Ik}(y,z)}{\partial k}e^{ikx}+ix\mathcal{B}_{Ik}(\boldsymbol{r})\right),\label{eq:work1}
\end{align}
where 
\begin{align*}
 & \boldsymbol{C}_{1}(\boldsymbol{r})=\frac{\boldsymbol{\nabla}\times\left(\frac{\partial\boldsymbol{b}_{Ik}(y,z)}{\partial k}e^{ikx}\right)}{\varepsilon_{1}(y,z;\omega_{Ik})},\\
 & \boldsymbol{C}_{2}(\boldsymbol{r})=\frac{\boldsymbol{\nabla}\times\left(ix\mathcal{B}_{Ik}(\boldsymbol{r})\right)}{\varepsilon_{1}(y,z;\omega_{Ik})}=\frac{i\boldsymbol{\hat{x}}\times\mathcal{B}_{Ik}(\boldsymbol{r})}{\varepsilon_{1}(y,z;\omega_{Ik})}+\frac{ix\boldsymbol{\nabla}\times\mathcal{B}_{Ik}(\boldsymbol{r})}{\varepsilon_{1}(y,z;\omega_{Ik})},\\
 & \boldsymbol{C}_{3}(\boldsymbol{r})=\frac{\partial}{\partial\omega_{Ik}}\left(\frac{1}{\varepsilon_{1}(y,z;\omega_{Ik})}\right)\boldsymbol{\nabla}\times\mathcal{B}_{Ik}(\boldsymbol{r}).
\end{align*}
Now write the curl of $\boldsymbol{C}_{2}(\boldsymbol{r})$ as 
\begin{align}
 & \boldsymbol{\nabla}\times\boldsymbol{C}_{2}(\boldsymbol{r})=\boldsymbol{\nabla}\times\boldsymbol{C}_{4}(\boldsymbol{r})+ix\boldsymbol{\nabla}\times\left(\frac{\boldsymbol{\nabla}\times\mathcal{B}_{Ik}(\boldsymbol{r})}{\varepsilon_{1}(y,z;\omega_{Ik})}\right)+\boldsymbol{C}_{6}(\boldsymbol{r}),\label{eq:curlC2}
\end{align}
where 
\begin{align*}
 & \boldsymbol{C}_{4}(\boldsymbol{r})=\frac{i\boldsymbol{\hat{x}}\times\mathcal{B}_{Ik}(\boldsymbol{r})}{\varepsilon_{1}(y,z;\omega_{Ik})},\\
 & \boldsymbol{C}_{6}(\boldsymbol{r})=i\hat{\boldsymbol{x}}\times\left(\frac{\boldsymbol{\nabla}\times\mathcal{B}_{Ik}(\boldsymbol{r})}{\varepsilon_{1}(y,z;\omega_{Ik})}\right).
\end{align*}
Using the expression (\ref{eq:curlC2}) for $\boldsymbol{\nabla}\times\boldsymbol{C}_{2}(\boldsymbol{r})$
in (\ref{eq:work1}), we see that the terms proportional to $x$ cancel
out by virtue of the master equation (\ref{eq:masterwork}). Further,
we can combine the $\boldsymbol{\nabla}\times\boldsymbol{C}_{1}(\boldsymbol{r})$
term on the left-hand-side of (\ref{eq:work1}) with the term involving
$\partial\boldsymbol{b}_{Ik}(y,z)/\partial k$ on the right-hand-side
of that equation to write (\ref{eq:work1}) as 
\begin{align}
 & \left(\mathcal{\overline{M}}-\frac{\omega_{Ik}^{2}}{c^{2}}\right)\left(\frac{\partial\boldsymbol{b}_{Ik}(y,z)}{\partial k}e^{ikx}\right)+\boldsymbol{\nabla}\times\boldsymbol{C}_{4}(\boldsymbol{r})+\boldsymbol{C}_{6}(\boldsymbol{r})\label{eq:work2}\\
 & =\frac{2\omega_{Ik}}{c^{2}}v_{Ik}\mathcal{B}_{Ik}(\boldsymbol{r})-v_{Ik}\boldsymbol{\nabla}\times\boldsymbol{C}_{3}(\boldsymbol{r}),\nonumber 
\end{align}
where we have introduced the Hermitian operator $\mathcal{\overline{M}}$,
\begin{align*}
 & \mathcal{\overline{M}}\boldsymbol{g}(\boldsymbol{r})\equiv\boldsymbol{\nabla}\times\left(\frac{\boldsymbol{\nabla}\times\boldsymbol{g}(\boldsymbol{r})}{\varepsilon_{1}(y,z;\omega_{Ik})}\right)
\end{align*}
for any vector function $\boldsymbol{g}(\boldsymbol{r}).$ 

Now choose a particular $k$, and introduce any length $L$. The derivation
presented here can be easily extended to a photonic crystal structure
as well, where instead of $\boldsymbol{b}_{Ik}(y,z)$ we would have
$\boldsymbol{b}_{Ik}(x;y,z),$with $\boldsymbol{b}_{Ik}(x;y,z)=\boldsymbol{b}_{Ik}(x+a;y,z)$,
where $a$ is the length of the unit cell; in such a case we would
take $L=a$., but in a channel waveguide the choice of $L$ is ireelevant.
 Then dot each side of (\ref{eq:work2}) into $\mathcal{B}_{Ik}^{*}(\boldsymbol{r})$
and integrate over all $y$ and $z$, and over $x$ from $-L/2$ to
$L/2.$ From the first term on the left-hand-side of (\ref{eq:work2})
we find 
\begin{align*}
 & \int_{-L/2}^{L/2}dx\int dydz\mathcal{B}_{Ik}^{*}(\boldsymbol{r})\cdot\left(\mathcal{\overline{M}}-\frac{\omega_{Ik}^{2}}{c^{2}}\right)\left(\frac{\partial\boldsymbol{b}_{Ik}(y,z)}{\partial k}e^{ikx}\right)\\
 & =\int_{-L/2}^{L/2}dx\int dydz\left(\left(\mathcal{\overline{M}}-\frac{\omega_{Ik}^{2}}{c^{2}}\right)\mathcal{B}_{Ik}^{*}(\boldsymbol{r})\right)\cdot\left(\frac{\partial\boldsymbol{b}_{Ik}(y,z)}{\partial k}e^{ikx}\right)\\
 & =0,
\end{align*}
where in the second line we have partially integrated and used the
fact that the terms in the integrand are the same at $z=-L/2$ and
$z=L/2$, and in the third line we have used the master equation (\ref{eq:masterwork}).
From the second term on the left-hand-side of (\ref{eq:work2}) we
find 
\begin{align*}
 & \int_{-L/2}^{L/2}dx\int dydz\mathcal{B}_{Ik}^{*}(\boldsymbol{r})\cdot\boldsymbol{\nabla}\times\boldsymbol{C}_{4}(\boldsymbol{r})\\
 & =i\int_{-L/2}^{L/2}dx\int dydz\left(\boldsymbol{\nabla}\times\mathcal{B}_{Ik}^{*}(\boldsymbol{r})\right).\left(\frac{\boldsymbol{\hat{x}}\times\mathcal{B}_{Ik}(\boldsymbol{r})}{\varepsilon_{1}(y,z;\omega_{Ik})}\right)\\
 & =-\mu_{0}\omega_{Ik}\int_{-L/2}^{L/2}dx\int dydz\mathcal{D}_{Ik}^{*}(\boldsymbol{r}).\left(\frac{\boldsymbol{\hat{x}}\times\mathcal{B}_{Ik}(\boldsymbol{r})}{\varepsilon_{1}(y,z;\omega_{Ik})}\right)\\
 & =\mu_{0}\omega_{Ik}\int_{-L/2}^{L/2}dx\int dydz\boldsymbol{\hat{x}}\cdot\left(\frac{\mathcal{D}_{Ik}^{*}(\boldsymbol{r})\times\mathcal{B}_{Ik}(\boldsymbol{r})}{\varepsilon_{1}(y,z;\omega_{Ik})}\right).
\end{align*}
In the second line we have used the result 
\begin{align}
 & \int_{-L/2}^{L/2}dx\int dydz\boldsymbol{U}(\boldsymbol{r})\cdot\left(\boldsymbol{\nabla}\times\boldsymbol{V}(\boldsymbol{r})\right)=\int_{-L/2}^{L/2}dx\int dydz\boldsymbol{V}(\boldsymbol{r})\cdot\left(\boldsymbol{\nabla}\times\boldsymbol{U}(\boldsymbol{r})\right),\label{eq:curl_identity}
\end{align}
which holds for functions $\boldsymbol{U}(\boldsymbol{r})$ and $\boldsymbol{V}(\boldsymbol{r})$
that vanish at the infinite limits of integration for $dy$ and $dz$,
and such that they yield no net contribution from the remaining limit
of integration (for $dx$); in the third line we have used the result
\begin{align}
 & \mathcal{D}_{Ik}(\boldsymbol{r})=\frac{i}{\mu_{0}\omega_{Ik}}\boldsymbol{\nabla}\times\mathcal{B}_{Ik}(\boldsymbol{r)},\label{eq:Maxwell_result1}
\end{align}
which holds since $\mathcal{D}_{Ik}(\boldsymbol{r})\exp(-i\omega_{Ik}t)$
and $\mathcal{B}_{Ik}(\boldsymbol{r})\exp(-i\omega_{Ik}t)$ satisfy
the linear Maxwell equations; in the fourth line we have used 
\begin{align}
 & \boldsymbol{u}\cdot\left(\boldsymbol{v}\times\boldsymbol{w}\right)=\boldsymbol{v}\cdot\left(\boldsymbol{w}\times\boldsymbol{u}\right),\label{eq:triple_vector}
\end{align}
for any three vectors $\boldsymbol{u}$, $\boldsymbol{v}$, and $\boldsymbol{w}$.
Finally, from the last term on the right-hand-side of (\ref{eq:work2})
we find 
\begin{align*}
 & \int_{-L/2}^{L/2}dx\int dydz\mathcal{B}_{Ik}^{*}(\boldsymbol{r})\cdot\boldsymbol{C}_{6}(\boldsymbol{r})\\
 & =i\int_{-L/2}^{L/2}dx\int dydz\boldsymbol{\hat{x}}\cdot\left(\left(\frac{\boldsymbol{\nabla}\times\mathcal{B}_{Ik}(\boldsymbol{r})}{\varepsilon_{1}(y,z;\omega_{Ik})}\right)\times\mathcal{B}_{Ik}^{*}(\boldsymbol{r})\right)\\
 & =\mu_{0}\omega_{Ik}\int_{-L/2}^{L/2}dx\int dydz\boldsymbol{\hat{x}}\cdot\left(\frac{\mathcal{D}_{Ik}(\boldsymbol{r})\times\mathcal{B}_{Ik}^{*}(\boldsymbol{r})}{\varepsilon_{1}(y,z;\omega_{Ik})}\right),
\end{align*}
where in the second line we have used (\ref{eq:triple_vector}) and
in the third line we have used (\ref{eq:Maxwell_result1}). Thus from
(\ref{eq:work2}) in all we have 
\begin{align}
 & \mu_{0}\omega_{Ik}\int_{-L/2}^{L/2}dx\int dydz\boldsymbol{\hat{x}}\cdot\left(\frac{\mathcal{D}_{Ik}(\boldsymbol{r})\times\mathcal{B}_{Ik}^{*}(\boldsymbol{r})+\mathcal{D}_{Ik}^{*}(\boldsymbol{r})\times\mathcal{B}_{Ik}(\boldsymbol{r})}{\varepsilon_{1}(y,z;\omega_{Ik})}\right)\label{eq:work3}\\
 & =v_{Ik}\mathcal{T},\nonumber 
\end{align}
where 
\begin{align}
 & \mathcal{T}=\frac{2\omega_{Ik}}{c^{2}}\int_{-L/2}^{L/2}dx\int dydz\mathcal{B}_{Ik}^{*}(\boldsymbol{r})\cdot\mathcal{B}_{Ik}(\boldsymbol{r})\label{eq:Tscript}\\
 & -\int_{-L/2}^{L/2}dx\int dydz\mathcal{B}_{Ik}^{*}(\boldsymbol{r})\cdot\nabla\times\boldsymbol{C}_{3}(\boldsymbol{r}).\nonumber 
\end{align}
Look first at the first term on the right-hand-side. Since $\mathcal{D}_{Ik}(\boldsymbol{r})\exp(-i\omega_{Ik}t)$
and $\mathcal{B}_{Ik}(\boldsymbol{r})\exp(-i\omega_{Ik}t)$ satisfy
the linear Maxwell equations, from Faraday's law we have 
\begin{align}
 & \mathcal{B}_{Ik}(\boldsymbol{r})=\frac{1}{i\omega_{Ik}}\boldsymbol{\nabla}\times\left(\frac{\mathcal{D}_{Ik}(\boldsymbol{r})}{\epsilon_{0}\varepsilon_{1}(y,z;\omega_{Ik})}\right),\label{eq:Maxwell_result2}
\end{align}
so 
\begin{align}
 & \frac{2\omega_{Ik}}{c^{2}}\int_{-L/2}^{L/2}dx\int dydz\mathcal{B}_{Ik}^{*}(\boldsymbol{r})\cdot\mathcal{B}_{Ik}(\boldsymbol{r})\label{eq:Tscript_work1}\\
 & =\frac{2}{ic^{2}}\int_{-L/2}^{L/2}dx\int dydz\mathcal{B}_{Ik}^{*}(\boldsymbol{r})\cdot\left(\boldsymbol{\nabla}\times\left(\frac{\mathcal{D}_{Ik}(\boldsymbol{r})}{\epsilon_{0}\varepsilon_{1}(y,z;\omega_{Ik})}\right)\right)\nonumber \\
 & =\frac{2}{ic^{2}}\int_{-L/2}^{L/2}dx\int dydz\left(\boldsymbol{\nabla}\times\mathcal{B}_{Ik}^{*}(\boldsymbol{r})\right)\cdot\left(\frac{\mathcal{D}_{Ik}(\boldsymbol{r})}{\epsilon_{0}\varepsilon_{1}(y,z;\omega_{Ik})}\right)\nonumber \\
 & =\frac{2\mu_{0}\omega_{Ik}}{c^{2}}\int_{-L/2}^{L/2}dx\int dydz\left(\frac{\mathcal{D}_{Ik}^{*}(\boldsymbol{r})\cdot\mathcal{D}_{Ik}(\boldsymbol{r})}{\epsilon_{0}\varepsilon_{1}(y,z;\omega_{Ik})}\right),\nonumber 
\end{align}
where in the second expression on the right-hand-side we have used
(\ref{eq:curl_identity}) and in the last expression we have used
(\ref{eq:Maxwell_result1}). Looking at the second term on the right-hand-side
of (\ref{eq:Tscript}) we have 
\begin{align*}
 & -\int_{-L/2}^{L/2}dx\int dydz\mathcal{B}_{Ik}^{*}(\boldsymbol{r})\cdot\nabla\times\boldsymbol{C}_{3}(\boldsymbol{r})\\
 & =-\int_{-L/2}^{L/2}dx\int dydz\mathcal{B}_{Ik}^{*}(\boldsymbol{r})\cdot\nabla\times\left(\frac{\partial}{\partial\omega_{Ik}}\left(\frac{1}{\varepsilon_{1}(y,z;\omega_{Ik})}\right)\boldsymbol{\nabla}\times\mathcal{B}_{Ik}(\boldsymbol{r})\right)\\
 & =-\int_{-L/2}^{L/2}dx\int dydz\boldsymbol{\nabla}\times\mathcal{B}_{Ik}^{*}(\boldsymbol{r})\cdot\left(\frac{\partial}{\partial\omega_{Ik}}\left(\frac{1}{\varepsilon_{1}(y,z;\omega_{Ik})}\right)\boldsymbol{\nabla}\times\mathcal{B}_{Ik}(\boldsymbol{r})\right)\\
 & =-\mu_{0}^{2}\omega_{Ik}^{2}\int_{-L/2}^{L/2}dx\int dydz\left(\mathcal{D}_{Ik}^{*}(\boldsymbol{r})\cdot\mathcal{D}_{Ik}(\boldsymbol{r})\right)\frac{\partial}{\partial\omega_{Ik}}\left(\frac{1}{\varepsilon_{1}(y,z;\omega_{Ik})}\right)\\
 & =-\frac{2\mu_{0}\omega_{Ik}}{c^{2}}\int_{-L/2}^{L/2}dx\int dydz\left(\mathcal{D}_{Ik}^{*}(\boldsymbol{r})\cdot\mathcal{D}_{Ik}(\boldsymbol{r})\right)\frac{\omega_{Ik}}{2}\frac{\partial}{\partial\omega_{Ik}}\left(\frac{1}{\epsilon_{0}\varepsilon_{1}(y,z;\omega_{Ik})}\right)
\end{align*}
or
\begin{align}
 & -\int_{-L/2}^{L/2}dx\int dydz\mathcal{B}_{Ik}^{*}(\boldsymbol{r})\cdot\nabla\times\boldsymbol{C}_{3}(\boldsymbol{r})\label{eq:Tscript_work2}\\
 & =-\int_{-L/2}^{L/2}dx\int dydz\boldsymbol{\nabla}\times\mathcal{B}_{Ik}^{*}(\boldsymbol{r})\cdot\left(\frac{\partial}{\partial\omega_{Ik}}\left(\frac{1}{\varepsilon_{1}(y,z;\omega_{Ik})}\right)\boldsymbol{\nabla}\times\mathcal{B}_{Ik}(\boldsymbol{r})\right)\nonumber \\
 & =-\frac{2\mu_{0}\omega_{Ik}}{c^{2}}\int_{-L/2}^{L/2}dx\int dydz\left(\mathcal{D}_{Ik}^{*}(\boldsymbol{r})\cdot\mathcal{D}_{Ik}(\boldsymbol{r})\right)\frac{\omega_{Ik}}{2}\frac{\partial}{\partial\omega_{Ik}}\left(\frac{1}{\epsilon_{0}\varepsilon_{1}(y,z;\omega_{Ik})}\right),\nonumber 
\end{align}
where in the first term on the right-hand-side we have used (\ref{eq:curl_identity}),
and in the second we have used (\ref{eq:Maxwell_result1}). Combining
(\ref{eq:Tscript_work1},\ref{eq:Tscript_work2}) in (\ref{eq:Tscript})
we find 
\begin{align*}
 & \mathcal{T}=\frac{2\mu_{0}\omega_{Ik}}{c^{2}}L,
\end{align*}
where we have used (\ref{eq:script_defs}) and (\ref{eq:norm_D_waveguide_disp}).
Using this in (\ref{eq:work3}), and using (\ref{eq:script_defs},\ref{eq:little_e_and_h})
we indeed find (\ref{eq:vIk}).

\section{Interaction picture}\label{app:interaction_picture}
In this appendix we show that it is always possible to go to an interaction picture where the terms proportional to $a_i^\dagger a_j$ in~\eqref{eq:quad_hamil} can be eliminated if the coefficients $\Delta_{i,j}$ are time independent. We split the Hamiltonian as
\begin{align}
H(t)=  \underbrace{ \hbar \sum_{i,j =1}^\ell \Delta_{i,j} a_i^\dagger a_j}_{\equiv H_0} + \underbrace{\hbar i \frac12 \sum_{k,l=1}^\ell \left[ \zeta_{k,l}(t) a_k^{\dagger } a_{l}^{\dagger} -\text{H.c.} \right] }_{\equiv V(t)},
\end{align}
To study the dynamics of the problem let us derive the equation of motion obeyed by a modified time evolution operator $\mathcal{ U}_I$ related to the original time evolution operator $\mathcal{ U}$ via
\begin{align}
\mathcal{ U}_I(t,t_0)\equiv e^{i  H_0 (t-t_0)/\hbar} \mathcal{ U}(t,t_0).
\end{align}
Using the Schr\"{o}dinger equation for $\mathcal{ U}$ in \eqref{eq:schrodinger} we find that 
\begin{align}\label{ch2:scheqi}
i \hbar \frac{d}{dt} \mathcal{ U}_I(t,t_0)=\left( e^{i  H_0(t-t_0)/\hbar} V(t) e^{-i  H_0(t-t_0)/\hbar} \right) \mathcal{ U}_I(t,t_0)=V_I(t) \mathcal{ U}_I(t,t_0).
\end{align}
In the last line we introduced the interaction picture Hamiltonian $V_I(t)$. For the explicit form of our Hamiltonian we can first obtain
\begin{align}
e^{i  H_0(t-t_0)/\hbar} \bm{a} e^{-i  H_0(t-t_0)/\hbar} = \exp\left( i \bm{\Delta} (t-t_0) \right) \bm{a} = \bm{U}(t)\bm{a},
\end{align}
where $\bm{U}(t)$ is a time-dependent unitary matrix. With this solution we can write
\begin{align}
V_I(t) = e^{i  H_0(t-t_0)/\hbar} V(t) e^{-i  H_0(t-t_0)/\hbar}  &=\hbar i \frac12 \sum_{k,l=1}^\ell \left[ \zeta_{k,l}(t) e^{i  H_0(t-t_0)/\hbar} a_k^{\dagger } a_{l}^{\dagger} e^{-i  H_0(t-t_0)/\hbar}  -\text{H.c.} \right] \\
&=\hbar i \frac12 \sum_{i,j,k,l=1}^\ell \left[ \zeta_{k,l}(t) U_{l,i}^*(t) a_i^\dagger U_{k,j}^*(t)a_j^\dagger  -\text{H.c.} \right] \\
&=\hbar i \frac12 \sum_{i,j=1}^\ell \left[ \tilde{\zeta}_{i,j}(t)  a_i^\dagger a_j^\dagger  -\text{H.c.} \right] ,
\end{align}
where we introduced the symmetric matrix $\tilde{\bm{\zeta}}(t) = \bm{U}^\dagger(t) \bm{\zeta}(t) \bm{U}^*(t)$. This completes the proof, since now we need to solve for the interaction picture evolution operator which is solely driven by terms of the form $a_i^\dagger a_j^\dagger$. 

\section{The Symplectic group}\label{app:symplectic}
In Sec.~\ref{sec:bogoliubov} we showed that the solution of the Heisenberg equations of motion had the form~\eqref{eq:symplectic}, and respected the equal time commutation relations. We can write the canonical commutation relation of the ladder operators more succinctly as
\begin{align}
[z_i, z_j^\dagger] = Z_{i,j}, \quad \bm{z} = \begin{pmatrix} 
\bm{a} \\
\bm{a}^\dagger
\end{pmatrix},  \quad \bm{Z} = \begin{pmatrix} 
\mathbb{I}_\ell & \bm{0} \\
\bm{0} & -\mathbb{I}_\ell
\end{pmatrix},
\end{align}
and then the fact that the Heisenberg transformation induced by $\bm{K}$ respects these commutation relations is equivalent to
\begin{align}\label{eq:symplectic_complex}
\bm{K} \bm{Z} \bm{K}^\dagger  = \bm{Z}.
\end{align}
To connect this equation to the usual definition of the symplectic group one introduced quadratures
\begin{align}
\bm{r} \equiv \begin{pmatrix}
\bm{q} \\
\bm{p}
\end{pmatrix}
= \sqrt{\hbar}\frac{1}{\sqrt{2}} \begin{pmatrix}
\mathbb{I}_\ell & \mathbb{I}_\ell \\
-i \mathbb{I}_\ell & i \mathbb{I}_\ell 
\end{pmatrix}
\begin{pmatrix}
	\bm{a} \\
	\bm{a}^\dagger
\end{pmatrix} = \sqrt{\hbar}\bm{R}\bm{z}.
\end{align} 
The canonical commutation relation for the operators in $\bm{r}$ follows from the one in $\bm{z}$ as
\begin{align}
[r_i,r_j] = i \hbar \Omega_{i,j} \quad \bm{\Omega} = \begin{pmatrix} 
0  & \mathbb{I}_\ell \\
 -\mathbb{I}_\ell & \bm{0}
\end{pmatrix}.
\end{align}
Now it is easy to state the transformation for quadratures
\begin{align}
\bm{z} \to \bm{z'} = \bm{K} \bm{z} \Longleftrightarrow \bm{r} \to \bm{r'} = \bm{S} \bm{r},
\end{align}
where
\begin{align}
\bm{S} = \bm{R} \bm{K} \bm{R}^\dagger = \begin{pmatrix}
\Re(\bm{V}) + \Re(\bm{W}) & \Im(\bm{W}) -\Im(\bm{V}) \\
\Im(\bm{V}) + \Im(\bm{W}) & \Re(\bm{V}) - \Re(\bm{W}),
\end{pmatrix}
\end{align}
The condition in~\eqref{eq:symplectic_complex} implies that the matrix $\bm{S}$ needs to satisfy
\begin{align}
\bm{S} \bm{\Omega} \bm{S}^T = \bm{\Omega}
\end{align}
which implies that $\bm{S}$ is a symplectic matrix \cite{serafini2017quantum,simon1988gaussian,dutta1995real,adesso2014continuous,weedbrook2012gaussian} .

\section{Form of the evolution operator in the high-gain regime}\label{app:form_of_U_0}
At the end of Sec.~\ref{sec:theket} we argued that the evolution operator associated with a general quadratic Hamiltonian can be written as the product of two exponentials
\begin{align}
\mathcal{U} = \mathcal{U}_1 \mathcal{U}_0, \quad \mathcal{U}_1  = \exp\left[ \frac12 \sum_{k,l=1}^\ell J_{k,l} a^\dagger_{k} a^\dagger_l  - \text{H.c.}\right],
\end{align}
where $\mathcal{U}_0$ is a unitary operator that satisfies $\mathcal{U}_0 \ket{\text{vac}} = \ket{\text{vac}}$. In this appendix we determine the form of the generator of $\mathcal{U}_0$. 
To pin down $\mathcal{U}_0$ we calculate 
\begin{align}
 \mathcal{U}_0^\dagger \mathcal{U}_1^\dagger \bm{a} \mathcal{U}_1 \mathcal{U}_0 = \bm{F} \left[ \oplus_{\lambda=1}^\ell \cosh r_\lambda \right] \bm{F}^*  \left\{ \mathcal{U}_0^\dagger \bm{a}  \mathcal{U}_0 \right\} + \bm{F} \left[ \oplus_{\lambda=1}^\ell \sinh r_\lambda \right] \bm{F}  \left\{ \mathcal{U}_0^\dagger \bm{a}^\dagger  \mathcal{U}_0 \right\},
\end{align}
and compare with the solution in~\eqref{eq:singular-values} which tell use that we want
\begin{align}
\bm{F}^*  \left\{ \mathcal{U}_0^\dagger \bm{a}  \mathcal{U}_0 \right\}  = \bm{G} \bm{a}	\Longleftrightarrow \bm{G}^\dagger\bm{F}^*  \left\{ \mathcal{U}_0^\dagger \bm{a}  \mathcal{U}_0 \right\}  =  \bm{a}	,
\end{align}
To satisfy this constraint we recall the results from Appendix~\ref{app:interaction_picture} and write 
\begin{align}\label{eq:U0}
\mathcal{U}_0 = \exp\left[ -i \sum_{j,k} \phi_{j,k}a^\dagger_j a_k  \right],
\end{align}
such that 
\begin{align}
\mathcal{U}_0^\dagger \bm{a}  \mathcal{U}_0 = \exp\left[ - i\bm{\phi} \right] \bm{a},
\end{align}
and comparing the two equations above tell us that we want $\bm{G}^\dagger\bm{F}^* \exp\left[ - i\bm{\phi} \right] = \mathbb{I}_{\ell}$
or equivalently $\bm{\phi} = -i \log_e \bm{G}^\dagger \bm{F}^*$. Note that since both $\bm{G}^\dagger$ and  $\bm{F}$ are unitaries their logarithm is $i$ times a hermitian matrix which upon multiplication by $(-i)$ give that $\bm{\phi}$ is a purely hermitian matrix as it should. Also note that as expected $\mathcal{U}_0 $ in~\eqref{eq:U0}  has the vacuum as an eigenket with eigenvalue 1.
It is interesting to note that Heisenberg transformation generated by a general quadratic Hamiltonian requires the exponentiaition of \emph{two} generators. This happens because the exponential map from the Lie algebra $\mathbf{sp}(2n, \mathbb{R})$ to the symplectic group Sp$(2n,\mathbb{R})$ is not surjective. However, any element of the group may be generated by the group multiplication of two elements of the group.

\section{Waveguide Hamiltonians}\label{sec:WG_Hamiltonians}
Concerning waveguides in which a second-order nonlinearity is dominant, the process with $2k_{S}=k_{P}$ leads to
\begin{equation}
H_{\text{DSV}}^{\text{SPDC}}=-\frac{\zeta_{\text{chan}}^{SSP}}{2}\int\dd x\,\psi_{S}^{\dagger}\left(x\right)\psi_{S}^{\dagger}\left(x\right)\psi_{P}\left(x\right)+\hc,
\end{equation}
and the process with $k_{S}+k_{I}=k_{P}$ leads to
\begin{equation}
H_{\text{NDSV}}^{\text{SPDC}}=-\zeta_{\text{chan}}^{SIP}\int\dd x\,\psi_{S}^{\dagger}\left(x\right)\psi_{I}^{\dagger}\left(x\right)\psi_{P}\left(x\right)+\hc,
\end{equation}
where
\begin{align}
\zeta_{\text{chan}}^{J_{1}J_{2}J_{3}}&=\frac{2}{\epsilon_{0}}\sqrt{\frac{\hbar^{3}\omega_{J_{1}}\omega_{J_{2}}\omega_{J_{3}}}{2^{3}}}\int\dd y\dd z\,\Gamma_{2}^{ijk}\left(\bm{r}\right)\left[d_{J_{1}k_{J_{1}}}^{i}\left(y,z\right)d_{J_{2}k_{J_{2}}}^{j}\left(y,z\right)\right]^{*}d_{J_{3}k_{J_{3}}}^{k}\left(y,z\right)\\
&=\hbar\sqrt{\frac{\hbar\omega_{J_{1}}\omega_{J_{2}}\omega_{J_{3}}}{2\epsilon_{0}\overline{n}_{J_{1}}\overline{n}_{J_{2}}\overline{n}_{J_{3}}A_{\text{chan}}^{J_{1}J_{2}J_{3}}}}\overline{\chi}_{2},\nonumber
\end{align}
with
\begin{equation}\label{eq:chi2_A}
\frac{1}{\sqrt{A_{\text{chan}}^{J_{1}J_{2}J_{3}}}}=\frac{\int\dd y\dd z\frac{\chi_{2}^{ijk}\left(y,z\right)}{\overline{\chi}_{2}}\left[e_{J_{1}}^{i}\left(y,z\right)e_{J_{2}}^{j}\left(y,z\right)\right]^{*}e_{J_{3}}^{k}\left(y,z\right)}{\mathcal{N}_{J_{1}}\mathcal{N}_{J_{2}}\mathcal{N}_{J_{3}}\sqrt{c^{3}/\left(v_{J_{1}}v_{J_{2}}v_{J_{3}}\right)}},
\end{equation}
and we have used~\eqref{eq:eh_from_db}, \eqref{eq:big_gammas}, and~\eqref{eq:N_norm}.

Concerning waveguides in which a third-order nonlinearity is dominant, the single-pump DSV process leads to
\begin{equation}\label{eq:H_SP-SFWM_pre}
H_{\text{DSV}}^{\text{SP-SFWM}}=-\frac{\gamma_{\text{chan}}^{PPPP}}{2}\hbar^{2}\omega_{P}v_{P}^{2}\int\dd x\,\psi_{P}^{\dagger}\left(x\right)\psi_{P}^{\dagger}\left(x\right)\psi_{P}\left(x\right)\psi_{P}\left(x\right),
\end{equation}
the process with $k_{S}+k_{I}=2k_{P}$ leads to~\eqref{eq:H_SP-SFWM}, the process with $2k_{S}=k_{P_{1}}+k_{P_{2}}$ leads to
\begin{align}\label{eq:H_DP-SFWM_pre}
H_{\text{DSV}}^{\text{DP-SFWM}}	&=-\gamma_{\text{chan}}^{SSP_{1}P_{2}}\hbar^{2}\overline{\omega}_{SSP_{1}P_{2}}\overline{v}_{SSP_{1}P_{2}}^{2}\int\dd x\,\psi_{S}^{\dagger}\left(x\right)\psi_{S}^{\dagger}\left(x\right)\psi_{P_{1}}\left(x\right)\psi_{P_{2}}\left(x\right)+\text{H.c.}\\
&\quad-\frac{\hbar^{2}}{2}\sum_{J}\gamma_{\text{chan}}^{JJJJ}\omega_{J}v_{J}^{2}\int\dd x\,\psi_{J}^{\dagger}\left(x\right)\psi_{J}^{\dagger}\left(x\right)\psi_{J}\left(x\right)\psi_{J}\left(x\right)\nonumber\\
&\quad-2\hbar^{2}\sum_{J\neq S}\gamma_{\text{chan}}^{JSJS}\sqrt{\omega_{J}\omega_{S}}v_{J}v_{S}\int\dd x\,\psi_{J}^{\dagger}\left(x\right)\psi_{S}^{\dagger}\left(x\right)\psi_{J}\left(x\right)\psi_{S}\left(x\right)\nonumber\\
&\quad-2\gamma_{\text{chan}}^{P_{1}P_{2}P_{1}P_{2}}\hbar^{2}\sqrt{\omega_{P_{1}}\omega_{P_{2}}}v_{P_{1}}v_{P_{2}}\int\dd x\,\psi_{P_{1}}^{\dagger}\left(x\right)\psi_{P_{2}}^{\dagger}\left(x\right)\psi_{P_{1}}\left(x\right)\psi_{P_{2}}\left(x\right),\nonumber
\end{align}
and the process with $k_{S}+k_{I}=k_{P_{1}}+k_{P_{2}}$ leads to
\begin{align}
H_{\text{NDSV}}^{\text{DP-SFWM}}&=-2\gamma_{\text{chan}}^{SIP_{1}P_{2}}\hbar^{2}\overline{\omega}_{SIP_{1}P_{2}}\overline{v}_{SIP_{1}P_{2}}^{2}\int\dd x\,\psi_{S}^{\dagger}\left(x\right)\psi_{I}^{\dagger}\left(x\right)\psi_{P_{1}}\left(x\right)\psi_{P_{2}}\left(x\right)+\hc\nonumber \\
&\quad-\frac{\hbar^{2}}{2}\sum_{J}\gamma_{\text{chan}}^{JJJJ}\omega_{J}v_{J}^{2}\int\dd x\,\psi_{J}^{\dagger}\left(x\right)\psi_{J}^{\dagger}\left(x\right)\psi_{J}\left(x\right)\psi_{J}\left(x\right)\nonumber \\
&\quad-2\hbar^{2}\sum_{J\neq S}\gamma_{\text{chan}}^{JSJS}\sqrt{\omega_{J}\omega_{S}}v_{J}v_{S}\int\dd x\,\psi_{J}^{\dagger}\left(x\right)\psi_{S}^{\dagger}\left(x\right)\psi_{J}\left(x\right)\psi_{S}\left(x\right)\nonumber \\
&\quad-2\hbar^{2}\sum_{J\neq I}\gamma_{\text{chan}}^{JIJI}\sqrt{\omega_{J}\omega_{I}}v_{J}v_{I}\int\dd x\,\psi_{J}^{\dagger}\left(x\right)\psi_{I}^{\dagger}\left(x\right)\psi_{J}\left(x\right)\psi_{I}\left(x\right)\nonumber \\
&\quad-2\gamma_{\text{chan}}^{P_{1}P_{2}P_{1}P_{2}}\hbar^{2}\sqrt{\omega_{P_{1}}\omega_{P_{2}}}v_{P_{1}}v_{P_{2}}\int\dd x\,\psi_{P_{1}}^{\dagger}\left(x\right)\psi_{P_{2}}^{\dagger}\left(x\right)\psi_{P_{1}}\left(x\right)\psi_{P_{2}}\left(x\right),\nonumber
\end{align}
where we have used~\eqref{eq:gamma_chan_general} and~\eqref{eq:v_bar_and_omega_bar}.

\section{Heisenberg equations of motion for waveguides}\label{sec:WG_equations}
Concerning waveguides in which a second-order nonlinearity is dominant, and using~\eqref{eq:psi_overline} along with $2\omega_{S}=\omega_{P}$, the Heisenberg equations of motion yield
\begin{align}\label{eq:SPDC_SSP_eqs}
\left(\frac{\partial}{\partial t}+v_{P}\frac{\partial}{\partial x}-i\frac{v_{P}^{\prime}}{2}\frac{\partial^{2}}{\partial x^{2}}\right)\left\langle \overline{\psi}_{P}\left(x,t\right)\right\rangle &=0,\\
\left(\frac{\partial}{\partial t}+v_{S}\frac{\partial}{\partial x}-i\frac{v_{S}^{\prime}}{2}\frac{\partial^{2}}{\partial x^{2}}\right)\overline{\psi}_{S}\left(x,t\right)&=\frac{i}{\hbar}\zeta_{\text{chan}}^{SSP}\left(x\right)\left\langle \overline{\psi}_{P}\left(x,t\right)\right\rangle \overline{\psi}_{S}^{\dagger}\left(x,t\right),\nonumber    
\end{align}
for the process with $2k_{S}=k_{P}$, and, using $\omega_{S}+\omega_{I}=\omega_{P}$
\begin{align}\label{eq:SPDC_SIP_eqs}
\left(\frac{\partial}{\partial t}+v_{P}\frac{\partial}{\partial x}-i\frac{v_{P}^{\prime}}{2}\frac{\partial^{2}}{\partial x^{2}}\right)\left\langle \overline{\psi}_{P}\left(x,t\right)\right\rangle &=0,\\
\left(\frac{\partial}{\partial t}+v_{S}\frac{\partial}{\partial x}-i\frac{v_{S}^{\prime}}{2}\frac{\partial^{2}}{\partial x^{2}}\right)\overline{\psi}_{S}\left(x,t\right)&=\frac{i}{\hbar}\zeta_{\text{chan}}^{SIP}\left(x\right)\left\langle \overline{\psi}_{P}\left(x,t\right)\right\rangle \overline{\psi}_{I}^{\dagger}\left(x,t\right), \nonumber\\
\left(\frac{\partial}{\partial t}+v_{I}\frac{\partial}{\partial x}-i\frac{v_{I}^{\prime}}{2}\frac{\partial^{2}}{\partial x^{2}}\right)\overline{\psi}_{I}\left(x,t\right)&=\frac{i}{\hbar}\zeta_{\text{chan}}^{SIP}\left(x\right)\left\langle \overline{\psi}_{P}\left(x,t\right)\right\rangle \overline{\psi}_{S}^{\dagger}\left(x,t\right),\nonumber
\end{align}
for the process with $k_{S}+k_{I}=k_{P}$. Note how the second of~\eqref{eq:SPDC_SSP_eqs} is of the form of~\eqref{eq:DSV_COEs}, and the second and third of~\eqref{eq:SPDC_SIP_eqs} are of the form of~\eqref{eq:NDSV_COEs}.

Concerning waveguides in which a third-order nonlinearity is dominant, and again using~\eqref{eq:psi_overline}, the Heisenberg equations of motion yield
\begin{align}\label{eq:SFWM_PPPP_eqs}
\left(\frac{\partial}{\partial t}+v_{P}\frac{\partial}{\partial x}-i\frac{v_{P}^{\prime}}{2}\frac{\partial^{2}}{\partial x^{2}}\right)\left\langle \overline{\psi}_{P}\left(x,t\right)\right\rangle &=i\gamma_{\text{chan}}^{PPPP}\hbar\omega_{P}v_{P}^{2}\left|\left\langle \overline{\psi}_{P}\left(x,t\right)\right\rangle \right|^{2}\left\langle \overline{\psi}_{P}\left(x,t\right)\right\rangle,\\
\left(\frac{\partial}{\partial t}+v_{P}\frac{\partial}{\partial x}-i\frac{v_{P}^{\prime}}{2}\frac{\partial^{2}}{\partial x^{2}}\right)\delta\overline{\psi}_{P}\left(x,t\right)&=i\gamma_{\text{chan}}^{PPPP}\hbar\omega_{P}v_{P}^{2}\left[\left\langle \overline{\psi}_{P}\left(x,t\right)\right\rangle ^{2}\delta\overline{\psi}_{P}^{\dagger}\left(x,t\right)+2\left|\left\langle \overline{\psi}_{P}\left(x,t\right)\right\rangle \right|^{2}\delta\overline{\psi}_{P}\left(x,t\right)\right]\nonumber,
\end{align}
where we have put $\overline{\psi}_{P}\left(x,t\right)\rightarrow\left\langle\overline{\psi}_{P}\left(x,t\right)\right\rangle+\delta\overline{\psi}_{P}\left(x,t\right)$, for the single-pump DSV process. Similarly, using $\omega_{S}+\omega_{I}=2\omega_{P}$, they yield \eqref{eq:NDSV_SP_SFWM_pump} and~\eqref{eq:NDSV_SP_SFWM_gen} for the process with $k_{S}+k_{I}=2k_{P}$, using $2\omega_{S}=\omega_{P_{1}}+\omega_{P_{2}}$ they yield
\begin{align}\label{eq:SFWM_SSP1P2_eqs}
\left(\frac{\partial}{\partial t}+v_{P_{1}}\frac{\partial}{\partial x}-i\frac{v_{P_{1}}^{\prime}}{2}\frac{\partial^{2}}{\partial x^{2}}\right)\left\langle \overline{\psi}_{P_{1}}\left(x,t\right)\right\rangle &=i\gamma_{\text{chan}}^{P_{1}P_{1}P_{1}P_{1}}\hbar\omega_{P_{1}}v_{P_{1}}^{2}\left|\left\langle \overline{\psi}_{P_{1}}\left(x,t\right)\right\rangle \right|^{2}\left\langle \overline{\psi}_{P_{1}}\left(x,t\right)\right\rangle\\
&\quad+2i\gamma_{\text{chan}}^{P_{1}P_{2}P_{1}P_{2}}\hbar\sqrt{\omega_{P_{1}}\omega_{P_{2}}}v_{P_{1}}v_{P_{2}}\left|\left\langle \overline{\psi}_{P_{2}}\left(x,t\right)\right\rangle \right|^{2}\left\langle \overline{\psi}_{P_{1}}\left(x,t\right)\right\rangle,\nonumber\\
\left(\frac{\partial}{\partial t}+v_{P_{2}}\frac{\partial}{\partial x}-i\frac{v_{P_{2}}^{\prime}}{2}\frac{\partial^{2}}{\partial x^{2}}\right)\left\langle \overline{\psi}_{P_{2}}\left(x,t\right)\right\rangle &=i\gamma_{\text{chan}}^{P_{2}P_{2}P_{2}P_{2}}\hbar\omega_{P_{2}}v_{P_{2}}^{2}\left|\left\langle \overline{\psi}_{P_{2}}\left(x,t\right)\right\rangle \right|^{2}\left\langle \overline{\psi}_{P_{2}}\left(x,t\right)\right\rangle\nonumber\\
&\quad+2i\gamma_{\text{chan}}^{P_{1}P_{2}P_{1}P_{2}}\hbar\sqrt{\omega_{P_{1}}\omega_{P_{2}}}v_{P_{1}}v_{P_{2}}\left|\left\langle \overline{\psi}_{P_{1}}\left(x,t\right)\right\rangle \right|^{2}\left\langle \overline{\psi}_{P_{2}}\left(x,t\right)\right\rangle,\nonumber\\
\left(\frac{\partial}{\partial t}+v_{S}\frac{\partial}{\partial x}-i\frac{v_{S}^{\prime}}{2}\frac{\partial^{2}}{\partial x^{2}}\right)\overline{\psi}_{S}\left(x,t\right)	&=2i\gamma_{\text{chan}}^{SSP_{1}P_{2}}\hbar\overline{\omega}_{SSP_{1}P_{2}}\bar{v}_{SSP_{1}P_{2}}^{2}\left\langle \overline{\psi}_{P_{1}}\left(x,t\right)\right\rangle \left\langle \overline{\psi}_{P_{2}}\left(x,t\right)\right\rangle \overline{\psi}_{S}^{\dagger}\left(x,t\right)\nonumber\\
&\quad+2i\gamma_{\text{chan}}^{P_{1}SP_{1}S}\hbar\sqrt{\omega_{P_{1}}\omega_{S}}v_{S}v_{P_{1}}\left|\left\langle \overline{\psi}_{P_{1}}\left(x,t\right)\right\rangle \right|^{2}\overline{\psi}_{S}\left(x,t\right)\nonumber\\
&\quad+2i\gamma_{\text{chan}}^{P_{2}SP_{2}S}\hbar\sqrt{\omega_{P_{2}}\omega_{S}}v_{S}v_{P_{2}}\left|\left\langle \overline{\psi}_{P_{2}}\left(x,t\right)\right\rangle \right|^{2}\overline{\psi}_{S}\left(x,t\right),\nonumber
\end{align}
for the process with $2k_{S}=k_{P_{1}}+k_{P_{2}}$, and using $\omega_{S}+\omega_{I}=\omega_{P_{1}}+\omega_{P_{2}}$ they yield
\begin{align}\label{eq:SFWM_SIP1P2_eqs}
\left(\frac{\partial}{\partial t}+v_{P_{1}}\frac{\partial}{\partial x}-i\frac{v_{P_{1}}^{\prime}}{2}\frac{\partial^{2}}{\partial x^{2}}\right)\left\langle \overline{\psi}_{P_{1}}\left(x,t\right)\right\rangle &=i\gamma_{\text{chan}}^{P_{1}P_{1}P_{1}P_{1}}\hbar\omega_{P_{1}}v_{P_{1}}^{2}\left|\left\langle \overline{\psi}_{P_{1}}\left(x,t\right)\right\rangle \right|^{2}\left\langle \overline{\psi}_{P_{1}}\left(x,t\right)\right\rangle\\
&\quad+2i\gamma_{\text{chan}}^{P_{1}P_{2}P_{1}P_{2}}\hbar\overline{\omega}_{P_{1}P_{2}P_{1}P_{2}}v_{P_{1}}v_{P_{2}}\left|\left\langle \overline{\psi}_{P_{2}}\left(x,t\right)\right\rangle \right|^{2}\left\langle \overline{\psi}_{P_{1}}\left(x,t\right)\right\rangle,\nonumber\\
\left(\frac{\partial}{\partial t}+v_{P_{2}}\frac{\partial}{\partial x}-i\frac{v_{P_{2}}^{\prime}}{2}\frac{\partial^{2}}{\partial x^{2}}\right)\left\langle \overline{\psi}_{P_{2}}\left(x,t\right)\right\rangle &=i\gamma_{\text{chan}}^{P_{2}P_{2}P_{2}P_{2}}\hbar\omega_{P_{2}}v_{P_{2}}^{2}\left|\left\langle \overline{\psi}_{P_{2}}\left(x,t\right)\right\rangle \right|^{2}\left\langle \overline{\psi}_{P_{2}}\left(x,t\right)\right\rangle\nonumber\\
&\quad+2i\gamma_{\text{chan}}^{P_{1}P_{2}P_{1}P_{2}}\hbar\overline{\omega}_{P_{1}P_{2}P_{1}P_{2}}v_{P_{1}}v_{P_{2}}\left|\left\langle \overline{\psi}_{P_{1}}\left(x,t\right)\right\rangle \right|^{2}\left\langle \overline{\psi}_{P_{2}}\left(x,t\right)\right\rangle,\nonumber\\
\left(\frac{\partial}{\partial t}+v_{S}\frac{\partial}{\partial x}-i\frac{v_{S}^{\prime}}{2}\frac{\partial^{2}}{\partial x^{2}}\right)\overline{\psi}_{S}\left(x,t\right)	&=2i\gamma_{\text{chan}}^{SIP_{1}P_{2}}\hbar\overline{\omega}_{SIP_{1}P_{2}}\bar{v}_{SIP_{1}P_{2}}^{2}\left\langle \overline{\psi}_{P_{1}}\left(x,t\right)\right\rangle \left\langle \overline{\psi}_{P_{2}}\left(x,t\right)\right\rangle \overline{\psi}_{I}^{\dagger}\left(x,t\right)\nonumber\\
&\quad+2i\gamma_{\text{chan}}^{SP_{1}SP_{1}}\hbar\overline{\omega}_{SP_{1}SP_{1}}v_{S}v_{P_{1}}\left|\left\langle \overline{\psi}_{P_{1}}\left(x,t\right)\right\rangle \right|^{2}\overline{\psi}_{S}\left(x,t\right)\nonumber\\
&\quad+2i\gamma_{\text{chan}}^{SP_{2}SP_{2}}\hbar\overline{\omega}_{SP_{2}SP_{2}}v_{S}v_{P_{2}}\left|\left\langle \overline{\psi}_{P_{2}}\left(x,t\right)\right\rangle \right|^{2}\overline{\psi}_{S}\left(x,t\right),\nonumber\\
\left(\frac{\partial}{\partial t}+v_{I}\frac{\partial}{\partial x}-i\frac{v_{I}^{\prime}}{2}\frac{\partial^{2}}{\partial x^{2}}\right)\overline{\psi}_{I}\left(x,t\right)	&=2i\gamma_{\text{chan}}^{SIP_{1}P_{2}}\hbar\overline{\omega}_{SIP_{1}P_{2}}\bar{v}_{SIP_{1}P_{2}}^{2}\left\langle \overline{\psi}_{P_{1}}\left(x,t\right)\right\rangle \left\langle \overline{\psi}_{P_{2}}\left(x,t\right)\right\rangle \overline{\psi}_{S}^{\dagger}\left(x,t\right)\nonumber\\
&\quad+2i\gamma_{\text{chan}}^{P_{1}IP_{1}I}\hbar\overline{\omega}_{P_{1}IP_{1}I}v_{P_{1}}v_{I}\left|\left\langle \overline{\psi}_{P_{1}}\left(x,t\right)\right\rangle \right|^{2}\overline{\psi}_{I}\left(x,t\right)\nonumber\\
&\quad+2i\gamma_{\text{chan}}^{P_{2}IP_{2}I}\hbar\overline{\omega}_{SP_{2}SP_{2}}v_{P_{2}}v_{I}\left|\left\langle \overline{\psi}_{P_{2}}\left(x,t\right)\right\rangle \right|^{2}\overline{\psi}_{I}\left(x,t\right),\nonumber
\end{align}
for the process with $k_{S}+k_{I}=k_{P_{1}}+k_{P_{2}}$. Note how the second of~\eqref{eq:SFWM_PPPP_eqs} and the third of~\eqref{eq:SFWM_SSP1P2_eqs} are both of the form of~\eqref{eq:DSV_COEs}, and the third and fourth of~\eqref{eq:SFWM_SIP1P2_eqs} are of the form of~\eqref{eq:NDSV_COEs}.

\section{Obtaining the matrix product form of an arbitrary tensor}\label{app:MPS}
Consider an arbitrary tensor with $\ell$ indices $R_{i_0\ldots i_{\ell-1}} \in \mathbb{C}^{c_0 \times c_1 \times \ldots \times c_{\ell-1}}$ where the index $i_k$ can take $c_k$ values. In this appendix we show how one can construct its matrix product representation by leveraging the singular value decomposition of matrices, which we recall now.
For a $c_0 \times c_1$ rectangular matrix with entries $M_{i j}$ one can construct its singular value decomposition as
\begin{align}
M_{ij} = \sum_{k = 0}^{\min\{c_0,c_1\}-1} U_{ik} \lambda_{kk} V_{kj}.
\end{align}
In the last equation one can take $\bm{U}$ to be a $c_0 \times c_0$ unitary matrix, $\bm{V}$ to be  $c_1 \times c_1$ unitary matrix and $\bm{\lambda}$ to be a $c_0 \times c_1$ diagonal matrix with non-negative entries. Equivalently, one can take $\bm{U}$ to be a rectangular $c_0 \times \min\{c_0,c_1\}$ semi-unitary matrix, $\bm{V}$ to be  $\min\{c_0,c_1\} \times D_1$ semi-unitary matrix and $\bm{\lambda}$ to be $\min\{c_0,c_1\} \times \min\{c_0,c_1\} $ square non-negative matrix.

A second useful property of a multidimensional array is that we are free to reshape it as a matrix by grouping indices together. Consider for example a four-index tensor $R_{i_0 i_1 i_2 i_3}$ where each index takes two possible values, $i_k \in \{0,1\}$.
We can group the last three indices $i_1 i_2 i_3$ into a common index that takes $2^3 = 8$ values. This allows us to consider the tensor $R$ as a $2 \times 8$ matrix. We can emphasize this partition by putting a semicolon between the indices we want to separate writing
\begin{align}
R_{i_0;i_1 i_2 i_3} \in \mathbb{C}^{2 \times 8}.
\end{align}
We can instead group the first and last two indices to now obtain a $2^2 \times 2^2 = 4 \times 4$ matrix that we write as
\begin{align}
R_{i_0 i_1 ; i_2 i_3} \in \mathbb{C}^{4 \times 4}.
\end{align}
Having introduced this useful notation and recalled the singular value decomposition for matrices we are now ready to decompose an arbitrary tensor.

To this end we split our arbitrary input tensor as $R_{i_0;i_1\ldots i_{\ell-1}}$ and then write the singular decomposition for this $c_0 \times c_{1:\ell-1}$ (with $c_{1:\ell-1} = c_1 \times \ldots \times c_{\ell-1}$) matrix
\begin{align}
R_{i_0;i_1\ldots i_{\ell-1}} = \sum_{\alpha_1}  U_{i_0; \alpha_1}^{[0]} \lambda^{[0]}_{\alpha_1 ; \alpha_1} V_{\alpha_1;i_1\ldots i_{\ell-1}}^{[0]},
\end{align}
where we do not write explicitly the possible values of the dummy index $\alpha_1$. This can be easily inferred by looking at the dimensions of the two index vectors separated by the semi-colon in the right-hand side. For this first step in our derivation $\alpha_1$ can take $\min\{c_0, c_1 \times \ldots c_{\ell-1}\}$ values.
We can now look at the tensor $V_{\alpha_1,i_1\ldots i_{\ell-1}}$ and apply once more the singular value decompositions grouping the first two indices together
\begin{align}
V_{\alpha_1,i_1;i_2\ldots i_{\ell-1}}^{[0]} = \sum_{\alpha_2} U_{\alpha_1,i_1;\alpha_2}^{[1]} \lambda^{[1]}_{\alpha_2;\alpha_2} V_{\alpha_2;i_2\ldots i_{\ell-1}}^{[1]},
\end{align}
which we can use to write
\begin{align}\label{eq:svd_step2}
R_{i_0 i_1\ldots i_{\ell-1}} = \sum_{\alpha_1,\alpha_2 } U_{i_0; \alpha_1}^{[0]} \lambda^{[0]}_{\alpha_1 ; \alpha_1} U_{\alpha_1,i_1;\alpha_2}^{[1]} \lambda^{[1]}_{\alpha_2;\alpha_2} V_{\alpha_2;i_2\ldots i_{\ell-1}}^{[1]}.
\end{align}
Now note that the leftover higher rank tensor $V^{[1]}$ has $\ell-1$ indices, one less than our original tensor $R$. We can continue as before and write the SVD of $V^{[1]}$ separating $\alpha_2$ and $i_2$ from the rest of indices
\begin{align}
V_{\alpha_2,i_2;i_3\ldots i_{\ell-1}}^{[1]} = \sum_{\alpha_3} U_{\alpha_2,i_2;\alpha_3}^{[2]} \lambda^{[2]}_{\alpha_3;\alpha_3} V_{\alpha_3;i_3\ldots i_{\ell-1}}^{[2]}.
\end{align}
We can then plug this decomposition into \eqref{eq:svd_step2} to obtain
\begin{align}
R_{i_0 i_1\ldots i_{\ell-1}} = \sum_{\alpha_1,\alpha_2 } U_{i_0; \alpha_1}^{[0]} \lambda^{[0]}_{\alpha_1 ; \alpha_1} U_{\alpha_1,i_1;\alpha_2}^{[1]} \lambda^{[1]}_{\alpha_2;\alpha_2} V_{\alpha_2;i_2\ldots i_{\ell-1}}^{[1]},
\end{align}
where we note that the leftover higher order tensor $V^{[2]}$ has $\ell-2$ indices. By this point it is easy to guess that in the $k^{\text{th}}$ step of this recursive decomposition we will have a tensor $V^{[k]}$ with $\ell-k$ indices $\{\alpha_{k+1},i_{k+1}\ldots,i_{\ell-1}\}$ on which we will apply the SVD to obtain
\begin{align}
V_{\alpha_{k+1},i_{k+1};i_{k+2}\ldots i_{\ell-1}}^{[k]} = \sum_{\alpha_{k+2}} U^{[k+1]}_{\alpha_{k+1},i_{k+1}; \alpha_2} \lambda^{[k+1]}_{\alpha_{k+2};\alpha_{k+2}} V^{[k+1]}_{\alpha_{k+2};i_{k+2}\ldots i_{\ell-1}}.
\end{align}
Based on this iterative decomposition we finally arrive at the first matrix product form of the tensor $R$,
\begin{align}
R_{i_0 i_1\ldots i_{\ell-1}} = \sum_{\alpha_1,\alpha_2 ,\ldots, \alpha_{\ell-1}} U_{i_0; \alpha_1}^{[0]} \lambda^{[0]}_{\alpha_1 ; \alpha_1} U_{\alpha_1,i_1;\alpha_2}^{[1]} \lambda^{[1]}_{\alpha_2;\alpha_2} \ldots U_{\alpha_{\ell-2},i_{\ell-2};\alpha_{\ell-1}}^{[1]} \lambda^{[1]}_{\alpha_{\ell-1};\alpha_{\ell-1}} V_{\alpha_{\ell-1},i_{\ell-1}}
\end{align}
which is the form used by Yanagimoto et al.~\cite{yanagimoto2021efficient}.
A second useful form can be obtained by introducing $(S^{[k]})^{\alpha_k}_{i_k,\alpha_{k+1}} \equiv U_{\alpha_k,i_k;\alpha_{k+1}}^{[k]} \lambda^{[k]}_{\alpha_{k+1};\alpha_{k+1}}$ as used in the main text.

\end{document}